\documentclass[11pt]{article}
\usepackage[utf8]{inputenc}
\usepackage[T1]{fontenc}
\pdfoutput = 1

\usepackage{amsmath,amssymb,amsthm}
\usepackage{graphicx,subfig}
\usepackage{changepage}
\usepackage{cite,url}
\usepackage{mathtools,xparse,MnSymbol,slashed}
\usepackage{multirow}
\usepackage{stackrel}
\allowdisplaybreaks

\usepackage{bbm}
\usepackage{bbold}
\usepackage{caption}
\usepackage{cite}
\usepackage{color}
    \definecolor{darkgreen}{rgb}{0,0.5,0}
    \definecolor{darkred}{rgb}{0.5,0,0}
    \definecolor{darkblue}{rgb}{0,0,0.6}
    \definecolor{purple}{rgb}{0.4,.2,0.7}
\usepackage{float}
\usepackage[hyperfootnotes = true, colorlinks = true, linkcolor = darkblue, citecolor = purple]{hyperref}

\numberwithin{equation}{section}

\usepackage[margin = 2.2cm]{geometry}
\usepackage[ragged]{footmisc}
\newcommand{\qeq}{\stackrel{?}{=}}
\newcommand{\qpropto}{\stackrel{?}{\propto}}
\renewcommand{\d}{\mathrm{d}}
\renewcommand{\i}{\mathrm{i}}

\newcommand{\vev}[1]{ \langle {#1} \rangle }

\DeclareMathOperator{\Tr}{Tr}

\renewcommand{\Re}[1]{{\mathrm{Re}}\left[ #1 \right]}
\renewcommand{\Im}[1]{{\mathrm{Im}}\left[ #1 \right]}

\def\bra#1{\langle #1 |}
\def\ket#1{| #1 \rangle}
\def\inner#1#2{\langle #1 | #2 \rangle}

\def\ketc#1{\| #1 \rrangle}
\def\innerc#1#2{\llangle #1 \| #2 \rrangle}

\def\Lbrac#1{\bigl\llangle #1 \bigr\|}
\def\Lketc#1{\bigl\| #1 \bigr\rrangle}

\DeclarePairedDelimiter{\norm}{\lVert}{\rVert}
\def\braR#1{\,{}_{\rm\scriptscriptstyle R}\!\langle #1 |}
\def\ketR#1{| #1 \rangle_{\rm\scriptscriptstyle R}}
\def\braL#1{\,{}_{\rm\scriptscriptstyle L}\!\langle #1 |}
\def\ketL#1{| #1 \rangle_{\rm\scriptscriptstyle L}}

\def\ketcR#1{\| #1 \rrangle_{\rm\scriptscriptstyle R}}

\def\ketcL#1{\| #1 \rrangle_{\rm\scriptscriptstyle L}}

\def\ketN#1{| #1 \rangle_{\rm\scriptscriptstyle N}}

\def\ketS#1{| #1 \rangle_{\rm\scriptscriptstyle S}}

\def\smg{{\scalebox{0.4}{$-\gamma$}}}
\def\ssmg{{\scalebox{0.3}{$-\gamma$}}}
\def\spg{{\scalebox{0.4}{$+\gamma$}}}
\def\sspg{{\scalebox{0.3}{$+\gamma$}}}

\begin{document}

\begin{titlepage}
\thispagestyle{empty}

\begin{flushright}
\end{flushright}

\vspace{1.2cm}
\begin{center}
\noindent{\bf \LARGE Black Hole and de~Sitter Microstructures \\ from a Semiclassical Perspective}

\vspace{0.4cm}

{\bf \large Chitraang Murdia$^{a,b}$, Yasunori Nomura$^{a,b,c}$, and Kyle Ritchie$^{a,b}$}
\vspace{0.3cm}\\

{\it $^a$ Berkeley Center for Theoretical Physics, Department of Physics, \\
University of California, Berkeley, CA 94720, USA}\\

{\it $^b$ Theoretical Physics Group, Lawrence Berkeley National Laboratory, \\ Berkeley, CA 94720, USA}\\

{\it $^c$ Kavli Institute for the Physics and Mathematics of the Universe (WPI), \\
UTIAS, The University of Tokyo, Kashiwa, Chiba 277-8583, Japan}\\

\vspace{0.3cm}
\end{center}

\begin{abstract}
We describe two different, but equivalent semiclassical views of black hole physics in which the equivalence principle and unitarity are both accommodated.
In one, unitarity is built-in, while the black hole interior emerges only effectively as a collective phenomenon involving horizon (and possibly other) degrees of freedom.
In the other, more widely studied approach, the existence of the interior is manifest, while the unitarity of the underlying dynamics can be captured only indirectly by incorporating certain nonperturbative effects of gravity.
These two pictures correspond to a distant description and the description based on entanglement islands/replica wormholes, respectively.
We also present a holographic description of de~Sitter spacetime based on the former approach, in which the holographic theory is located on the stretched horizon of a static patch.
We argue that the existence of these two approaches is rooted in the two formulations of quantum mechanics:\ the canonical and path integral formalisms.
\end{abstract}

\end{titlepage}

\tableofcontents
\newpage

\section{Introduction and Summary}
\label{sec:intro}

For a long time, the puzzle of black hole information loss has caused confusion about how gravity works at the quantum level~\cite{Hawking:1976ra,Mathur:2009hf,Almheiri:2012rt}.
This confusion arises mostly due to the following features of the semiclassical description of black holes:
\begin{itemize}
\item[(i)] Despite the fact that spacetime is fundamentally quantum mechanical, it is described as a classical object.
\item[(ii)] While the fundamental theory has a preferred class of time foliations for spacetimes with a horizon, general relativity seems to treat all the coordinates equally.
\end{itemize}
The purpose of this paper is to elucidate these points and present a coherent picture in which the results of semiclassical theory are consistently interpreted to address issues related to the information puzzle.
This picture builds on recent developments of our understanding of quantum gravity.

The first hint that spacetime consists of quantum degrees of freedom came from the discovery that black holes have entropy~\cite{Bekenstein:1973ur,Hawking:1974sw}.
In the standard statistical mechanical interpretation, this implies that the horizon, a region which general relativity describes as empty space, contains a quantum mechanical substance.
This interpretation is indeed supported by the anti-de~Sitter (AdS)/conformal field theory (CFT) correspondence~\cite{Maldacena:1997re}---a concrete realization of holography~\cite{tHooft:1993dmi,Susskind:1994vu,Banks:1996vh,Maldacena:1997re,Bousso:2002ju}.

Recent progress in understanding the AdS/CFT correspondence, and holography more generally, has shown that while general relativity is diffeomorphism invariant, there appear to be a preferred class of coordinates at the quantum level.
From the viewpoint of the boundary theory, these coordinates correspond to descriptions based on quantum operators constructed by a simple procedure or operators which are not exponentially complex in fundamental degrees of freedom~\cite{Hamilton:2005ju,Hamilton:2006az,Nomura:2018kji,Murdia:2020iac,Brown:2019rox,Engelhardt:2021mue}, and they cover only a portion of spacetime when there is a horizon.
In a system with a black hole, these coordinates are associated with Schwarzschild time slicing, or any other time foliation associated with an observer located outside the horizon.
This conforms to the earlier idea that black hole evolution obeys the standard rules of quantum mechanics---and hence is unitary---when viewed by an external observer~\cite{tHooft:1984kcu,tHooft:1990fkf,Susskind:1993if,Page:1993wv}.
Indeed, we see that there is nothing unusual with the results of semiclassical theory if a black hole is described using only external frames.

The issue, then, is how to interpret the existence of the black hole interior which arises when the external coordinates are analytically extended in general relativity.
We take the view that the picture of the black hole interior arises only effectively at the semiclassical level.
In particular, we adopt the construction in Refs.~\cite{Nomura:2018kia,Nomura:2019qps,Nomura:2019dlz,Langhoff:2020jqa,Nomura:2020ska} in which the interior emerges because of the special, chaotic nature of the horizon dynamics.%
\footnote{
 A detailed description of relations of this construction to earlier work~\cite{Papadodimas:2012aq,Verlinde:2012cy,Nomura:2012ex,Maldacena:2013xja,Hayden:2018khn} is given in Ref.~\cite{Nomura:2020ska} and throughout this paper.
}
A key idea is that when the black hole is described in an external quasi-static reference frame, its large acceleration with respect to the free falling frame makes the dynamics at the horizon---more precisely the stretched horizon~\cite{Susskind:1993if}---appear string theoretic, which is chaotic across all low energy species.
This makes the black hole vacuum microstate generic in the relevant microcanonical ensemble, allowing us to erect the effective theory of the interior.

The erected effective theory is defined only up to errors of order $e^{-(S_{\rm BH} + S_{\rm rad})/2}$, where $S_{\rm BH}$ and $S_{\rm rad}$ are the Bekenstein-Hawking entropy and the coarse-grained entropy of the degrees of freedom entangled with the black hole, respectively.
Operators describing the interior are state dependent~\cite{Papadodimas:2012aq,Papadodimas:2013jku,Papadodimas:2015jra}, though only weakly in the sense of Refs.~\cite{Nomura:2020ska,Hayden:2018khn}.
Reflecting the fact that the Hilbert space associated with the black hole system is finite dimensional, the effective theory can be used only for a finite time interval; the existence of the black hole singularity is consistent with this~\cite{Nomura:2018kia}.
Similarly, weak cosmic censorship can be viewed as a statement that a distant description, which corresponds to a simple boundary description in holography, can be used for an arbitrarily long time, to the extent that the theory is well defined in the infrared.

In this paper, we review the construction described above and refine it, including subtle evolutionary effects for a dynamically formed black hole.
We present the whole framework in a coherent manner and address various issues associated with it, including its realization in the boundary description of holography.
In doing so, we also discuss the relation of the present construction to other recent related works.
In particular, we discuss the relation between the picture described here and that based on quantum extremal surfaces and entanglement wedge reconstruction~\cite{Penington:2019npb,Almheiri:2019psf,Almheiri:2019hni,Penington:2019kki,Almheiri:2019qdq}.
We argue that the latter emerges through coarse graining necessary to describe a semiclassical black hole without specifying its microscopic structure~\cite{Langhoff:2020jqa}.
We also see that the two are closely tied, respectively, to the canonical and path integral formulations of quantum mechanics~\cite{Nomura:2020ewg}.

We expect that at the macroscopic level, generic (quasi-)static horizons, including cosmic horizons, have locally the same statistical features.s
Indeed, the horizon of de~Sitter spacetime has the same entropy per area as that of a black hole~\cite{Gibbons:1977mu}, and so is the temperature of the Hawking cloud at the stretched horizon.
In fact, building on the analyses in Refs.~\cite{Nomura:2019qps,Nomura:2017fyh}, we see that the holographic description of a static patch of de~Sitter spacetime is very much an ``inside-out'' version of that of a black hole.
We present this description in the context of more general holography for cosmological spacetimes~\cite{Sanches:2016sxy,Nomura:2016ikr}.
We also discuss the relationship of the present description with other recent proposals~\cite{Alishahiha:2004md,Alishahiha:2005dj,Susskind:2021dfc,Susskind:2021esx,Shaghoulian:2021cef,Shaghoulian:2022fop}.

\subsection{Overall picture and outline of the paper}

In the rest of this section, we present an overview of the picture presented in this paper, pointing to where the details of each subject are covered.
The assumptions about the setup which we adopt throughout the paper will be summarized at the end of this section.

\subsubsection*{No puzzle for a black hole when viewed from the exterior}

Consider a non-rotating, non-charged black hole in a 4-dimensional asymptotically flat spacetime.
At the classical level, the black hole is uniquely specified by one continuous parameter:\ its mass $M$.
We view this system from a distance, i.e., we describe it using Schwarzschild time slicing or something related to it in a simple manner.

When quantum effects are included, the black hole has a finite entropy
\begin{equation}
  S_{\rm BH}(M) = 4\pi \frac{G_{\rm N} M^2}{\hbar c},
\label{eq:S_BH}
\end{equation}
where $G_{\rm N}$ is Newton's constant.
This means that at the quantum level the black hole is characterized by discrete---albeit exponentially dense---states, rather than a continuous, classical number.%
\footnote{
 Of course, quantum mechanics allows for a superposition of these independent states, so that the expectation value of the energy, or $M$, can take continuous values.
}
Specifically, the number of independent states in the energy interval between $M-\delta M/2$ and $M+\delta M/2$ is given by
\begin{equation}
  {\cal N}(M) \sim e^{S_{\rm BH}(M)} \frac{\delta M}{M}.
\end{equation}
Note that this is a general phenomenon in quantum mechanics.
Like other physical systems, the entropy in Eq.~\eqref{eq:S_BH} diverges in the limit $\hbar \rightarrow 0$.

The semiclassical theory treats the black hole as a classical object while retaining the finite nature of the entropy.
This is not inconsistent:\ the concept of entropy can be defined at the level of thermodynamics without explicitly taking into account the fundamental discreteness.
The disadvantage of this treatment, however, is that one can no longer resolve each microstate, hence requiring a statistical, or thermodynamic, treatment of the system~\cite{Bekenstein:1973ur,Hawking:1974sw,Bardeen:1973gs,Bekenstein:1974ax}.

At the semiclassical level, a black hole is described as having a definite mass $M$ but with the Hawking cloud around it, which is in a thermal mixed state of temperature
\begin{equation}
  T_{\rm H} = \frac{\hbar c^3}{8\pi k G_{\rm N} M},
\end{equation}
where $k$ is the Boltzmann constant.
This is a proxy for an ensemble of black hole microstates with energies between $M-\delta M/2$ and $M+\delta M/2$, which dominates the corresponding microcanonical ensemble of states associated with the spacetime region near the black hole.
Below, we adopt natural units $\hbar = c = k = 1$.

The formation and evaporation of a black hole described in an external frame is a process in which quantum information in the initial collapsing matter is dispersed among spacetime degrees of freedom, represented by the Bekenstein-Hawking entropy; Hawking emission then transfers it back to matter degrees of freedom in the semiclassical description.
For an external observer, this entire process is unitary if all the microscopic degrees of freedom are accounted for.
An important point is that with quantum effects, the instantaneously-defined apparent horizon is stretched, at which the local (Tolman) Hawking temperature is the string scale~\cite{Susskind:1993if}.%
\footnote{
 By the string scale, we mean the scale at which the low energy effective field theory description breaks down.
 The appearance of this scale is associated with the nonzero Newton's constant, which controls quantum corrections to the system since it always appears with $\hbar$.
}
The trajectory of this stretched horizon is timelike, so an object falling into a black hole reaches there within finite time.
This object is then absorbed by the black hole, whose information will eventually be sent back to ambient space as Hawking radiation.

This implies that in the external frame description, the stretched horizon behaves as a regular material surface as far as the flow of quantum information is concerned.
The semiclassical theory, however, treats the horizon degrees of freedom appearing in the middle of the process to be classical, or at best an ensemble of quantum states represented by the Hawking cloud.
Therefore, in any semiclassical calculation, the quantum information in the initial matter {\it must} appear to be lost in the final state; there is no way to completely describe the microscopic information on the horizon while staying in the semiclassical regime.
In other words, Hawking's calculation~\cite{Hawking:1974sw} must have led to information loss, and indeed it did~\cite{Hawking:1976ra}.

\subsubsection*{Emergence of the interior}

A puzzling feature of the picture described above is that general relativity seems to imply that the black hole horizon is smooth.
Namely, when an object freely falls into the horizon, it does not experience anything special there.
This is obviously not the case when an object falls into a regular material surface.

As discussed in Refs.~\cite{Nomura:2018kia,Nomura:2019qps,Nomura:2019dlz,Langhoff:2020jqa,Nomura:2020ska}, the stretched horizon is distinguished from other, regular material surfaces by its chaotic~\cite{Maldacena:2015waa}, fast-scrambling~\cite{Hayden:2007cs,Sekino:2008he} dynamics across {\it all} low energy species.
Here, by all low energy species, we mean all quantum fields appearing in the low energy effective theory below the string scale.
These features arise because of the large relative acceleration, of order the string scale, between the quasi-static frame (a natural frame in holography) and the free falling frame at the stretched horizon.
It is this aspect of the horizon dynamics that allows us to erect a description in which an object falls freely through the stretched horizon.
This is done by making the state take a fully generic form in the relevant microcanonical ensemble.%
\footnote{
For a charged or rotating black hole, the relevant ensemble consists of microstates with charge or angular momentum constrained to lie within a small window dictated by quantum uncertainties.
}

Specifically, when described in a quasi-static reference frame, quantum degrees of freedom of a black hole consist of modes in a spatial region near the stretched horizon (called the zone) as well as those at the stretched horizon.
We call them zone and horizon modes, respectively.
While the dynamics of the former is described by the low energy effective field theory, that of the latter is not.
Note that these modes are defined for each time interval in which the black hole can be viewed as quasi-static.

Let us now focus on a (small) subset of the zone modes which is relevant for describing an infalling physical object.
In particular, the object will be described by excitations of these modes over the black hole vacuum.
Let us call these modes hard modes and all other black hole (zone and horizon) modes soft modes.
From the assumption that the horizon degrees of freedom obey chaotic dynamics, we can then conclude that a black hole vacuum microstate takes the form
\begin{equation}
  \ket{\Psi(M)} = \sum_n \sum_{i_n = 1}^{e^{S_{\rm bh}(M-E_n)}}\!\!\! c_{n i_n} \ket{\{ n_\alpha \}} \ket{\psi^{(n)}_{i_n}},
\label{eq:BH-intro}
\end{equation}
where $\ket{\{ n_\alpha \}}$ represent states of the hard modes, specified by the occupation numbers $n_\alpha$ for each mode $\alpha$, and $\ket{\psi^{(n)}_{i_n}}$ are the states of the soft modes that have energy $M - E_n$, where $E_n$ is the energy carried by $\ket{\{ n_\alpha \}}$.
Note that since the total energy of the black hole system is constrained to be $M$, the number of independent horizon-mode states $\ket{\psi^{(n)}_{i_n}}$ coupling to $\ket{\{ n_\alpha \}}$ in Eq.~\eqref{eq:BH-intro} is given by $S_{\rm bh}(M-E_n)$, the density of states at energy $M-E_n$. 
Here,
\begin{equation}
  S_{\rm bh}(E) = 4\pi E^2 l_{\rm P}^2
\end{equation}
is the Bekenstein-Hawking entropy density at energy $E$ with $l_{\rm P}$ being the Planck
length.
(The contribution from the hard modes to the black hole entropy is negligible.)
Also note that because of the chaotic nature of the horizon dynamics, the coefficients $c_{n i_n}$ take random values across all low energy species.
In fact, the quasi-static form of Eq.~\eqref{eq:BH-intro} is achieved quickly after any disturbance, i.e.\ within the scrambling timescale of order $M l_{\rm P}^2 \ln(M l_{\rm P})$~\cite{Hayden:2007cs,Sekino:2008he}.

The universality of the form of the state in Eq.~\eqref{eq:BH-intro} allows us to erect the effective theory of the interior.
Specifically, we can define the normalized state coupling to $\ket{\{ n_\alpha \}}$ in Eq.~\eqref{eq:BH-intro}
\begin{equation}
  \ketc{\{ n_\alpha \}} = \varsigma_n \sum_{i_n = 1}^{e^{S_{\rm bh}(M-E_n)}} c_{n i_n} \ket{\psi^{(n)}_{i_n}},
\end{equation}
where
\begin{equation}
  \varsigma_n = \frac{1}{\sqrt{\sum_{i_n = 1}^{e^{S_{\rm bh}(M-E_n)}} |c_{n i_n}|^2}}
  = e^{\frac{E_n}{2T_{\rm H}}} \sqrt{\sum_m e^{-\frac{E_m}{T_{\rm H}}}}\, \left[ 1 + O\Bigl(e^{-\frac{1}{2}S_{\rm bh}(M)}\Bigr) \right].
\end{equation}
Plugging this into Eq.~\eqref{eq:BH-intro}, we obtain the standard thermofield double form~\cite{Unruh:1976db,Israel:1976ur}
\begin{equation}
  \ket{\Psi(M)} = \frac{1}{\sqrt{\sum_m e^{-\frac{E_m}{T_{\rm H}}}}} \sum_n e^{-\frac{E_n}{2T_{\rm H}}} \ket{\{ n_\alpha \}} \ketc{\{ n_\alpha \}},
\label{eq:BH-coarse-intro}
\end{equation}
up to exponentially small corrections of order $e^{-S_{\rm bh}/2}$.
We can thus evolve an object in the zone (generated by acting creation/annihilation operators on $\ket{\{ n_\alpha \}}$'s) using the time evolution operator associated with the proper time of the falling object, which is different from the original time evolution operator in the boundary theory but still local in bulk spacetime~\cite{Nomura:2018kia,Nomura:2019qps,Nomura:2019dlz,Langhoff:2020jqa,Nomura:2020ska}.
This implies the existence of a description in which the falling object passes the horizon and enters the black hole interior smoothly.

We emphasize that the criterion for a state to take the universal form in Eq.~\eqref{eq:BH-intro} is much stronger than that for regular thermalization; in particular, it must exhibit universal thermalization across all low energy species.
It is reasonable to expect that such strong thermalization is achieved within a reasonable timescale only by the string scale dynamics, which singles out the stretched horizon.
We also stress that the prescription of obtaining interior spacetime described here does not require a detailed knowledge about microscopic dynamics of quantum gravity; only some basic assumptions are sufficient.

As the evaporation of a black hole progresses, information stored in the horizon degrees of freedom is gradually transferred to Hawking radiation.
After the Page time~\cite{Page:1993wv}, entanglement between hard modes and early Hawking radiation becomes nonnegligible, so that the interior description must involve this Hawking radiation~\cite{Maldacena:2013xja}.
The way this occurs is that the black hole interior, more precisely $\ketc{\{ n_\alpha \}}$'s in Eq.~\eqref{eq:BH-coarse-intro}, must involve early Hawking radiation {\it in addition to} the black hole degrees of freedom.

This last statement may seem to contradict the recent claim about entanglement wedge reconstruction that the interior of an old black hole can be reconstructed using only early Hawking radiation~\cite{Penington:2019npb,Almheiri:2019psf,Almheiri:2019hni}.
This is, however, not the case~\cite{Langhoff:2020jqa}.
A key point is that such entanglement wedge reconstruction makes crucial use of boundary time evolution, which allows for reconstructing states of the black hole and radiation modes at an earlier time.
On the other hand, the effective theory of the interior uses these modes directly at the time relevant for describing an infalling object.
This understanding resolves an apparent puzzle associated with causality in entanglement wedge reconstruction.

In this paper, we first perform a detailed analysis of the near horizon mode structure in Sections~\ref{sec:QFT}~and~\ref{sec:stretched}, and study its relation to the holographic boundary picture in Section~\ref{sec:holo}.
The construction of the effective theory of the interior is presented in Sections~\ref{sec:stretched}~and~\ref{sec:extension}, where we also discuss the refinement of the construction needed to accommodate the effects of evaporation on the instantaneous state of the zone modes.
The relationship of this picture to entanglement wedge reconstruction is discussed in Section~\ref{sec:extension}.

\subsubsection*{de~Sitter holography}

The similarity between the static patch description of de~Sitter spacetime and the external description of black hole spacetime has been studied for a long time~\cite{Gibbons:1977mu}.
Based on this similarity, it was suggested that de~Sitter spacetime may admit a holographic description with finite-dimensional Hilbert space~\cite{Banks:2000fe,Witten:2001kn}.
In this paper, building on the earlier analysis in Refs.~\cite{Nomura:2019qps,Nomura:2017fyh}, we develop a holographic description of de~Sitter spacetime based on the (quasi-)static picture.

First, we note that the analogue of a collapse formed, single-sided black hole is cosmological de~Sitter spacetime, which arises approximately in the middle of a cosmological history, e.g.\ at late times in a bubble universe with positive cosmological constant.
In this case, there is a single static patch as viewed by an observer (timelike geodesic), which is the inside-out analogue of the exterior of the black hole.
We define the zone and horizon modes as those inside (on the observer side of) the stretched horizon and those on the stretched horizon, respectively.
As in the case of a black hole, a microstate for the de~Sitter vacuum then takes the form
\begin{equation}
  \ket{\Psi(E)} = \sum_n \sum_{i_n = 1}^{e^{S_{\rm dS}(E-E_n)}}\!\!\! c_{n i_n} \ket{\{ n_\alpha \}} \ket{\psi^{(n)}_{i_n}},
\label{eq:dS-intro}
\end{equation}
where $\ket{\{ n_\alpha \}}$ represent states of the hard modes and $\ket{\psi^{(n)}_{i_n}}$ are the states of the soft modes that have energy $E - E_n$, where $E_n$ is the energy of $\ket{\{ n_\alpha \}}$ measured at the location of the observer.
Here,
\begin{equation}
  S_{\rm dS}(E) = \pi E^2 l_{\rm P}^2
\end{equation}
represents the de~Sitter entropy, with $E$ being the ``energy'' of the de~Sitter vacuum related to the Hubble radius $\alpha$ by
\begin{equation}
  E = \frac{\alpha}{l_{\rm P}^2}.
\end{equation}
Again, the coefficients $c_{n i_n}$ take random values across all low energy species because of universally chaotic dynamics at the horizon.
This randomization is achieved rather quickly, within the scrambling timescale of order $\alpha \ln(\alpha/l_{\rm P})$~\cite{Susskind:2011ap}.

Using a microstate in Eq.~\eqref{eq:dS-intro}, the effective theory describing a region outside the horizon (as viewed from the observer) can be constructed analogously to the black hole case.
By identifying the normalized state $\ketc{\{ n_\alpha \}}$ using Eq.~\eqref{eq:dS-intro}, we can rewrite the full state as
\begin{equation}
  \ket{\Psi(E)} = \frac{1}{\sqrt{\sum_m e^{-\frac{E_m}{T_{\rm H}}}}} \sum_n e^{-\frac{E_n}{2T_{\rm H}}} \ket{\{ n_\alpha \}} \ketc{\{ n_\alpha \}},
\label{eq:dS-coarse-intro}
\end{equation}
where $T_{\rm H}$ is the de~Sitter temperature.
This state represents the semiclassical vacuum state of global de~Sitter spacetime at time when the effective theory is erected.

We thus find that the effective global de~Sitter picture emerges from a cosmological de~Sitter spacetime as a collective phenomenon involving horizon modes.
Like the black hole case, the effective theory of global de~Sitter spacetime is intrinsically semiclassical in that there is an intrinsic ambiguity of order $e^{-S_{\rm GH}/2}$ in the definition of the theory, where $S_{\rm GH}$ is the Gibbons-Hawking entropy.
This, therefore, addresses the issue identified as a puzzling feature in Ref.~\cite{Goheer:2002vf} that symmetries of classical (global) de~Sitter spacetime cannot be implemented exactly in a finite-dimensional Hilbert space.

The holographic theory of de~Sitter spacetime described here is presented in Sections~\ref{sec:stretched}~and~\ref{sec:extension}.
Its relation to holography in more general spacetimes is discussed in Section~\ref{sec:holo}, where we will see how static patch de~Sitter holography arises naturally from holography in more general cosmological spacetimes.
In Section~\ref{sec:extension}, we will describe the construction of the effective theory and see that this theory is sufficient to provide a semiclassical description of future measurements of the observer.

\subsubsection*{Intrinsically two-sided systems}

While we find that the analytically extended Schwarzschild and global de~Sitter spacetimes emerge from the more physical, single-sided spacetimes via collective dynamics involving horizon modes, there is nothing theoretically wrong with considering an ``intrinsically two-sided'' system, which comprises two copies of the holographic theory for the single-sided system.

In Section~\ref{sec:extended}, we analyze such two-sided systems using our framework.
We will see that these two-sided systems lead to physics similar to the corresponding single-sided systems at the semiclassical level, despite the fact that the microscopic structures of relevant states are significantly different.
We also comment on possible relations of our framework to other proposals for holographic theories developed in the context of intrinsically two-sided systems.
In particular, we discuss the relationship of our description of de~Sitter spacetime with the DS/dS correspondence~\cite{Alishahiha:2004md,Alishahiha:2005dj} and the Shaghoulian-Susskind proposal~\cite{Susskind:2021dfc,Susskind:2021esx,Shaghoulian:2021cef,Shaghoulian:2022fop}.

\subsubsection*{Gravitational path integral:\ ensemble from coarse graining}

The approach described so far is formulated most naturally in the canonical formalism of quantum mechanics.
For example, we have introduced the concept of horizon modes and their associated states.
However, quantum mechanics must also be formulable using path integral.
What does the picture look like in this case?

The path integral formalism has a very different starting point.
In the context of quantum gravity, it is the collection of classical field configurations on classical geometries, which are then integrated over to obtain physical results.
This implies, however, that a black hole (or de~Sitter) spacetime appearing in the path integral must be treated as a (semi)classical object, so its detailed microscopic structure cannot be discriminated.
In fact, since the energy differences between different black hole microstates are suppressed by $e^{-S_{\rm BH}(M)}$, these states cannot be discriminated by direct measurement.

From the microscopic point of view, this means that a black hole appearing in gravitational path integral is a coarse-grained object.
This mandatory coarse graining introduces the concept of ensemble averaging in interpreting the result of gravitational path integral.
This interpretation, in fact, reproduces many features which are attributed to the ensemble nature of holographic theories in lower dimensional quantum gravity~\cite{Penington:2019kki} using an ensemble of microscopic states in a single theory~\cite{Langhoff:2020jqa}.
However, if the classical black hole spacetime necessarily represents an ensemble of microstates, how can the underlying unitarity of a black hole evolution be captured by the quantum extremal surface method~\cite{Penington:2019npb,Almheiri:2019psf,Almheiri:2019hni} which uses such classical spacetime?

A key observation is that while gravitational path integral can only calculate ensemble averages, it can still do so for many different quantities.
In particular, adopting the replica method, the gravitational path integral can calculate the traces of powers of the density matrix of Hawking radiation $\overline{\Tr \rho_R^n}$.
This is the replica wormhole calculation of Refs.~\cite{Penington:2019kki,Almheiri:2019qdq}.
From this, the ensemble average of the von~Neumann entropy of the radiation can be obtained
\begin{equation}
  \overline{S_R} = - \lim_{n \rightarrow 1} \frac{\partial}{\partial n} \overline{\Tr \rho_R^n} \sim {\rm min}\{ S_{\rm rad}, S_{\rm bh} \},
\end{equation}
which follows the Page curve.
The reason why we could obtain the Page curve here is because we have calculated the ensemble average of the microscopic von~Neumann entropy, which obeys the Page curve for all members of the ensemble.
This is unlike Hawking's calculation which gives the von~Neumann entropy of the averaged state.

This replica method prescription involving gravitational path integrals is virtually equivalent to the entanglement island prescription for calculating entropies, adopted in Refs.~\cite{Penington:2019npb,Almheiri:2019psf,Almheiri:2019hni}.
This, therefore, provides the following interpretation of the results in Refs.~\cite{Penington:2019npb,Almheiri:2019psf,Almheiri:2019hni}:\ while any formalism treating a black hole as a classical object, including the gravitational path integral, must involve an ensemble average over microstates, microscopic information about some quantities can still be deduced by computing such ensemble averages.
The entanglement island prescription (implicitly) adopts this for von~Neumann entropies.

The issue described here is discussed in Section~\ref{sec:path-int}.
It is based on the picture outlined in Refs.~\cite{Langhoff:2020jqa,Nomura:2020ewg}.
The idea that the semiclassical description involves an ensemble of microstates was also discussed in Refs.~\cite{Pollack:2020gfa,Belin:2020hea,Liu:2020jsv,Freivogel:2021ivu,Cotler:2022rud}, and the understanding of the Page curve presented there is based on the developments in Refs.~\cite{Bousso:2020kmy,Engelhardt:2020qpv,Renner:2021qbe,Qi:2021oni,Almheiri:2021jwq,Blommaert:2021fob,Chandra:2022fwi}.

\subsubsection*{Quantum mechanics vs general relativity in quantum gravity}

It is important that the extremization procedure involved in the calculation of an entanglement island is performed on global spacetime of general relativity.
In particular, for a black hole spacetime, it must be performed on the whole spacetime including the interior of the black hole.

This picture, therefore, is complementary to that based on the external view of the black hole.
Here the existence of the black hole interior is manifest, while understanding the unitarity of black hole evolution requires a method incorporating nonperturbative effects of quantum gravity, such as replica wormholes.
On the other hand, in the framework based on the external view, the unitarity of the evolution is built-in, and the interior emerges only effectively as a collective phenomenon involving horizon (and possibly other, entangled) degrees of freedom.

Despite the fact that the two pictures appear very different, they give the same physical conclusions.
In particular, a black hole evolves unitarily and has a smooth horizon.
The origin of historical confusions about black hole physics come from the fact that only one of these features is manifest in a given low energy description; the other appears in a highly nontrivial manner.
It is interesting that the description in which unitarity of quantum mechanics is manifest naturally comes with the canonical/Hamiltonian formulation of quantum mechanics, while the one in which the interior predicted by general relativity is manifest is naturally associated with the path integral/Lagrangian formulation.
It is an interesting question if a microscopic formulation of quantum gravity can make both these features manifest.

\subsection{List of assumptions}

Here we list out the assumptions that we use throughout the paper.
We focus on spacetimes which are spherically symmetric at the semiclassical level, although excitations on them are not restricted to be spherically symmetric.
We take the number of spacetime dimensions to be $3+1$, although our arguments can be generalized straightforwardly to $d+1$ dimensions with $d \geq 3$.

When discussing black holes, we mostly consider a spherically symmetric, non-charged black hole in an asymptotically flat spacetime (or an analogous object such as a spherically symmetric, non-charged small black hole in an asymptotically AdS spacetime).
It is straightforward to include the effect of a charge or rotation unless the black hole is extremal or near extremal.
An extension to (near) extremal and lower dimensional black holes is expected to be nontrivial, since these black holes have different structures for the densities of states than ``generic'' black holes considered in this paper; see, e.g., Refs.~\cite{Maxfield:2020ale,Heydeman:2020hhw}.

\section{Quantum Field on Background Spacetime}
\label{sec:QFT}

In this section, we discuss the behavior of quantum fields in spherically symmetric background spacetimes.
The calculation presented here is elementary.
The results, however, are used in later sections, so we include them for completeness.

We consider a 4-dimensional spacetime with the metric
\begin{equation}
  \d s^2 = - f(r) \d t^2 + \frac{1}{f(r)} \d r^2 + r^2 \d\Omega^2,
\label{eq:metric-gen}
\end{equation}
where $\d \Omega^2 = \d \theta^2 + \sin^2\!\theta \d \phi^2$.
For simplicity, we consider a minimally-coupled real scalar field $\Phi$ in this spacetime, whose action is given by
\begin{equation}
  I = \frac{1}{2} \int \d^4 x \sqrt{-g} \left( -g^{\mu\nu} \partial_{\mu} \Phi \partial_{\nu} \Phi - m_\Phi^2 \Phi^2 \right).
\label{eq:I_Phi}
\end{equation}
After changing the radial coordinate to the tortoise coordinate $r_*$ defined by
\begin{equation}
  \d r_* = \frac{\d r}{f(r)},
\end{equation}
the metric becomes
\begin{equation}
  \d s^2 = f(r) \left( -\d t^2 + \d r_*^2 \right) + r^2 \d\Omega^2.
\end{equation}
Here, $r$ should be regarded as a function of $r_*$.
The equation of motion derived from Eq.~\eqref{eq:I_Phi} is then given by
\begin{equation}
    \left( -\partial_t^2 + \partial_{r_*}^2 - \frac{f(r) f'(r)}{r} + \frac{f(r)}{r^2} \partial_\Omega^2 - m_{\Phi}^2 f(r) \right)(r \Phi) = 0,
\end{equation}
where $\partial_\Omega^2$ is defined by $\partial_\Omega^2 \chi = (1/\sin\theta) \partial_\theta(\sin\theta \partial_\theta \chi) + (1/\sin^2\!\theta) \partial_\phi^2 \chi$.

We look for positive frequency solutions of the form
\begin{equation}
  \Phi(t,r_*,\theta,\phi) = e^{-i \omega t} \frac{\varphi_{\ell m}(r_*)}{r} Y_{\ell m}(\theta,\phi)
\label{eq:Phi_varphi}
\end{equation}
with $\omega \geq 0$, where $Y_{\ell m}(\theta,\phi)$ represent real spherical harmonics, satisfying $\partial_\Omega^2 Y_{\ell m} = -\ell(\ell+1) Y_{\ell m}$.
This results in a linear, second-order  differential equation for $\varphi_{\ell m}(r)$
\begin{equation}
  -\frac{\d^2}{(\d r_*)^2} \varphi_{\ell m}(r_*) + \left\{ V_\ell(r_*) - \omega^2 \right\} \varphi_{\ell m}(r_*) = 0,
\label{eq:de_varphi}
\end{equation}
where the effective potential $V_\ell(r_*)$ is given by
\begin{equation}
  V_\ell(r) = f(r) \left(  \frac{f'(r)}{r} +  \frac{\ell(\ell+1)}{r^2} + m_\Phi^2 \right).
\end{equation}
The corresponding equations for higher spin fields are generally more complicated, but they can also be derived using a semiclassical method~\cite{Regge:1957td,Price:1971fb,Teukolsky:1973ha,Hatsuda:2020iql}.

We are interested in spacetime that has a horizon at radius $r_+$ determined by
\begin{equation}
  f(r_+) = 0.
\end{equation}
The location, $r_{\rm s}$, of the stretched horizon~\cite{Susskind:1993if} is determined by the condition that the proper distance between $r = r_+$ and $r_{\rm s}$ is the string length $l_{\rm s}$:
\begin{equation}
  \left| \int_{r_+}^{r_{\rm s}} \frac{dr}{\sqrt{f(r)}} \right| \approx l_{\rm s},
\label{eq:stretch-def}
\end{equation}
giving
\begin{equation}
  |r_{\rm s} - r_+| \approx \frac{|f'(r_+)|\, l_{\rm s}^2}{4}.
\end{equation}
Here, we have assumed that $|f'(r_+)|$ is not much suppressed compared with the natural size determined by dimensional analysis, $|f'(r_+)| \sim 1/r_+$, which is indeed the case for the spacetimes that we consider in this paper.
The Hawking temperature, as measured at $r$ satisfying $f(r) = 1$, is given by
\begin{equation}
  T_{\rm H} = \frac{|f'(r_+)|}{4\pi},
\end{equation}
so that the location of the stretched horizon also coincides with the place where the local (Tolman) Hawking temperature $T_{\rm loc}(r) = T_{\rm H}/\sqrt{f(r)}$ becomes the string scale, $\approx 1/2\pi l_{\rm s}$~\cite{Langhoff:2020jqa}.

\subsection{Schwarzschild black hole}
\label{subsec:BH}

\begin{figure}[t]
\centering
  \includegraphics[width=0.7\textwidth]{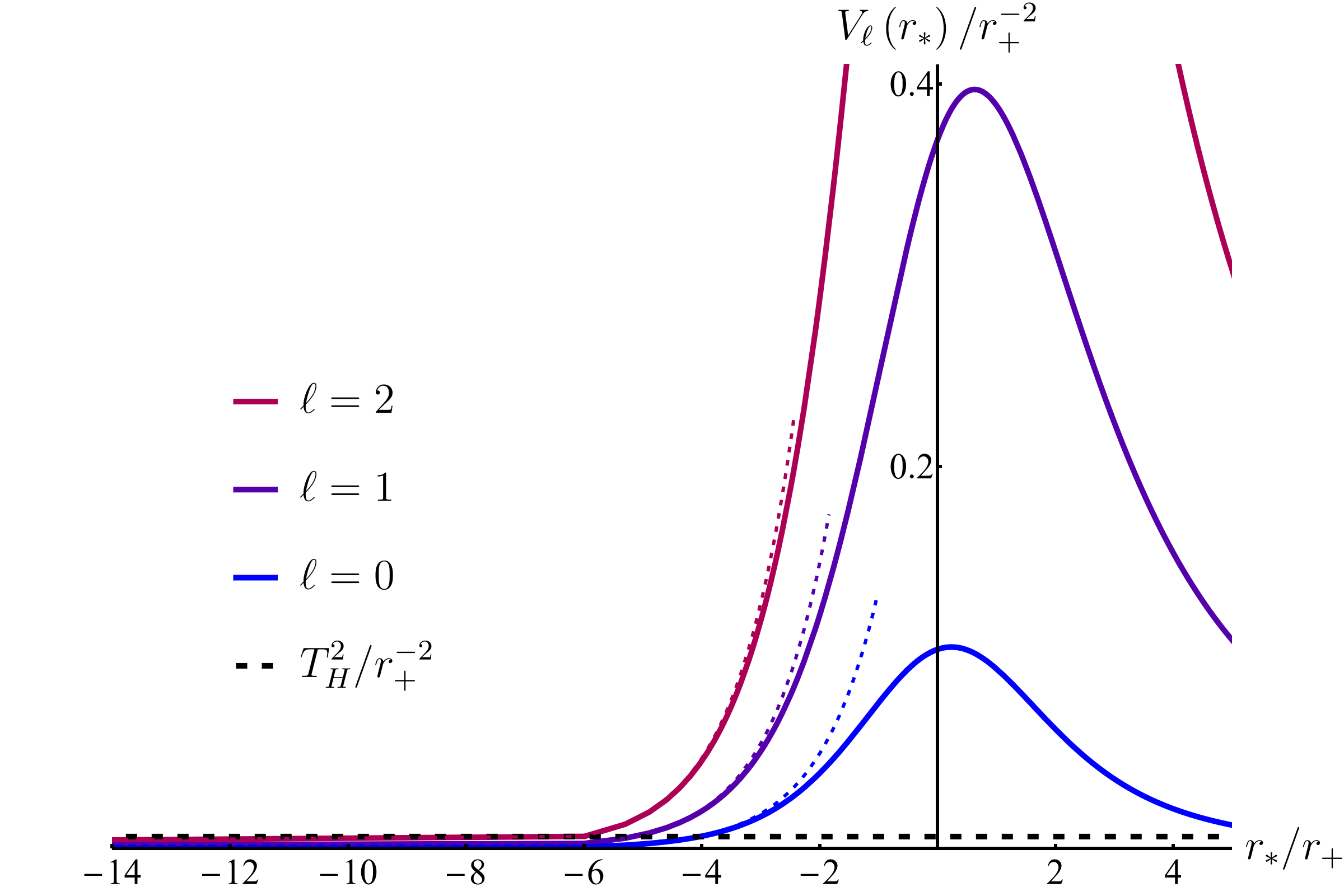}
\caption{
 The effective potential $V_{\ell}$ of a Schwarzschild black hole (solid lines) for a minimally-coupled massless scalar field with $\ell = 0,1,2$ as a function of the tortoise coordinate $r_*$.
 The Hawking temperature $T_{\rm H} = 1/4\pi r_+$ is indicated by the horizontal dashed line.
 The approximated curves in Eq.~\eqref{eq:BH-V-approx} are also plotted (dotted lines) in the region where they are relevant.}
\label{fig:BH_Veff}
\end{figure}

Let us consider a Schwarzschild black hole of mass $M$.
The metric is given by
\begin{equation}
  f(r) = 1 - \frac{r_+}{r},
\label{eq:metric-Sch}
\end{equation}
where $r_+ = 2Ml_{\rm P}^2$.
The tortoise coordinate is given by
\begin{equation}
  r_* = r + r_+ \ln\frac{r-r_+}{r_+},
\end{equation}
which maps $r:(r_+,\infty)$ to $r_*:(-\infty,\infty)$, and the effective potential is
\begin{equation}
  V_\ell(r_*) = \left( 1 - \frac{r_+}{r} \right) \left( \frac{r_+}{r^3} + \frac{\ell(\ell+1)}{r^2} + m_\Phi^2 \right),
\end{equation}
which is plotted in Fig.~\ref{fig:BH_Veff} for $m_\Phi = 0$.%
\footnote{
 For a general, not necessarily a scalar, field of $m_\Phi=0$, we have
 \begin{equation}
   V_\ell(r_*) = \left( 1 - \frac{r_+}{r} \right) \left( \frac{(1-s^2) r_+}{r^3} + \frac{\ell(\ell+1)}{r^2} \right),
 \label{eq:V-spin}
 \end{equation}
 where $s$ is the spin-weight parameter and $\ell \geq |s|$.
 In the near horizon limit, this leads to the approximate potential of Eq.~\eqref{eq:BH-V-approx} but with $\lambda_\ell = (\ell^2 + \ell + 1 - s^2)/r_+^2$, giving a higher potential barrier for larger $|s|$ for a fixed value of $\ell-|s|$.
\label{ft:V-spin}}
The stretched horizon is located at
\begin{equation}
  r_{\rm s} - r_+ \approx \frac{l_{\rm s}^2}{4r_+}
\quad\Longleftrightarrow\quad
  r_{*{\rm s}} - r_+ \approx -2 r_+ \ln\frac{2r_+}{l_{\rm s}},
\label{eq:r_st_s-bh}
\end{equation}
and the Hawking temperature is given by
\begin{equation}
  T_{\rm H} = \frac{1}{4\pi r_+}.
\end{equation}

While we cannot solve Eq.~\eqref{eq:de_varphi} analytically, we can study the behavior of the field in the near horizon and far regions by making appropriate approximations.
We are mostly interested in the near horizon ``zone'' region
\begin{equation}
  r_{\rm s} < r < r_{\rm z},
\qquad
  r_{\rm z} \approx \frac{3}{2} r_+,
\end{equation}
where $r_{\rm z}$ indicates the location of the potential barrier.
The near horizon limit corresponds to $r-r_+ \ll O(r_+)$, where $r_* \approx r_+ + r_+ \ln[(r-r_+)/r_+] \,\Longleftrightarrow\, r \approx r_+ + r_+ e^{(r_*-r_+)/r_+}$.
The effective potential in this region is given by
\begin{equation}
  V_\ell(r_*) \approx \lambda_\ell\, e^{\frac{r_*-r_+}{r_+}},
\qquad
  \lambda_\ell = \frac{\ell^2 + \ell + 1}{r_+^2} + m_{\Phi}^2.
\label{eq:BH-V-approx}
\end{equation}
This approximation is valid if $r_*$ is negative with $|r_*|$ sufficiently larger than $r_+$; see Fig.~\ref{fig:BH_Veff}.
This implies that we can trust solutions obtained using Eq.~\eqref{eq:BH-V-approx} if $\omega^2 \ll V_\ell(r_* = -r_+) \sim \lambda_\ell$.

Two independent real solutions of Eq.~\eqref{eq:de_varphi} with Eq.~\eqref{eq:BH-V-approx} can be taken as
\begin{equation}
  \Re{I_{2i r_+ \omega}\Bigl(2 \sqrt{\lambda_\ell}\, r_+ e^{\frac{r_*-r_+}{2r_+}}\Bigr)},
\quad
  \Im{I_{2i r_+ \omega}\Bigl(2 \sqrt{\lambda_\ell}\, r_+ e^{\frac{r_*-r_+}{2r_+}}\Bigr)},
\end{equation}
where $I_\nu(x)$ is the modified Bessel function of the first kind.
The first solution is exponentially increasing in $r_*$ at large $r_*$, so it does not correspond to modes that are localized in the zone region; it corresponds to decaying modes for signals sent from the far region.
We thus focus on the second solution, which is exponentially damped at large $r_*$.
This solution is approximated by a trigonometric function in the near horizon region
\begin{equation}
  \varphi_{\ell m}(r_*) \,\,\,\propto\,\,\, \Im{I_{2i r_+ \omega}\Bigl(2 \sqrt{\lambda_\ell}\, r_+ e^{\frac{r_*-r_+}{2r_+}}\Bigr)}
\,\,\xrightarrow[r_* < 0]{|r_*| \gg r_+}\,\,
  \sin\left[ \omega r_* + \omega r_+ \bigl(\ln(\lambda_\ell r_+^2) - 1\bigr) \right].
\end{equation}
Given that the effective potential at the stretched horizon has the value $V_\ell(r_{*{\rm s}}) = \lambda_\ell l_{\rm s}^2/4 r_+^2$, we find
\begin{equation}
  \omega \gtrsim \frac{\sqrt{\lambda_\ell}\, l_{\rm s}}{2 r_+}.
\end{equation}
We also find that imposing a boundary condition at the stretched horizon makes the spectrum discrete, with the gap between adjacent levels given by
\begin{equation}
  \varDelta\omega \equiv \omega_n - \omega_{n-1} \approx \frac{\pi}{2 r_+ \left|\ln\frac{2}{\sqrt{\lambda_\ell}\, l_{\rm s}}\right|},
\label{eq:vardelta-omega_BH}
\end{equation}
where we have imposed a simple Dirichlet boundary condition $\varphi_{\ell m}(r_{*{\rm s}}) = 0$ for illustrative purposes.
Note that for a given $\ell$, each level has ($2\ell+1$)-fold degeneracy corresponding to $m = -\ell, -\ell+1, \cdots, \ell$.
While the details of the spectrum do depend on the boundary condition, its basic structure---the discreteness and the scale characterizing the gaps---does not.

\subsection{de~Sitter spacetime}
\label{subsec:dS}

\begin{figure}[t]
\centering
  \includegraphics[width=0.7\textwidth]{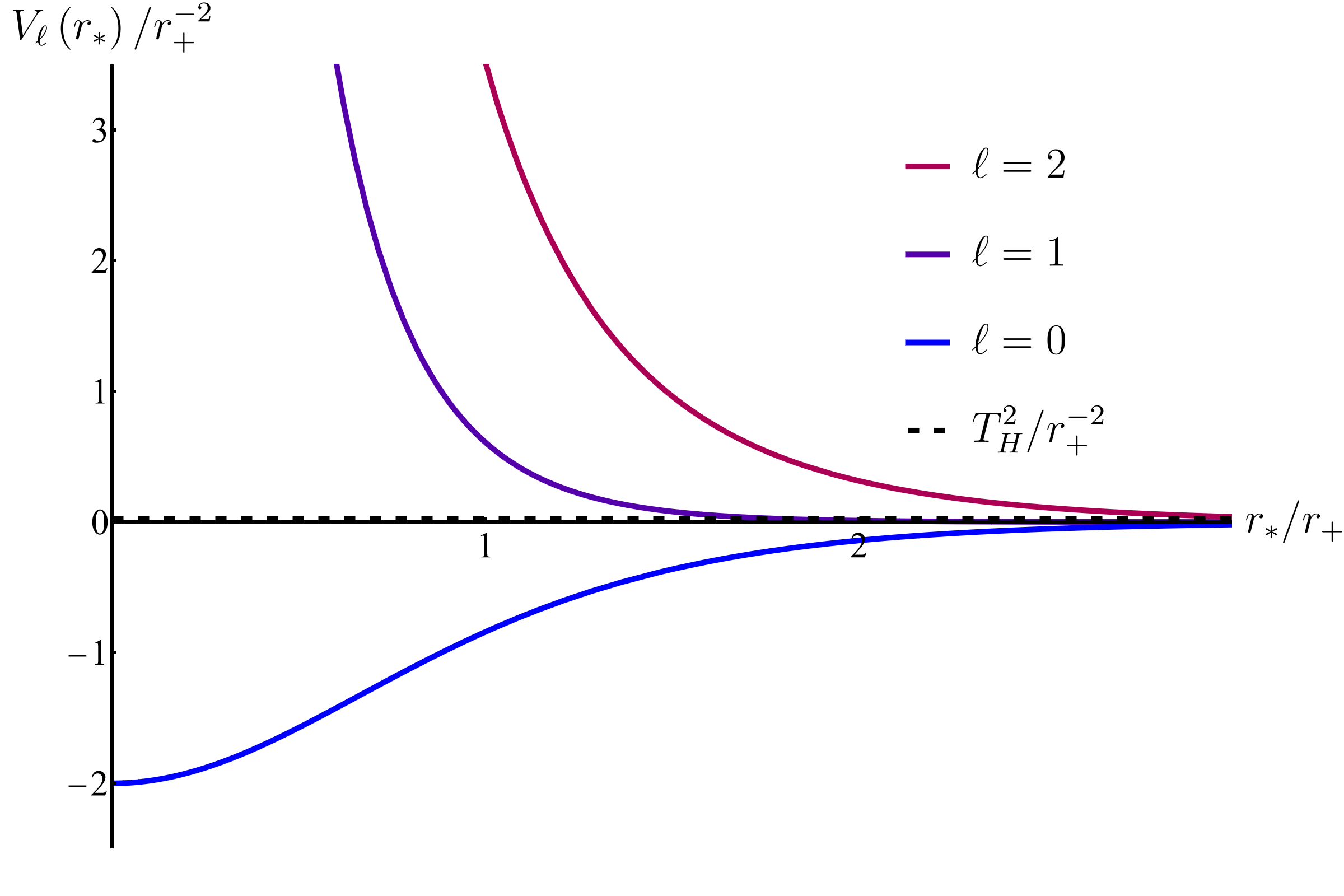}
\caption{
 The effective potential $V_{\ell}$ of de~Sitter spacetime for a minimally-coupled massless scalar field with $\ell = 0,1,2$ as a function of the tortoise coordinate $r_*$.
 The Hawking temperature $T_{\rm H} = 1/2\pi r_+$ is indicated by the horizontal dashed line.}
\label{fig:dS_Veff}
\end{figure}

A similar analysis can be performed for de~Sitter spacetime in static coordinates
\begin{equation}
  f(r) = 1 - \frac{r^2}{r_+^2},
\end{equation}
leading to the same basic conclusion.
Here, $r_+ = \alpha$ with $\alpha$ being the Hubble radius.
The tortoise coordinate is given by
\begin{equation}
  r_* = \frac{r_+}{2} \ln\frac{1+\frac{r}{r_+}}{1-\frac{r}{r_+}},
\end{equation}
which maps $r:(0,r_+)$ to $r_*:(0,\infty)$, and the effective potential is
\begin{equation}
  V_\ell(r_*) = \left( 1 - \frac{r^2}{r_+^2} \right) \left( -\frac{2}{r_+^2} + \frac{\ell(\ell+1)}{r^2} + m_{\Phi}^2 \right),
\end{equation}
which is plotted in Fig.~\ref{fig:dS_Veff} for $m_{\Phi} = 0$.
The stretched horizon is located at
\begin{equation}
  r_+ - r_{\rm s} \approx \frac{l_{\rm s}^2}{2r_+}
\quad\Longleftrightarrow\quad
  r_{*{\rm s}} \approx r_+ \ln\frac{2r_+}{l_{\rm s}},
\label{eq:r_st_s-dS}
\end{equation}
and the Hawking temperature is given by
\begin{equation}
  T_{\rm H} = \frac{1}{2\pi r_+}.
\end{equation}

For $\ell \neq 0$, the effective potential blows up at small values of $r_*$.
Therefore, if $\omega$ is small, we can restrict to $r_* > r_+$ which corresponds to the region near the cosmological horizon.
For $\ell = 0$, the effective potential blows up at small $r_*$ only when $m_\Phi > \sqrt{2} /r_+$; otherwise, the potential is confining, leading to a small number of bound states.
Below, we focus on the case $m_\Phi > \sqrt{2} /r_+$ for $\ell = 0$.

With this restriction, the effective potential in the near horizon region is given by
\begin{equation}
  V_\ell(r_*) \approx 4\lambda_\ell\, e^{-2\frac{r_*}{r_+}},
\qquad
  \lambda_\ell = \frac{\ell^2 + \ell - 2}{r_+^2} + m_\Phi^2,
\label{eq:dS-V-approx}
\end{equation}
except for $\ell = 1$ with $m_\Phi = 0$, in which case
\begin{equation}
  V_{\ell=1}(r_*) \approx \frac{32}{r_+^2} e^{-4\frac{r_*}{r_+}}
\quad \mbox{for }\, m_\Phi = 0.
\label{eq:dS-V-approx-2}
\end{equation}
This approximation is valid only for $r_* \gtrsim r_+$.
We can thus trust the solutions of Eq.~\eqref{eq:de_varphi} obtained using these $V_\ell(r_*)$'s only if $\omega^2 \ll V(r_* = r_+) \sim \lambda_\ell$ (or $\ll 1/r_+^2$ for $\ell=1$, $m_\Phi=0$).

The most general solution corresponding to the approximate potential in Eq.~\eqref{eq:dS-V-approx} is given by
\begin{equation}
  \varphi_{\ell m}(r_*) = \Re{A\, I_{i r_+ \omega}\Bigl(2 \sqrt{\lambda_\ell}\, r_+ e^{-\frac{r_*}{r_+}}\Bigr)},
\end{equation}
where $A \in \mathbb{C}$ is an arbitrary constant.
In order for the original field $\Phi(t,r_*,\theta,\phi)$ in Eq.~\eqref{eq:Phi_varphi} to be well-defined, the exact solution of $\varphi_{\ell m}(r_*)$ must vanish at least as fast as $r_*$ when $r_* \approx r \to 0$.
This condition fixes the phase of $A$ in the approximate solution, leading to a single solution for $\varphi_{\ell m}(r_*)$ in the near horizon region:
\begin{equation}
  \varphi_{\ell m}(r_*) \,\xrightarrow[]{r_* \gg r_+}\,
  \sin\left[ \omega r_* - \frac{1}{2} \omega r_+ \ln(\lambda_\ell r_+^2) + \delta \right],
\end{equation}
where $\delta$ is determined by the phase of $A$.
For $\ell = 1$ and $m_\Phi = 0$, a similar analysis gives
\begin{equation}
  \varphi_{\ell m}(r_*)  \,\,\propto\,\, \Re{A\, I_{\frac{i}{2} r_+ \omega}\Bigl(2 \sqrt{2}\, e^{-\frac{2r_*}{r_+}}\Bigr)} 
\,\xrightarrow[]{r_* \gg r_+}\,
  \sin\left[ \omega r_* - \frac{\ln 2}{4} \omega r_+  + \delta \right].
\end{equation}

Given that the effective potential in Eq.~\eqref{eq:dS-V-approx} has the value $V_\ell(r_{*{\rm s}}) = \lambda_\ell l_{\rm s}^2/r_+^2$, we find
\begin{equation}
  \omega \gtrsim \frac{\sqrt{\lambda_\ell}\, l_{\rm s}}{r_+}
\qquad \mbox{for }\, (\ell,m_\Phi) \neq (1,0),
\end{equation}
and by imposing the boundary condition $\varphi_{\ell m}(r_{*{\rm s}}) = 0$, we obtain a discrete spectrum with
\begin{equation}
  \varDelta\omega \approx \frac{\pi}{r_+ \left|\ln\frac{2}{\sqrt{\lambda_\ell}\, l_{\rm s}}\right|}
\qquad \mbox{for }\, (\ell,m_\Phi) \neq (1,0).
\end{equation}
For $(\ell,m_\Phi) = (1,0)$, the corresponding quantities are given instead by $V_\ell(r_{*{\rm s}}) = 2 l_{\rm s}^4/r_+^6$ and
\begin{equation}
  \omega \gtrsim \frac{\sqrt{2} l_{\rm s}^2}{r_+^3},
\quad
  \varDelta\omega \approx \frac{\pi}{r_+ \ln\frac{2^{3/4} r_+}{l_{\rm s}}}
\qquad \mbox{for }\, (\ell,m_\Phi) = (1,0).
\end{equation}
Once again, the basic structure of the spectrum found here does not depend on the boundary condition at the stretched horizon.

\subsection{General near horizon limit}
\label{subsec:gen-hor}

We now see that the structure of the spectrum for
\begin{equation}
  \omega \ll \sqrt{\frac{\ell^2}{r_+^2} + m_\Phi^2}, \qquad \ell \gg 1
\label{eq:omega-gen}
\end{equation}
in the near horizon region is universal for generic horizons, beyond the black hole and de~Sitter spacetimes discussed so far.
This reflects the universality of the near horizon limit, giving Rindler spacetime.

To see this, let us consider a spacetime with the metric in Eq.~\eqref{eq:metric-gen} which has a horizon at $r = r_+$.
We assume that $f(r) > 1$ in the region $r > r_+$, which we call the allowed region.
For a given horizon, this can always be arranged.
Specifically, if $f(r) > 1$ in $r < r_+$, as in de~Sitter spacetime, we redefine $r \to -r$ (and $r_+ \to -r_+$) to make the allowed region $r > r_+$.
Note that this makes $r$ and $r_+$ negative.

In the near horizon region, we then have $f(r) \approx f'(r_+) (r-r_+)$ with $f'(r_+) > 0$.
Here, we have assumed that $f'(r_+)$ is not too suppressed compared with its natural size $f'(r_+) \sim 1/|r_+|$.
(This excludes the horizon of a near extremal black hole from our consideration.)
The tortoise coordinate in the near horizon region is then
\begin{equation}
  r_* \approx \frac{1}{f'(r_+)} \ln\frac{r-r_+}{c |r_+|},
\end{equation}
where $c > 0$ is an unimportant $O(1)$ number defining the origin of $r_*$,%
\footnote{
 In sections~\ref{subsec:BH}~and~\ref{subsec:dS}, $c$ was taken to be $1/e$ and $2$, respectively.
}
and the effective potential in this region is given by
\begin{equation}
  V_\ell(r_*) \approx c |r_+| f'(r_+) \lambda_\ell\, e^{f'(r_+) r_*} \approx \lambda_\ell\, e^{f'(r_+) r_*},
\qquad
  \lambda_\ell = \frac{f'(r_+)}{r_+} + \frac{\ell(\ell+1)}{r_+^2} + m_\Phi^2.
\label{eq:V-approx-gen}
\end{equation}
We want $\lambda_\ell > 0$ for this potential to trap modes in the near horizon region.
(Here we exclude the nongeneric case of $\lambda_\ell = 0$ from consideration.)
If $r_+ > 0$, this condition is satisfied for any values of $\ell$ and $m_{\Phi}$.
If $r_+ < 0$, we restrict our treatment to modes with large enough $\ell$ or $m_\Phi$ such that $\lambda_\ell > 0$.

For the above approximation to hold, we need $\omega^2 \ll V(r_* \sim -1/f'(r_+)) \sim \lambda_\ell$, which for $\ell \gg 1$ gives the condition in Eq.~\eqref{eq:omega-gen}.
Assuming that $\omega$ is in this range, we can obtain the general solution to Eq.~\eqref{eq:de_varphi} as
\begin{equation}
  \varphi_{\ell m}(r_*) = \Re{A\, I_{\!\frac{2i \omega}{f'(r_+)}}\!\biggl(2 \sqrt{\frac{c|r_+|\lambda_\ell}{f'(r_+)}}\, e^{\frac{f'(r_+) r_*}{2}}\biggr)},
\end{equation}
where $A \in \mathbb{C}$ is an arbitrary constant.
The relevant solution is selected by a boundary condition at $r_* \approx O(r_+)$, which depends on spacetime under consideration.
For Schwarzschild black hole and de~Sitter spacetimes, these were given at $r \approx r_{\rm z}$, and $r=0$, respectively.
This fixes the phase of $A$ and gives us a single solution for each $\omega$.
In the near horizon limit, it takes the form
\begin{equation}
  \varphi_{\ell m}(r_*) \approx \sin\biggl(\omega r_* + \frac{\omega}{f'(r_+)} \ln\frac{c|r_+|\lambda_\ell}{f'(r_+)} + \delta\biggr)
\qquad \mbox{for }\, r_* < 0,\, |r_*| \gg |r_+|,
\end{equation}
where $\delta$ is an $O(1)$ phase determined by the phase of $A$.

The existence of the stretched horizon, at $r_{*{\rm s}} \approx -(1/f'(r_+)) \ln(4c|r_+|/l_{\rm s}^2 f'(r_+))$, has two effects.
First, since $V_\ell(r_{*{\rm s}}) \approx \lambda_\ell l_{\rm s}^2/|r_+|^2$, frequency $\omega$ for given $\lambda_\ell$ is bounded from below:
\begin{equation}
  \omega \gtrsim \frac{\sqrt{\lambda_\ell}\, l_{\rm s}}{|r_+|}.
\end{equation}
Second, imposing a boundary condition at $r_* = r_{*{\rm s}}$ quantizes the spectrum:
\begin{equation}
  \varDelta\omega \sim \frac{\pi f'(r_+)}{2 \left|\ln\frac{2}{\sqrt{\lambda_\ell}\, l_{\rm s}}\right|}.
\end{equation}
We find that these are universal features of the spectrum in the regime of Eq.~\eqref{eq:omega-gen}.

\section{The Absence of Spacetime below the String Length}
\label{sec:stretched}

A key element of the analysis in the previous section is the hypothesis that spacetime, as we usually perceive, does not exist below the string length.
In normal circumstances, this hardly affects low energy physics because of Wilsonian decoupling.
In the presence of a horizon, however, large gravitational red/blue shift makes this fact relevant for low energy physics.
Its significance can be best seen in the tortoise coordinate, in which the wavelength of a massless mode is preserved while propagating; in terms of this coordinate, the removal of the spacetime region within the string length from the horizon leads to excising a half line.

The proper distance, used in defining the stretched horizon in Eq.~\eqref{eq:stretch-def}, is associated with the particular time foliation that leads to the static form of the metric, Eq.~\eqref{eq:metric-gen}.
In the case of an evaporating black hole, we apply our treatment to a sufficiently small time window, e.g.\ $\varDelta t \lesssim r_+$, in which the system can be viewed as approximately static.
The issue of time slicing will be discussed further in Section~\ref{sec:holo}.

In this section, we demonstrate that the Bekenstein-Hawking entropy, $S_{\rm BH}$, as well as Hawking temperature, $T_{\rm H}$, follow only from two inputs from the ultraviolet (UV) physics (or two assumptions from the point of view of low energy theory):
\begin{itemize}
\item Spacetime does not exist below the string length, which introduces the ``boundary'' of space---the stretched horizon---at a proper distance $l_{\rm s}$ away from the horizon.
\item Physics associated with the boundary is (maximally) quantum chaotic~\cite{Maldacena:2015waa} across all low energy species~\cite{Nomura:2019qps}, so the black hole and de~Sitter vacuum microstates (or the microstates associated with any horizon) are typical in a suitable microcanonical ensemble.
\end{itemize}
In particular, these reproduce $S_{\rm BH}$ and $T_{\rm H}$ up to incalculable $O(1)$ coefficients.
In the situation where the state evolves in time at the semiclassical level, for example when the black hole is evaporating or the system is perturbed by excitations, we assume that the typicality described above is reached quickly.
In fact, it is believed that the dynamics associated with horizons is fast scrambling~\cite{Hayden:2007cs,Sekino:2008he}.%
\footnote{
 As discussed in Ref.~\cite{Nomura:2019qps}, this assumption implies the absence of fundamental global symmetries (see, e.g., Refs.~\cite{Bekenstein:1971hc,Zeldovich:1977be,Banks:2010zn,Harlow:2018jwu}); specifically, any linearly-realized global symmetry is explicitly broken by an $O(1)$---or not exponentially suppressed---amount at the string scale.
}

Next we discuss the UV sensitivities of the various modes, classifying them into modes that can be reliably described in the semiclassical theory and those that are intrinsically quantum gravitational.
This provides a refinement of the concept of hard and soft modes introduced in Refs.~\cite{Nomura:2018kia,Nomura:2019qps} to describe an analytic extension of spacetime at the microscopic level; the meaning of this analytic extension in quantum gravity will be discussed in more detail in Section~\ref{sec:extension}.
Finally, we will discuss how the vacuum and excited states of semiclassical theory are related to the microscopic description given here.

\subsection{Entropy, temperature, and microstates for a semiclassical vacuum}
\label{subsec:vacuum}

As stated above, we assume that the unknown UV physics of quantum gravity appears in low energy physics as a lack of space below the proper length of order $l_{\rm s}$.
As we have seen in Section~\ref{sec:QFT}, this makes the spectrum discrete:
\begin{equation}
  \omega^{(\ell)}_n = \omega^{(\ell)}_0 + n \varDelta\omega^{(\ell)}
\qquad
  n = 0,1,2,\cdots,
\label{eq:gen-omega}
\end{equation}
where
\begin{equation}
  \omega^{(\ell)}_0 \approx \frac{\sqrt{\lambda_\ell} l_{\rm s}}{|r_+|},
\qquad
  \varDelta\omega^{(\ell)} \sim \left|\frac{\pi f'(r_+)}{2\ln\frac{2}{\sqrt{\lambda_\ell}l_{\rm s}}}\right|.
\label{eq:gen-omega-2}
\end{equation}
Following Refs.~\cite{Nomura:2018kia,Nomura:2019qps}, we take the view that the energy of the system, which is usually attributed to the background, is carried by the quanta that fill these energy levels.

For concreteness, let us consider a Schwarzschild black hole.
The case of de~Sitter spacetime can be analyzed similarly, which we will discuss at the end of this subsection.
Suppose that there are $N^{(\ell)}_n$ quanta of field $\Phi$ at the $n$-th level of angular momentum $\ell$, which has ($2\ell+1$)-fold degeneracy.
The total energy carried by $\Phi$ is then
\begin{equation}
  E_\Phi = g_\Phi \sum_{\ell=0}^{\infty} \sum_{n=0}^{\infty} \omega^{(\ell)}_n N^{(\ell)}_n,
\end{equation}
where $g_\Phi$ is the number of degrees of freedom for field $\Phi$.
($g_\Phi = 1$ for a real scalar field.)
The picture of Refs.~\cite{Nomura:2018kia,Nomura:2019qps} is that the sum of this energy for all low energy fields represents the total energy of the system (as measured at $r$ satisfying $f(r) = 1$), which is determined self-consistently by the spacetime background used in calculating $\omega^{(\ell)}_n$.
In the present case
\begin{equation}
  E_{\rm bh} = \sum_\Phi E_\Phi = M,
\label{eq:const}
\end{equation}
where the sum runs over all low energy fields that can be viewed as elementary in the effective field theory at a scale slightly below $1/l_{\rm s}$.%
\footnote{
 We ignore the kinetic energy, which is not important for a black hole that is spherically symmetric at the classical level.
}

While the expressions for $\omega^{(\ell)}_0$ and $\varDelta\omega^{(\ell)}$ are different for a field with spin, their values are of the same order as those of a scalar field.
In particular, using Eq.~\eqref{eq:V-spin} in footnote~\ref{ft:V-spin}, one finds that $\lambda_\ell$ in Eq.~\eqref{eq:gen-omega-2} is simply replaced as
\begin{equation}
  \lambda_\ell \,\rightarrow\, \lambda_{\ell,s} = \frac{\ell^2 + \ell + 1 - s^2}{r_+^2} \quad (\ell \geq |s|)
\end{equation}
for $m_\Phi = 0$.
This does not change the values of $\omega^{(\ell)}_0$ and $\varDelta\omega^{(\ell)}$ much, as long as $|s| \approx O(1)$ which we assume here.
In any case, the precise numbers for these quantities are not important, or trustable, as we will discuss in Section~\ref{subsec:UV}.

Because of the assumption of chaotic and fast scrambling dynamics at the stretched horizon, especially those across all low energy species, the distribution of energy among various species and levels is determined purely by the content of low energy fields.
In particular, the distribution of quanta in each degree of freedom is given by maximizing the combinatorial numbers
\begin{equation}
  C_\Phi\bigl(\{ N^{(\ell)}_n \}\bigr) = \begin{cases}
    \prod_{\ell=0}^{\infty} \prod_{n=0}^{\infty} \frac{(N^{(\ell)}_n+2\ell)!}{N^{(\ell)}_n! (2\ell)!}
      \quad (N^{(\ell)}_n \geq 0) & \mbox{for } \Phi\mbox{: boson} \\
    \prod_{\ell=0}^{\infty} \prod_{n=0}^{\infty} \frac{(2\ell+1)!}{N^{(\ell)}_n! (2\ell+1-N^{(\ell)}_n)!}
      \quad (0 \leq N^{(\ell)}_n \leq 2\ell+1) & \mbox{for } \Phi\mbox{: fermion}
  \end{cases}
\label{eq:C_Phi}
\end{equation}
under the constraint of Eq.~\eqref{eq:const}.%
\footnote{
 As in the standard statistical mechanics, this gives the most probable configuration of quanta.
 The probability of finding other configurations satisfying the constraint is not zero but exponentially suppressed.
}
Following the standard analysis in statistical mechanics, we find
\begin{equation}
  N^{(\ell)}_n = \frac{2\ell+1}{e^{\beta \omega^{(\ell)}_n} \mp 1},
\label{eq:N_BH}
\end{equation}
where $\mp$ takes the minus and plus signs for bosonic and fermionic degrees of freedom, respectively, and $\beta$ is determined by the condition
\begin{equation}
  E_{\rm bh} = \sum_\Phi g_\Phi \sum_{\ell=0}^{\infty} \sum_{n=0}^{\infty} \frac{(2\ell+1) \omega^{(\ell)}_n}{e^{\beta \omega^{(\ell)}_n} \mp 1}.
\label{eq:E_BH}
\end{equation}
Note that $\beta$ in Eq.~\eqref{eq:E_BH} does not depend on $\Phi$, and there is no ``chemical potential'' for any $\Phi$, since the dynamics at the stretched horizon is chaotic across all low energy species.

We can calculate the right-hand side of Eq.~\eqref{eq:E_BH} by using Eq.~\eqref{eq:gen-omega} and replacing the sum over $\ell$ and $n$ with the corresponding integrals.
Assuming that
\begin{equation}
  \ell \gg 1,
\qquad
  \frac{\ell^2}{r_+^2} \gg m_\Phi^2,
\end{equation}
which is justified a posteriori as the integral is dominated by $\ell \sim r_+/l_{\rm s}$, we may use
\begin{equation}
  \omega^{(\ell)}_0 \approx \frac{\ell\, l_{\rm s}}{r_+^2},
\qquad
  \varDelta\omega^{(\ell)} \sim \frac{1}{\left|r_+ \ln\frac{r_+}{\ell\, l_{\rm s}}\right|}.
\end{equation}
This gives
\begin{equation}
  E_{\rm bh} \approx N_{\rm dof} \frac{r_+^5}{\beta^4 l_{\rm s}^2} \left( \left|\ln\frac{\beta}{r_+}\right| + O(1) \right),
\end{equation}
where $N_{\rm dof} = \sum_\Phi g_\Phi$ is the total number of degrees of freedom of low energy fields.
Using
\begin{equation}
  \frac{1}{l_{\rm P}^2} \approx \frac{N_{\rm dof}}{l_{\rm s}^2}
\label{eq:lP-ls}
\end{equation}
(see, e.g., Ref.~\cite{Dvali:2007hz}) as well as $E_{\rm bh} = M$ and $r_+ = 2Ml_{\rm P}^2$, we find
\begin{equation}
  \beta \approx M l_{\rm P}^2,
\label{eq:beta-BH}
\end{equation}
which is the inverse Hawking temperature.
We also find that
\begin{equation}
  S = \sum_\Phi g_\Phi \ln C_\Phi \approx M^2 l_{\rm P}^2,
\label{eq:S-BH}
\end{equation}
as indicated by the Bekenstein-Hawking entropy.
Note that since the contribution to
\begin{equation}
  \ln C_\Phi \approx \sum_{\ell=0}^{\infty} 2\ell \sum_{n=0}^{\infty} \beta \omega^{(\ell)}_n e^{-\beta \omega^{(\ell)}_n}
\label{eq:log-C_Phi}
\end{equation}
comes predominantly from the $\ell \approx O(r_+/l_{\rm s})$ modes, the entropy is given primarily by the number of different independent states that these large $\ell$ modes can take.

In the present method based on a low energy description of the system, the coefficients in Eqs.~\eqref{eq:beta-BH} and \eqref{eq:S-BH} cannot be obtained.
This is because the quantities are dominated by the contributions from the $\ell \sim r_+/l_{\rm s}$ modes that are localized near the stretched horizon, where the effect of unknown UV physics dominates.
While this implies that the calculation is UV sensitive, its agreement with the results of Bekenstein and Hawking still gives us information about the UV physics; in particular, this unknown physics does not drastically increase the degrees of freedom compared to what is suggested by naively cutting off the spacetime at $r_* \approx r_{*{\rm s}}$.
Note that this calculation is equivalent to that in Refs.~\cite{Nomura:2018kia,Nomura:2019qps}, in which the mass and entropy of a black hole are obtained by integrating appropriate powers of the local Hawking temperature.
The logic, however, is reversed here; once the geometry is given (within a time window of order $r_+$) by Eqs.~\eqref{eq:metric-gen} and \eqref{eq:metric-Sch}, and correspondingly the energy of the system by $M$, then a typical state represents a black hole vacuum microstate with the temperature and entropy of the black hole given by Eqs.~\eqref{eq:beta-BH} and \eqref{eq:S-BH}.
Microstates of the black hole vacuum correspond to different ways in which the energy levels in Eq.~\eqref{eq:gen-omega} are occupied under the energy constraint.

As in the standard thermal system, the microscopic state of a black hole changes generically in a timescale of order the inverse temperature $\beta \approx M l_{\rm P}^2$.
This implies that the energy of the system, i.e.\ the mass of the black hole, can be specified only up to the precision of $\varDelta E \sim 1/\beta$; the state of a black hole comprises a superposition of energy eigenstates with the spread of eigenvalues of order $1/\beta$ or larger.
In the following, we always assume that the state of the system of interest, e.g.\ a black hole or de~Sitter spacetime, is specified with this maximal precision.
A similar comment also apples to other quantities, such as the momentum of a black hole, whose minimal uncertainty is of order $\varDelta p \sim \varDelta(\sqrt{2 M E}) \sim 1/Ml_{\rm P}^2$.
The number of independent states consistent with this specification is given by the Bekenstein-Hawking entropy of Eq.~\eqref{eq:S-BH}.
If the state involves superpositions of wider ranges of energy, momentum, and so on, e.g.\ as a result of backreaction of Hawking emission~\cite{Page:1979tc,Nomura:2012cx}, then our discussion below applies to each branch specified with the maximal precision for these quantities.

Note that a typical state in the space spanned by the independent microstates specified by $E = M$ and $p = 0$ within the minimal uncertainties of $\varDelta E$ and $\varDelta p$ has angular momentum of order the uncertainty $\varDelta J \sim \sqrt{N_{\rm dof}}\, r_+/l_{\rm s} \sim r_+/l_{\rm P}$.
This is consistent with the maximal precision with which angles can be specified consistently with the UV cutoff:\ $\varDelta\theta \sim l_{\rm s}/r_+$ for each species.
We can thus regard these microstates as those of a non-rotating black hole in semiclassical theory.

We finally mention that we can discuss de~Sitter spacetime in a similar manner.
One difference is that we do not have a well-established notion of energy attributed to the spacetime in this case.
However, by taking
\begin{equation}
  E_{\rm dS} = \frac{\alpha}{l_{\rm P}^2},
\label{eq:E_dS}
\end{equation}
as implied by the Bekenstein-Hawking (or Gibbons-Hawking~\cite{Gibbons:1977mu}) entropy $S_{\rm GH} = \pi \alpha^2/l_{\rm P}^2$ and Hawking temperature $T_{\rm H} = 1/2\pi \alpha$, we reproduce all the properties associated with a static patch of the de~Sitter spacetime~\cite{Nomura:2019qps}.
(Physically, this energy can be specified only up to the precision of order $T_{\rm H}$.)
While the relationship of this energy to more conventionally defined energies is not clear, it represents some ``energy'' defined at $r = 0$, at which $f(r) = 1$, rather than at asymptotic infinity.%
\footnote{
 It is interesting to note that if one considers the quantity $E = \int_\Sigma\! \sqrt{-g}\, \rho\, d^3x$, along the lines of Ref.~\cite{Aoki:2020prb}, then one would get $E = \alpha/l_{\rm P}^2$.
 Here, $\Sigma$ is the $t = 0$ surface of {\it global} de~Sitter spacetime comprising two static patches, $g = -r^4 \sin^2\!\theta$ is the determinant of the {\it spacetime} metric (in the static coordinates), $\rho = 3/8\pi l_{\rm P}^2 \alpha^2$ is the energy density, and $d^3x = dr d\theta d\phi$.
}

\subsection{UV (in)sensitivity: zone and horizon modes}
\label{subsec:UV}

Even though the density of states given by Eqs.~\eqref{eq:gen-omega} and \eqref{eq:gen-omega-2} reproduces the entropy and temperature associated with the horizon, we do not expect that the precise spectrum is given by these expressions.
This is because interactions near the stretched horizon, with the effective coupling given by $T_{\rm loc}(r_{\rm s})^2 l_{\rm s}^2 \sim O(1)$, deform the spectrum significantly from that of free theories (though the density of states will not change when coarse grained at a scale of order $r_+$ in the $r_*$ coordinate).

To analyze this issue in more detail, let us consider the effective potential $V_\ell(r_*)$ for a fixed $\ell$ and negligible $m_\Phi$.
From Eq.~\eqref{eq:gen-omega-2}, the average gap between adjacent energy levels is given by
\begin{equation}
  \varDelta\omega^{(\ell)} \sim \frac{1}{r_+ \ln\frac{r_+}{\ell\, l_{\rm s}}}.
\end{equation}
For a black hole, the height of the barrier is given by
\begin{equation}
  \omega^{(\ell)}_{\rm barrier} \sim \frac{\ell}{r_+}.
\end{equation}
As we have seen, the temperature of the system is $\sim 1/r_+$, and the modes with
\begin{equation}
  \omega \lesssim \frac{1}{r_+}
\,\mbox{ and }\;
  \ell \lesssim \frac{r_+}{l_{\rm s}}
\end{equation}
are significantly occupied.
For each of these modes, the wavefunction is oscillatory between $r_* = r_{*{\rm s}}$ and
\begin{equation}
  r_* \approx -\frac{2}{f'(r_+)} \ln\ell \equiv r^{(\ell)}_{*{\rm b}},
\end{equation}
outside of which it is exponentially damped.
Thus the size of the region supporting these modes is given, in units of the wavelength of Hawking radiation, as
\begin{equation}
  \frac{\bigl|r^{(\ell)}_{*{\rm b}} - r_{*{\rm s}}\bigr|}{r_+} \approx \left|\frac{1}{r_+ f'(r_+)} \ln\frac{r_+}{\ell^2 l_{\rm s}^2 |f'(r_+)|}\right| 
  \sim \left|\ln\frac{r_+}{\ell\, l_{\rm s}}\right|,
\label{eq:zone-width}
\end{equation}
where we have used $|f'(r_+)| \sim 1/r_+$ in the last expression.

\begin{figure}[t]
\centering
  \subfloat[\centering $\ell \approx O(1)$]{{\includegraphics[width=0.45\textwidth]{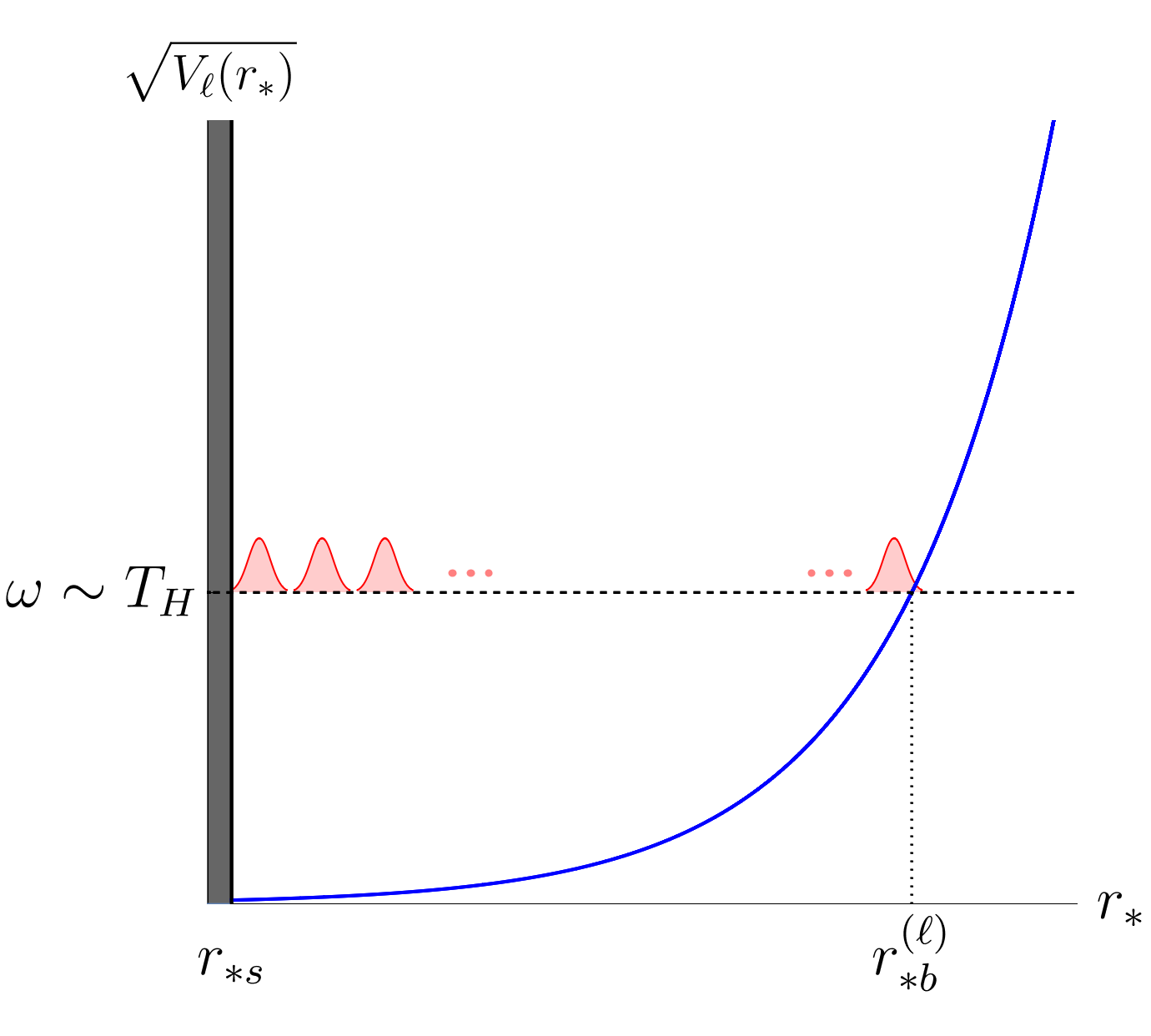} }}
\hspace{1cm}
  \subfloat[\centering $\ell \approx O(r_{+}/l_{s})$]{{\includegraphics[width=0.45\textwidth]{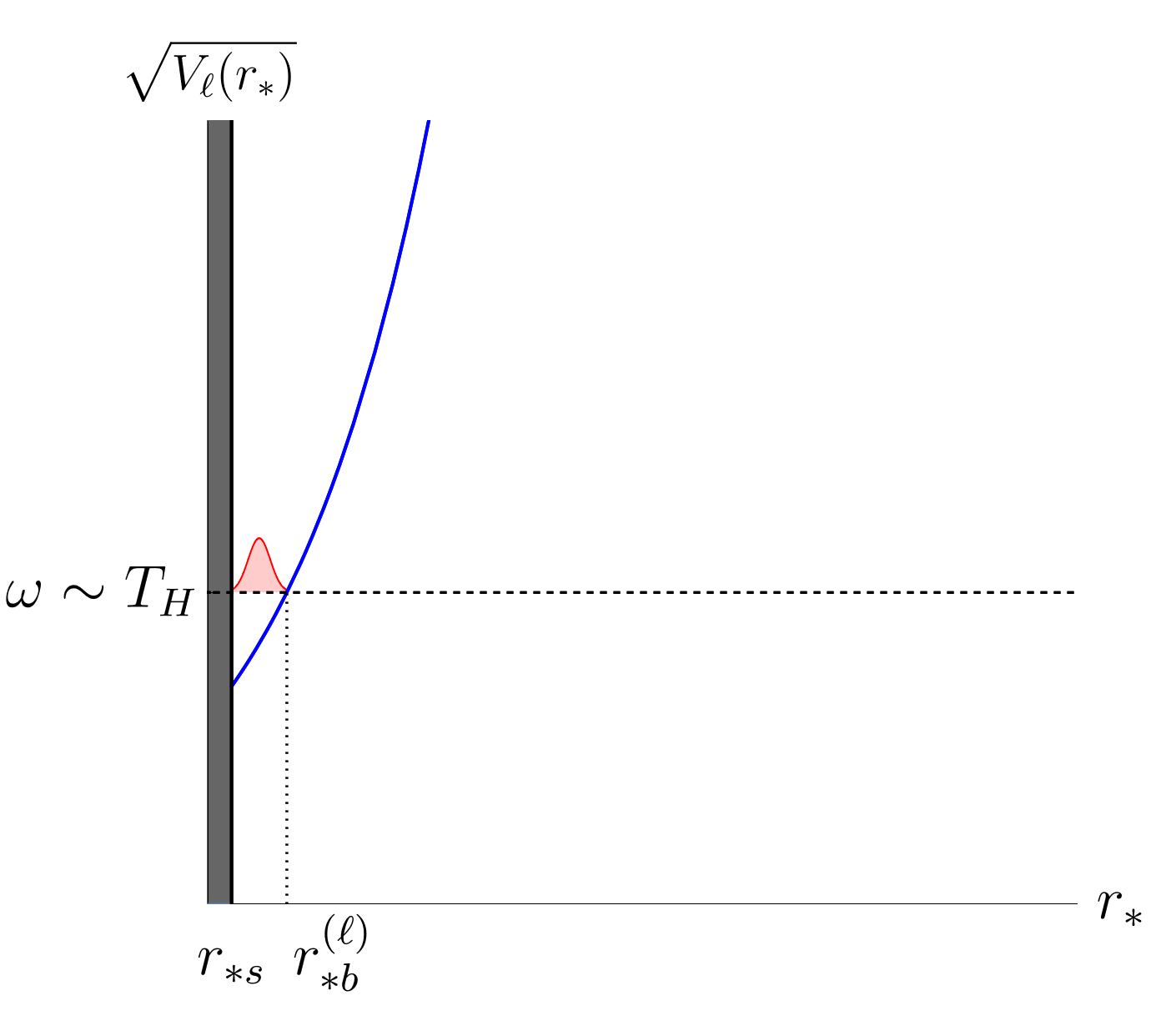} }}
\caption{
 A schematic depiction of approximately orthogonal, independent modes having $\omega \approx O(T_{\rm H})$ and localized in the zone for a field of negligible mass.
 (a) For $\ell \approx O(1)$, there are $\sim \ln(r_+/l_{\rm s})$ independent ingoing and outgoing modes.
 Their approximate basis can be taken to be wavepackets of width $\sim r_+$ in $r_*$, distributed uniformly over the classically allowed region between $r_{*{\rm s}}$ and $r^{(\ell)}_{*{\rm b}}$, where $r^{(\ell)}_{*{\rm b}} - r_{*{\rm s}} \sim r_+ \ln(r_+/l_{\rm s})$.
 (b) For modes with $\ell \approx O(r_{+}/l_{s})$, the classically allowed region is near the stretched horizon with $r^{(\ell)}_{*{\rm b}} - r_{*{\rm s}} \sim r_+$, so all these modes must be regarded as horizon modes.}
\label{fig:modes}
\end{figure}

For illustration, let us fist consider the two extreme cases of $\ell \approx O(1)$ and $O(r_+/l_{\rm s})$.
For $\ell \approx O(1)$, there are $1/\varDelta\omega^{(\ell)} r_+ \sim \ln(r_+/l_{\rm s})$ independent modes that have $\omega \approx O(1/r_+)$ for each orbital and magnetic quantum numbers $\ell$ and $m$.
We may take them to be wavepackets of width $\approx r_+$ in $r_*$ distributed uniformly between $r_{*{\rm s}}$ and $r^{(\ell)}_{*{\rm b}}$ without having a significant overlap with each other; see Fig.~\ref{fig:modes}.
In order for these to form a basis, we need to prepare two sets of wavepackets, moving toward larger and smaller values of $r_*$.
Among these (approximately orthogonal) wavepackets, the ones closest to the stretched horizon, i.e.\ those located within $\sim r_+$ in $r_*$ from the stretched horizon, are special in that their dynamics cannot be described by a semiclassical theory.
This is because the interaction strength of the unknown UV dynamics is strong there, which can also be seen from the fact that the unknown $O(1)$ coefficient in the definition of the stretched horizon in Eq.~\eqref{eq:stretch-def} translates into the ambiguity of the location of the stretched horizon of order $r_+$ in $r_*$.
We call modes corresponding to these wavepackets, i.e.\ the wavepackets ``next to'' the stretched horizon, {\it horizon modes}.
On the other hand, the dynamics of other $\sim \ln(r_+/l_{\rm s})$ wavepackets can be described by a semiclassical theory (at least) in the relevant timescale of order $1/T_{\rm H} \sim r_+$.
Modes associated with these semiclassically describable wavepackets are called {\it zone modes}.

For $\ell \approx O(r_+/l_{\rm s})$, the situation is different.
In this case, there are only $1/\varDelta\omega^{(\ell)} r_+ \approx O(1)$ independent modes having $\omega \approx O(1/r_+)$ for each $\ell$ and $m$, which, given Eq.~\eqref{eq:zone-width}, are all supported within $\sim r_+$ from the stretched horizon.
Therefore, they are all horizon modes.
As stated earlier, the entropy of a black hole (or de~Sitter spacetime) is dominated by the number of independent states of these high $\ell$ modes
\begin{equation}
  S \,\sim\, N_{\rm dof} \sum_{\ell=0}^{O(r_+/l_{\rm s})} \sum_{m=-\ell}^\ell O(1) \,\sim\, \frac{r_+^2}{l_{\rm P}^2},
\end{equation}
so the dynamics of the black hole (de~Sitter) microstates cannot be described by a semiclassical theory.
Indeed, the dynamics of these modes is expected to be nonlocal in the spatial directions along the horizon~\cite{Hayden:2007cs,Sekino:2008he}.

In general, we call modes localized near the stretched horizon, i.e.\ within $\sim r_+$ in $r_*$, horizon modes while those away from it we call zone modes.
We use the term ``zone modes'' also in spacetimes other than the black hole spacetime, including de~Sitter spacetime, even though there may not be real zones in these spacetimes.
In the case where the region near the horizon is connected to another (ambient/bath) region across a barrier of the effective potential, we restrict the use of the term zone modes to those modes that are contained in the region near the horizon.
For example, for a Schwarzschild black hole in asymptotically flat spacetime, zone modes only refer to modes within the zone $r \lesssim r_{\rm z}$ of the black hole.
Modes located outside the barrier, $r \gtrsim r_{\rm z}$, are not directly involved in the near horizon dynamics, and we call them {\it far modes}.

It is important to realize that the terminologies introduced above are associated with the spatial position of modes at a given time, or more precisely a time interval of width $\varDelta t \lesssim O(r_+)$ within which the system can be regarded as approximately static.
This implies, for example, that a zone mode according to the classification at time $t_1$ can be a horizon, zone, or far mode (or a superposition of them) according to the classification at another time $t_2$.
In particular, if the mode is well localized in the zone at time $t_1$ and is propagating toward the stretched horizon, then it will be a horizon mode according to the classification at $t_2$, a time after this mode has reached the stretched horizon.
Generally, a horizon mode remains as a horizon mode for a long time, although it occasionally becomes an outgoing zone mode through interactions at the stretched horizon.

\subsection{Vacuum microstates of semiclassical theories}
\label{subsec:excitation}

Consider a state specified by the occupation numbers of all the levels in Eq.~\eqref{eq:gen-omega} (with the precise values of $\omega^{(\ell)}_n$ modified by interactions).
This should be understood as a state of full quantum gravity.
In particular, a typical pure state given in this way represents a microstate of a black hole or de~Sitter vacuum.
We now connect this description to the picture in semiclassical theory, wherein a vacuum state takes the form of a mixed, (approximately) thermal state.

For this purpose, we first consider a subset of zone modes which is relevant for describing a physical object (or objects) falling into the horizon.
In the context of de~Sitter spacetime, this means an object accelerating away from the observer toward the cosmological horizon.
We can choose a subset of modes to accommodate {\it any} semiclassical object; the object is then described as an excitation of this set of modes over the semiclassical vacuum state.%
\footnote{
 This subset can be chosen to represent bound or meta-stable states (or an object made out of them).
 We may even choose it to represent a microscopic black hole, although such a construction would have to be made more precise.
}
We call these modes {\it hard modes}.
They comprise only a tiny subset of the zone modes, since even the most entropic configuration of semiclassical matter has the entropy suppressed by powers of $l_{\rm P}/r_+$ compared with the entropy of the Hawking cloud~\cite{tHooft:1993dmi,Nomura:2013lia}, of which the entropy of the zone modes comprises an $O(1)$ fraction.
The rest of the zone modes and the horizon modes are together called {\it soft modes}.

Similar to the zone and horizon modes, the hard and soft modes are defined at a specific time $t$.
In particular, a mode defined as a hard (or soft) mode at time $t$ may not be a hard (soft) mode at another time.
This issue, however, is less relevant for these modes than for the zone and horizon modes, since the concept is used almost exclusively to construct an effective theory describing the region behind the horizon, which is erected at a given instantaneous time $t$ as we will see below.

We label hard modes collectively by $\alpha$, which includes all possible quantum numbers like species, level, and orbital and spin angular momenta.
The state of these modes is specified by giving the occupation number $n_\alpha$ ($\geq 0$) for each $\alpha$, which we denote by $\ket{\{ n_\alpha \}}$ and normalize such that
\begin{equation}
  \inner{\{ m_\alpha \}}{\{ n_\alpha \}} = \delta_{\{ m_\alpha \}, \{ n_\alpha \}} = \delta_{mn},
\label{eq:zone-ortho}
\end{equation}
where $m$ and $n$ are shorthand notations of $\{ m_\alpha \}$ and $\{ n_\alpha \}$.
We assume that different hard-mode states are observationally distinguishable in that two different states do not have identical quantum numbers within the uncertainties $\varDelta E$, $\varDelta p$, and so on.

Suppose that the system consists only of zone and horizon modes; examples include de~Sitter spacetime.
\footnote{
 This is always the case if the effective potential increases monotonically as $r_*$ moves away from the stretched horizon.
 Another system exhibiting a similar behavior is Rindler spacetime, although in this case the system has a planar rather than spherical symmetry.
}
In this class of systems, the state of the entire system is generally given as an entangled state of hard and soft modes.
Because of the energy constraint, the state of soft modes that comes with the hard-mode state $\ket{\{ n_\alpha \}}$ must have energy $E - E_n$, up to the uncertainty $\varDelta E$ required by quantum mechanics, which is typically of the order of the Hawking temperature $T_{\rm H}$.
Here, $E$ is the total energy of the system, e.g.\ $E_{\rm dS}$ in Eq.~\eqref{eq:E_dS} for de~Sitter spacetime, and
\begin{equation}
  E_n = E_{\{ n_\alpha \}} = \sum_\alpha n_\alpha \omega_\alpha
\end{equation}
is the energy of the hard-mode state $\ket{\{ n_\alpha \}}$.

The relevant Hilbert space of the system is then given by
\begin{equation}
  {\cal H}(E) = \bigoplus_n \left( \ket{\{ n_\alpha \}} \otimes {\cal H}^{(n)}_{\rm soft} \right),
\label{eq:H-E}
\end{equation}
where ${\cal H}^{(n)}_{\rm soft}$ is the Hilbert space spanned by the soft-mode states that carry energy $E - E_n$ within the uncertainty $\varDelta E$.
Given that the number of hard modes is much smaller than the number of relevant soft modes, the effective dimension of the Hilbert space ${\cal H}^{(n)}_{\rm soft}$ is given by
\begin{equation}
  \ln {\rm dim}\,{\cal H}^{(n)}_{\rm soft} = S(E-E_n),
\end{equation}
where $S(E)$ is the entropy density of the system at energy $E$.
For de~Sitter spacetime, it is given by the Gibbons-Hawking entropy
\begin{equation}
  S(E) = S_{\rm dS}(E) = \pi E^2 l_{\rm P}^2,
\label{eq:S_dS}
\end{equation}
which gives the standard expression of $S_{\rm GH} = \pi \alpha^2 / l_{\rm P}^2$ for $E = E_{\rm dS} = \alpha/l_{\rm P}^2$.

A typical state described in Section~\ref{subsec:vacuum} corresponds to a typical state in the Hilbert space ${\cal H}(E)$ of Eq.~\eqref{eq:H-E}.
For our purposes, we need not be very precise about what we mean by typical, but for concreteness one might imagine a state that is typical in ${\cal H}(E)$ under the Haar measure.
Denoting a set of generic orthonormal basis states of ${\cal H}^{(n)}_{\rm soft}$ by $\ket{\psi^{(n)}_{i_n}}$ ($i_n = 1,\cdots,e^{S(E-E_n)}$), the state we are interested in can be written as
\begin{equation}
  \ket{\Psi(E)} = \sum_n \sum_{i_n = 1}^{e^{S(E-E_n)}}\!\!\! c_{n i_n} \ket{\{ n_\alpha \}} 
    \ket{\psi^{(n)}_{i_n}},
\label{eq:sys-state_dS}
\end{equation}
where the real and imaginary parts of complex coefficients $c_{n i_n}$'s can be viewed as taking random values following independently the Gaussian distributions with
\begin{equation}
  \vev{{\rm Re}\,c_{n i_n}} = \vev{{\rm Im}\,c_{n i_n}} = 0,
\qquad
  \sqrt{\vev{({\rm Re}\,c_{n i_n})^2}} = \sqrt{\vev{({\rm Im}\,c_{n i_n})^2}} = \frac{1}{\sqrt{2 e^{S_{\rm sys}}}}.
\label{eq:c-distr}
\end{equation}
Here, the brackets represent the ensemble average over (a sufficiently large portion of) the $(n,i_n)$ space, and
\begin{equation}
  e^{S_{\rm sys}} = \sum_n e^{S(E-E_n)} \equiv z\, e^{S(E)}
\label{eq:S_sys}
\end{equation}
is the number of independent states of the form of Eq.~\eqref{eq:sys-state_dS} with
\begin{equation}
  z = \sum_n e^{-\frac{E_n}{T_{\rm H}}},
\label{eq:def-z}
\end{equation}
where we have used $\partial S(E)/\partial E = 1/T_{\rm H}$ and $E_n \ll E$.%
\footnote{
 We have used the equal sign for a relation that becomes exact in the thermodynamic limit.
}
In other words, we can say that $c_{n i_n}$ is a complex Gaussian random variable, which implies that the phases of $c_{n i_n}$'s are distributed uniformly.

Note that by taking the basis of ${\cal H}^{(n)}_{\rm soft}$ for each $n$, we find that soft-mode states are orthogonal
\begin{equation}
  \inner{\psi^{(m)}_{i_m}}{\psi^{(n)}_{j_n}} = \delta_{mn} \delta_{i_m j_n}.
\label{eq:hor-ortho}
\end{equation}
This is because states $\ket{\psi^{(n)}_{i_n}}$ with different $n$ can be observationally discriminated, which follows from the distinguishability of different hard-mode states $\ket{\{ n_\alpha \}}$ as well as the fact that the state of combined hard and soft mode system is specified with the minimal uncertainty.
In particular, even if some observables (e.g.\ $E$ and $p$) may have to be intrinsically coarse grained (by $\varDelta E$ and $\varDelta p$), $\ket{\psi^{(n)}_{i_n}}$'s with different $n$ can still be discriminated at the semiclassical level, and hence are orthogonal.%
\footnote{
 We assume that such coarse grainings would be performed using smoothing functions which damp very rapidly outside the windows of order $\varDelta E$ and $\varDelta p$ so that Eq.~\eqref{eq:zone-ortho} is valid with sufficient accuracy.
 With this assumption, Eq.~\eqref{eq:hor-ortho} is also valid at the same level of accuracy.
}

We also note that $S_{\rm sys} = S(E) + \ln z$ is equal to $S(E)$ at the leading order in $l_{\rm P}/r_+$ and $1/E_n r_+$.
This implies that the standard interpretation of the Gibbons-Hawking entropy as the entropy of de~Sitter spacetime persists.
A similar comment applies to a black hole, for which the density of soft-mode states is given by the Bekenstein-Hawking entropy
\begin{equation}
  S(E) = S_{\rm bh}(E) = 4\pi E^2 l_{\rm P}^2,
\end{equation}
although in this case the entire system also has degrees of freedom outside the zone, and $S_{\rm sys} = S(E) + \ln z$ represents only the entropy of the black hole system (i.e.\ the zone and horizon modes) without including the contribution from the far modes.

We now take a complete set of orthonormal states of the form of Eq.~\eqref{eq:sys-state_dS}, i.e.\ states having the energy $E$ within $\varDelta E$, in a generic basis:
\begin{equation}
  \ket{\Psi_A(E)} = \sum_n \sum_{i_n = 1}^{e^{S(E-E_n)}} c^A_{n i_n} \ket{\{ n_\alpha \}} 
    \ket{\psi^{(n)}_{i_n}}
\quad
  (A = 1,\cdots,e^{S_{\rm sys}}),
\label{eq:sys-microstate_dS}
\end{equation}
where
\begin{equation}
  \inner{\Psi_A(E)}{\Psi_B(E)} = \delta_{AB}
  \quad\Longleftrightarrow\quad
  \sum_n \sum_{i_n = 1}^{e^{S(E-E_n)}}\!\!\! c^{A*}_{n i_n} c^{B}_{n i_n} = \delta_{AB}.
\label{eq:norm}
\end{equation}
In general, these states provide a basis for microstates of a semiclassical vacuum.
For example, if the spectrum $\omega_\alpha$ is given such that it represents hard modes inside the stretched de~Sitter horizon, then the set $\{\, \ket{\Psi_A(E)} \,\}$ forms a basis for the microstates of the de~Sitter vacuum.

We stress that in order for the states of Eq.~\eqref{eq:sys-microstate_dS} to be microstates of the spacetime, they need to be taken generically in the space of ${\cal H}(E)$.
For example, if we took $e^{S_{\rm sys}}$ orthonormal states that are, or approximately are, product states
\begin{equation}
  \ket{\Psi(E)} \approx \ket{\{ n_\alpha \}} \ket{\psi^{(n)}_{i_n}},
\end{equation}
then these states would still form a ``basis'' of the microstates in Eq.~\eqref{eq:sys-microstate_dS} in the sense that all the states of the form in Eq.~\eqref{eq:sys-microstate_dS} can be obtained by superposing them, although none of them is by itself a microstates of the spacetime under consideration---these special,  exponentially rare, states (states whose entanglement structure is significantly different from generic states) are ``firewall''~\cite{Almheiri:2012rt} states which do not represent the spacetime under consideration.
This is because spacetime is a manifestation of the entanglement structure of a holographic boundary state~\cite{Ryu:2006bv,Hubeny:2007xt,Lewkowycz:2013nqa,Faulkner:2013ana,Engelhardt:2014gca,Dong:2016hjy}, and entanglement cannot be represented as a linear operator---the concept of a linear vector space comprising the microstates of a spacetime is only an approximate one~\cite{Nomura:2017fyh,Almheiri:2016blp}.
This issue will be discussed further in the next section.

Since a semiclassical theory can describe only the dynamics of hard modes, it concerns only about the state of these modes.
Thus, the vacuum state $\rho_{\rm vac}(E)$ in a semiclassical theory is obtained by taking a typical vacuum microstate and then tracing out the soft modes.
Specifically, it is given by
\begin{align}
  \rho_{\rm vac}(E) &= \Tr_{\rm soft} \ket{\Psi_A(E)} \bra{\Psi_A(E)} \nonumber\\
  &= \sum_n \left( \sum_{i_n = 1}^{e^{S(E-E_n)}}\!\! |c^{A}_{n i_n}|^2 \right) \ket{\{ n_\alpha \}} \bra{\{ n_\alpha \}} \nonumber\\
  &= \sum_n \frac{e^{-\frac{E_n}{T_{\rm H}}}}{z} \ket{\{ n_\alpha \}} \bra{\{ n_\alpha \}} + O\Bigl(e^{-\frac{1}{2}S(E)}\Bigr).
\label{eq:rho_vac}
\end{align}
Here, in the last equation we have used
\begin{equation}
  \varepsilon_n^{AB} \equiv z\, e^{\frac{E_n}{T_{\rm H}}} \sum_{i_n = 1}^{e^{S(E-E_n)}} c^{A*}_{n i_n} c^B_{n i_n} - \delta_{AB} 
  \approx O\Bigl(e^{-\frac{1}{2}S(E-E_n)}\Bigr),
\label{eq:varepsilon}
\end{equation}
which can be derived from Eq.~\eqref{eq:c-distr}.
More precisely, when we vary $A$ and $B$, $\varepsilon_n^{AB}$ behave as complex Gaussian random variables with mean $0$ and variance $e^{-S(E-E_n)}$, but obeying
\begin{equation}
  (\varepsilon_n^{AB})^* = \varepsilon_n^{BA},
\qquad
  \sum_n \frac{e^{-\frac{E_n}{T_{\rm H}}}}{z} \varepsilon_n^{AB} = 0,
\label{eq:epsilon-sum}
\end{equation}
where the second relation follows from Eq.~\eqref{eq:norm}.
In the case of de~Sitter spacetime, the thermal state in Eq.~\eqref{eq:rho_vac} gives the vacuum state describing a static patch of the de~Sitter spacetime with Hubble radius $\alpha = E\, l_{\rm P}^2$.

The hard and soft modes described here provide a refinement of the hard and soft modes defined in Refs.~\cite{Nomura:2018kia,Nomura:2019qps,Nomura:2019dlz,Langhoff:2020jqa,Nomura:2020ska}.
In Refs.~\cite{Nomura:2018kia,Nomura:2019qps,Nomura:2019dlz,Langhoff:2020jqa,Nomura:2020ska} a simple frequency space criterion was used to define the hard and soft modes, while in our new definition here, the hard modes are chosen to be a subset of the zone modes whose dynamics we intend to describe at the semiclassical level.
This makes it possible, for example, to describe the dynamics of zone modes with $\omega \sim O(T_{\rm H})$ using a semiclassical theory, which was not possible with the previous definition.
For many practical purposes, however, the two definitions are interchangeable.
For a small object falling into the horizon, for example, the difference between the two definitions is not significant if we choose the frequency cutoff to be sufficiently larger than $T_{\rm H}$.
We can then employ the same construction for spacetime beyond the horizon, which we will discuss in Section~\ref{sec:extension}.

\subsubsection*{Evaporating black hole}

The situation is more complicated if the region near the horizon is coupled to an ambient/bath system.
An important example of this is a black hole in asymptotically flat spacetime; other examples include a small black hole in asymptotically AdS spacetime and a large AdS black hole coupled to a separate bath system.
In this case, the near horizon system, consisting of the zone and horizon, evolves in time, and this evolution modifies the vacuum states.

For concreteness, let us focus on a black hole in asymptotically flat spacetime.
The state of the entire system then involves the hard, soft, and far (located outside the zone, $r > r_{\rm z}$) modes.
Thus, denoting orthonormal basis states of the far modes by $\ket{\phi_a}$, one would consider the state of the system to be given by%
\footnote{
 We assume that the standard issues for a factorization of Hilbert space in quantum field theory, such as those associated with short distance divergences and constraints from gauge invariance, are dealt with appropriately.
}
\begin{equation}
  \ket{\Psi_A(M)} \,\qeq\,\, \sum_n \sum_{i_n = 1}^{e^{S_{\rm bh}(M-E_n)}} \sum_{a = 1}^{S_{\rm rad}} c^A_{n i_n a} \ket{\{ n_\alpha \}} \ket{\psi^{(n)}_{i_n}} \ket{\phi_a},
\label{eq:Psi-no-evol}
\end{equation}
where we have assumed that the black hole system, comprising the hard and soft modes, is at rest and has energy $M$ (up to the uncertainty of order $T_{\rm H}$).
Here, $e^{S_{\rm rad}}$ is the number of independent far-mode states relevant here, i.e.\ those significantly entangled with the black hole system (typically Hawking radiation emitted earlier from the black hole), and $A$ is the index for microstates running over
\begin{equation}
  A = 1,\cdots,e^{S_{\rm s+r}},
\qquad
  S_{\rm s+r} = S_{\rm sys} + S_{\rm rad},
\end{equation}
with $S_{\rm sys}$ given by Eq.~\eqref{eq:S_sys} with $S(E) = S_{\rm bh}(E)$.
The coefficients $c^A_{n i_n a}$ satisfy the properties analogous to those in Eq.~\eqref{eq:c-distr}
\begin{equation}
  \vev{{\rm Re}\,c^A_{n i_n a}} = \vev{{\rm Im}\,c^A_{n i_n a}} = 0,
\qquad
  \sqrt{\vev{({\rm Re}\,c^A_{n i_n a})^2}} = \sqrt{\vev{({\rm Im}\,c^A_{n i_n a})^2}} = \frac{1}{\sqrt{2 e^{S_{\rm s+r}}}},
\label{eq:c-distr-BH}
\end{equation}
with brackets representing the ensemble average over (a sufficiently large portion of) the $(n,i_n,a)$ space.
Note that $\ket{\Psi_A(M)}$ represent microstates of the system with the black hole put in the semiclassical vacuum, and a generic state in the Hilbert space of dimension $e^{S_{\rm s+r}}$ has the black hole of mass $M$.
Since black hole evaporation is a thermodynamically irreversible process~\cite{Zurek:1982zz,Page:1983ug}, most of these microstates do not become a state with a larger black hole in empty space when evolved backward in time---there is some junk radiation around it.
This, however, does not change the fact that there are $e^{S_{\rm s+r}}$ independent microstates relevant for the discussion here.

The fact that the height of the potential barrier is finite, however, implies that only modes with $\omega < \omega^{(\ell)}_{\rm barrier}$ are thermalized in the zone.
Given that the dynamics at the stretched horizon is strongly coupled, outgoing modes (i.e.\ modes moving toward larger $r$) with $\omega > \omega^{(\ell)}_{\rm barrier}$ can still be viewed as obeying the thermal distribution, but this is not the case for ingoing modes with $\omega > \omega^{(\ell)}_{\rm barrier}$.
At the microscopic level, this implies that the microcanonical ensemble in Section~\ref{subsec:vacuum} is taken with the extra constraint that the occupation numbers of ingoing modes with $\omega > \omega^{(\ell)}_{\rm barrier}$ are zero, leading to%
\footnote{
 Strictly speaking, there are small but nonzero amplitudes for outgoing hard modes with $\omega > \omega^{(\ell)}_{\rm barrier}$ to be reflected back from the potential barrier.
 We mostly ignore this effect because it is not essential for our discussion.
 Including it, however, is straightforward; instead of taking the terms with $\exists \alpha', n_{\alpha'} \neq 0$ to be exactly absent, we keep these terms with small coefficients (compared to those of the terms with $\forall \alpha', n_{\alpha'} = 0$).
 Note that the size of these coefficients in general depends strongly on $\{ n_{\alpha'} \}$, reflecting the $\omega_{\alpha'}$ dependence of the reflection amplitudes.
\label{ft:nonzero-ref}}
\begin{equation}
  \ket{\Psi_A(M)} \,\qpropto\,\, \sum_n\! \vphantom{\frac{1}{1+\frac{1}{1+\frac{1}{1+\frac{1}{1+\frac{1}{1+\frac{1}{x}}}}}}} \smash{\prod_{\vphantom{\frac{1+\frac{1}{x}}{1+\frac{1}{x}}} \smash{\alpha' \in \begin{array}{l}{\scriptstyle\rm ingoing}\\[-7pt] {\scriptstyle \!\!\! \omega > \omega^{(\ell)}_{\rm barrier}}\end{array}}}}\!\!\!\!\!\!\!\!\!\!\! \delta_{n_{\alpha'},0} \sum_{i_n = 1}^{e^{S_{\rm bh}(M-E_n)}} \sum_{a = 1}^{S_{\rm rad}} c^A_{n i_n a} \ket{\{ n_\alpha \}} \ket{\psi^{(n)}_{i_n}} \ket{\phi_a},
\label{eq:sys-micro_BH}
\end{equation}
where the $\propto$ symbol represents ``up to a normalization constant,'' and the question mark above it indicates that this relation is tentative and will be updated momentarily.
(The effect of the lack of ingoing soft modes with $\omega > \omega^{(\ell)}_{\rm barrier}$ is negligible for our purpose.)
This corresponds to taking the Hartle-Hawking~\cite{Hartle:1976tp} and Unruh~\cite{Unruh:1976db} vacua for hard modes with $\omega < \omega^{(\ell)}_{\rm barrier}$ and $> \omega^{(\ell)}_{\rm barrier}$, respectively.
The effect of the constraint on soft modes is negligible, since the vast majority of the relevant modes have $\omega \sim T_{\rm H}$ and $\ell \gg 1$, and hence $\omega \ll \omega^{(\ell)}_{\rm barrier}$.

The story, however, does not end here.
Because of the coupling between the black hole system (zone $+$ horizon) and the asymptotically flat spacetime around the region $r \sim r_{\rm z}$, thermal quanta in the zone ``leak'' into the latter.
This occurs mostly via $s$-wave modes that tunnel through the potential barrier.
We emphasize that this process, occurring near the edge of the zone~\cite{Nomura:2014woa,Nomura:2014voa} (for related discussions, see Refs.~\cite{Unruh:1977ga,Israel:2015ava,Giddings:2015uzr}), is governed by semiclassical physics---it does not involve strongly coupled, intrinsically quantum gravitational physics in any significant way.
Given a black hole microstate in Eq.~\eqref{eq:sys-micro_BH} (slightly modified due to backreaction; see below), the emission of quanta into the asymptotic region occurs unitarily following the dynamics of standard quantum field theory.
The apparent violation of unitarity in Hawking's analysis~\cite{Hawking:1976ra} occurs because we cannot calculate the configuration of zone mode quanta using semiclassical theory due to the strong dynamics {\it near the stretched horizon}.
It is this incalculability that makes the semiclassical description of Hawking radiation, obtained after tracing out the soft modes, intrinsically thermal and hence leading to a mixed final state.
Physics away from the stretched horizon can indeed be fully semiclassical, even within and at the edge of the zone.%
\footnote{
 This implies that the process of black hole mining~\cite{Unruh:1982ic,Brown:2012un} also occurs unitarily, which is governed by semiclassical physics if it is performed away from the stretched horizon.
 It also implies that the semiclassical calculation of the gray body factor, such as that in Ref.~\cite{Page:1976df}, is valid as long as $r_+/l_{\rm s}$ is sufficiently large; see below.
}

\begin{figure}[t]
\centering
  \includegraphics[width=0.7\textwidth]{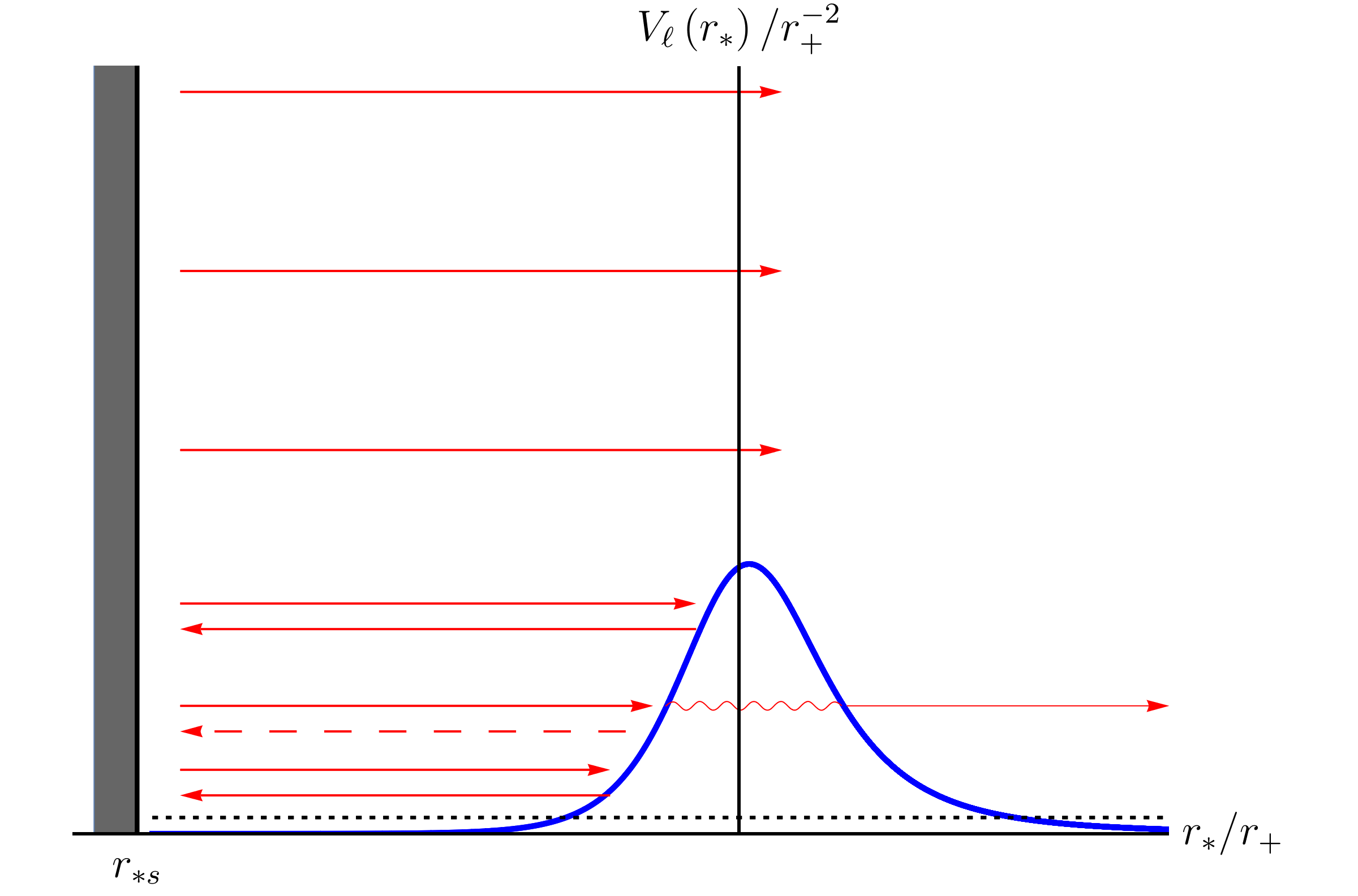}
\caption{
 A schematic depiction of quanta occupying various modes.
 Solid arrows indicate that the occupation numbers of the modes are determined by the thermodynamic consideration as described in Section~\ref{subsec:vacuum}.
 For zone modes, this applies to the outgoing mode as well as the ingoing modes with $\omega \lesssim \omega^{(\ell)}_{\rm barrier}$, except that backreaction of Hawking emission at $r \sim r_{\rm z}$ leads to a deficit for some of these ingoing modes as indicated by the dashed arrow.}
\label{fig:zone}
\end{figure}

The emission of Hawking particles to the ambient space at $r \sim r_{\rm z}$ gives a backreaction to the state of the black hole.
In this region, quanta of the zone region is leaked into the ambient space through tunneling the potential barrier (and also via thermal hopping to some extent).
This removes some of the quanta that would be reflected back to the zone by the potential, producing a deficit in ingoing zone mode quanta relative to those in the states of Eq.~\eqref{eq:sys-micro_BH}, i.e.\ with the Hartle-Hawking vacuum.
Note that the process occurs only for low energy fields of mass $m_\Phi \lesssim T_{\rm H}$.
Given that an $O(1)$ number of quanta are emitted within each time interval of order $1/T_{\rm H} \sim r_+$, there are $\sim \ln(r_+/l_{\rm s})$ quanta missing throughout the zone for each low energy field $\Phi$ of $m_\Phi \lesssim T_{\rm H}$.
Denoting the annihilation operators for hard mode quanta by $b_{\bar{\alpha}}$ (see below for more detail), we finally find that the microstates of an evaporating black hole are given by
\begin{equation}
  \ket{\Psi_A(M)} \,\,\propto\!\!\!\! \vphantom{\frac{1}{1+\frac{1}{1+\frac{1}{1+\frac{1}{1+\frac{1}{1+\frac{1}{x}}}}}}} \smash{\prod_{\vphantom{\frac{1+\frac{1}{x}}{1+\frac{1}{x}}} \smash{\bar{\alpha} \in \begin{array}{l}{\scriptstyle\rm \,\,\,\,\,\, ingoing}\\[-7pt] {\scriptstyle \!\!\!\! m_\Phi \lesssim T_{\rm H},\, \omega \sim T_{\rm H}}\end{array}}}}\!\!\!\!\!\!\!\!\!\!\!\!\! b_{\bar{\alpha}}\;\;\; \sum_n\! \vphantom{\frac{1}{1+\frac{1}{1+\frac{1}{1+\frac{1}{1+\frac{1}{1+\frac{1}{x}}}}}}} \smash{\prod_{\vphantom{\frac{1+\frac{1}{x}}{1+\frac{1}{x}}} \smash{\alpha' \in \begin{array}{l}{\scriptstyle\rm ingoing}\\[-7pt] {\scriptstyle \!\!\! \omega > \omega^{(\ell)}_{\rm barrier}}\end{array}}}}\!\!\!\!\!\!\!\!\!\!\! \delta_{n_{\alpha'},0} \sum_{i_n = 1}^{e^{S_{\rm bh}(M-E_n)}} \sum_{a = 1}^{S_{\rm rad}} c^A_{n i_n a} \ket{\{ n_\alpha \}} \ket{\psi^{(n)}_{i_n}} \ket{\phi_a},
\label{eq:sys-micro_BH-evol}
\end{equation}
where the number of annihilation operators $b_{\bar{\alpha}}$ in the product is at most of order $\ln(r_+/l_{\rm s})$ for each field; the precise number depends on the choice of the hard modes.
A schematic depiction of the occupation of various modes is given in Fig.~\ref{fig:zone}.

The deficit of ingoing modes described above implies that there is a negative energy flux carrying negative entropy for each field of $m_\Phi \lesssim T_{\rm H}$.
Note that these energy and entropy are measured with respect to the thermal, Hartle-Hawking vacuum state in Eq.~\eqref{eq:sys-micro_BH} for modes with $\omega \sim T_{\rm H}$.
The flux has negative entropy because with the lack of some of the zone mode quanta, the number of independent states realizing the most probable configuration is smaller by $e^{O(1) \ln(r_+/l_{\rm s})}$ for each field.
Thus, the microstate index $A$ in Eq.~\eqref{eq:sys-micro_BH-evol} runs effectively only for
\begin{equation}
  A = 1,\cdots,A_{\rm max},
\qquad
  \ln A_{\rm max} - S_{\rm tot} \sim -\ln\frac{r_+}{l_{\rm s}},
\end{equation}
where $S_{\rm tot}$ is the coarse-grained entropy of the total system ignoring the backreaction
\begin{equation}
  e^{S_{\rm tot}} = \sum_n \! \vphantom{\frac{1}{1+\frac{1}{1+\frac{1}{1+\frac{1}{1+\frac{1}{1+\frac{1}{x}}}}}}} \smash{\prod_{\vphantom{\frac{1+\frac{1}{x}}{1+\frac{1}{x}}} \smash{\alpha' \in \begin{array}{l}{\scriptstyle\rm ingoing}\\[-7pt] {\scriptstyle \!\!\! \omega > \omega^{(\ell)}_{\rm barrier}}\end{array}}}}\!\!\!\!\!\!\!\!\!\!\! \delta_{n_{\alpha'},0}\,\, e^{S_{\rm bh}(M-E_n)} e^{S_{\rm rad}} 
  \equiv z' e^{S_{\rm bh}(M)+S_{\rm rad}},
\label{eq:S_tot}
\end{equation}
and we have taken the number of low energy degrees of freedom relevant for the emission process to be of $O(1)$.%
\footnote{
 This implies that the range of $i_n$ in Eq.~(\ref{eq:sys-micro_BH-evol}) is, strictly speaking, smaller than that in Eqs.~(\ref{eq:Psi-no-evol}) or (\ref{eq:sys-micro_BH}).
 For the lack of a better notation, we interpret the sum in Eq.~(\ref{eq:sys-micro_BH-evol}), and analogous expressions later, to include this minor effect.
}
As discussed in Refs.~\cite{Nomura:2014woa,Nomura:2014voa}, this negative entropy is essential for the unitarity of the black hole evolution, in particular for the black hole to keep relaxing into a lower mass black hole while absorbing the negative energy flux.
Again, we emphasize that semiclassical physics is sufficient to understand the unitary emission process at $r \sim r_{\rm z}$ and the resulting emergence of a negative energy and entropy flux.
The unknown UV physics enters only in the process occurring at the stretched horizon, in which the ingoing negative energy-entropy flux is absorbed into horizon modes and the black hole relaxes into its semiclassical vacuum state.

So far, we have assumed that our black hole is large, in particular $\ln(r_+/l_{\rm s}) \gg 1$.
In this limit, the difference of energies between adjacent discrete levels for modes relevant for Hawking emission, $\sim r_+/\ln(r_+/l_{\rm s})$, is much smaller than the uncertainty of energy of Hawking quanta, $\sim r_+$, so that the effect of the discreteness of energy levels is negligible in calculating the spectrum of Hawking radiation.
In particular, the semiclassical calculation of the spectrum, including the gray body factor, persists with high precision.
If the value of $\ln(r_+/l_{\rm s})$ is reduced, however, the effect of the discreteness of levels may become important.
In particular, if the size of the classically allowed region $r^{(\ell)}_{*{\rm b}} - r_{*{\rm s}} \approx 2 r_+ \ln(r_+/l_{\rm s})$, becomes smaller than a half wavelength of a Hawking quantum $\lambda_{\rm H}/2 \approx \pi/T_{\rm H} \approx 4\pi^2 r_+$, i.e.
\begin{equation}
  r_+ \lesssim e^{2\pi^2} l_{\rm s},
\label{eq:breaking}
\end{equation}
then the effect may become non-negligible.
(This condition can also be obtained by requiring $\varDelta\omega$ in Eq.~\eqref{eq:vardelta-omega_BH} to be larger than $T_{\rm H}$.)
A naive guess is that in this regime, the energy of each Hawking quantum is larger than that obtained by the semiclassical calculation, since the frequency of the lowest energy level is expected to become larger than $T_{\rm H}$.
An interesting point is that the black hole enters the regime of Eq.~\eqref{eq:breaking} before its mass is reduced to the Planck mass, since this condition can be written as
\begin{equation}
  M \lesssim \frac{e^{2\pi^2} l_{\rm s}}{2 l_{\rm P}^2}.
\end{equation}
For $l_{\rm s}/l_{\rm P} \approx 20$, as we might expect in our universe, the right-hand side is $4 \times 10^9$ times larger than the Planck mass (corresponding to the black hole of $T_{\rm H} \approx 1 \times 10^8~{\rm GeV}$).
We leave further discussion of this issue, including its possible phenomenological implications, for the future.

\subsection{Excited states}

Before concluding this section, let us discuss semiclassical excitations in the zone.
The states we have considered so far are typical states in a suitably defined microcanonical ensemble.
For a black hole, for example, the relevant ensemble consists of the states that contain a fixed energy $E$ in a spatial region $r \lesssim r_{\rm z}$ within an uncertainty $\varDelta E$.
These states are all vacuum states from the point of view of semiclassical theory as indicated by the fact that the Bekenstein-Hawking entropy is associated with the background (vacuum) spacetime in semiclassical theory.

We can, however, consider atypical states obtained by acting annihilation and/or creation operators of hard modes
\begin{align}
  b_\gamma &= \sum_n \sqrt{n_\gamma}\, 
    \ket{\{ n_\alpha - \delta_{\alpha\gamma} \}} \bra{\{ n_\alpha \}},
\label{eq:ann}\\*
  b_\gamma^\dagger &= \sum_n \sqrt{n_\gamma + 1}\, 
    \ket{\{ n_\alpha + \delta_{\alpha\gamma} \}} \bra{\{ n_\alpha \}}
\label{eq:cre}
\end{align}
A state obtained in this way is not typical as it does not have the most probable configuration among the states in the ensemble.
on these vacuum states, where $\gamma$ specifies the mode which is annihilated/created by the operator.
(States obtained by acting annihilation operators can become relevant when one considers backreaction of the Hawking emission or black hole mining process.
In particular, some of the operators $b^{({\rm z})}_{\bar{\alpha}}$ in Eq.~\eqref{eq:sys-micro_BH-evol} can be superpositions of the $b_\gamma$ operators, if we choose the hard modes to include the relevant zone modes.)

We are interested in semiclassical excitations whose backreaction to the geometry is negligible, or regarded as being small, for example a baseball falling into an astronomical black hole.
This implies that the number of creation/annihilation operators that can be acted on a vacuum state is limited, so that their algebra defined on a vacuum state does not close in a strict mathematical sense.
This type of structure is in fact common in erecting a semiclassical theory in quantum gravity, and here we simply treat the space of states obtained in this way as a Hilbert space; for a mathematically more rigorous treatment, see, e.g., Ref.~\cite{Ghosh:2017gtw}.
In holography, one can think of this space as a code subspace embedded in a physical Hilbert space~\cite{Almheiri:2014lwa,Pastawski:2015qua,Hayden:2016cfa}, although we do not discuss the error correcting nature of operators in this paper.

Strictly speaking, the space of semiclassical excitations generated by $b_\gamma$ and $b_\gamma^\dagger$ acted on each vacuum state is not orthogonal to the space of vacuum microstates; namely, the Hilbert space cannot be strictly written as ${\cal H}_{\rm exc} \otimes {\cal H}_{\rm vac}$~\cite{Nomura:2020ska}.
One can see this by calculating inner products between states obtained by exciting black hole microstates in Eq.~\eqref{eq:sys-micro_BH}:
\begin{align}
  \bra{\Psi_{A}(M)} b_\beta^\dagger b_\gamma \ket{\Psi_{B}(M)} 
  &= \delta_{\beta\gamma} \sum_n\! \vphantom{\frac{1}{1+\frac{1}{1+\frac{1}{1+\frac{1}{1+\frac{1}{1+\frac{1}{x}}}}}}} \smash{\prod_{\vphantom{\frac{1+\frac{1}{x}}{1+\frac{1}{x}}} \smash{\alpha' \in \begin{array}{l}{\scriptstyle\rm ingoing}\\[-7pt] {\scriptstyle \!\!\! \omega > \omega^{(\ell)}_{\rm barrier}}\end{array}}}}\!\!\!\!\!\!\!\!\!\!\! \delta_{n_{\alpha'},0} \sum_{i_n = 1}^{e^{S_{\rm bh}(M-E_n)}} 
    \sum_{a = 1}^{e^{S_{\rm rad}}} n_\gamma\, c^{A*}_{n i_n a} c^{B}_{n i_n a},
\label{eq:non-ortho-1}\\*
  \bra{\Psi_{A}(M)} b_\beta b_\gamma^\dagger \ket{\Psi_{B}(M)} 
  &= \delta_{\beta\gamma} \sum_n\! \vphantom{\frac{1}{1+\frac{1}{1+\frac{1}{1+\frac{1}{1+\frac{1}{1+\frac{1}{x}}}}}}} \smash{\prod_{\vphantom{\frac{1+\frac{1}{x}}{1+\frac{1}{x}}} \smash{\alpha' \in \begin{array}{l}{\scriptstyle\rm ingoing}\\[-7pt] {\scriptstyle \!\!\! \omega > \omega^{(\ell)}_{\rm barrier}}\end{array}}}}\!\!\!\!\!\!\!\!\!\!\! \delta_{n_{\alpha'},0} \sum_{i_n = 1}^{e^{S_{\rm bh}(M-E_n)}} 
    \sum_{a = 1}^{e^{S_{\rm rad}}} (n_\gamma + 1)\, c^{A*}_{n i_n a} c^{B}_{n i_n a}.
\label{eq:non-ortho-2}
\end{align}
These are not proportional to $\delta_{AB}$ in general, so that excited states built on different vacuum microstates are not necessarily orthogonal.
However, for $A\neq B$, the right-hand sides of the above equations are exponentially suppressed by a factor of $e^{-S_{\rm tot}/2}$, where we have taken
\begin{equation}
  \inner{\Psi_A(M)}{\Psi_B(M)} = \delta_{AB}
  \quad\Longleftrightarrow\quad
  \sum_n\! \vphantom{\frac{1}{1+\frac{1}{1+\frac{1}{1+\frac{1}{1+\frac{1}{1+\frac{1}{x}}}}}}} \smash{\prod_{\vphantom{\frac{1+\frac{1}{x}}{1+\frac{1}{x}}} \smash{\alpha' \in \begin{array}{l}{\scriptstyle\rm ingoing}\\[-7pt] {\scriptstyle \!\!\! \omega > \omega^{(\ell)}_{\rm barrier}}\end{array}}}}\!\!\!\!\!\!\!\!\!\!\! \delta_{n_{\alpha'},0} \sum_{i_n = 1}^{e^{S_{\rm bh}(M-E_n)}} \sum_{a = 1}^{e^{S_{\rm rad}}}  c^{A*}_{n i_n a} c^{B}_{n i_n a} = \delta_{AB}.
\label{eq:norm_s+r}
\end{equation}
Therefore, the deviation from the product space structure is exponentially small.

Incidentally, the annihilation and creation operators in Eqs.~\eqref{eq:ann} and \eqref{eq:cre} satisfy the standard commutation relations
\begin{equation}
  [b_\beta, b_\gamma^\dagger] = \delta_{\beta\gamma} \sum_n \ket{\{ n_\alpha \}} \bra{\{ n_\alpha \}},
\qquad
  [b_\beta, b_\gamma] = [b_\beta^\dagger, b_\gamma^\dagger] = 0
\label{eq:ann-cre_alg}
\end{equation}
as operators, without having an exponentially small correction.
In erecting a semiclassical theory, one regards all the microstates $\ket{\Psi_A(M)}$ as representing the same geometry ${\cal M}$.
This allows us to define field operators
\begin{equation}
  \Phi_\Gamma(x) = \sum_{\gamma'} \left( b_\gamma u_{\gamma'}(x) + b_\gamma^\dagger v_{\gamma'}(x) \right),
\end{equation}
where we have split the index $\gamma$ into the index for species $\Gamma$ and that for other quantum numbers $\gamma'$, e.g.\ a component of spin, whose structure may depend on $\Gamma$:\ $\gamma = (\Gamma,\gamma')$.
Here, $u_{\gamma'}(x)$ and $v_{\gamma'}(x)$ are mode functions defined in the allowed region of ${\cal M}$.
Because of Eq.~\eqref{eq:ann-cre_alg}, these field operators and their conjugate momenta obey the standard equal-time commutation relations, so that the resulting semiclassical theory respects causality exactly.
For black hole and de~Sitter spacetimes, this theory can be used to describe physics in the regions $r > r_{\rm s}$ and $< r_{\rm s}$, respectively.

\section{Holographic Description}
\label{sec:holo}

So far, we have assumed that the fundamental description of a system represents spacetime in the allowed region (with the horizon degrees of freedom included).
For example, this corresponds to taking a distant view for a black hole and a static patch view for de~Sitter spacetime.
Why is this the case?

In this section, we discuss this issue from the perspective of holography.
We assume that these spacetimes arise in setups in which the holographic description has only a single boundary; in particular, the black hole is formed by a gravitational collapse, and de~Sitter spacetime arises in a cosmological context, for example as a universe created by bubble nucleation~\cite{Coleman:1980aw} which is filled with a positive cosmological constant.
These setups are, arguably, more ``realistic'' than those with multiple boundaries, which will be discussed in Section~\ref{sec:extended}.
We conclude the section with a discussion on how the (effective) boundary Hilbert spaces describing the spacetimes considered here are related to the infinite-dimensional ``fundamental'' Hilbert space.

\subsection{Black hole spacetime}
\label{subsec:holo-BH}

We begin by considering a collapse formed black hole in an asymptotically AdS spacetime.
We would like to know what spacetime picture can be obtained from the boundary CFT by reconstructing the bulk in a simple manner.
By simple reconstruction, we mean reconstruction of the bulk using only low-complexity, causally-propagating operators and sources in the CFT~\cite{Engelhardt:2021mue}.
In particular, we are interested in obtaining a ``gauge-fixed'' bulk description where the spacetime is foliated by equal-time hypersurfaces, using which the canonical formulation of quantum mechanics can be employed.

One way to obtain such a description is to ``pull'' the boundary into the bulk by coarse graining boundary degrees of freedom~\cite{Nomura:2018kji,Murdia:2020iac,Bao:2018pvs}.
Based on intuition from tensor networks~\cite{Pastawski:2015qua,Hayden:2016cfa,Swingle:2009bg}, one can consider a series of states~\cite{Miyaji:2015yva} defined on successfully ``renormalized'' boundaries obtained by moving the original boundary inward to the bulk.
The coarse-grained degrees of freedom are distributed locally on a renormalized boundary, although the dynamics they obey are not necessarily local.

This procedure is expected to work beyond the AdS/CFT context, which we assume to be the case, and henceforth we will not necessarily assume that the spacetime is asymptotically AdS.
A particular way of performing this renormalization is to ``continuously'' coarse grain boundary degrees of freedom uniformly throughout the renormalized boundary space.
This leads to a specific spacelike or null hypersurface swept by a series of renormalized boundaries called a {\it holographic slice}~\cite{Nomura:2018kji}, which we will discuss in more detail below.
This surface plays the role of an equal-time hypersurface in the bulk, providing a gauge fixing necessary for the canonical formulation.

\begin{figure}[t]
\centering
  \subfloat[\centering classical]{{\includegraphics[width=0.3\textwidth]{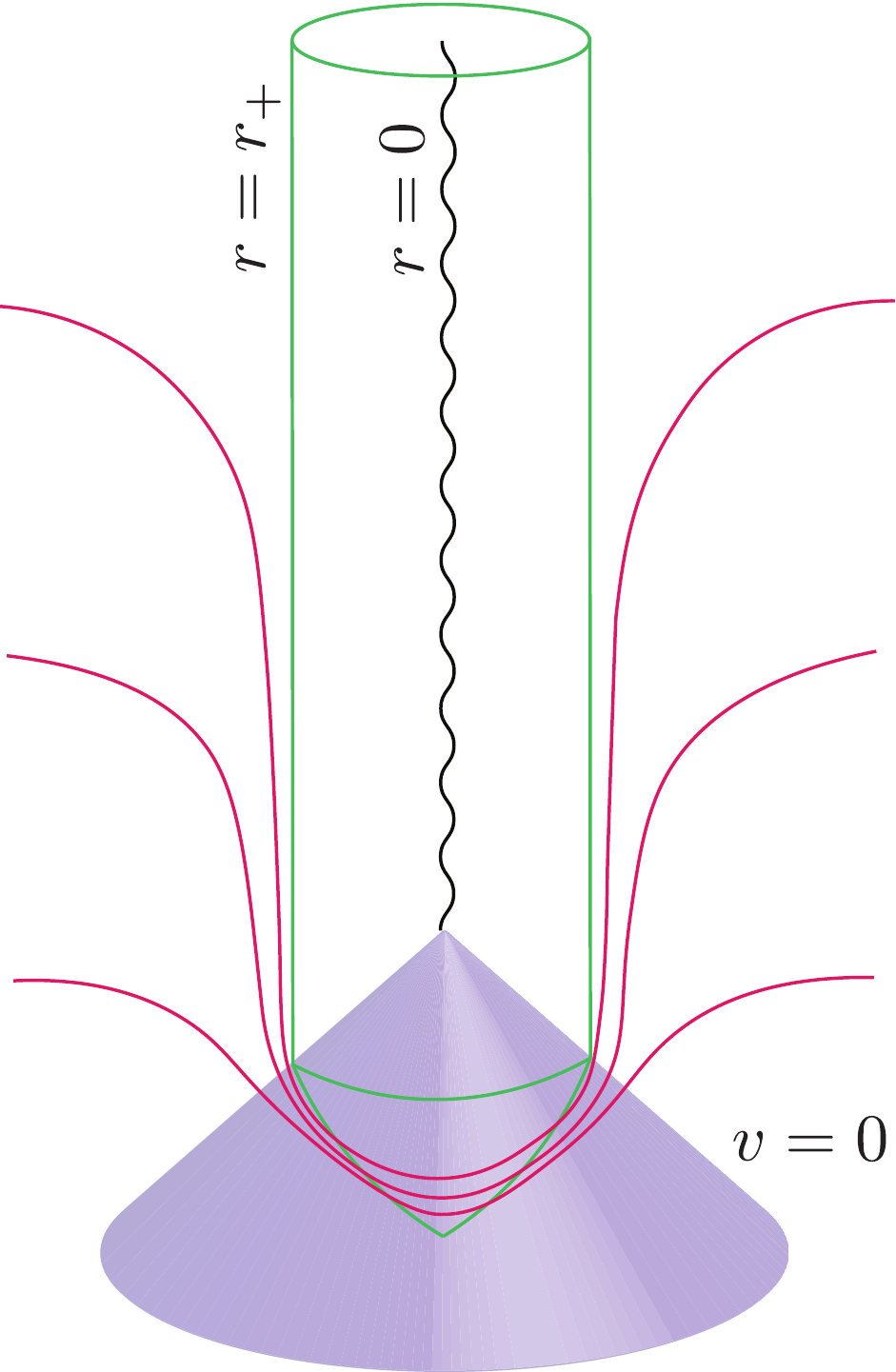} }}
\hspace{2.5cm}
  \subfloat[\centering quantum]{{\includegraphics[width=0.3\textwidth]{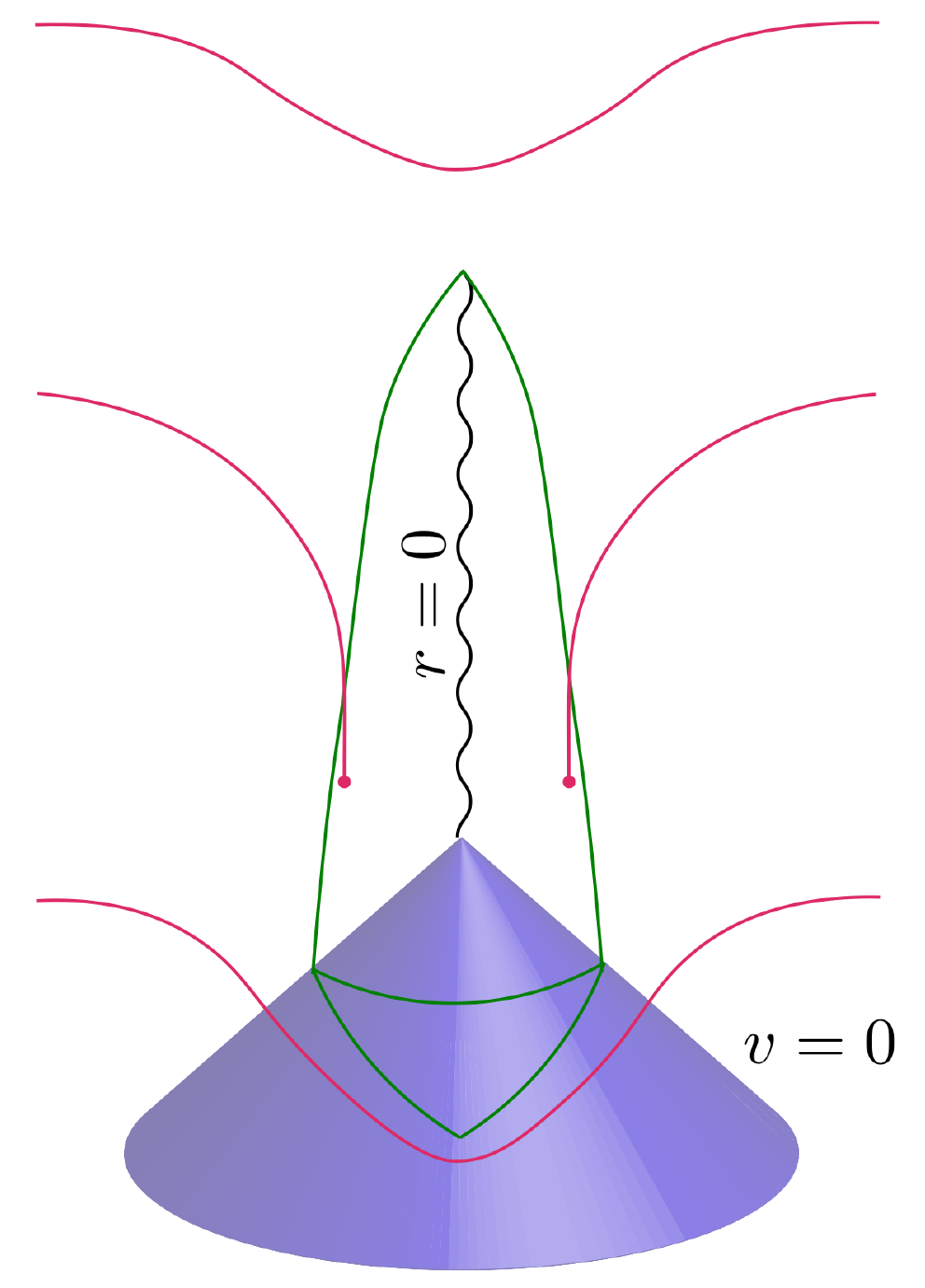} }}
\caption{
 Holographic slices in a black hole spacetime in ingoing Eddington-Finkelstein coordinates $(v,r)$ obtained (a) without and (b) with quantum effects in the bulk.
 These slices correspond to equal-time hypersurfaces in the bulk, which approach Schwarzschild time slices near the black hole.
 The figures are taken from Refs.~\cite{Nomura:2018kji} and \cite{Murdia:2020iac}.}
\label{fig:slices}
\end{figure}

In Fig.~\ref{fig:slices}(a), we depict holographic slices in a black hole spacetime obtained without including quantum effects in the bulk.
We find that the slices do not enter the black hole interior, hence providing a distant view of the black hole.
This is because the black hole horizon, sufficiently after the black hole has stabilized, plays a role of a barrier~\cite{Engelhardt:2013tra} for extremal surfaces, which are used to move renormalized boundaries.
In fact, the slices approach Schwarzschild time slices near the black hole; they then stay near the horizon for long time and are eventually capped off at $r = 0$.

In order to describe an evaporating black hole, quantum effects in the bulk must be included.
The procedure of forming a holographic slice can be extended to incorporate these effects~\cite{Murdia:2020iac}.
In this case, the holographic slice is smoothly capped off at $r=0$ initially, but at some (boundary) time it becomes asymptoting to a quantum extremal surface located (approximately) at the horizon; and as time progresses further, it again becomes a surface without a hole.
This is depicted in Fig.~\ref{fig:slices}(b).
This behavior of holographic slices, in fact, can be used to define what we mean by the ``formation'' and ``evaporation'' of a black hole.
The holographic slices approximate Schwarzschild time slices in this case as well, giving a distant description of the black hole.
Note that the region swept by the holographic slices is essentially the simple wedge~\cite{Engelhardt:2021mue}, which can be reconstructed in a simple manner using the Hamilton-Kabat-Lifschytz-Lowe (HKLL) procedure~\cite{Hamilton:2005ju,Hamilton:2006az,Heemskerk:2012mn} together with boundary time evolutions with sources.

In any event, the description of a black hole that naturally results from a holographic theory with a single boundary is an exterior/distant one.
In this description, the unitarity of black hole evolution is not an issue, since the stretched horizon behaves as a regular material surface from the point of view of quantum information flow.
The question, rather, is in what sense the near empty interior region exists for an infalling observer as predicted by general relativity.
We will come back to this issue in Section~\ref{sec:extension}.

\subsection{de~Sitter (or cosmological) spacetime}
\label{subsec:holo-dS}

To discuss holographic descriptions beyond AdS/CFT, we need to introduce a ``boundary'' in spacetime on which a holographic state can be defined.
One way to do this is to consider a renormalized boundary deep in the bulk---called a renormalized leaf~\cite{Nomura:2018kji,Murdia:2020iac} in this context---and ``unrenormalize'' it by successively integrating in relevant degrees of freedom, making the holographic slice grow  toward the ``non-renormalized'' boundary.
This procedure depends on the background spacetime, but we can perform it in each branch of a state representing a well-defined spacetime in the semiclassical limit.

\subsubsection*{General picture}

To be specific, let us adopt the scheme in Refs.~\cite{Nomura:2018kji,Murdia:2020iac}, in which successively renormalized boundaries $\sigma(\lambda)$, parameterized by ``renormalization scale'' $\lambda$, are obtained by the flow equation
\begin{equation}
  \frac{dx^\mu}{d\lambda} = s^\mu,
\label{eq:flow-eq}
\end{equation}
where $x^\mu$ are the embedding coordinates of the codimension-2 surface $\sigma(\lambda)$, on which the holographic states are defined, and $s^\mu$ is the evolution vector given by
\begin{equation}
  s^\mu = \frac{1}{2}(\Theta_k l^\mu + \Theta_l k^\mu).
\label{eq:s-mu}
\end{equation}
Here, $\{ k^\mu, l^\mu \}$ are the future-directed null vectors orthogonal to $\sigma(\lambda)$, normalized such that $k_\mu l^\mu = -2$,%
\footnote{
 This condition does not fix the normalizations of $k^\mu$ and $l^\mu$ separately, but it is sufficient to ensure the validity of the following treatment.
 In particular, $s^\mu$ is invariant under rescalings of $k^\mu$ and $l^\mu$ satisfying $k_\mu l^\mu = -2$.
 Below, we fix this freedom conveniently in each setup when we give explicit expressions for $k^\mu$, $l^\mu$, $\theta_{k,l}$, and $\Theta_{k,l}$.
}
and $\Theta_{k,l}$ represent quantum expansions in the corresponding directions~\cite{Bousso:2015mna}, which reduce to the classical expansions $\theta_{k,l}$ in the limit that bulk quantum effects are ignored.%
\footnote{
 More precisely, we must use a modified version of the quantum expansion which includes a bulk entropy contribution from an ``exterior'' region, as described in Ref.~\cite{Murdia:2020iac}.
}
We assume that $\sigma(\lambda)$ is neither quantum trapped nor antitrapped, so that the quantum expansion $\Theta_s$ associated with the evolution vector $s^\mu$ satisfies
\begin{equation}
  \Theta_s = \Theta_k \Theta_l \leq 0.
\end{equation}
This condition is needed for the consistency of the interpretation that the flow (toward larger $\lambda$) corresponds to coarse graining of the boundary degrees of freedom.

\begin{figure}[t]
\centering
  \includegraphics[width=0.9\textwidth]{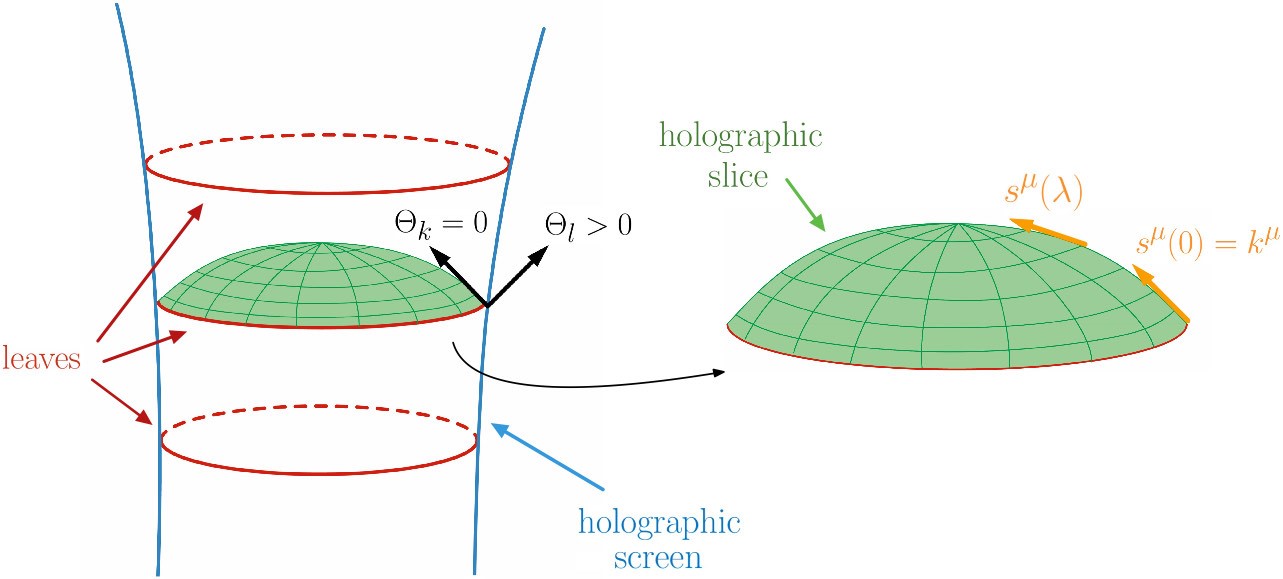}
\vspace{2mm}
\caption{
 A holographic theory resides on a holographic screen, which is a hypersurface foliated by quantum marginally antitrapped (or trapped) surfaces called leaves.
 At a given boundary time, a boundary state lives on a leaf, from which one can reconstruct a bulk equal-time hypersurface using the flow equation in Eq.~\eqref{eq:flow-eq}.}
\label{fig:holo}
\end{figure}

Since a holographic slice is nothing but a codimension-1 surface swept by $\sigma(\lambda)$, we can extend it ``outward'' using the flow equation.
For a generic spacetime, we can perform this extension up to the point where one of the $\Theta_{k,l}$ becomes zero, i.e.\ until $\sigma(\lambda)$ becomes a quantum marginally trapped or antitrapped surface.
Suppose that $\Theta_k = 0$ throughout this surface, which we call $\sigma(0)$.
Then, the dimension of the Hilbert space associated with states on the holographic slice is bounded~\cite{Bousso:2015mna,Bousso:1999xy} by the generalized entropy of $\sigma(0)$, so that the $\sigma(0)$ can be identified as an equal-time surface of a non-renormalized boundary---or a leaf---on which a non-renormalized boundary state is given.%
\footnote{
 If we follow the flow further beyond $\lambda = 0$, then we lose this property; in this sense, $\sigma(0)$ is the ``maximally unrenormalized'' leaf.
}
A codimension-1 surface foliated by such leaves is called a holographic screen~\cite{Bousso:1999cb}, or $Q$-screen in the quantum context~\cite{Bousso:2015eda}, on which a holographic theory for general spacetimes is supposed to live~\cite{Nomura:2017fyh,Sanches:2016sxy,Nomura:2016ikr}.
This framework is indeed consistent with the hypothetical relationship~\cite{Miyaji:2015yva} between bulk spacetime and the entanglement structure of a boundary state.
A sketch of a holographic screen, leaves, and a holographic slice is given in Fig.~\ref{fig:holo}.

\begin{figure}[t]
\centering
  \subfloat[\centering asymptotically AdS spacetime]{{\includegraphics[width=0.21\textwidth]{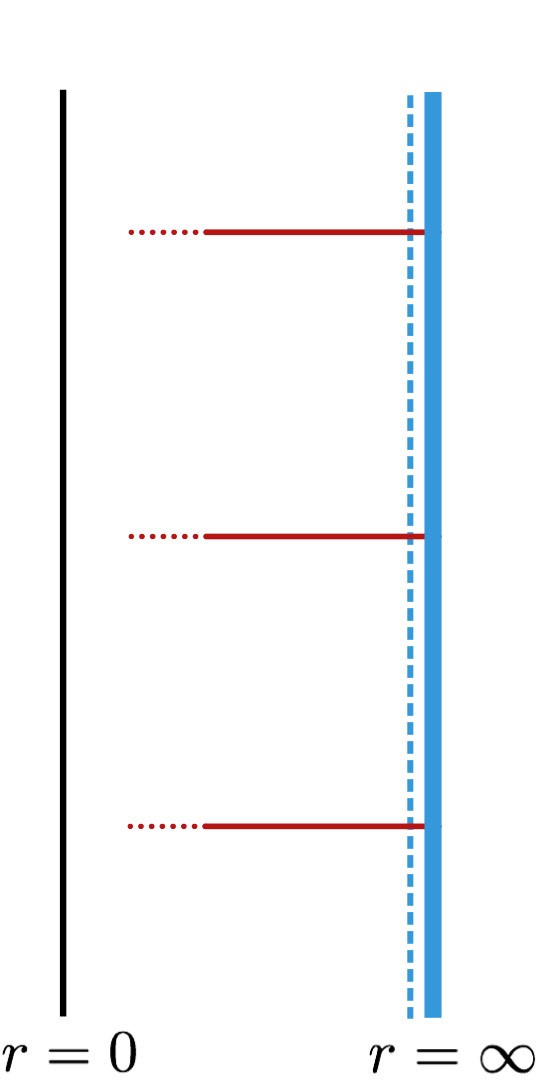}}}
\hspace{4cm}
  \subfloat[\centering asymptotically flat spacetime]{{\includegraphics[width=0.21\textwidth]{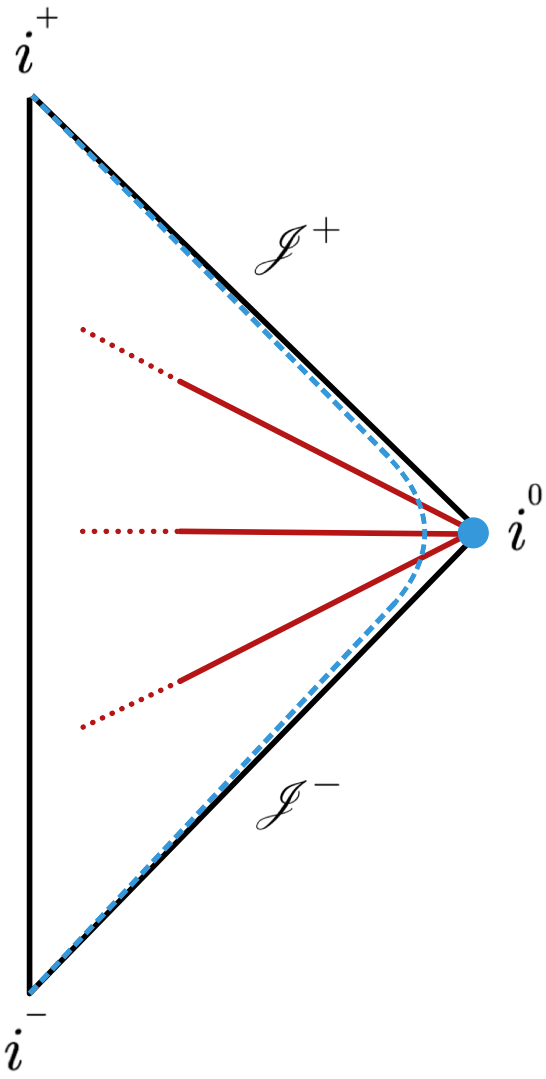}}}
\caption{
 Renormalized holographic screens, on which regularized holographic theories reside, are timelike hypersurfaces in an asymptotically AdS and flat spacetimes, which are depicted by the dotted blue lines in the Penrose diagrams in (a) and (b), respectively.
 In both cases, holographic slices agree with the time slices given by timelike Killing vectors in the asymptotic regions.
 In the non-renormalized limit, physics occurring in a finite time is described by holographic theories located at the conformal boundary ($r = \infty$) and spatial infinity ($i^0$) in the cases of asymptotically AdS and flat spacetimes, respectively.}
\label{fig:AdS-flat}
\end{figure}

In some cases, non-renormalized leaves cannot be reached with a finite evolution in $\lambda$.
In particular, this occurs in asymptotically AdS and flat spacetimes.
In an asymptotically AdS spacetime, the metric in the asymptotic region can be expanded in a Fefferman-Graham series~\cite{FG}
\begin{equation}
  ds^2 = \frac{L^2}{z^2} \bigl\{ g_{ab}(x^a,z) dx^a dx^b + dz^2 \bigr\},
\label{eq:FG-1}
\end{equation}
where $L$ is the AdS length scale, $a,b = 0,\cdots,d-1$, and $g_{ab}(x^a,z) = g_{ab}^{(0)}(x^a) + z^2 g_{ab}^{(2)}(x^a) + \cdots$ with $g_{ab}^{(0)}$ being the conformal boundary metric.
Suppose that $\sigma(\lambda)$ is taken to be a constant $t = x^0$ surface at a constant $z = \epsilon$.
The null normals are then given by $k_\mu = (L/\epsilon)(- dt + dz)$ and $l_\mu = (L/\epsilon)(-dt - dz)$, yielding $\Theta_k = -(d-1)/L$ and $\Theta_l = (d-1)/L$ up to higher order corrections in $\epsilon$.
Here, we have used the fact that the quantum expansions $\Theta_{k,l}$ approach classical expansions $\theta_{k,l}$ in the asymptotic region because of the lack of matter there.
This implies that a leaf of any regularized boundary located at $z > 0$ is, in fact, a renormalized one, and that the non-renormalized holographic screen can lie formally only at spacelike infinity,%
\footnote{
 We adopt a definition of the non-renormalized holographic screen such that it is a hypersurface foliated by leaves, on which {\it at least one} of the $\Theta_{k,l}$ vanishes.
}
where a nonregularized holographic CFT lives in AdS/CFT.
It also gives the evolution vector
\begin{equation}
  s^\mu = \frac{(d-1)\epsilon}{L^2} \frac{\partial}{\partial z} + O(\epsilon^2),
\end{equation}
showing that the holographic slice extending inward from a regularized boundary evolves initially in the $z$ direction, up to corrections suppressed by $\epsilon$; see Fig.~\ref{fig:AdS-flat}(a).
This behavior is the same as that of the conventional holographic renormalization group flow~\cite{Akhmedov:1998vf,Alvarez:1998wr,Balasubramanian:1999jd,Skenderis:1999mm,deBoer:1999tgo} in the AdS/CFT correspondence.
We also find that the flow freezes, $|dz/d\lambda| \rightarrow 0$, as $z \rightarrow 0$.

For an asymptotically flat spacetime, the metric in the asymptotic region can be expanded in the Bondi-Sachs form~\cite{Bondi:1962px,Sachs:1962wk} as
\begin{equation}
  ds^2 = -\frac{V}{r} e^{2\beta} du^2 - 2 e^{2\beta} du dr + r^2 h_{AB} \bigl(dx^A - U^A du\bigr) \bigl(dx^B - U^B du\bigr),
\end{equation}
where $A,B = 1,\cdots,d-1$, and each function admits a large $r$ expansion of the form $V = r + O(1)$, $\beta = O(r^{-2})$, $U^A = O(r^{-2})$, and $h_{AB} = O(1)$.
As in the case of an asymptotically AdS spacetime, let us consider $\sigma(\lambda)$ which is a constant time slice of the surface $r = R$.
The null normals are then given by $k_\mu = du$ and $l_\mu = (V/r) du + 2 dr$, leading to $\Theta_k = -2/R$ and $\Theta_l = 2/R$ at the leading order in $1/R$.
We thus find that a leaf of a regularized holographic screen is a renormalized one, and the flow vector is
\begin{equation}
  s^\mu = \frac{2}{R} \left( -\frac{\partial}{\partial u} + \frac{\partial}{\partial r} \right) + O\left(\frac{1}{R^2}\right),
\end{equation}
showing that the generated flow follows a hypersurface of equal Minkowski time in the inward radial direction, up to corrections suppressed by $1/R$, and that it freezes as $R \rightarrow \infty$.
This is illustrated in Fig.~\ref{fig:AdS-flat}(b), from which we see that any process occurring in the bulk within a finite time interval can be described by a boundary theory located at spatial infinity $i^0$.
By taking the time interval to infinity, one recovers an $S$-matrix description in the bulk.
This view is consistent with discussions of flat space holography advanced, e.g., in Refs.~\cite{deBoer:2003vf,Strominger:2013jfa,Cheung:2016iub,Pasterski:2016qvg,Laddha:2020kvp}.

\subsubsection*{de~Sitter spacetime}

Let us now discuss de~Sitter spacetime.
As we will see below, care is needed to consider holography of de~Sitter spacetime in a static patch.
Our view is that in this context, the concept of exact de~Sitter spacetime arises only as a result of idealization, very much analogous to an eternal single-sided black hole in an asymptotically flat spacetime (which does not exist because of Hawking radiation).%
\footnote{
 This is consistent with the expectation from string theory that there is no absolutely stable de~Sitter vacuum~\cite{Dyson:2002pf,Kachru:2003aw}.
}

For now, we bypass this issue by considering de~Sitter spacetime in a cosmological setup.
In particular, we consider an empty bubble universe in which there is a positive vacuum energy density $\rho_\Lambda$.
The interior of the bubble is an open Friedmann-Lema\^{i}tre-Robertson-Walker (FLRW) universe described by the metric
\begin{equation}
  ds^2 = -d\tau^2 + a(\tau)^2 \left[ d\chi^2 + \left\{ \frac{1}{\sqrt{-\kappa}} \sinh\bigl(\sqrt{-\kappa}\chi\bigr) \right\}^2 d\Omega^2 \right].
\label{eq:dS-FLRW}
\end{equation}
Here,
\begin{equation}
  a(\tau) = \sqrt{\frac{-\kappa}{\tilde{\Lambda}}} \sinh\Bigl(\sqrt{\tilde{\Lambda}}\,\tau\Bigr)
\end{equation}
is the scale factor, where $\tilde{\Lambda}$ is related to the vacuum energy density by $\tilde{\Lambda} = 8\pi \rho_\Lambda l_{\rm P}^2/3$, and $\kappa < 0$ is a curvature parameter related to the physical curvature radius by $r_{\rm curv}(\tau) = a(\tau)/\sqrt{-\kappa}$.
The holographic screen is located at the apparent horizon, where $\theta_k = 0$ and $\theta_l > 0$:
\begin{equation}
  \chi = \chi_{\rm sc}(\tau) = \frac{1}{\sqrt{-\kappa}} \ln\left[\coth\frac{\sqrt{\tilde{\Lambda}}\,\tau}{2}\right],
\end{equation}
where we have taken $k^\mu$ and $l^\mu$ to be the future-directed ingoing and outgoing null vector orthogonal to a leaf, respectively, and we have indexed leaves by FLRW times at their locations.
Here, we have ignored the difference between classical and quantum expansions, which is valid because we consider the semiclassical vacuum state, implying that Gibbons-Hawking radiation is not extracted as semiclassical radiation.

\begin{figure}[t]
\centering
\vspace{2mm}
  \includegraphics[width=0.4\textwidth]{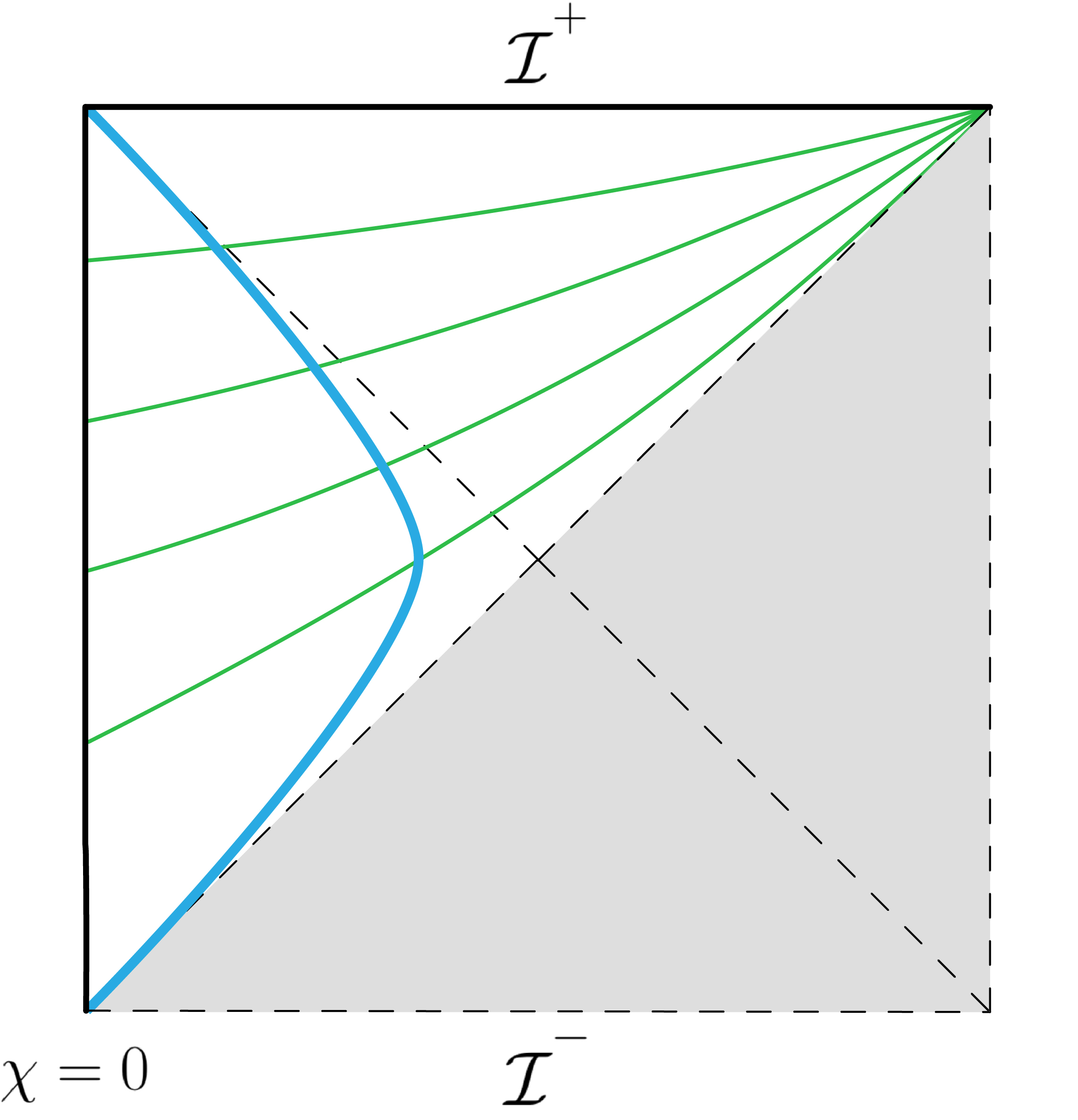}
\caption{
 Penrose diagram of de~Sitter spacetime.
 A de~Sitter bubble corresponds to the upper-left half of this diagram, where FLRW equal-time hypersurfaces are drawn by green lines.
 The holographic screen is located along the trajectory of the apparent horizon, which is depicted by the solid blue curve.}
\label{fig:FLRW-dS}
\end{figure}

At late times, the universe behaves like a de~Sitter spacetime:
\begin{equation}
  a(\tau) \approx \frac{\sqrt{-\kappa}}{2\sqrt{\tilde{\Lambda}}}\, e^{\sqrt{\tilde{\Lambda}}\,\tau}
\quad\mbox{for }
  \tau \gg \frac{1}{\sqrt{\tilde{\Lambda}}}.
\end{equation}
The metric in Eq.~\eqref{eq:dS-FLRW} then becomes flat slicing of the de~Sitter spacetime; see Fig.~\ref{fig:FLRW-dS}.
On the other hand, the physical distance to the holographic screen approaches the Hubble radius $1/\sqrt{\tilde{\Lambda}}$:
\begin{equation}
  a(\tau) \chi_{\rm sc}(\tau) \,\,=\,\, \frac{1}{\sqrt{\tilde{\Lambda}}} \sinh\Bigl(\sqrt{\tilde{\Lambda}}\,\tau\Bigr) \ln\left[\coth\frac{\sqrt{\tilde{\Lambda}}\,\tau}{2}\right]
\,\,\,\xrightarrow[\tau \gg \frac{1}{\sqrt{\tilde{\Lambda}}}]{}\,\,\,
  \frac{1}{\sqrt{\tilde{\Lambda}}} \left( 1 - \frac{2}{3} e^{-2\sqrt{\tilde{\Lambda}}\,\tau} \right).
\end{equation}
The holographic theory thus describes a static patch of the late-time de~Sitter spacetime, i.e.\ the patch inside the holographic screen.

The holographic slices approach static de~Sitter time slices at late times.
To see this, we can calculate $k^\mu$ and $l^\mu$ for arbitrary $(\tau,\chi)$
\begin{equation}
  k^\mu = \begin{pmatrix} 1 \\ -\sqrt{\frac{\tilde{\Lambda}}{-\kappa}}\frac{1}{\sinh\bigl(\sqrt{\tilde{\Lambda}}\,\tau\bigr)} \end{pmatrix},
\qquad
  l^\mu = \begin{pmatrix} 1 \\ \sqrt{\frac{\tilde{\Lambda}}{-\kappa}}\frac{1}{\sinh\bigl(\sqrt{\tilde{\Lambda}}\,\tau\bigr)} \end{pmatrix}
\end{equation}
and their associated quantum expansions
\begin{equation}
\begin{cases}
  \Theta_k \approx \theta_k = 2 \sqrt{\tilde{\Lambda}} \left[ \coth\Bigl(\sqrt{\tilde{\Lambda}}\,\tau\Bigr) - \frac{\coth(\sqrt{-\kappa}\,\chi)}{\sinh\bigl(\sqrt{\tilde{\Lambda}}\,\tau\bigr)} \right], \\
  \Theta_l \approx \theta_l = 2 \sqrt{\tilde{\Lambda}} \left[ \coth\Bigl(\sqrt{\tilde{\Lambda}}\,\tau\Bigr) + \frac{\coth(\sqrt{-\kappa}\,\chi)}{\sinh\bigl(\sqrt{\tilde{\Lambda}}\,\tau\bigr)} \right].
\end{cases}
\end{equation}
The evolution vector is then
\begin{equation}
  \begin{pmatrix} s^\tau \\ s^\chi \end{pmatrix} = 2\sqrt{\tilde{\Lambda}} \begin{pmatrix} \coth\Bigl(\sqrt{\tilde{\Lambda}}\,\tau\Bigr) \\ -\sqrt{\frac{\tilde{\Lambda}}{-\kappa}}\frac{\coth(\sqrt{-\kappa}\,\chi)}{\sinh^2\bigl(\sqrt{\tilde{\Lambda}}\,\tau\bigr)} \end{pmatrix}.
\end{equation}

We are interested in the spacetime region well inside the holographic screen
\begin{equation}
  \chi \ll \chi_{\rm sc}(\tau) \,\,\xrightarrow[e^{\sqrt{\tilde{\Lambda}}\,\tau} \gg 1]{}\,\, \sqrt{-\kappa}\,\chi \ll 2 e^{-\sqrt{\tilde{\Lambda}}\,\tau}
\end{equation}
at late times
\begin{equation}
  \alpha = \frac{1}{\sqrt{\tilde{\Lambda}}} \ll r_{\rm curv}(\tau) \,\,\longrightarrow\,\, e^{\sqrt{\tilde{\Lambda}}\,\tau} \gg 1,
\end{equation}
where $\alpha$ is the Hubble radius of the late time universe.
In this region, the relation between the FLRW coordinates $(\tau,\chi)$ and the de~Sitter static coordinates $(t,r)$ is given by
\begin{equation}
\begin{cases}
  \alpha\, e^{\frac{\tau}{\alpha}} = \sqrt{\alpha^2 - r^2}\, e^{\frac{t}{\alpha}}, \\
  \frac{\alpha}{2} \sqrt{-\kappa}\, \chi\, e^{\frac{\tau}{\alpha}} = r,
\end{cases}
\end{equation}
leading to the $s^\mu$ vector in the static coordinates
\begin{equation}
  \begin{pmatrix} s^t \\ s^r \end{pmatrix} \approx \begin{pmatrix} -\frac{8}{3\alpha^3} \frac{r^2}{(1-r^2/\alpha^2)^2} e^{-\frac{2t}{\alpha}} \\ -\frac{2}{r} \Bigl( 1 - \frac{r^2}{\alpha^2} \Bigr) \end{pmatrix}
  \approx \begin{pmatrix} -\frac{8 r^2}{3 \alpha^3} e^{-\frac{2t}{\alpha}} \\ -\frac{2}{r} \end{pmatrix}.
\end{equation}
We thus find that both $s^t$ and $s^r$ are negative, and $|s^t| \ll |s^r|$ in the relevant region.

From the above analysis, we conclude that in the holographic description of de~Sitter spacetime, the region swept by holographic slices is the interior of the static patch.
Thus, what is analogous to the exterior of a single-sided black hole is the interior of a static patch in de~Sitter spacetime.
This indicates that the interior of the static patch is the region in which semiclassical field operators can be reconstructed in a simple manner.
We note that the same conclusion can also be obtained by regularizing de~Sitter spacetime in different ways, for example by considering a big-bang universe filled with two fluid components with the equation-of-state parameters $w = -1$ and $w > -1$, or with a single fluid component of $w = -1+\epsilon$, where $\epsilon$ ($> 0$) is taken sufficiently small that the system can be viewed as in a de~Sitter vacuum~\cite{Nomura:2017fyh,Sanches:2016sxy}.
This provides a justification for the description of de~Sitter spacetime adopted in Sections~\ref{sec:QFT}~and~\ref{sec:stretched}, focusing on a single static patch.%
\footnote{
 It is also comforting that there is a perturbative positive energy theorem in a static patch of de~Sitter spacetime~\cite{Abbott:1981ff}.
}

It is important that the holographic description based on a single static patch, which we may call ``single-sided'' de~Sitter spacetime, assumes an appropriate physical regularization.
Had we started with exact de~Sitter spacetime, then the location of a leaf would be on the bifurcation surface or the future horizon.
In this case, a subregion of the leaf would have degenerate extremal surfaces, all of which are located on the future horizon and have areas equal to the volume of the subregion.
This would imply that holographic slices sweep only a codimension-1 surface in the bulk, i.e.\ the future horizon, failing to reconstruct the codimension-0 spacetime~\cite{Nomura:2017fyh}.%
\footnote{
 More precisely, if we use global information of the boundary state, the interior of the static patch may be reconstructed.
 However, the entanglement wedge of a boundary subregion in this case contains either the entirety or none of the interior, depending on the size of the subregion.
 This indicates a ``singular'' nature of the limit of de~Sitter spacetime.
}
For $\Theta_k = 0$ and $\Theta_l \neq 0$, this can also be seen from the fact that $s^\mu \,\propto\, k^\mu$.

\begin{figure}[t]
\centering
\vspace{2mm}
  \includegraphics[width=0.4\textwidth]{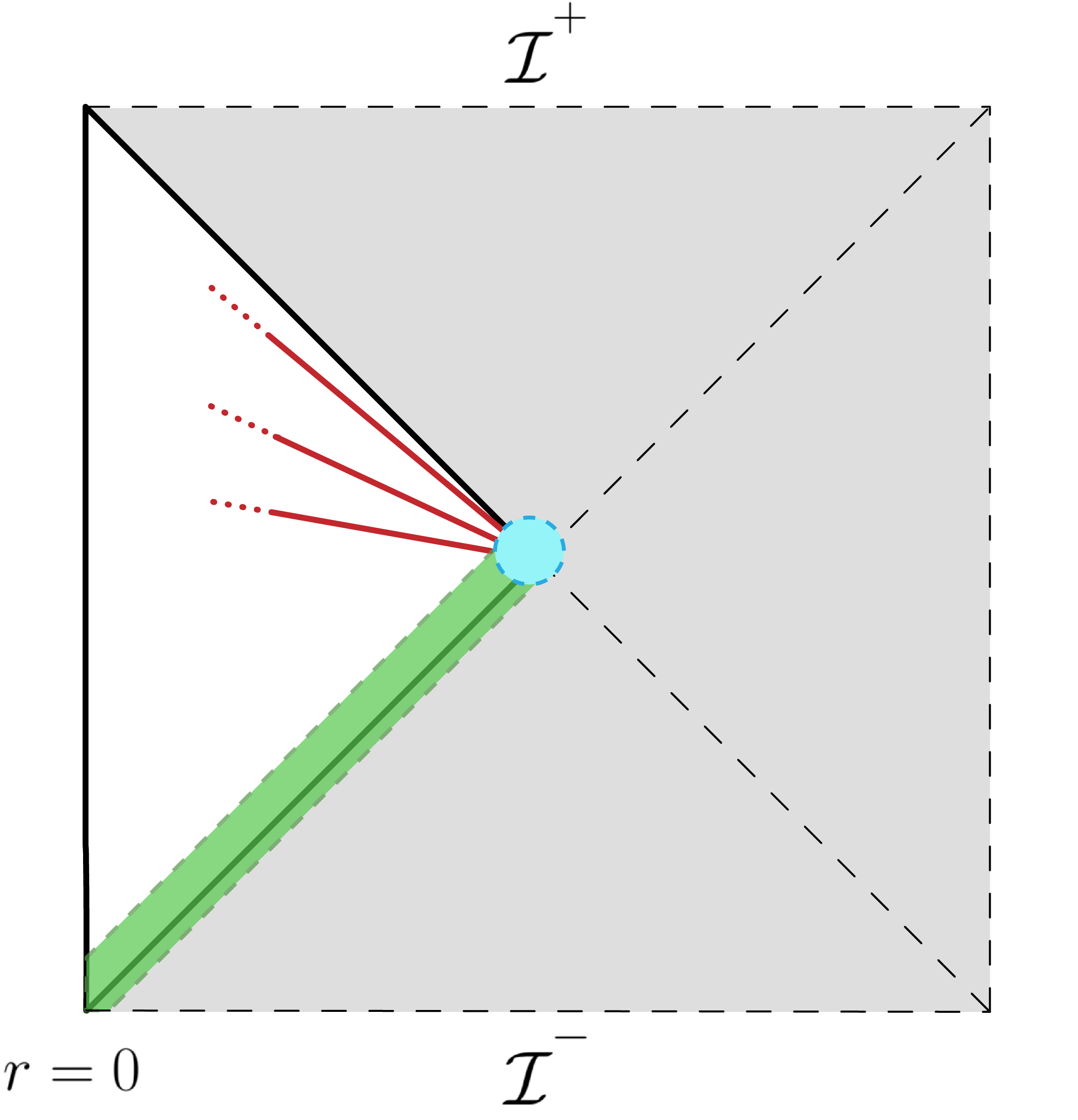}
\caption{
 A holographic description of ``single-sided'' de~Sitter spacetime.
 Holographic slices agree (approximately) with equal-time hypersurfaces in static coordinates.
 The smaller the amount of UV renormalization becomes, the closer renormalized leaves are to the bifurcation surface.
 The geometry in the vicinity of the past horizon (indicated by light green) and the bifurcation surface (light blue) is regularization dependent.}
\label{fig:dS-static}
\end{figure}

The picture of a single-sided de~Sitter spacetime is shown in Fig.~\ref{fig:dS-static}, in which the region near the bifurcation surface and the past horizon should be viewed as regularization dependent.
This picture can also be obtained if we begin with a renormalized holographic screen foliated by renormalized leaves located deep inside the static patch and then push the screen outward by unrenormalizing it using the flow equation.
Renormalized leaves then approach the bifurcation surface (or the future horizon if the assumed deviation from exact de~Sitter spacetime is significant), but they never get there.
Given the non-decoupling of bulk gravity on these leaves, we expect that the holographic theory on the screen is gravitational, but we do not make any further speculation about this theory here.

\subsubsection*{de~Sitter entropy}

The fact that holographic states in a single-sided de~Sitter spacetime are analogous to those in a single-sided black hole suggests that we can interpret the de~Sitter entropy in an analogous manner to the black hole case~\cite{Nomura:2014yka}.
At the classical level, de~Sitter spacetime is parameterized by one continuous number:\ the Hubble radius $\alpha$.
At the quantum level, this freedom leads to the corresponding independent quantum states, which are discretized.
The number of independent states corresponding to the value of the Hubble radius between $\alpha$ and $\alpha + \delta\alpha$ is
\begin{equation}
  {\cal N} \sim e^{S_{\rm GH}} \frac{\delta\alpha}{\alpha},
\label{eq:N_dS}
\end{equation}
where
\begin{equation}
  S_{\rm GH} = \frac{\pi \alpha^2}{l_{\rm P}^2}
\label{eq:S_GH}
\end{equation}
is the Gibbons-Hawking entropy.
As in standard statistical mechanics, this result does not depend on the detailed choice of $\delta\alpha$ (unless $\delta\alpha$ is taken to be exponentially small in $S_{\rm GH}$).

Now, suppose that the time scale for the evolution of a de~Sitter microstate is given by $\varDelta t$.
Using the energy of Eq.~\eqref{eq:E_dS}, we find that the required uncertainty of $\alpha$, determined by $\varDelta E \varDelta t \sim 1$, is
\begin{equation}
  \varDelta\alpha \sim \frac{l_{\rm P}^2}{\varDelta t}.
\end{equation}
We expect that $\varDelta t \sim 1/T_{\rm H} \sim \alpha$ because of large redshift between the location of the stretched horizon, where everything is controlled by the string scale, and the location at which the time $t$ is measured, $r = 0$.
This gives
\begin{equation}
  \varDelta\alpha \sim \frac{l_{\rm P}^2}{\alpha}.
\label{varDelta-alpha}
\end{equation}
In the semiclassical regime $\alpha \gg l_{\rm P}$, this uncertainty is smaller than the Planck length, $\varDelta\alpha \ll l_{\rm P}$, which is consistent with the fact that the Hubble radius can be precisely specified (treated classically) in a semiclassical theory.

Note that the situation described above is analogous to the black hole case.
For a black hole, $E = M$ and $r_+ \sim M l_{\rm P}^2$ give $\varDelta r_+ \sim l_{\rm P}^2/\varDelta t$.
Assuming $\varDelta t \sim 1/T_{\rm H} \sim M l_{\rm P}^2$, this leads to
\begin{equation}
  \varDelta r_+ \sim \frac{1}{M}.
\end{equation}
In the semiclassical regime $M \gg 1/l_{\rm P}$, we find $\varDelta r_+ \ll l_{\rm P}$.

There is one apparent difference between the de~Sitter and black hole cases.
Since the de~Sitter spacetime is an ``inside-out'' version of the black hole spacetime, simple operators can cause excitations inside the stretched horizon, $r < r_{\rm s}$, and with many such excitations the geometry will be backreacted.
In particular, we can form a black hole inside the de~Sitter horizon, leading to a different semiclassical geometry.
We will now try to understand microscopic entropies of such geometries and their relations to the de~Sitter entropy.%
\footnote{
 A similar issue was also discussed in Ref.~\cite{Susskind:2021dfc}.
}

We naturally expect that the original theory on the holographic screen can accommodate all such solutions.
This interpretation is consistent if we regard a holographic theory of de~Sitter spacetime to be associated with a fixed vacuum energy $\rho_\Lambda$ (within uncertainty), and not a fixed horizon radius.
In this case, the solution with a black hole has an entropy smaller than the solution with no black hole~\cite{Gibbons:1977mu}, so the dimension of the Hilbert space of the holographic theory is
\begin{equation}
  {\rm dim}\,{\cal H} \sim \sum_{\varDelta S \leq \varDelta S_{\rm max}} e^{S_{\rm GH} - \varDelta S}
\quad\Rightarrow\quad
  \ln {\rm dim}\,{\cal H} \sim S_{\rm GH}
\label{eq:inclusion}
\end{equation}
even including all the spacetimes with varying sizes of black holes.
Here, $\varDelta S$ is the entropy deficit of a spacetime with a black hole(s), $S_{\rm GH}$ is the Gibbons-Hawking entropy expressed in terms of the vacuum energy $\rho_\Lambda$ or the Hubble radius $\alpha_0$ of the de~Sitter spacetime without a black hole
\begin{equation}
  S_{\rm GH} = \frac{\pi \alpha_0^2}{l_{\rm P}^2}= \frac{3}{8 \rho_\Lambda l_{\rm P}^4},
\end{equation}
and $\varDelta S_{\rm max}$ is the entropy deficit when the cosmic and black hole horizons have the same radius, which occurs with $r_{+,{\rm bh}} = r_{+,{\rm dS}} = \alpha_0/\sqrt{3}$:
\begin{equation}
  \varDelta S_{\rm max} = S_{\rm GH} - 2 \times \frac{\pi\alpha_0^2}{3 l_{\rm P}^2} = \frac{\pi \alpha_0^2}{3 l_{\rm P}^2} = \frac{1}{8 \rho_{\Lambda} l_{\rm P}^4},
\end{equation}
where the factor of $2$ comes from the fact that we have both cosmological and black hole horizons.
The fact that $\varDelta S_{\rm max}$ is positive serves as a consistency check for our interpretation.%
\footnote{
 In $d+1$ dimensions, this occurs when the black hole mass becomes
 \begin{equation}
   M = \frac{d-1}{d-2} \left(\frac{d-2}{d}\right)^{d/2} \frac{{\rm vol}(\Omega_{d-1})}{8\pi l_{\rm P}^{d-1}}\, \alpha_0^{d-2}
 \end{equation}
 with
 \begin{equation}
    r_{+,{\rm bh}} = r_{+,{\rm dS}} = \sqrt{\frac{d-2}{d}}\, \alpha_0.
 \end{equation}
 Here, ${\rm vol}(\Omega_{d-1}) = 2 \pi^{d/2}/\Gamma(d/2)$ is the volume (area) of the $(d-1)$-dimensional unit sphere.
 The maximal entropy deficit is thus
 \begin{equation}
   \varDelta S_{\rm max} = \frac{{\rm vol}(\Omega_{d-1}) \alpha_0^{d-1}}{4 l_{\rm P}^{d-1}}\, \left[ 1 - 2\times \left(\frac{d-2}{d}\right)^{\frac{d-1}{2}} \right],
 \end{equation}
 which is indeed positive for all $d \geq 2$.
}

In fact, because of the inside-out nature, forming black holes inside the de~Sitter horizon is analogous to forming (small) black holes outside the black hole horizon, specifically in the zone, while keeping the total energy fixed (though the latter necessarily breaks the spherical symmetry).
Such excited states can be included in the Hilbert space associated with the central black hole without changing its dimensions, $\ln {\rm dim}\,{\cal H} = S_{\rm bh}$, at the leading order.

\subsection{Holographic Hilbert spaces}

Before concluding this section, we discuss the structure of holographic boundary Hilbert spaces in a more general setting.
The picture presented here builds on the analyses performed in Refs.~\cite{Nomura:2018kji,Nomura:2017fyh} and is suggested by relations between bulk geometries and boundary entanglement entropies.

\subsubsection*{Effective boundary Hilbert spaces {\boldmath ${\cal H}_{\rm eff}({\cal A})$}}

Consider a set of states associated with a leaf $\sigma$ characterized by its volume ${\cal A}$ on the boundary.
(${\cal A}$ is an area from the point of view of bulk spacetime.)
Holography implies that the number of degrees of freedom on the leaf is given by
\begin{equation}
  {\cal N} = \frac{{\cal A}}{4 G_{\rm N}},
\label{eq:cal-N}
\end{equation}
which we will assume to be uniformly distributed over the leaf.
Here, we have used $G_{\rm N}$ instead of $l_{\rm P}$, since discussion in this subsection does not depend on the number of spacetime dimensions.
We denote the Hilbert space comprising states of these ${\cal N}$ degrees of freedom by ${\cal H}_{\rm eff}({\cal A})$, so that
\begin{equation}
  \ln {\rm dim}\, {\cal H}_{\rm eff}({\cal A}) = \frac{{\cal A}}{4 G_{\rm N}}.
\label{eq:dim-H_eff-A}
\end{equation}
The reason for the subscript ``eff'' will become clear later.

We expect that states of these ${\cal N}$ degrees of freedom can represent various different spacetimes, or more precisely the domain of dependence $D_\sigma$ of a spacelike hypersurface bounded by $\sigma$ in these spacetimes.
In addition, for a given bulk spacetime, there may be many independent states in ${\cal H}_{\rm eff}({\cal A})$ that span the space of microstates for the spacetime, as in the case of black hole and de~Sitter spacetimes.
How can these happen?

Let us assume that the entanglement entropy of subregions of a boundary state $\ket{\psi}$ dual to a semiclassical geometry can be calculated via the Hubeny-Rangamani-Ryu-Takayanagi (HRRT) prescription~\cite{Ryu:2006bv,Hubeny:2007xt,Lewkowycz:2013nqa} (or its quantum extension~\cite{Faulkner:2013ana,Engelhardt:2014gca,Dong:2016hjy}).
In particular, we assume that the boundary and bulk Hilbert spaces can be appropriately factorized, which may involve a gauge choice or the introduction of edge modes~\cite{Buividovich:2008gq,Donnelly:2011hn,Casini:2013rba,Ghosh:2015iwa,Donnelly:2016auv,Takayanagi:2019tvn}.
Here we consider the ``classical limit,'' meaning that all the subregions we consider contain $O({\cal N})$ degrees of freedom.
Given a bulk spacetime, one can then find the corresponding entanglement entropies for all subregions of the boundary.
The collection of all boundary subregions and their corresponding entanglement entropies will be referred to as the entanglement structure of the state, which we denote by ${\cal S}(\ket{\psi})$.

First, we note that for a given entanglement structure ${\cal S}_0$, we can always find a basis of the Hilbert space ${\cal H}_{\rm eff}({\cal A})$ in which all basis states have the specified entanglement structure.
This is because by applying local unitaries to a state, one can generate $e^{\cal N}$ orthogonal states while preserving the entanglement structure of the original state.
This fact, however, does not mean that these $e^{\cal N}$ states span an $e^{\cal N}$-dimensional space of microstates for bulk spacetime corresponding to the entanglement structure ${\cal S}_0$.
Indeed, by generically superposing $e^{O({\cal N})}$ of these states, one would obtain a state whose entanglement structure is drastically different from ${\cal S}_0$, so that the resulting state is dual to a completely different spacetime, if it represents bulk spacetime at all.

Of course, given an entanglement structure, there exists a subspace of dimension $e^{O({\cal N}^p)}$ with $p < 1$ in which generic states have this same entanglement structure up to $O({\cal N}^p)$ corrections.
This is because we generally have
\begin{equation}
  {\cal S}\biggl( \sum_{i=1}^{e^{M}} c_i \ket{\psi_i} \biggr) = {\cal S}_0 + O(M),
\label{eq:EE_triv}
\end{equation}
where ${\cal S}(\ket{\psi_i}) = {\cal S}_0$ for all $i$, so that for $M = O({\cal N}^p)$ the corrections are suppressed by powers of ${\cal N}$ compared to ${\cal S}_0$, which is of order ${\cal N}$.
The subspace obtained in this way, however, comprises only an exponentially small subset of ${\cal H}_{\rm eff}({\cal A})$; in particular, it is a measure zero subset of ${\cal H}_{\rm eff}({\cal A})$ in the classical limit.

A nontrivial thing is that for ${\cal N} \gg 1$, there exist subspaces {\it of dimension $e^{O({\cal N})}$} in which generic states have the same entanglement structure ${\cal S}_0$ up to small corrections.
Specifically, for such a subspace spanned by basis states $\ket{\psi_i}$ ($i = 1, \cdots, e^{Q {\cal N}}$), we have
\begin{equation}
  {\cal S}\biggl( \sum_{i=1}^{e^{Q {\cal N}}} c_i \ket{\psi_i} \biggr) = {\cal S}_0 + O({\cal N}^p;\, p<1),
\label{eq:EE_nontriv}
\end{equation}
where $Q \leq 1$ does not scale with ${\cal N}$.
The existence of these subspaces with entanglement structures invariant under superpositions is expected from canonical typicality (also referred to as the general canonical principle)~\cite{Goldstein:2005aib,Popescu:2005}, which states that generic states in such a subspace have the same reduced density matrix for small subsystems (up to small corrections).
This is the case despite the fact that the size of the subspace is large enough that one would naively think that superpositions would ruin the entanglement structure at $O({\cal N})$.
The proof of this statement is purely kinematical and hence applies generally.
In fact, according to canonical typicality the correction term in Eq.~(\ref{eq:EE_nontriv}) is exponentially small, $O(e^{-Q {\cal N}/2})$.

\begin{figure}[t]
\centering
  \includegraphics[height=0.4\textwidth]{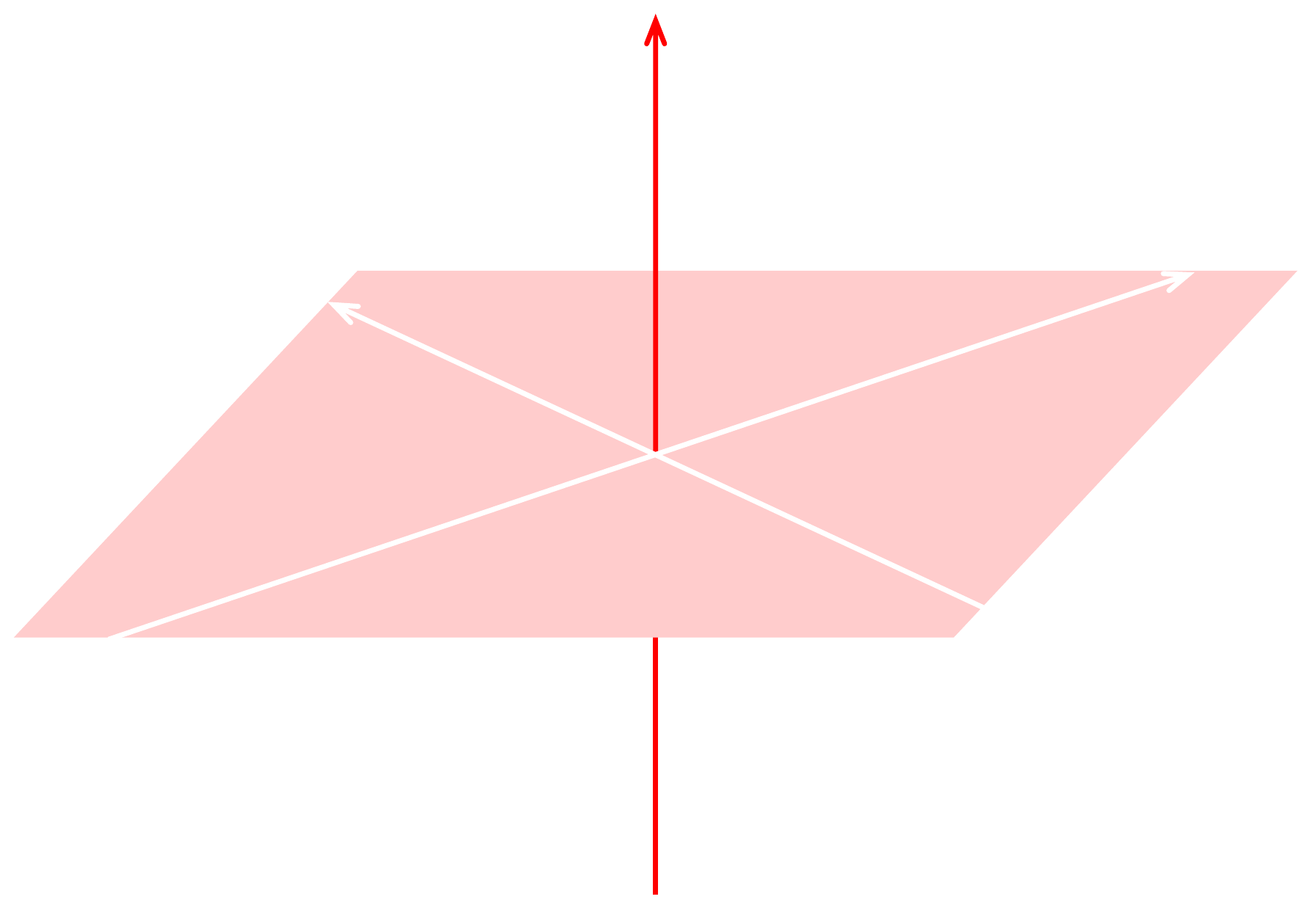}
\caption{
 A sketch of the collection of states in holographic Hilbert space ${\cal H}_{\rm eff}({\cal A})$ which have the same entanglement structure at leading order in $1/{\cal N}$.
 It forms an $e^{Q{\cal N}}$-dimensional subspace of ${\cal H}_{\rm eff}({\cal A})$ (represented by the pink plane) except that exponentially rare nongeneric states are excluded (white arrows), and that $e^{\cal N}-e^{Q{\cal N}}$ isolated states orthogonal to it are added (red arrow).}
\label{fig:Hilbert}
\end{figure}

In Fig.~\ref{fig:Hilbert}, we show a sketch of the collection ${\cal C}$ of states in ${\cal H}_{\rm eff}({\cal A})$ which have the same entanglement structure ${\cal S}_0$ up to corrections higher order in $1/{\cal N}$
\begin{equation}
  {\cal C} = \bigl\{ \ket{\psi} \,\big|\, {\cal S}(\ket{\psi}) = {\cal S}_0  + O({\cal N}^p;\, p<1) \bigr\}
\end{equation}
in the case that they form a subspace of dimension $e^{Q{\cal N}}$ with $Q < 1$.
In addition to this subspace represented by the pink plane, ${\cal C}$ contains $e^{\cal N}-e^{Q{\cal N}}$ states (slightly ``thickened'' in ${\cal H}_{\rm eff}({\cal A})$) orthogonal to it, which are schematically represented by the red arrow.
Furthermore, in the $e^{Q{\cal N}}$-dimensional subspace, there are exponentially rare states that do not have the entanglement structure ${\cal S}_0$.
These states can be obtained by fine-tuning coefficients when we expand $\ket{\psi}$ in terms of the basis states $\ket{\psi_i}$ of the subspace and are represented by white arrows.

In Ref.~\cite{Nomura:2017fyh}, it was argued that it is this $e^{Q{\cal N}}$-dimensional subspace that comprises the space of microstates for a spacetime.
In particular, for $Q < 1$ the corresponding entanglement structure ${\cal S}_0$ can be non-maximal,%
\footnote{
 By the entanglement structure being maximal, we mean that for any subregion $A$ its entanglement entropy $S_A$ is maximal, i.e.\ $S_A = \norm{A}/4G_{\rm N}$, at the leading order in $1/\norm{A}$.
 Here $\norm{A}$ is the volume of $A$ on the boundary.
}
and generic states in this subspace can have dual bulk spacetimes which are simply reconstructable.
In this case, even if one considers an exponentially large superposition of microstates, geometric operators are effectively linear so long as the state is generic within the subspace~\cite{Almheiri:2016blp}.
On the other hand, if $Q=1$, the ``subspace'' is the whole Hilbert space ${\cal H}_{\rm eff}({\cal A})$, so applying Page's analysis~\cite{Page:1993df}, we see that the only entanglement structure consistent with Eq.~(\ref{eq:EE_nontriv}) is that of maximal entropy.
In this case, the resulting spacetime is not in a simple wedge, and reconstruction of the bulk requires some level of nonlinearity, or state dependence.%
\footnote{
 This is indeed the case for the interior of a black hole; see Section~\ref{subsec:int-BH}.
}

The structure discussed above allows for a single holographic Hilbert space ${\cal H}_{\rm eff}({\cal A})$ to harbor effective subspaces dual to different geometries.
In fact, similarly to Eq.~\eqref{eq:inclusion}, one can show that ${\cal H}_{\rm eff}({\cal A})$ satisfying Eq.~\eqref{eq:dim-H_eff-A} can support a number of $e^{Q{\cal N}}$-dimensional subspaces with $Q < 1$.
Geometric operators are approximately linear in each of these subspaces, which gives the effective linear space of microstates for a fixed semiclassical geometry.

\subsubsection*{Fundamental boundary Hilbert space {\boldmath ${\cal H}_{\rm UV}$}}

So far, we have considered boundary Hilbert space with fixed volume ${\cal A}$: ${\cal H}_{\rm eff}({\cal A})$.
However, general spacetime involves boundary evolution in which the volume of a leaf (a boundary equal-time surface) changes~\cite{Nomura:2016ikr}.
Such an evolution can occur in the ``fundamental'' Hilbert space ${\cal H}_{\rm UV}$ which contains, at least effectively, ${\cal H}_{\rm eff}({\cal A})$'s with different ${\cal A}$'s:%
\footnote{
 A semiclassical description is valid only when ${\rm dim}{\cal H}_{\rm eff}({\cal A}) \gg 1$, but we will be sloppy about it in writing Eq.~\eqref{eq:H_UV}.
}
\begin{equation}
  {\cal H}_{\rm UV} \supset \{ {\cal H}_{\rm eff}({\cal A}) \,|\, {\rm dim}{\cal H}_{\rm eff}({\cal A}) \in \mathbb{N} \}.
\label{eq:H_UV}
\end{equation}
A naive possibility is to literally have ${\cal H}_{\rm UV} \supset \bigoplus_{\cal A} {\cal H}_{\rm eff}({\cal A})$, but this need not be the case.
In fact, motivated by the relation between geometric objects in the bulk and quantum information theoretic quantities on the boundary, which has been learned in AdS/CFT and is expected to apply beyond, one can imagine that ${\cal H}_{\rm eff}({\cal A})$'s are contained in ${\cal H}_{\rm UV}$ in a more intricate manner.

In Refs.~\cite{Nomura:2018kji,Miyaji:2015yva}, it was envisioned that ${\cal H}_{\rm eff}({\cal A})$ is embedded in ${\cal H}_{\rm UV}$ as an effective subspace (in the sense discussed before; see, e.g., Fig.~\ref{fig:Hilbert}) in which a generic state has the property
\begin{equation}
  \sum_i S_i = \frac{\cal A}{4 G_{\rm N}},
\label{eq:eff-dim}
\end{equation}
where $S_i$ represents the entanglement entropy of the state in a sufficiently small subregion, $A_i$, of the holographic space $\Omega$, on which $\mathcal{H}_{\rm UV}$ is defined; the sum runs over all of these small subregions such that $\Omega = \cup_i A_i$ and $A_i \cap A_j = \emptyset$ ($i \neq j$).
This allows us to consider holographic states of all bulk spacetimes, and also their dynamics, in a single Hilbert space $\mathcal{H}_{\rm UV}$.

The Hilbert space $\mathcal{H}_{\rm UV}$ will be defined by introducing a short distance cutoff $\delta$ in $\Omega$ and then sending $\delta$ to zero, so that
\begin{equation}
  {\rm dim} {\cal H}_{\rm UV} \rightarrow \infty.
\end{equation}
For ($d+1$)-dimensional asymptotically AdS and flat bulk spacetimes, $S_i$ will behave as
\begin{equation}
  S_i \sim \frac{\norm{\partial A_i}}{\delta^{d-2}}
\quad\mbox{and}\quad
  S_i = f(A_i) \frac{\norm{A_i}}{\delta^{d-1}},
\end{equation}
respectively.
Here, $\norm{x}$ represents the volume of the object $x$, $\partial A_i$ is the boundary of $A_i$, and $f(A_i)$ is a function of $O(1)$.%
\footnote{
 In asymptotically flat spacetime, for example, the ratio of $f(A_i)$ for $A_i$ being a half of $\Omega$, $A_{1/2}$, to that for $A_i$ being a cutoff size region, $A_\delta$, is $f(A_{1/2})/f(A_\delta) = (1/\sqrt{\pi}) \{ \Gamma[d/2]/\Gamma[(d+1)/2] \}$.
}

According to this picture, states representing a cosmological spacetime with finite leaf area comprise a tiny effective subspace of ${\cal H}_{\rm UV}$, obtained from a generic state in ${\cal H}_{\rm UV}$ by an infinite number (in the limit $\delta \rightarrow 0$) of fine-tunings.%
\footnote{
 A similar picture was discussed for de~Sitter spacetime in Ref.~\cite{Maltz:2016iaw}.
}
It will be interesting to study the dynamics in $\mathcal{H}_{\rm UV}$ in the thermodynamic limit, using information from semiclassical theory.

\section{Analytic Extension of Spacetime in Quantum Gravity}
\label{sec:extension}

In a single-sided system, the spacetime region obtained from analytic extension in general relativity---e.g.\ the interior of a black hole and the region outside a static patch in de~Sitter spacetime---emerges as a collective phenomenon involving horizon (and possibly other) modes.
This can occur because a huge gravitational red/blueshift at the stretched horizon makes the string dynamics relevant in a static description, which makes the state take a generic, universal form across all low energy species.
In this section, we review this construction~\cite{Nomura:2018kia,Nomura:2019qps,Nomura:2019dlz,Langhoff:2020jqa,Nomura:2020ska} and refine it to include the effect of black hole evolution analyzed in Section~\ref{subsec:excitation}.

\subsection{The interior of a black hole}
\label{subsec:int-BH}

Consider a state of a black hole of mass $M$ at some time $t = t_*$.
We assume that it is in the semiclassical vacuum state, which is achieved typically more than one scrambling time $t_{\rm scr}$ after the last perturbation, where%
\footnote{
 This expression is obtained from the fact that the scrambling time represented in ingoing Eddington-Finkelstein time $v = t + r_*$ is~\cite{Penington:2019npb}
 \begin{equation}
   v_{\rm scr} = \frac{1}{2\pi T_{\rm H}} \ln S_{\rm bh} \approx 4 r_+ \ln\frac{r_+}{l_{\rm P}}.
 \label{eq:v_scr}
 \end{equation}
 The scrambling time in Schwarzschild time is then related to this time by~\cite{Langhoff:2020jqa}
 \begin{equation}
   t_{\rm scr} = v_{\rm scr} + r_{*{\rm s}} \approx 2 r_+ \ln\frac{r_+}{l_{\rm P}},
 \end{equation}
 where the $r_{*{\rm s}}$ term in the middle expression comes from the fact that the scrambling time in Eq.~\eqref{eq:v_scr} is defined as the minimal time needed to recover information about an object falling into the stretched horizon (at $r_* = r_{*{\rm s}}$) at a location sufficiently far from the horizon, $r_* \approx O(r_+)$.
 In the last equation, we have used $r_{*{\rm s}} \approx -2r_+ \ln(r_+/l_{\rm P})$, obtained from Eq.~\eqref{eq:r_st_s-bh} by identifying $l_{\rm s}$ with $l_{\rm P}$.
}
\begin{equation}
  t_{\rm scr} = 2r_+ \biggl[ \ln\frac{r_+}{l_{\rm P}} + O(1) \biggr].
\label{eq:t_scr}
\end{equation}
As discussed in Section~\ref{subsec:vacuum}, we assume that the mass of the black hole (as well as other quantities such as the momentum) is specified with the maximal precision allowed by the uncertainty principle, $\varDelta M \sim T_{\rm H}$.
If the state of the system involves a superposition of a black hole of a wider range of masses, then our consideration below applies to each of the branches containing a black hole with minimal uncertainties.%
\footnote{
 In practice, different branches decohere with environment, so focusing on a single branch is phenomenologically forced on us when we discuss the physics of the black hole itself, such as its interior, using the pure state language.
 Of course, when discussing more global aspects, such as the full unitarity of a black hole formation and evaporation process, we need to take into account all these branches~\cite{Page:1979tc,Nomura:2012cx}.
 However, the effect from such a superposition, i.e.\ a superposition of ``macroscopically distinguishable'' black holes, is subdominant in entropic consideration, compared with the Bekenstein-Hawking entropy associated with a black hole with minimal uncertainties.
}

Ignoring evolution effects for now, a black hole vacuum state is given as in Eq.~\eqref{eq:Psi-no-evol}.
This state picks out a set of special states of the combined system of soft and far modes:\ those multiplied by the hard-mode states $\ket{\{ n_\alpha \}}$ in Eq.~\eqref{eq:Psi-no-evol}.
We denote these states using the double-ket symbol
\begin{equation}
  \ketc{\{ n_\alpha \}_A} = \varsigma^A_n \sum_{i_n = 1}^{e^{S_{\rm bh}(M-E_n)}} \sum_{a = 1}^{e^{S_{\rm rad}}} c^A_{n i_n a} \ket{\psi^{(n)}_{i_n}} \ket{\phi_a}.
\label{eq:ketc}
\end{equation}
Here, $A$ is the index for the microstate, specified by the coefficients $c^A_{n i_n a}$ in Eq.~\eqref{eq:Psi-no-evol}, and $\varsigma_n^A$ is the normalization constant
\begin{equation}
  \varsigma^A_n = \frac{1}{\sqrt{\sum_{i_n = 1}^{e^{S_{\rm bh}(M-E_n)}} \sum_{a = 1}^{e^{S_{\rm rad}}} c^{A*}_{n i_n a} c^A_{n i_n a}}}
  = \sqrt{z}\,\, e^{\frac{E_n}{2T_{\rm H}}} \left( 1 - \frac{1}{2}\tilde{\varepsilon}_n^{AA} \right),
\label{eq:alpha_nA}
\end{equation}
where $z$ is given by Eq.~\eqref{eq:def-z}.
The last expression is obtained by using statistical properties of $c^A_{n i_n a}$ in Eq.~\eqref{eq:c-distr-BH}, and $\tilde{\varepsilon}_n^{AB}$ is the quantity analogous to $\varepsilon_n^{AB}$ in Eq.~\eqref{eq:varepsilon}:
\begin{equation}
  \tilde{\varepsilon}_n^{AB} \equiv z\, e^{\frac{E_n}{T_{\rm H}}} \sum_{i_n = 1}^{e^{S_{\rm bh}(M-E_n)}} \sum_{a = 1}^{e^{S_{\rm rad}}} c^{A*}_{n i_n a} c^B_{n i_n a} - \delta_{AB} 
  \approx O\Bigl( e^{-\frac{1}{2}\{ S_{\rm bh}(M-E_n)+S_{\rm rad} \}} \Bigr),
\end{equation}
which satisfies
\begin{equation}
  (\tilde{\varepsilon}_n^{AB})^* = \tilde{\varepsilon}_n^{BA},
\qquad
  \sum_n \frac{e^{-\frac{E_n}{T_{\rm H}}}}{z} \tilde{\varepsilon}_n^{AB} = 0.
\end{equation}

Substituting Eq.~\eqref{eq:ketc} into Eq.~\eqref{eq:Psi-no-evol}, we see that the state can be written in the thermofield double form
\begin{equation}
  \ket{\Psi_A(M)} = \frac{1}{\sqrt{z}} \sum_n e^{-\frac{E_n}{2T_{\rm H}}} \ket{\{ n_\alpha \}} \ketc{\{ n_\alpha \}_A},
\label{eq:TFD}
\end{equation}
up to exponentially suppressed corrections of order $e^{-\{S_{\rm bh}(M)+S_{\rm rad}\}/2}$.
We can also check that the states $\ketc{\{ n_\alpha \}_A}$ are orthonormal in $\{ n_\alpha \}$ as well as the microstate index $A$, up to exponentially small corrections~\cite{Nomura:2020ska}:
\begin{equation}
  \innerc{\{ m_\alpha \}_A}{\{ n_\alpha \}_B} = \delta_{m n}\, \eta_n^{AB},
\label{eq:inner-c}
\end{equation}
where
\begin{equation}
  \eta_n^{AB} \equiv \begin{cases} 
    1 & \mbox{for } A = B \\
    \tilde{\varepsilon}_n^{AB} \approx O\Bigl( e^{-\frac{1}{2}\{ S_{\rm bh}(M-E_n)+S_{\rm rad} \}} \Bigr) & \mbox{for } A \neq B.
  \end{cases}
\label{eq:eta}
\end{equation}
Since the spectrum of the states $\ket{\{ n_\alpha \}}$ represents semiclassical physics in the zone of the single-sided black hole, the corresponding states $\ketc{\{ n_\alpha \}_A}$ of the soft and far modes can be identified as the states in the second exterior of an effective two-sided black hole, given by Eq.~\eqref{eq:TFD}.

With this identification, one can construct annihilation and creation operators for modes in the second exterior
\begin{align}
  \tilde{b}_\gamma^A &= \sum_n \sqrt{n_\gamma}\, \Lketc{\{ n_\alpha - \delta_{\alpha\gamma} \}_A} \Lbrac{\{ n_\alpha \}_A}
\label{eq:ann-m-eff}\\
  &= \sum_n \sqrt{n_\gamma}\, \varsigma^A_{n_\smg} \varsigma^{A*}_n\! \sum_{i_{n_\ssmg} = 1}^{e^{S_{\rm bh}(M-E_{n_\ssmg})}} \sum_{j_n = 1}^{e^{S_{\rm bh}(M-E_n)}} \sum_{a,b = 1}^{e^{S_{\rm rad}}} c^A_{n_\smg i_{n_\ssmg} a} c^{A*}_{n j_n b} \ket{\psi^{(n_\smg)}_{i_{n_\ssmg}}} \ket{\phi_a} \bra{\psi^{(n)}_{j_n}} \bra{\phi_b},
\label{eq:ann-m-orig}\\*
  \tilde{b}_\gamma^{A\dagger} &= \sum_n \sqrt{n_\gamma + 1}\, \Lketc{\{ n_\alpha + \delta_{\alpha\gamma} \}_A} \Lbrac{\{ n_\alpha \}_A}
\label{eq:cre-m-eff}\\
  &= \sum_n \sqrt{n_\gamma + 1}\, \varsigma^A_{n_\spg} \varsigma^{A*}_n\! \sum_{i_{n_\sspg} = 1}^{e^{S_{\rm bh}(M-E_{n_\sspg})}} \sum_{j_n = 1}^{e^{S_{\rm bh}(M-E_n)}} \sum_{a,b = 1}^{e^{S_{\rm rad}}} c^A_{n_\spg i_{n_\sspg} a} c^{A*}_{n j_n b} \ket{\psi^{(n_\spg)}_{i_{n_\sspg}}} \ket{\phi_a} \bra{\psi^{(n)}_{j_n}} \bra{\phi_b},
\label{eq:cre-m-orig}
\end{align}
in addition to those in the first exterior, Eqs.~\eqref{eq:ann} and \eqref{eq:cre}.
Here, $n_{\pm\gamma} \equiv \{ n_\alpha \pm \delta_{\alpha\gamma} \}$, and we have used the same symbol $\gamma$ to specify both the first and second exterior modes.
From these operators, one can then construct annihilation and creation operators for interior modes, as well as the infalling time evolution operator, through an appropriate Bogoliubov transformation~\cite{Nomura:2018kia,Nomura:2019qps,Nomura:2019dlz,Langhoff:2020jqa,Nomura:2020ska}.%
\footnote{
 This construction differs from a similar construction in Refs.~\cite{Papadodimas:2012aq,Papadodimas:2013jku,Papadodimas:2015jra}, in which the degrees of freedom that are identified as those in the first exterior of the effective two-sided black hole increase as the black hole evaporates; in particular, Hawking radiation emitted earlier composes degrees of freedom in the first exterior.
 In the construction here, the number of degrees of freedom composing the first exterior (zone modes) decreases as the evaporation progresses; in particular, early Hawking radiation is identified as a part of the degrees of freedom in the second exterior.

 The idea that the construction of interior operators involves early Hawking radiation was promoted in Ref.~\cite{Maldacena:2013xja}, but its specific realization is different here.
 In contrast to the picture laid out in Ref.~\cite{Maldacena:2013xja}, the second exterior of the effective two-sided geometry arises primarily from degrees of freedom directly associated with the black hole (soft modes), and the involvement of Hawking radiation is indirect (although it is significant for an old black hole, i.e.\ a black hole that is nearly maximally entangled with the rest of the system).
 In particular, the structure of entanglement is not bipartite between the first exterior and early Hawking radiation degrees of freedom as envisioned in Ref.~\cite{Maldacena:2013xja}.
}
We will see this construction more explicitly below for an evaporating black hole.

In this picture, the second exterior of the black hole emerges effectively as a collective phenomenon associated with the soft and far modes.
Note that while the energy of each soft or early Hawking mode may be tiny, $\lesssim T_{\rm H}$, their collective excitation can have a much larger energy, $\gg T_{\rm H}$.
In particular, a ``quasi particle'' created by $\tilde{b}_\gamma^{A\dagger}$ in the second exterior has energy $E_\gamma$, which is negative and $|E_\gamma| > T_{\rm H}$.
The fact that this energy is negative implies that the Hamiltonian $H$ of the original single-sided black hole is mapped to the generator of the timelike isometry in the effective two-sided picture
\begin{equation}
  H \mapsto H_{\rm R} - H_{\rm L},
\label{eq:H_HR-HL}
\end{equation}
where $H_{\rm R}$ and $H_{\rm L}$ are the Hamiltonian operators acting on the first and second exteriors, respectively.
As we will see below, excitation in the black hole interior is a superposition of quasi-particle excitations built by $\tilde{b}_\gamma^A$'s and $\tilde{b}_\gamma^{A\dagger}$'s and original excitations in the zone.

It is worth mentioning that we have no freedom in choosing the basis in the space of $\ketc{\{ n_\alpha \}_A}$'s.
In other words, the interpretation that $\ketc{\{ n_\alpha \}_A}$ is a state in which $n_\alpha$ of the mode corresponding to $\alpha$ in the first exterior is excited in the second exterior, is not invariant under unitary transformation
\begin{equation}
  \ketc{\{ n_\alpha \}_A} \,\rightarrow\, \ketc{\{ n_\alpha \}_A} = \sum_{n'} U_{nn'} \ketc{\{ n'_\alpha \}_A},
\end{equation}
so that a state obtained by acting $\tilde{b}_\gamma^A$ or $\tilde{b}_\gamma^{A\dagger}$ on $\ket{\Psi_A(M)}$ in Eq.~\eqref{eq:Psi-no-evol} cannot be regarded as the vacuum state at the semiclassical level.
This is because the minimal uncertainty condition imposed on $\ket{\Psi_A(M)}$ allows us to physically distinguish between different $\ket{\{ n_\alpha \}}$'s, and hence there is no ambiguity in defining the corresponding $\ketc{\{ n_\alpha \}_A}$'s.
Our framework, therefore, does not suffer from the ``frozen vacuum'' problem posed in Ref.~\cite{Bousso:2013ifa}.

We also comment that $\ket{\phi_a}$ in Eq.~\eqref{eq:ketc} may not consist of only the early Hawking radiation; if a major portion of the radiation interacts with other degrees of freedom, such degrees of freedom must also be included in $\ket{\phi_a}$.
This is because $\ket{\phi_a}$'s represent states of all the degrees of freedom that are entangled significantly with the hard and soft modes.
It follows that the degrees of freedom that were once the second-exterior degrees of freedom keep playing that role even if they no longer take the form of Hawking radiation.
This situation, however, does not last forever.
The role as a second exterior emerges only relationally with respect to the hard modes.
Thus, after the black hole is evaporated completely, the Hawking radiation as well as any other degrees of freedom that have interacted with it become regular matter that does not have any spacetime interpretation.

Finally, we emphasize that the construction described here is performed in a theory with gravity in the bulk, implying that the precise spectrum of the horizon and zone modes (represented by the boundary conditions on the stretched horizon in the low energy field theory) are not arbitrary; rather, they are determined by a consistent UV theory.
In other words, the construction here should be viewed as being performed in a holographic boundary theory (see, e.g., footnote~\ref{ft:holo-const} below).
The physics in a nongravitational bulk can be reproduced if we send $l_{\rm P} \rightarrow 0$ (with fixed $r_+ = 2Ml_{\rm P}^2$), which implies $l_{\rm s} \rightarrow 0$ because of Eq.~(\ref{eq:lP-ls}).
The change from the distant description to the infalling description considered here is then reduced to the change of reference frames in quantum field theory (from the Rindler to Minkowski frame for $r_+ \rightarrow \infty$).

\subsubsection*{Effective theory of the interior of an evaporating black hole}

As we have seen in Section~\ref{subsec:excitation}, for an evaporating black hole some of the ingoing zone modes are not excited in the black hole vacuum; see, e.g., Eq.~\eqref{eq:sys-micro_BH-evol} and Fig.~\ref{fig:zone}.
Strictly speaking, however, the coefficients of the terms involving zone-mode states in which these modes are excited are not exactly zero (though they are exponentially suppressed); see footnote~\ref{ft:nonzero-ref}.
We can therefore define the double-ket states as in Eq.~\eqref{eq:ketc} for all $\{ n_\alpha \}$ and construct annihilation and creation operators, Eqs.~\eqref{eq:ann-m-eff}--\eqref{eq:cre-m-orig}, for the second exterior mode corresponding to any hard mode $\gamma$ in the first exterior.

Nevertheless, it is true that in the black hole vacuum state, the hard-mode states in which these ingoing modes are excited do not have sizable coefficients.
We thus have to use Eq.~\eqref{eq:ketc} in Eq.~\eqref{eq:sys-micro_BH-evol} instead of Eq.~\eqref{eq:Psi-no-evol}, and we obtain
\begin{equation}
  \ket{\Psi_A(M)} \,\,\propto\!\!\!\! \vphantom{\frac{1}{1+\frac{1}{1+\frac{1}{1+\frac{1}{1+\frac{1}{1+\frac{1}{x}}}}}}} \smash{\prod_{\vphantom{\frac{1+\frac{1}{x}}{1+\frac{1}{x}}} \smash{\bar{\alpha} \in \begin{array}{l}{\scriptstyle\rm \,\,\,\,\,\, ingoing}\\[-7pt] {\scriptstyle \!\!\!\! m_\Phi \lesssim T_{\rm H},\, \omega \sim T_{\rm H}}\end{array}}}}\!\!\!\!\!\!\!\!\!\!\!\!\! b_{\bar{\alpha}}\;\;\; \sum_n\! \vphantom{\frac{1}{1+\frac{1}{1+\frac{1}{1+\frac{1}{1+\frac{1}{1+\frac{1}{x}}}}}}} \smash{\prod_{\vphantom{\frac{1+\frac{1}{x}}{1+\frac{1}{x}}} \smash{\alpha' \in \begin{array}{l}{\scriptstyle\rm ingoing}\\[-7pt] {\scriptstyle \!\!\! \omega > \omega^{(\ell)}_{\rm barrier}}\end{array}}}}\!\!\!\!\!\!\!\!\!\!\! \delta_{n_{\alpha'},0}\, e^{-\frac{E_n}{2T_{\rm H}}} \ket{\{ n_\alpha \}} \ketc{\{ n_\alpha \}_A},
\label{eq:TFD-evol}
\end{equation}
up to exponentially small corrections.
We thus find that the ingoing modes $\alpha'$ with the frequency larger than the barrier height are virtually not entangled with the corresponding second exterior modes in the vacuum microstate.

\begin{figure}[t]
\centering
  \includegraphics[height=0.4\textwidth]{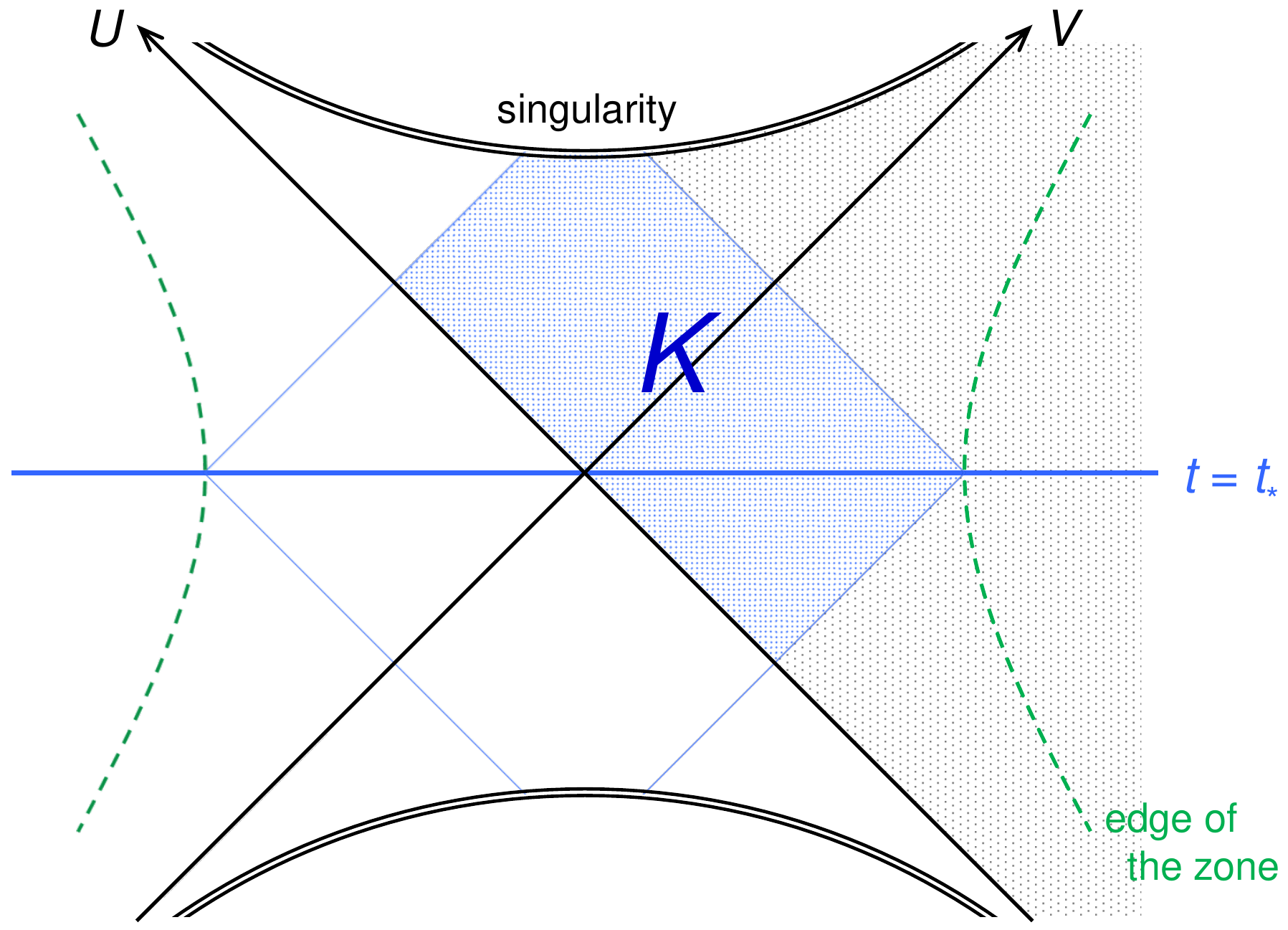}
\caption{
 The spacetime region $K$ described by the effective theory of the interior erected at a boundary (Schwarzschild) time $t = t_*$.}
\label{fig:eff-int}
\end{figure}

This implies that the spacetime is not smooth across the future horizon of the second exterior, or the past horizon of the first exterior, since the missing entanglement is essential for the connectedness of the spacetime there.
The spacetime region which the effective theory built on the state in Eq.~\eqref{eq:TFD-evol} describes is thus the shaded region in Fig.~\ref{fig:eff-int}, which we refer to as region $K$.
Here, $U$ and $V$ are the Kruskal-Szekeres coordinates erected at $t = t_*$:
\begin{equation}
  ds^2 = -dU dV + r^2 d\Omega_{d-1}^2,
\label{eq:KS-1}
\end{equation}
which are given in the near horizon region by
\begin{eqnarray}
  \begin{cases}
    U = -R\, e^{-\omega} \\
    V = R\, e^{\omega},
  \end{cases}
\qquad
  \begin{cases}
    U = R\, e^{-\omega} \\
    V = R\, e^{\omega},
  \end{cases}
\qquad
  \begin{cases}
    U = R\, e^{-\omega} \\
    V = -R\, e^{\omega},
  \end{cases}
\qquad
  \begin{cases}
    U = -R\, e^{-\omega} \\
    V = -R\, e^{\omega}
  \end{cases}
\label{eq:KS-2}
\end{eqnarray}
for Region~I ($U < 0, V > 0$), Region~II ($U, V > 0$), Region~III ($U > 0, V < 0$), and Region~IV ($U, V < 0$), respectively, with
\begin{equation}
  R = 2\sqrt{r_+|r-r_+|},
\qquad
  \omega = \frac{1}{2r_+}(t-t_*).
\label{eq:R-omega}
\end{equation}
With the state given by Eq.~\eqref{eq:TFD-evol}, the interior of the black hole is in the semiclassical vacuum in the infalling frame.

We now study in more detail how the interior region is described in this effective theory.
Let us first take the Schr\"{o}dinger picture (as we have implicitly been doing).
We want to understand what an object located in the zone and falling toward the black hole will experience as it crosses the horizon.
An excited state with $N$ particles in the zone can be obtained by applying appropriate superpositions of creation operators $b_\gamma^\dagger$ on a black hole vacuum state:
\begin{equation}
  \ket{\Psi(t=t_*)} \equiv \prod_{i=1}^{N} \left( \sum_\gamma f^{(i)}_\gamma b_\gamma^\dagger \right) \ket{\Psi_A(M)},
\end{equation}
where $\ket{\Psi_A(M)}$ is the black hole vacuum microstate in Eq.~\eqref{eq:sys-micro_BH-evol}, and $f^{(i)}$ is the wavefunction of the $i$-th particle represented in the $\gamma$ space.
This state can be straightforwardly mapped to that in the effective theory---we simply have to take $\ket{\Psi_A(M)}$ to be that in Eq.~\eqref{eq:TFD-evol} instead of Eq.~\eqref{eq:sys-micro_BH-evol}.
We denote this state in the effective theory by $\ket{\Psi(\tau = 0)}$:
\begin{equation}
  \ket{\Psi(t=t_*)} \mapsto \ket{\Psi(\tau = 0)},
\end{equation}
by choosing the origin of the infalling time $\tau$ to match the boundary time $t = t_*$.

The operators used to interpret a state in the effective theory is given by $b_\gamma$ and $b_\gamma^\dagger$ in Eqs.~\eqref{eq:ann} and \eqref{eq:cre} and $\tilde{b}_\gamma^A$ and $\tilde{b}_\gamma^{A\dagger}$ in Eqs.~\eqref{eq:ann-m-eff} and \eqref{eq:cre-m-eff}.
In particular, we can form infalling mode operators out of superpositions of these operators
\begin{align}
  a_\xi^A &= \sum_\gamma \bigl( \alpha_{\xi\gamma} b_\gamma + \beta_{\xi\gamma} b_\gamma^\dagger + \zeta_{\xi\gamma} \tilde{b}_\gamma^A + \eta_{\xi\gamma} \tilde{b}_\gamma^{A\dagger} \bigr),
\label{eq:a_xi}\\*
  a_\xi^{A\dagger} &= \sum_\gamma \bigl( \beta_{\xi\gamma}^* b_\gamma + \alpha_{\xi\gamma}^* b_\gamma^\dagger + \eta_{\xi\gamma}^* \tilde{b}_\gamma^A + \zeta_{\xi\gamma}^* \tilde{b}_\gamma^{A\dagger} \bigr),
\label{eq:a_xi-dag}
\end{align}
where $\xi$ is the label in which the frequency $\omega$ with respect to boundary time $t$ is traded with the frequency $\Omega$ associated with infalling time $\tau$, and $\alpha_{\xi\gamma}$, $\beta_{\xi\gamma}$, $\zeta_{\xi\gamma}$, and $\eta_{\xi\gamma}$ are the Bogoliubov coefficients calculable using the standard field theory method~\cite{Unruh:1976db,Israel:1976ur}.%
\footnote{
 For a massless scalar field, for example, Eq.~\eqref{eq:a_xi} takes the form
 \begin{equation}
   a_\xi^A = \pm \frac{i}{2\pi \sqrt{\Omega_\xi T_{\rm H}}} \int_0^\infty\! d\omega_\gamma \left[ \frac{\Xi}{\sqrt{1 - e^{-\frac{\omega_\gamma}{T_{\rm H}}}}}\, b_\gamma + \frac{\Xi^*}{\sqrt{e^{\frac{\omega_\gamma}{T_{\rm H}}} - 1}}\, b_\gamma^\dagger - \frac{\Xi^*}{\sqrt{1 - e^{-\frac{\omega_\gamma}{T_{\rm H}}}}}\, \tilde{b}_\gamma^A - \frac{\Xi}{\sqrt{e^{\frac{\omega_\gamma}{T_{\rm H}}} - 1}}\, \tilde{b}_\gamma^{A\dagger} \right]
 \end{equation}
 in the near horizon limit.
 Here, we have adopted the continuum notation for the sum over the frequency, and $\Xi = \left(\Omega_\xi/2\pi T_{\rm H}\right)^{\pm\frac{i\omega_\gamma}{2\pi T_{\rm H}}} \Gamma\left(1 \pm \frac{\omega_\gamma}{2\pi i T_{\rm H}}\right)/\left|\Gamma\left(1 \pm \frac{\omega_\gamma}{2\pi i T_{\rm H}}\right)\right|$ is a pure phase.
 The $\pm$ symbol in these equations takes $+$ and $-$ for ingoing and outgoing modes, respectively.
}

To obtain the generator of infalling time evolution $\tilde{H}$, we start from the generator of boundary time evolution
\begin{equation}
  H = H_{\rm soft} + \sum_\gamma \omega_\gamma b_\gamma^\dagger b_\gamma + H_{\rm int}\bigl( \{ b_\gamma \}, \{ b_\gamma^\dagger \} \bigr),
\label{eq:H}
\end{equation}
where $H_{\rm soft}$ gives time evolution of the soft modes as well as interactions between the hard and soft modes.
The infalling time evolution generator is then given by
\begin{equation}
  \tilde{H} = \sum_\xi \Omega_\xi a_\xi^{A\dagger} a_\xi^A + \tilde{H}_{\rm int}\bigl( \{ a_\xi^A \}, \{ a_\xi^{A\dagger} \} \bigr),
\label{eq:tilde-H}
\end{equation}
where $\tilde{H}_{\rm int}\bigl( \{ a_\xi^A \}, \{ a_\xi^{A\dagger} \} \bigr)$ is determined by matching it with $H_{\rm int}(\{ b_\gamma \}, \{ b_\gamma^\dagger \})$.
The state in the effective theory evolves as
\begin{equation}
  \ket{\Psi(\tau)} = e^{-i\tilde{H}\tau} \ket{\Psi(\tau = 0)}.
\end{equation}
Note that the generator in Eq.~\eqref{eq:tilde-H} contains a term that involves both $b_\gamma$ and $\tilde{b}_\gamma^A$ operators, e.g.\ $b_\gamma^\dagger \tilde{b}_{\gamma'}^A$, even in the free part.
Thus, it acts on the hard and soft (and far) modes simultaneously, as can be seen in Eqs.~\eqref{eq:ann-m-orig} and \eqref{eq:cre-m-orig}.

The effective theory is defined with the initial state on $\Sigma$, the zone at $t=t_*$, and its mirror in the second exterior $\tilde{\Sigma}$, and the spacetime is not smoothly connected across the $U$ axis (or formally has an infinite energy shock wave along the $U$ axis).
We can, therefore, trust results obtained using the theory only in the region
\begin{equation}
  K = D(\Sigma \cup \tilde{\Sigma}) \,\cap\, \{ (U,V) | V > 0 \},
\label{eq:ET-region}
\end{equation}
shown in Fig.~\ref{fig:eff-int}, where $D(X)$ represents the domain of dependence of $X$.
This is, however, all we need to describe the fate of the fallen object.
Note that here we consider the fate of an object falling into a black hole in its semiclassical vacuum.
Also, if an object leaves the region $D(\Sigma \cup \tilde{\Sigma})$ before hitting the singularity, we can choose to erect an effective theory at a different time such that the entire trajectory of the object (until it hits the singularity) is contained in the region described by the effective theory; such a choice is always possible~\cite{Nomura:2018kia}.

The description of the interior which is more along the lines of a conventional quantum field theory treatment can be obtained by adopting the Heisenberg picture~\cite{Nomura:2020ska}.
With the operators in Eqs.~\eqref{eq:a_xi} and \eqref{eq:a_xi-dag}, quantum field operators at $\tau = 0$ are given by
\begin{equation}
  \tilde{\Phi}_\Gamma({\bf x},0) = \sum_{s,\Omega,{\bf L}} \left( f_s(\Omega,{\bf L})\, \varphi_{\Omega,{\bf L}}({\bf x})\, a_\xi^A + g_s(\Omega,{\bf L})\, \varphi_{\Omega,{\bf L}}^*({\bf x})\, a_{\xi^c}^{A\dagger} \right),
\end{equation}
where we have decomposed index $\xi$ into $\Gamma$, $s$, $\Omega$, and ${\bf L}$ which represent species, spin, frequency, and orbital angular momentum quantum numbers, respectively, with $\xi^c$ representing the CPT conjugate of $\xi$.
Here, $f_s(\Omega,{\bf L})$ and $g_s(\Omega,{\bf L})$ are the standard factors providing Lorentz representation of the field (Dirac spinors, polarization vectors, etc), and $\varphi_{\Omega,{\bf L}}({\bf x})$ are the spatial wavefunctions, determined by matching $\tilde{\Phi}_\Gamma({\bf x},0)$ with quantum field operators of the original theory at $t = t_*$.%
\footnote{
 This match requires information of the coefficients in Eqs.~\eqref{eq:a_xi} and \eqref{eq:a_xi-dag}.
}

The Heisenberg picture field operators can then be defined as
\begin{equation}
  \tilde{\Phi}_\Gamma({\bf x},\tau) = e^{i\tilde{H}\tau} \tilde{\Phi}_\Gamma({\bf x},0) e^{-i\tilde{H}\tau},
\end{equation}
where $\tilde{H}$ is given in Eq.~\eqref{eq:tilde-H}.
The quantities we are interested in are the correlators
\begin{equation}
  \left\langle \tilde{\Phi}_{\Gamma_1}(x_1) \tilde{\Phi}_{\Gamma_2}(x_2) \,\cdots\, \tilde{\Phi}_{\Gamma_n}(x_n) \right\rangle = \Tr\left[ \tilde{\rho}(0) \tilde{\Phi}_{\Gamma_1}(x_1) \tilde{\Phi}_{\Gamma_2}(x_2) \,\cdots\, \tilde{\Phi}_{\Gamma_n}(x_n) \right],
\label{eq:correlators}
\end{equation}
where $x_i = \{ {\bf x}_i, \tau_i \}$, and $\tilde{\rho}(0) = \ket{\Psi(\tau = 0)}\bra{\Psi(\tau = 0)}$.
Since these are expectation values in the state at a fixed time $\tau = 0$, we must adopt the in-in formalism rather than the more conventional in-out formalism to calculate them.
This ultimately comes from the fact that the $S$-matrix cannot be defined at the semiclassical level for an object falling into a black hole.

Using the Schwinger-Keldysh method, Eq.~\eqref{eq:correlators} can be written as a path integral over an appropriate closed time contour with the boundary condition given by $\tilde{\rho}(0)$.
We can also calculate it using perturbation theory in the canonical in-in formalism.
Note that fields in Eq.~\eqref{eq:correlators} need not be in the interior of the black hole; they only need to be in the region $K$.
We can thus compute correlators between fields inside and outside the horizon using Eq.~\eqref{eq:correlators}.

In the construction of the effective theory described so far, we have used an input from semiclassical theory to determine the coefficients in Eqs.~\eqref{eq:a_xi} and \eqref{eq:a_xi-dag}.
However, we expect that this is ultimately not needed.
In particular, we expect that these coefficients are determined (though not uniquely) by the requirement that the generator $\tilde{H}$ can be written in the local form in terms of the original quantum fields~\cite{Nomura:2019qps}; namely, when $a_\xi^A$'s in $\tilde{H}$ are represented by $b_\gamma$'s and $\tilde{b}_\gamma^A$'s which in turn are represented by quantum fields and their canonical conjugates, $\tilde{H}$ takes a local form in the first and emergent second exteriors.
To find the coefficients directly from the boundary theory, one possibility is to use a physical probe to construct the infalling Hamiltonian, along the lines of Refs.~\cite{Jafferis:2020ora,Gao:2021tzr}.
We leave a detailed study of these issues for the future.

\subsubsection*{State dependence and intrinsic ambiguity}

The operators in the effective theory constructed so far, $\tilde{b}_\gamma^A$, $\tilde{b}_\gamma^{A\dagger}$, $a_\xi^A$, and $a_\xi^{A\dagger}$, depended on the vacuum microstate, indexed by $A$, on which the excited states are built.
This dependence, however, can be relaxed in such a way that a single set of operators can describe an object in the zone even if it is entangled arbitrarily with the vacuum microstates~\cite{Nomura:2020ska,Hayden:2018khn} (see also Ref.~\cite{Papadodimas:2015jra}).

The Hilbert space spanned by all the independent vacuum microstates of a black hole of mass $M$ is given by
\begin{equation}
  {\cal M} = \Biggl\{ \sum_{A=1}^{e^{S_{\rm tot}}} a_A \ket{\Psi_A(M)} \,\Bigg|\, a_A \in \mathbb{C},\, \sum_{A=1}^{e^{S_{\rm tot}}} |a_A|^2 = 1 \Biggr\},
\label{eq:cal-M}
\end{equation}
where $S_{\rm tot}$ is given by Eq.~\eqref{eq:S_tot}.
We consider a subspace of ${\cal M}$ spanned by $e^{S_{\rm eff}}$ independent microstates 
\begin{equation}
  \hat{\cal M} = \Biggl\{ \sum_{A'=1}^{e^{S_{\rm eff}}} a_{A'} \ket{\Psi_{A'}(M)} \,\Bigg|\, a_{A'} \in \mathbb{C},\, \sum_{A'=1}^{e^{S_{\rm eff}}} |a_{A'}|^2 = 1 \Biggr\},
\label{eq:hat-cal-M}
\end{equation}
where $S_{\rm eff} < S_{\rm tot}$.
By choosing the bases of ${\cal M}$ and $\hat{\cal M}$ appropriately, we can take $\{ \ket{\Psi_{A'}(M)} \}$ to be a subset of $\{ \ket{\Psi_A(M)} \}$, so $\inner{\Psi_{A'}(M)}{\Psi_{B'}(M)} = \delta_{A'B'}$.

We now define the following operators associated with the Hilbert subspace $\hat{\cal M}$
\begin{equation}
  \tilde{\cal B}_\gamma = \sum_{A'=1}^{e^{S_{\rm eff}}} \tilde{b}_\gamma^{A'},
\qquad
  \tilde{\cal B}_\gamma^\dagger = \sum_{A'=1}^{e^{S_{\rm eff}}} \tilde{b}_\gamma^{A'\dagger},
\label{eq:global-ann-cre}
\end{equation}
where $\tilde{b}_\gamma^{A'}$ and $\tilde{b}_\gamma^{A'\dagger}$ are given by Eqs.~\eqref{eq:ann-m-orig} and \eqref{eq:cre-m-orig}.
These operators will work as desired if the excited states obtained by acting $\tilde{b}_\gamma^{A'}$'s and $\tilde{b}_\gamma^{A'\dagger}$'s on $\ket{\Psi_{A'}(M)}$ are effectively orthogonal for different $A'$'s.
In fact, we can show that the algebra of these operators in the black hole Hilbert space built on $\hat{\cal M}$, i.e.\ the space spanned by the vacuum microstates of $\hat{\cal M}$ and the states in which these microstates have been excited, is the same as that of the mode operators in the second exterior of the corresponding two-sided black hole, up to correction of order~\cite{Nomura:2020ska}
\begin{equation}
  \epsilon = {\rm max}\left\{ \frac{e^{\frac{E_{\rm max}}{2T_{\rm H}}}}{e^{\frac{1}{2}\{ S_{\rm bh}(M)+S_{\rm rad} \}}}, \frac{e^{\frac{E_{\rm max}}{2T_{\rm H}}+S_{\rm eff}}}{e^{S_{\rm bh}(M)+S_{\rm rad}}} \right\},
\label{eq:epsilon}
\end{equation}
where $E_{\rm max}$ is the maximum energy which an excitation can carry in the semiclassical theory.
Given that $E_{\rm max} \ll M$, we find that the error is exponentially small for
\begin{equation}
  S_{\rm eff} \,\prec\, S_{\rm bh}(M) + S_{\rm rad} \,\approx\, S_{\rm tot},
\label{eq:S_eff-cond}
\end{equation}
where the symbol $\prec$ here means that $S_{\rm eff}$ is smaller than $S_{\rm bh}(M) + S_{\rm rad}$ and that the fractional difference between $S_{\rm eff}$ and $S_{\rm bh}(M) + S_{\rm rad}$ is not exponentially small, specifically
\begin{equation}
  S_{\rm bh}(M) + S_{\rm rad} - S_{\rm eff} \gg \frac{E_{\rm max}}{2T_{\rm H}}.
\end{equation}
Below, we use the symbols $\prec$ and $\succ$ to mean similar relations.

Similarly, we can define infalling mode operators
\begin{align}
  {\cal A}_\xi &= \sum_\gamma \bigl( \alpha_{\xi\gamma} b_\gamma + \beta_{\xi\gamma} b_\gamma^\dagger + \zeta_{\xi\gamma} \tilde{\cal B}_\gamma + \eta_{\xi\gamma} \tilde{\cal B}_\gamma^\dagger \bigr),
\label{eq:A_xi}\\*
  {\cal A}_\xi^\dagger &= \sum_\gamma \bigl( \beta_{\xi\gamma}^* b_\gamma + \alpha_{\xi\gamma}^* b_\gamma^\dagger + \eta_{\xi\gamma}^* \tilde{\cal B}_\gamma + \zeta_{\xi\gamma}^* \tilde{\cal B}_\gamma^\dagger \bigr),
\label{eq:A_xi-dag}
\end{align}
which act linearly in the black hole Hilbert space built on $\hat{\cal M}$.
Here, the coefficients $\alpha_{\xi\gamma}$, $\beta_{\xi\gamma}$, $\zeta_{\xi\gamma}$, and $\eta_{\xi\gamma}$ are the same as those in Eqs.~\eqref{eq:a_xi} and \eqref{eq:a_xi-dag}.
The matrix elements of products of ${\cal A}_\xi$ and ${\cal A}_\xi^\dagger$ in the black hole Hilbert space built on $\hat{\cal M}$, then, are the same as the corresponding field theory values on the two-sided black hole background, up to corrections suppressed by $\epsilon$ in Eq.~\eqref{eq:epsilon}.%
\footnote{
 Precisely speaking, the operators ${\cal A}_\xi$ and ${\cal A}_\xi^\dagger$ (and $\tilde{\cal B}_\gamma$ and $\tilde{\cal B}_\gamma^\dagger$) can be used for states built on a larger vacuum microstate space.
 Specifically, the algebra of these operators is the same as the corresponding semiclassical algebra in the space of black hole states built on a typical state in subspace $\hat{\cal M}'$ of ${\cal M}$ as long as ${\rm dim}(\hat{\cal M}' \cap \hat{\cal M}) \succ {\rm dim}\hat{\cal M}'/2$~\cite{Nomura:2020ska}.
}

The existence of operators ${\cal A}_\xi$ and ${\cal A}_\xi^\dagger$ allows us to erect the effective theory such that the dependence of operators on states is invisible in the effective theory.
Suppose that the state at $t = t_*$ is an entangled state between semiclassical excitations and black hole vacuum microstates
\begin{equation}
  \ket{\Psi(t_*)} = \sum_{A=1}^{S_{\rm tot}} \sum_I d_{A I}(t_*) \ket{\Psi_{A,I}(M)},
\label{eq:Psi-t*}
\end{equation}
where $\ket{\Psi_{A,I}(M)}$ represents the state in which the semiclassical excitation $I$ exists on the black hole vacuum state $\ket{\Psi_A(M)}$.
Even in this case, given that the logarithm of the dimension of the excitation Hilbert space, $S_{\rm exc}$, is much smaller than $S_{\rm tot}$, we can write the state using the Schmidt decomposition as
\begin{equation}
  \ket{\Psi(t_*)} = \sum_{I=1}^{\cal K} g_I \ket{\Psi_{A(I),I}(M)},
\label{eq:Psi-t*_Sch}
\end{equation}
where $\sum_{I=1}^{\cal K} |g_I|^2 = 1$, $g_I > 0$, and ${\cal K}$ is the Schmidt number.
The point is that since ${\cal K}$ satisfies
\begin{equation}
  {\cal K} \leq S_{\rm exc} \ll S_{\rm bh}(M) + S_{\rm rad},
\end{equation}
$\hat{\cal M}$ can always be taken to contain the space of vacuum microstates spanned by $\{ \ket{\Psi_{A(I)}(M)} | I=1,\cdots,{\cal K} \}$.
This guarantees that the effective theory respects the standard tenet of quantum mechanics that physical observables are given by linear operators acting on the Hilbert space of the theory.

Since $S_{\rm eff}$ only needs to satisfy Eq.~\eqref{eq:S_eff-cond}, one might think that a single, fixed set of ${\cal A}_\xi$, ${\cal A}_\xi^\dagger$ operators can cover the states built on most of the black hole vacuum microstates in ${\cal M}$ of Eq.~\eqref{eq:cal-M} by taking
\begin{equation}
  S_{\rm eff} = c\, \{ S_{\rm bh}(M) + S_{\rm rad} \}
\label{eq:local-Seff}
\end{equation}
with $c$ close to (but not exponentially close to) $1$.
This is, however, not the case.
The dimension of the space $\hat{\cal M}_\perp$ of vacuum microstates that are orthogonal to the states in $\hat{\cal M}$ is
\begin{equation}
  {\rm dim}\, \hat{\cal M}_\perp = e^{S_{\rm tot}}-e^{S_{\rm eff}},
\end{equation}
which is much larger than ${\rm dim}\, \hat{\cal M} = e^{S_{\rm eff}}$ even for $c$ close to $1$.
In fact, for $c > 1/2$, there is a simple relation between the fraction of ${\cal M}$ which a fixed set of operators can cover and the size of error for using these operators:
\begin{equation}
  \frac{{\rm dim}\, \hat{\cal M}}{{\rm dim}\, {\cal M}} \,\approx\, \epsilon.
\end{equation}
This makes it clear that we cannot use a fixed set of operators to cover a significant fraction of states in ${\cal M}$ while keeping the error $\epsilon$ small.
In fact, to cover all states in ${\cal M}$ by fixed sets of operators, we need double exponentially large number, $O(e^{e^{S_{\rm tot}-S_{\rm eff}}})$, of sets.%
\footnote{
 This is related to the well-known fact that in a Hilbert space of dimension $e^S \gg 1$, there are $O(e^{e^S})$ approximately orthogonal states with exponentially small overlaps of $O(e^{-S/2})$.
}

Finally, we note that the construction of the effective theory described here has an intrinsic ambiguity coming from the fact that the actions of infalling mode operators are not strictly orthogonal to $\hat{\cal M}$ in the space of black hole microstates.
Specifically, we find that the inner product between states obtained by operating mode operators on two vacuum microstates $\ket{\Psi_A(M)}$ and $\ket{\Psi_B(M)}$ reproduces the field theory value multiplied by $\delta_{AB}$, but with corrections of order $\epsilon$ in Eq.~\eqref{eq:epsilon} which are not proportional to $\delta_{AB}$.
The existence of these exponentially suppressed corrections means that an excited state cannot have an exact and unambiguous association with a unique vacuum microstate.

The fact that the corrections are only of order $\epsilon$, however, implies that up to these exponentially suppressed corrections, the mode operators $b_\gamma$, $b_\gamma^\dagger$, $\tilde{\cal B}_\gamma$, $\tilde{\cal B}_\gamma^\dagger$, ${\cal A}_\xi$, and ${\cal A}_\xi^\dagger$ act only on the excitation index $I$, and not on the vacuum index $A'$.
In other words, ignoring these corrections, the Hilbert space can be viewed as
\begin{equation}
  {\cal H} \approx {\cal H}_{\rm exc} \otimes ({\cal H}_{\rm vac} \cong \hat{\cal M}),
\label{eq:prod}
\end{equation}
where these mode operators act only on ${\cal H}_{\rm exc}$, and an excited state can be associated ``uniquely'' with a vacuum microstate.
The exponentially suppressed corrections discussed here constitute an intrinsic ambiguity of semiclassical physics, resulting from the fact that the black hole system (consisting of zone and horizon modes as well as the relevant degrees of freedom of far modes) is finite dimensional.

\subsubsection*{Consistency with the semiclassical expectation}

Suppose that a falling object hits the stretched horizon at $t = t_*$ when viewed from the exterior.
The object then disappears from the zone, but this does not mean that the state immediately becomes the quasi-equilibrium form of Eq.~\eqref{eq:sys-micro_BH-evol} with Eq.~\eqref{eq:c-distr-BH} at $t = t_*$; rather, it stays in a state with excited horizon modes for a while.
Now, consider that we erect an effective theory of the interior shortly after $t_*$:\ $t = t_* + \delta t$.
In this case, the effective theory can have semiclassical excitations in the interior reflecting the fact that there is an object that has fallen into the horizon at $t_*$.
These include everything that the object does to the spacetime region described by the effective theory; in particular, the excitations existing in the effective theory need not be the object itself.%
\footnote{
 Suppose, for example, that the object hits the stretched horizon at $t = t_*$ with almost the speed of light.
 Then, the object will not appear in the effective theory erected at $t = t_* + \delta t$, since it is squeezed into the $U$ axis so strongly that it cannot exist in the effective theory, a theory that cannot describe physics below the Planck length.
 Even in this case, however, the object may emit a high energy quantum in the outward direction shortly after it crosses the horizon.
 The quantum would then have to appear in the effective theory as a signal emerging from somewhere on the $U$ axis to the positive $V$ direction, if $\delta t$ is sufficiently small.
 (This modification of the boundary condition along the $U$ axis must occur though a map of horizon excitations in the microscopic theory to the effective theory, which requires a UV physics.)
}

On the other hand, if we erect an effective theory more than one scrambling time after the last disturbance to the stretched horizon, $t = t_* + \varDelta t$ with $\varDelta t > t_{\rm scr}$, then we expect that the effective theory finds the semiclassical vacuum in the interior, since the black hole state has already equilibrated by then.
(There can be zone mode excitations corresponding to semiclassical objects in the zone.)
Is this consistent with the semiclassical expectation?

\begin{figure}[t]
\centering
  \includegraphics[height=0.4\textwidth]{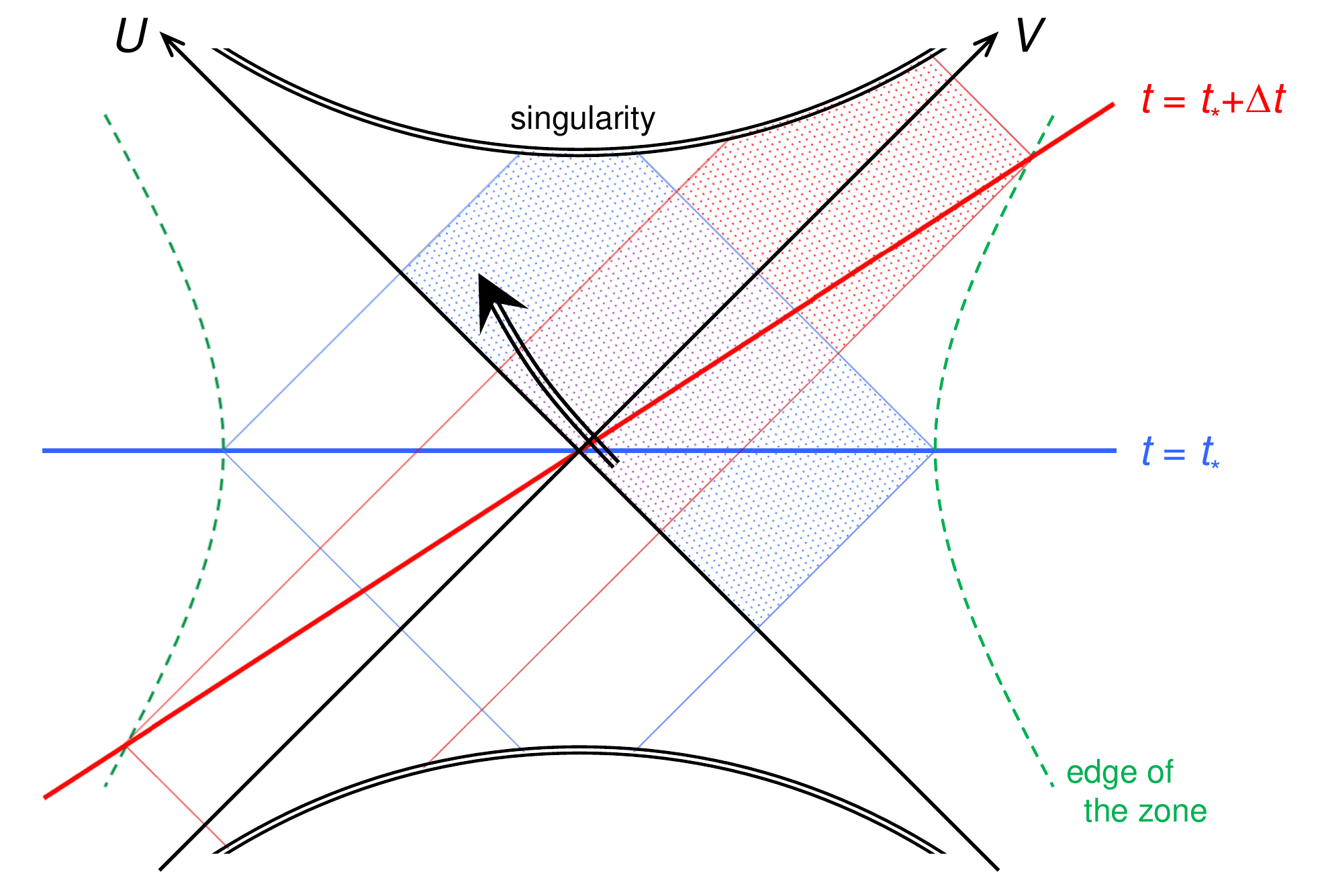}
\caption{
 The spacetime regions described by effective theories erected at $t = t_*$ (blue) and $t = t_* + \varDelta t$ with $\varDelta t > 0$ (red).
 The black arrow indicates a falling object that reaches the stretched horizon at $t = t_*$.}
\label{fig:eff-ths}
\end{figure}

In Fig.~\ref{fig:eff-ths}, we show by the central blue diamond the spacetime region described by the effective theory erected at $t = t_*$, at which the object reaches the stretched horizon.
The coordinates $U$ and $V$ are given by Eqs.~\eqref{eq:KS-2} and \eqref{eq:R-omega}.
Now, consider the Kruskal-Szekeres coordinates $\tilde{U}$ and $\tilde{V}$ adapted to the effective theory erected at $t = t_* + \varDelta t$, which are related to $U$ and $V$ by
\begin{equation}
  \tilde{U} = U e^{\frac{1}{2r_+}\varDelta t},
\qquad
  \tilde{V} = V e^{-\frac{1}{2r_+}\varDelta t}
\label{eq:KS-rel}
\end{equation}
(so that $\tilde{U} = U$ and $\tilde{V} = V$ for $\varDelta t = 0$).
The spacetime region that can be described by this theory, depicted by the red shaded region in Fig.~\ref{fig:eff-ths} for $\varDelta t > 0$, must thus satisfy
\begin{equation}
  \tilde{U} \lesssim r_{\rm z} = O(r_+)
\quad\Rightarrow\quad
  U \lesssim r_+ e^{-\frac{1}{2r_+}\varDelta t}.
\label{eq:U-bound}
\end{equation}
It is then clear that the object that fell into the horizon at $t = t_*$ spends only tiny proper time in this region, given by
\begin{equation}
  \varDelta \tau \lesssim r_+ e^{-\frac{1}{2r_+}\varDelta t}.
\end{equation}
Since holography limits the maximum amount of information an object can handle to be of $O(1)$ per Planck time, this implies that the object cannot cause any physical effect in an effective theory erected after $t = t_* + \varDelta t_{\rm max}$, with
\begin{equation}
  \varDelta t_{\rm max} = 2r_+ \ln\frac{r_+}{l_{\rm P}},
\label{eq:Dt_max}
\end{equation}
where the expression is reliable up to fractional corrections of order $1/\ln(r_+/l_{\rm P})$.
Comparing this with the scrambling time in Eq.~\eqref{eq:t_scr}, we find that the picture is indeed consistent with what we expect from semiclassical theory.

\subsubsection*{Young black hole and the role of the Page time}

The construction of the effective theory described so far applies to a black hole of any age.
However, if the black hole is young, i.e.\ if it is not maximally entangled with other systems, then we can have an alternative construction in which the $\tilde{b}_\gamma^A$ and $\tilde{b}_\gamma^{A\dagger}$ operators act only on soft-mode states, without involving far mode degrees of freedom~\cite{Nomura:2019dlz,Nomura:2020ska}.

This is done by projecting operators in Eqs.~\eqref{eq:ann-m-eff} and \eqref{eq:cre-m-eff} on the space of soft-mode states using the Petz map~\cite{Petz:1986tvy,Petz:1988usv}.
Specifically, we can take
\begin{align}
  \tilde{b}_\gamma^A &= \sigma^{-1/2} \left[ \sum_a \bra{\phi_a} \left( \sum_n \sqrt{n_\gamma}\, \Lketc{\{ n_\alpha - \delta_{\alpha\gamma} \}_A} \Lbrac{\{ n_\alpha \}_A} \right) \ket{\phi_a} \right] \sigma^{-1/2},
\label{eq:ann-young}\\*
  \tilde{b}_\gamma^{A\dagger} &= \sigma^{-1/2} \left[ \sum_a \bra{\phi_a} \left( \sum_n \sqrt{n_\gamma + 1}\, \Lketc{\{ n_\alpha + \delta_{\alpha\gamma} \}_A} \Lbrac{\{ n_\alpha \}_A} \right) \ket{\phi_a} \right] \sigma^{-1/2}.
\label{eq:cre-young}
\end{align}
Here,
\begin{equation}
  \sigma = \sum_a \bra{\phi_a} \left( \sum_n \Lketc{\{ n_\alpha \}_A} \Lbrac{\{ n_\alpha \}_A} \right) \ket{\phi_a},
\end{equation}
and $\ketc{\{ n_\alpha \}_A}$'s are given by Eq.~\eqref{eq:ketc}.
Note that $\tilde{b}_\gamma^A$, $\tilde{b}_\gamma^{A\dagger}$, and $\sigma$ are operators acting only on the space of soft-mode states.
Infalling mode operators acting only on the hard and soft modes can then be constructed by substituting these $\tilde{b}_\gamma^A$ and $\tilde{b}_\gamma^{A\dagger}$ in Eqs.~\eqref{eq:a_xi} and \eqref{eq:a_xi-dag}.

The algebra of the above mode operators in the black hole Hilbert space, i.e.\ the space obtained by acting mode operators on the vacuum state in Eq.~\eqref{eq:sys-micro_BH-evol}, follows that of semiclassical theory up to errors of order
\begin{equation}
  \epsilon_{\rm young} = {\rm max} \left\{ \frac{1}{e^{\frac{1}{2}S_{\rm bh}(M)}},\, \frac{e^{S_{\rm rad}}}{e^{S_{\rm bh}(M)}} \right\}.
\end{equation}
Therefore, if the black hole is young, i.e.\ $S_{\rm rad} < S_{\rm bh}(M)$, then the operators $\tilde{b}_\gamma^A$ and $\tilde{b}_\gamma^{A\dagger}$ in Eqs.~\eqref{eq:ann-young} and \eqref{eq:cre-young} as well as $a_\xi^A$ and $a_\xi^{A\dagger}$ constructed using them can be used to describe the interior of the black hole (up to corrections suppressed exponentially in a macroscopic entropy).
This elucidates the role of the Page time in constructing the effective theory:\ if a black hole is young, then interior operators {\it can be} represented purely using the black hole degrees of freedom (i.e.\ zone and horizon modes), while if it is old, then the operators {\it must} involve the early Hawking radiation (i.e.\ far modes).

The promotion of $\tilde{b}_\gamma^A$, $\tilde{b}_\gamma^{A\dagger}$, $a_\xi^A$, and $a_\xi^{A\dagger}$ operators to act linearly in $\hat{\cal M}$ can be made similarly as before.
In particular, errors of the promoted operators are of order
\begin{equation}
  \hat{\epsilon}_{\rm young} = {\rm max}\left\{ \frac{1}{e^{\frac{1}{2}S_{\rm bh}(M)}},\, \frac{e^{S_{\rm rad}+S_{\rm eff}}}{e^{S_{\rm bh}(M)}} \right\},
\label{eq:eps-psi}
\end{equation}
so that these operators work correctly as long as
\begin{equation}
  S_{\rm eff} \prec S_{\rm bh}(M) - S_{\rm rad}.
\end{equation}
This agrees with the result of the general analysis in Ref.~\cite{Hayden:2018khn}.

Incidentally, in the way of constructing an effective theory of the interior described here, operators $\tilde{b}_\gamma^A$ and $\tilde{b}_\gamma^{A\dagger}$ cannot be represented using only far modes, even if the black hole is old.
Namely, unlike Eqs.~\eqref{eq:ann-young} and \eqref{eq:cre-young}, projecting operators in Eqs.~\eqref{eq:ann-m-eff} and \eqref{eq:cre-m-eff} on the space of far-mode states does not work.
Technically, this is because of the energy constraint imposed on the black hole, i.e.\ the combined system of zone and horizon modes.
The relation of this statement to entanglement wedge reconstruction, in which the interior of an old black hole is reconstructed only using the early Hawking radiation, will be discussed below.

\subsubsection*{Relation to entanglement wedge reconstruction}

As we have seen, operators in our effective theory describing the interior must involve horizon degrees of freedom, regardless of the age of the black hole.
On the other hand, the analysis~\cite{Penington:2019npb,Almheiri:2019psf,Almheiri:2019hni,Penington:2019kki,Almheiri:2019qdq} based on holographic entanglement wedge reconstruction~\cite{Czech:2012bh,Wall:2012uf,Headrick:2014cta,Jafferis:2015del,Dong:2016eik,Cotler:2017erl} says that after the Page time operators acting on early radiation are sufficient to reconstruct a portion of the black hole interior.
What is the relation between these two statements?

A key ingredient to understanding this is the boundary time evolution~\cite{Langhoff:2020jqa}.
In general, entanglement wedge reconstruction assumes that we know the time evolution operator of the boundary theory; in models discussed in Refs.~\cite{Penington:2019npb,Almheiri:2019psf,Almheiri:2019hni,Penington:2019kki,Almheiri:2019qdq}, for example, the Hamiltonian of a system consists of boundary conformal field theory as well as an auxiliary theory coupling to it.
With this knowledge, one can reconstruct the state of some of the horizon modes at $t = t_{\rm h}$ from the state of the radiation at $t = t_R$ if
\begin{equation}
  t_R > t_{\rm h} + t_{\rm scr}.
\label{eq:th-cond}
\end{equation}
Note that as discussed in Section~\ref{subsec:UV}, the notion of zone, horizon, and far modes is associated with a specific time $t$, so that a component of a horizon mode at some time can become a far mode at a later time.

\begin{figure}[t]
\centering
  \subfloat[]{{\includegraphics[width=0.49\textwidth]{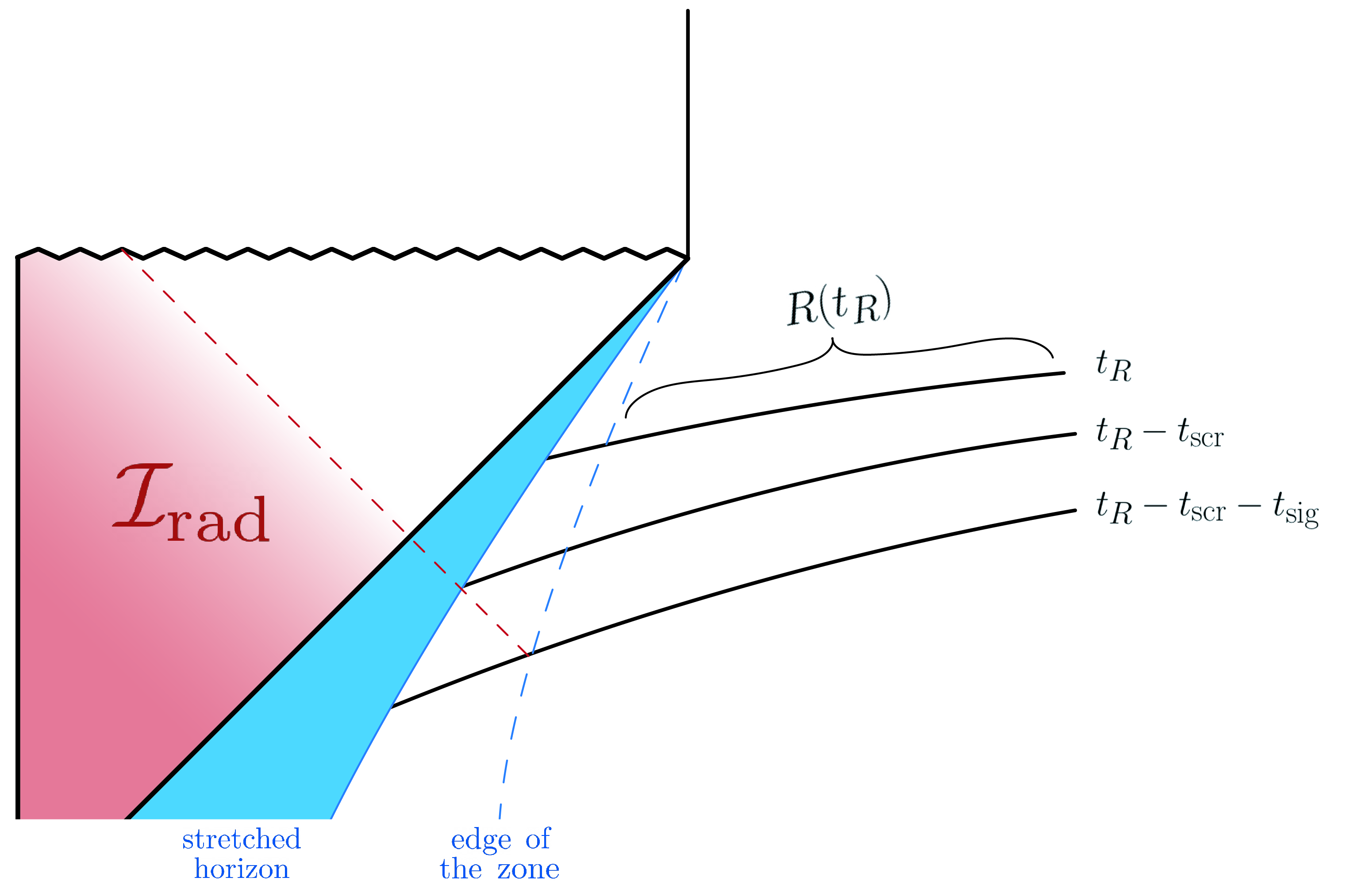}}}
\hspace{2mm}
  \subfloat[]{{\includegraphics[width=0.49\textwidth]{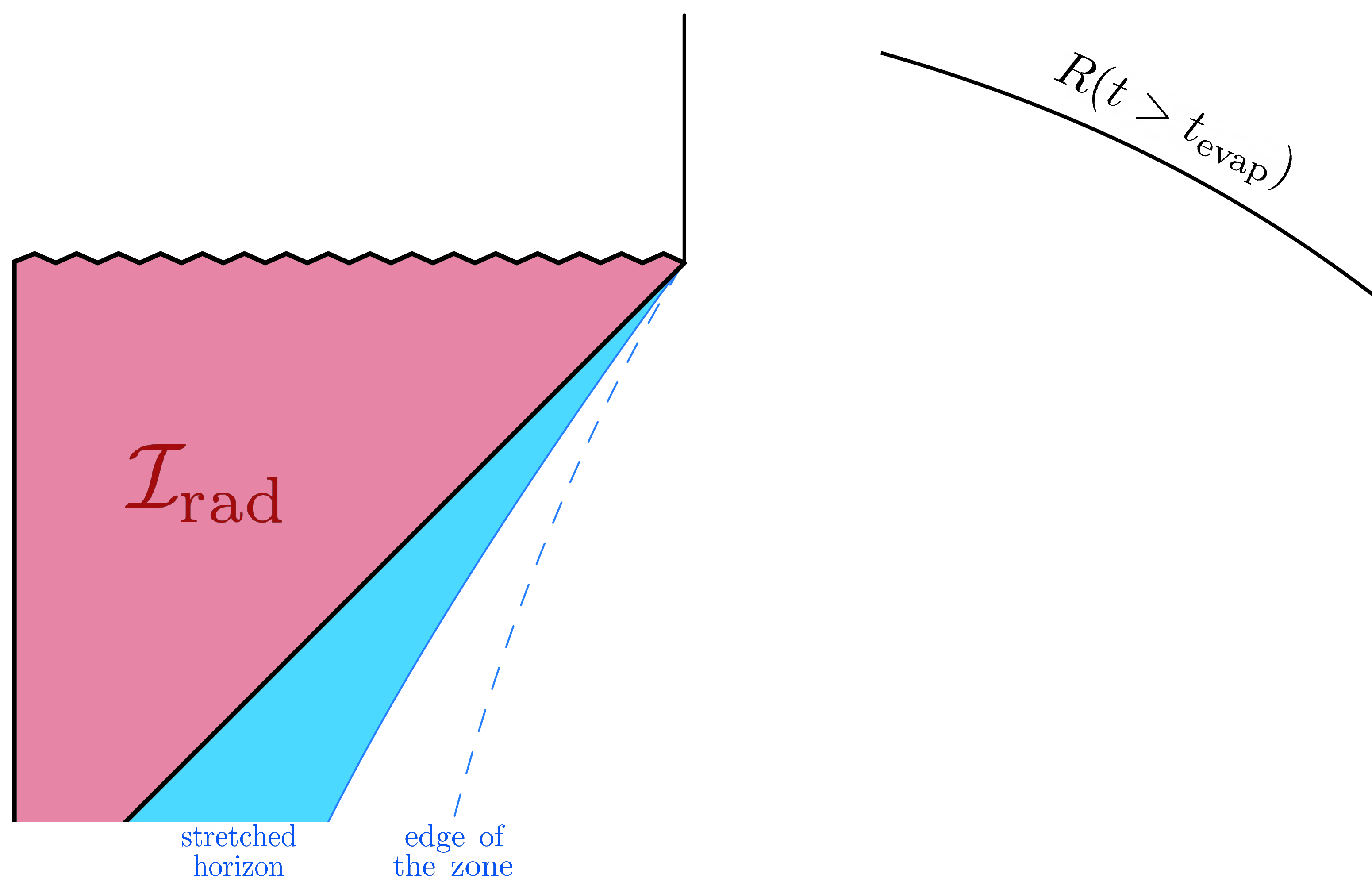}}}
\caption{
 (a) Radiation at $t = t_R$, $R(t_R)$, contains information about some of the horizon and zone modes at $t < t_R - t_{\rm scr} - t_{\rm sig}$.
 This allows us to reconstruct a portion ${\cal I}_{\rm rad}$ of the interior spacetime corresponding to the inner wedge of the stretched horizon at $t = t_R - t_{\rm scr}$.
 The amount of information that can be reconstructed, however, depends on the location of an object in ingoing time, which is indicated by the gradation of the red shade.
 (b) Radiation after the black hole is fully evaporated, $R(t > t_{\rm evap})$, allows for reconstruction of the full interior region.}
\label{fig:wedge}
\end{figure}

A detailed way in which the reconstruction described above works was discussed in Ref.~\cite{Langhoff:2020jqa}.
It is essentially the Hayden-Preskill protocol~\cite{Hayden:2007cs} applied to the horizon and zone modes.
Recall that zone modes at $t = t_{\rm h} - t_{\rm sig}$ either become horizon modes (for ingoing modes) or far modes (for outgoing modes) by $t = t_{\rm h}$.
Here, $t_{\rm sig} \approx |r_{*{\rm s}}|$ is the signal propagation time:\ the time it takes for a massless quantum to propagate from the stretched horizon to the edge of the zone.
Hence, radiation at $t_R$ satisfying Eq.~\eqref{eq:th-cond} can reconstruct some of the zone modes.
In fact, we can arbitrarily choose which zone modes to reconstruct, and we can take them to be hard modes.
This implies that radiation at $t = t_R$ can reconstruct an object in the interior of the effective theory erected at $t=t_*$ if
\begin{equation}
  t_* < t_R - t_{\rm scr} - t_{\rm sig}
\label{eq:reconst-1}
\end{equation}
and $t_* + t_{\rm scr} > t_{\rm Page}$.
In other words, an object in the black hole interior can be reconstructed from radiation at $t = t_{\rm R}$ if it exists in the inner wedge of the stretched horizon at
\begin{equation}
  t = t_R - t_{\rm scr},
\label{eq:reconst-2}
\end{equation}
assuming $t_R > t_{\rm Page}$.
This is sketched in Fig.~\ref{fig:wedge}(a).
The entanglement island, ${\cal I}_{\rm rad}$, of the radiation, $R(t_R)$, represents the spacetime region in which some information about the region is reconstructed.
Thus, its edge, i.e.\ the minimal quantum extremal surface for $R(t_R)$, is located near the stretched horizon at $t = t_R - t_{\rm scr}$.
Given that the discussion for reconstruction here is intrinsically semiclassical, and hence does not have a precision of resolving the Planck scale, this is consistent with what was found in Refs.~\cite{Penington:2019npb,Almheiri:2019psf,Almheiri:2019hni}.

We stress that while the entanglement island ${\cal I}_{\rm rad}$ represents the interior region which one can reconstruct from the radiation $R(t_R)$, the amount of information that can be reconstructed, or the size of code subspace~\cite{Almheiri:2014lwa,Pastawski:2015qua} one can construct, depends on the location in this region.
Suppose one wants to reconstruct an interior object that has a significant amount of information, say having (coarse-grained) entropy $S_{\rm obj}$.
Then, the radiation must contain the corresponding amount of information.
This modifies Eqs.~\eqref{eq:reconst-1} and \eqref{eq:reconst-2} to
\begin{equation}
  t_* < t_R - t_{\rm scr} - t_{\rm sig} - \frac{S_{\rm obj}}{T_{\rm H}}
\label{eq:tst-cond}
\end{equation}
and
\begin{equation}
  t = t_R - t_{\rm scr} - \frac{S_{\rm obj}}{T_{\rm H}},
\end{equation}
where $t_R$ must satisfy $t_R > t_{\rm Page} + S_{\rm obj}/T_{\rm H}$.
This feature inherits from the use of the Hayden-Preskill protocol, and more fundamentally boundary time evolution, in the reconstruction.
Because of the scrambling and quantum error correcting nature of black hole dynamics, one can arbitrarily choose {\it which} information to reconstruct~\cite{Hayden:2007cs}, but the {\it amount} of information to be reconstructed is bounded from above by
\begin{equation}
  S_{\rm obj} \leq T_{\rm H} (t_R - t_{\rm scr} - v_{\rm obj}),
\label{eq:S_obj-bound}
\end{equation}
where $v_{\rm obj}$ is the ingoing Eddington-Finkelstein time $v$ at the location of the object, chosen so that $v = t$ at the stretched horizon.
This is represented by the gradation of the red shade in Fig.~\ref{fig:wedge}(a).

The discussion above makes it clear that entanglement wedge reconstruction is nothing but the statement of unitarity as viewed from the exterior.
Indeed, the entanglement island of radiation after the black hole is fully evaporated can be viewed as the entire interior; see Fig.~\ref{fig:wedge}(b).%
\footnote{
 Strictly speaking, the existence of an island in this case is not rigorously established, since its edge cannot be obtained as a surface that is quantum extremized in the regime in which a semiclassical description is valid.
 However, from physical consideration as well as continuity of ${\cal I}_{\rm rad}$ in $t$, we expect that $R(t)$ ($t > t_{\rm evap}$) has the entire interior region as the ``island.''
}

The understanding of entanglement wedge reconstruction described above also elucidates why an operator acting only on radiation $R(t_R)$, in particular a unitary operator representing a physical manipulation of $R(t_R)$, can affect the interior of a black hole ${\cal I}_{\rm rad}$ in a seemingly acausal manner.
In the manifestly unitary picture adopted so far in this paper, based on an external view of the black hole, an excitation $X$ in the interior ${\cal I}_{\rm rad}$ is, in fact, ``located'' in the zone and/or stretched horizon at $t < t_R$ (and possibly in the early radiation as well for an old black hole), which is timelike separated from $R(t_R)$.
There is, therefore, no a~priori reason why the operator generating $X$ commutes with operators acting on the radiation at $t = t_R$.
In particular, an operator ${\cal O}_R$ acting on $R(t_R)$ can be a  ``precursor''~\cite{Polchinski:1999yd,Giddings:2001pt} of $X$; i.e., when evolved backward in time, ${\cal O}_R$ induces a change of the state of an object falling into the black hole at an earlier time, which changes the interior of the black hole.

On the other hand, in the global spacetime picture of general relativity, it is still true that radiation at $t=t_R$ is spacelike separated from the interior region ${\cal I}_{\rm rad}$.
This raises the question of why there can be any unitary operator acting on $R(t_R)$ which affects ${\cal I}_{\rm rad}$, without completely jeopardizing the intuition coming from the semiclassical, global spacetime picture.
This issue was studied in Ref.~\cite{Kim:2020cds}, in which it was argued that if the dynamics of a black hole is sufficiently complex (e.g.\ leading to a pseudorandom state), then the information about the interior ${\cal I}_{\rm rad}$ cannot be accessed by any simple operator acting on $R(t_R)$ which does not have exponential computational complexity.
This reproduces the causality of the semiclassical theory, assuming that it represents only a feature of simple operations performed on the system in each observer frame.

In the present context, the required complexity arises from the dynamics of the stretched horizon as viewed from the exterior of the black hole.
Note that the time $t_*$ at which the effective theory describing ${\cal I}_{\rm rad}$ is erected must satisfy Eq.~(\ref{eq:tst-cond}).
This implies that ingoing zone modes (as well as most of the horizon modes) at $t = t_*$ must go through the complex horizon dynamics before they become radiation modes at $t = t_R$.
In particular, this is true for the ingoing hard modes for any choice of hard modes describing the object carrying the entropy $S_{\rm obj}$.
The semiclassical picture emerging from our framework, therefore, respects causality, at least in the sense of Ref.~\cite{Kim:2020cds}.

In entanglement wedge reconstruction, reconstructing the interior on a Hawking radiation state involves backward time evolution, so that the resulting boundary operators acting on the radiation degrees of freedom are highly fragile; i.e., a small deformation of the operators destroys the success of the reconstruction.
In the case of the effective theory, on the other hand, interior operators are given simply by Eqs.~\eqref{eq:ann-m-orig}, \eqref{eq:cre-m-orig}, \eqref{eq:a_xi}, and \eqref{eq:a_xi-dag} in terms of hard, soft, and far mode operators at time $t_*$, which can be obtained easily in the boundary theory from the structure of entanglement of the whole state at $t = t_*$, if such a state is given.%
\footnote{
 A specific construction can go as follow.
 We can fist construct zone mode operators $b_\gamma$ and $b_\gamma^\dagger$ using the HKLL procedure~\cite{Hamilton:2005ju,Hamilton:2006az,Heemskerk:2012mn} or its extension~\cite{Engelhardt:2021mue}.
 We can then expand the full state $\ket{\Psi(t_*)}$ at $t = t_*$ in terms of the eigenstates of number operators $b_\alpha^\dagger b_\alpha$'s and identify the $\ketc{\{ n_\alpha \}}$ states associated with $\ket{\Psi(t_*)}$.
 This allows us to define $\tilde{b}_\gamma$ and $\tilde{b}_\gamma^\dagger$ operators acting on $\ketc{\{ n_\alpha \}}$'s, and hence infalling operators $a_\xi$ and $a_\xi^\dagger$ through Eqs.~\eqref{eq:a_xi} and \eqref{eq:a_xi-dag}.
 The effective theory of the interior (erected at $t = t_*$) can then be obtained using the generator $\tilde{H}$ of time evolution, given in Eq.~\eqref{eq:tilde-H}.
\label{ft:holo-const}}
In this ``equal-time'' conversion between the distant and infalling descriptions, which use modes at a given time $t_*$ directly, there is no upper limit on the amount of information about the infalling object described, such as that in Eq.~(\ref{eq:S_obj-bound}); the only restriction on the size of the object is the one coming from the validity of the semiclassical picture.

\subsection{The outside of a de~Sitter static patch}

A construction analogous to the black hole interior can be applied to a state describing de~Sitter spacetime~\cite{Nomura:2019qps}.
As we have argued in Section~\ref{subsec:holo-dS}, a microstate representing empty de~Sitter spacetime consists of the state of the ``zone modes'' inside a static patch as well as that of the horizon modes.
Thus, by choosing a subset of the zone modes as hard modes, it can be written in the form of Eq.~\eqref{eq:sys-microstate_dS} with $S(E-E_n) \rightarrow S_{\rm dS}(E-E_n)$.
Here, $S_{\rm dS}(E)$ is given by Eq.~\eqref{eq:S_dS}, and $E$ is the ``energy'' given in terms of the Hubble radius $\alpha$ by $E = \alpha/l_{\rm P}^2$; see Eq.~\eqref{eq:E_dS}.
As in the case of a black hole, we can define the ``double-ket'' states using soft modes
\begin{equation}
  \ketc{\{ n_\alpha \}_A} = \varsigma^A_n \sum_{i_n = 1}^{e^{S_{\rm dS}(E-E_n)}} c^A_{n i_n} \ket{\psi^{(n)}_{i_n}},
\label{eq:ketc-dS}
\end{equation}
where $A = 1,\cdots,e^{S_{\rm sys}}$, and $\varsigma_n^A$ is the normalization constant
\begin{equation}
  \varsigma^A_n = \frac{1}{\sqrt{\sum_{i_n = 1}^{e^{S_{\rm dS}(E-E_n)}} c^{A*}_{n i_n} c^A_{n i_n}}}
  = \sqrt{z}\,\, e^{\frac{E_n}{2T_{\rm H}}} \left[ 1 + O\Bigl( e^{-\frac{1}{2}S_{\rm dS}(E-E_n)} \Bigr) \right],
\end{equation}
where $T_{\rm H} = 1/2\pi\alpha$, and $z$ is given by Eq.~\eqref{eq:def-z}.
In terms of these states, the de~Sitter vacuum microstate can be rewritten as
\begin{equation}
  \ket{\Psi_A(E)} = \frac{1}{\sqrt{z}} \sum_n e^{-\frac{E_n}{2T_{\rm H}}} \ket{\{ n_\alpha \}} \ketc{\{ n_\alpha \}_A}.
\label{eq:TFD-dS}
\end{equation}

Since the state $\ket{\{ n_\alpha \}}$ is specified by the occupation numbers $n_\alpha$ of modes inside the stretched horizon of a static patch, which we refer to as the polar region here, the thermofield double state in Eq.~\eqref{eq:TFD-dS} represents a vacuum state at $\tau = 0$ of global de~Sitter spacetime of which the original static patch is a portion.
Here, $\tau$ is global de~Sitter time, or time associated with closed slicing.
This implies that $\ketc{\{ n_\alpha \}_A}$ should be identified as the state of the other hemisphere of the static patch at $\tau = 0$.
States in which modes in the polar region are excited or deexcited are those obtained by acting corresponding creation operators Eq.~\eqref{eq:cre} or annihilation operators Eq.~\eqref{eq:ann}, respectively, to a de~Sitter vacuum microstate.%
\footnote{
 Deexcited states become relevant if Gibbons-Hawking radiation is extracted by a material.
}
These states are naturally mapped to excited states in the effective theory built on Eq.~\eqref{eq:TFD-dS}.

As in the case of a black hole, we can introduce annihilation and creation operators for modes in the other hemisphere in the effective theory as
\begin{align}
  \tilde{b}_\gamma^A &= \sum_n \sqrt{n_\gamma}\, \varsigma^A_{n_\smg} \varsigma^{A*}_n\! \sum_{i_{n_\ssmg} = 1}^{e^{S_{\rm dS}(E-E_{n_\ssmg})}} \sum_{j_n = 1}^{e^{S_{\rm dS}(E-E_n)}} c^A_{n_\smg i_{n_\ssmg}} c^{A*}_{n j_n} \ket{\psi^{(n_\smg)}_{i_{n_\ssmg}}} \bra{\psi^{(n)}_{j_n}},
\label{eq:ann-m-orig-dS}\\*
  \tilde{b}_\gamma^{A\dagger} &= \sum_n \sqrt{n_\gamma + 1}\, \varsigma^A_{n_\spg} \varsigma^{A*}_n\! \sum_{i_{n_\sspg} = 1}^{e^{S_{\rm dS}(E-E_{n_\sspg})}} \sum_{j_n = 1}^{e^{S_{\rm dS}(E-E_n)}} c^A_{n_\spg i_{n_\sspg}} c^{A*}_{n j_n} \ket{\psi^{(n_\spg)}_{i_{n_\sspg}}} \bra{\psi^{(n)}_{j_n}},
\label{eq:cre-m-orig-dS}
\end{align}
which can be used to form annihilation and creation operators for global time slicing
\begin{align}
  a_\xi^A &= \sum_\gamma \bigl( \alpha_{\xi\gamma} b_\gamma + \beta_{\xi\gamma} b_\gamma^\dagger + \zeta_{\xi\gamma} \tilde{b}_\gamma^A + \eta_{\xi\gamma} \tilde{b}_\gamma^{A\dagger} \bigr),
\label{eq:a_xi-dS}\\*
  a_\xi^{A\dagger} &= \sum_\gamma \bigl( \beta_{\xi\gamma}^* b_\gamma + \alpha_{\xi\gamma}^* b_\gamma^\dagger + \eta_{\xi\gamma}^* \tilde{b}_\gamma^A + \zeta_{\xi\gamma}^* \tilde{b}_\gamma^{A\dagger} \bigr),
\label{eq:a_xi-dag-dS}
\end{align}
where $\xi$ is the label in which the frequency $\omega$ with respect to static time $t$ is traded with the frequency $\Omega$ associated with global time $\tau$, and $\alpha_{\xi\gamma}$, $\beta_{\xi\gamma}$, $\zeta_{\xi\gamma}$, and $\eta_{\xi\gamma}$ are the Bogoliubov coefficients (which, of course, differ from the black hole case).
The correspondence between the black hole and de~Sitter cases is summarized in Table~\ref{tab:corresp}. 
\begin{table}[t]
\begin{center}
\begin{tabular}{rc|c}
  & \mbox{Evaporating black hole} & \mbox{Cosmological de~Sitter spacetime} \\
\hline
  \multirow{ 3}{*}{\mbox{microscopic level} $\left\{ \begin{matrix} {} \\ {} \\ {} \end{matrix} \right.$} 
  & \mbox{zone region} & \mbox{polar region} \\
  & \mbox{stretched horizon} & \mbox{stretched horizon} \\
  & \mbox{far region} & \mbox{-----} \\
\hline
  \multirow{ 2}{*}{\mbox{effective theory} $\left\{ \begin{matrix} {} \\ {} \end{matrix} \right.$} 
  & \mbox{two-sided black hole} & \mbox{global de~Sitter spacetime} \\
  & \mbox{the second exterior} & \mbox{the other hemisphere}
\end{tabular}
\end{center}
\caption{
 Correspondence between an evaporating black hole and cosmological de~Sitter spacetime.}
\label{tab:corresp}
\end{table}
Note that for de~Sitter spacetime, there is no region corresponding to the region outside the zone of an evaporating black hole.

Physical quantities in global de~Sitter spacetime can be calculated using the global time evolution operator
\begin{equation}
  U(\tau) = e^{-i\tilde{H}\tau},
\end{equation}
where
\begin{equation}
  \tilde{H} = \sum_\xi \Omega_\xi a_\xi^{A\dagger} a_\xi^A + \tilde{H}_{\rm int}\bigl( \{ a_\xi^A \}, \{ a_\xi^{A\dagger} \} \bigr).
\label{eq:tilde-H-dS}
\end{equation}
In the Heisenberg picture, this can be done by evolving quantum fields $\tilde{\Phi}_a({\bf x},0)$ formed from $a_\xi^A$ and $a_\xi^{A\dagger}$ at $\tau = 0$ as
\begin{equation}
  \tilde{\Phi}_a({\bf x},\tau) = U(\tau)^\dagger \tilde{\Phi}_a({\bf x},0) U(\tau)
\end{equation}
and sandwiching their products by the state $\ket{\Psi(\tau = 0)}$ of the effective theory at $\tau = 0$ obtained by matching with the state of the microscopic theory at $t=0$.
($\ket{\Psi(\tau = 0)} = \ket{\Psi_A(E)}$ if the system is in the semiclassical vacuum.)
In the Schr\"{o}dinger picture, the state must be evolved with $U(\tau)$ while $\tilde{\Phi}_a({\bf x},0)$ are inserted at intermediate stages of the evolution.

The promotion of operators to a less state-dependent form can be made in a similar way to the back hole case, Eqs.~\eqref{eq:global-ann-cre}, \eqref{eq:A_xi} and \eqref{eq:A_xi-dag}.
The analysis of errors of operator algebras can be performed similarly.
In particular, the global description in the effective theory has an intrinsic error of order
\begin{equation}
  \epsilon = {\rm max}\left\{ \frac{e^{\frac{E_{\rm max}}{2T_{\rm H}}}}{e^{\frac{1}{2}S_{\rm GH}}}, \frac{e^{\frac{E_{\rm max}}{2T_{\rm H}}+S_{\rm eff}}}{e^{S_{\rm GH}}} \right\},
\label{eq:epsilon-dS}
\end{equation}
where $S_{\rm GH}$ is the Gibbons-Hawking entropy, $E_{\rm max}$ is the maximum energy that an excitation can carry in the effective theory, and $S_{\rm eff}$ represents the size of the microscopic Hilbert space covered by the promoted operators, as defined in Eq.~\eqref{eq:hat-cal-M}.

Like the case of a black hole, the effective theory of global de~Sitter spacetime is intrinsically semiclassical in that the algebra of operators in the theory is defined only up to an uncertainty of $e^{-S_{\rm GH}/2}$.
This is consistent with the observation in Ref.~\cite{Goheer:2002vf} that symmetries of classical de~Sitter spacetime cannot be implemented exactly in a finite-dimensional Hilbert space.

\subsubsection*{Consistency with the semiclassical expectation}

\begin{figure}[t]
\begin{center}
  \includegraphics[width=0.4\textwidth]{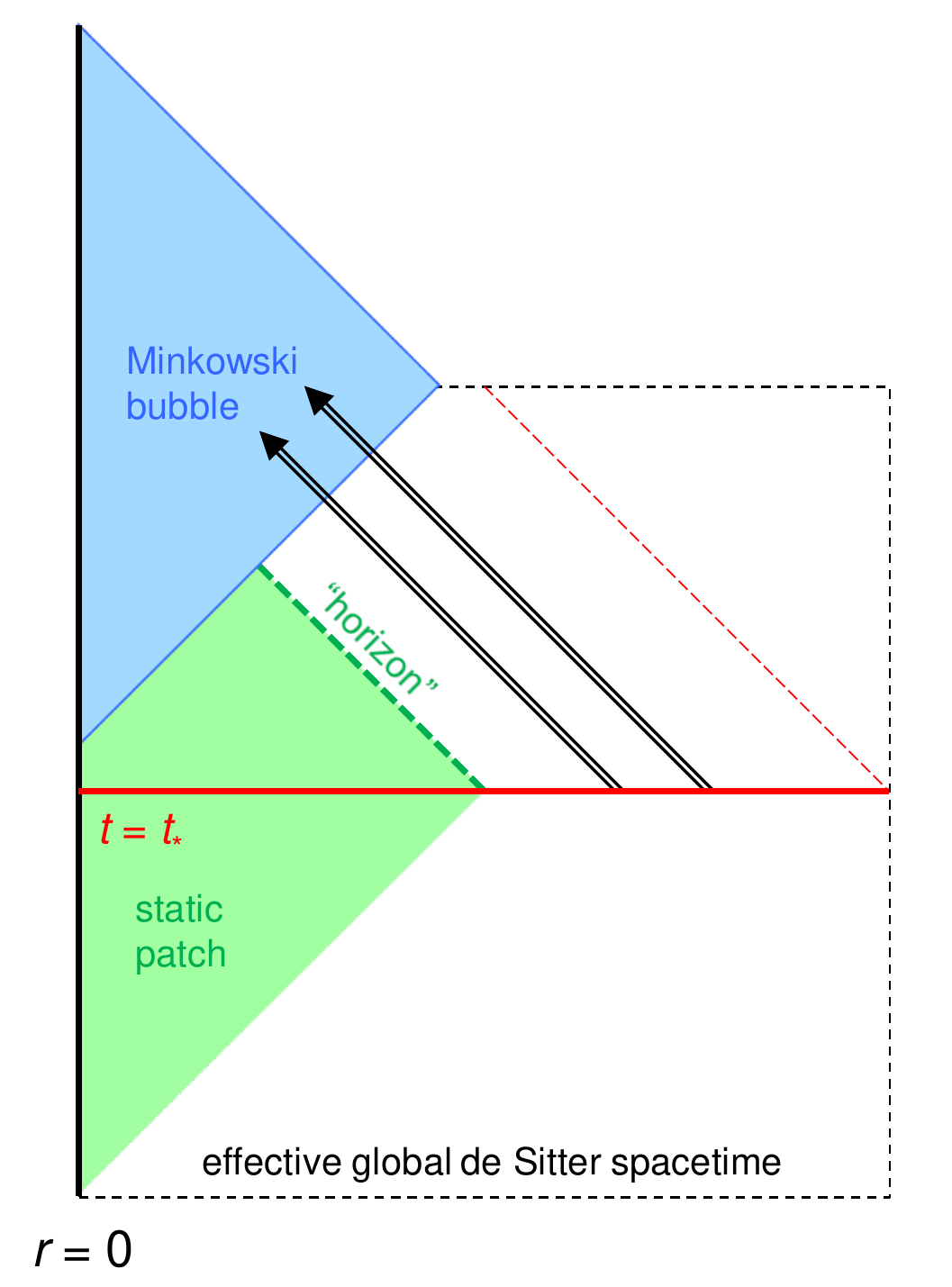}
\vspace{-2mm}
\end{center}
\caption{
 Information outside the static patch in the effective theory can be retrieved if the system leaves the de~Sitter phase at a later time, for example by tunneling into a Minkowski bubble universe.}
\label{fig:dS-Minkowski}
\end{figure}

Suppose an object hits the stretched horizon in the static patch description at some time $t = t_{\rm obj}$.
The state shortly after it will not take the form of Eq.~\eqref{eq:sys-microstate_dS} with random $c^A_{n i_n}$ which leads to the vacuum state of Eq.~\eqref{eq:TFD-dS}.
Instead, it stays in an excited state for a while, reflecting the existence of an object just outside the horizon.
In the effective theory, this is represented as the existence of an excitation in the other hemisphere at $\tau = 0$.
Such an excitation has a physical significance, since we can retrieve (a part of) information about it if the system leaves the de~Sitter phase after $t = t_{\rm obj}$, e.g.\ by tunneling into a Minkowski vacuum; see Fig.~\ref{fig:dS-Minkowski}.

According to the static patch description, the system is expected to relax into an equilibrium state of the form of Eq.~\eqref{eq:sys-microstate_dS} when one scrambling time%
\footnote{
 To obtain this expression, we can consider an analogue of ingoing Eddington-Finkelstein time in de~Sitter spacetime, $v = t - r_*$, and use the general expression for scrambling time to obtain
 \begin{equation}
   v_{\rm scr} = \frac{1}{2\pi T_{\rm H}} \ln S_{\rm dS} \approx 2 \alpha \ln\frac{\alpha}{l_{\rm P}}.
 \end{equation}
 The scrambling time in static time is then
 \begin{equation}
   t_{\rm scr} = v_{\rm scr} - r_{*{\rm s}} \approx \alpha \ln\frac{r_+}{l_{\rm P}},
 \end{equation}
 where we have used $r_{*{\rm s}} \approx \alpha \ln(r_+/l_{\rm P})$, obtained from Eq.~\eqref{eq:r_st_s-dS} by identifying $l_{\rm s}$ with $l_{\rm P}$.
}
\begin{equation}
  t_{\rm scr} = \alpha \biggl[ \ln\frac{\alpha}{l_{\rm P}} + O(1) \biggr]
\label{eq:t_scr-dS}
\end{equation}
has passed after the object hits the stretched horizon.
This implies that if we erect the effective theory at $t_* > t_{\rm obj} + t_{\rm scr}$, then the effective theory sees the vacuum in the other hemisphere.
Is this consistent with the semiclassical expectation?

\begin{figure}[t]
\begin{center}
  \includegraphics[width=0.5\textwidth]{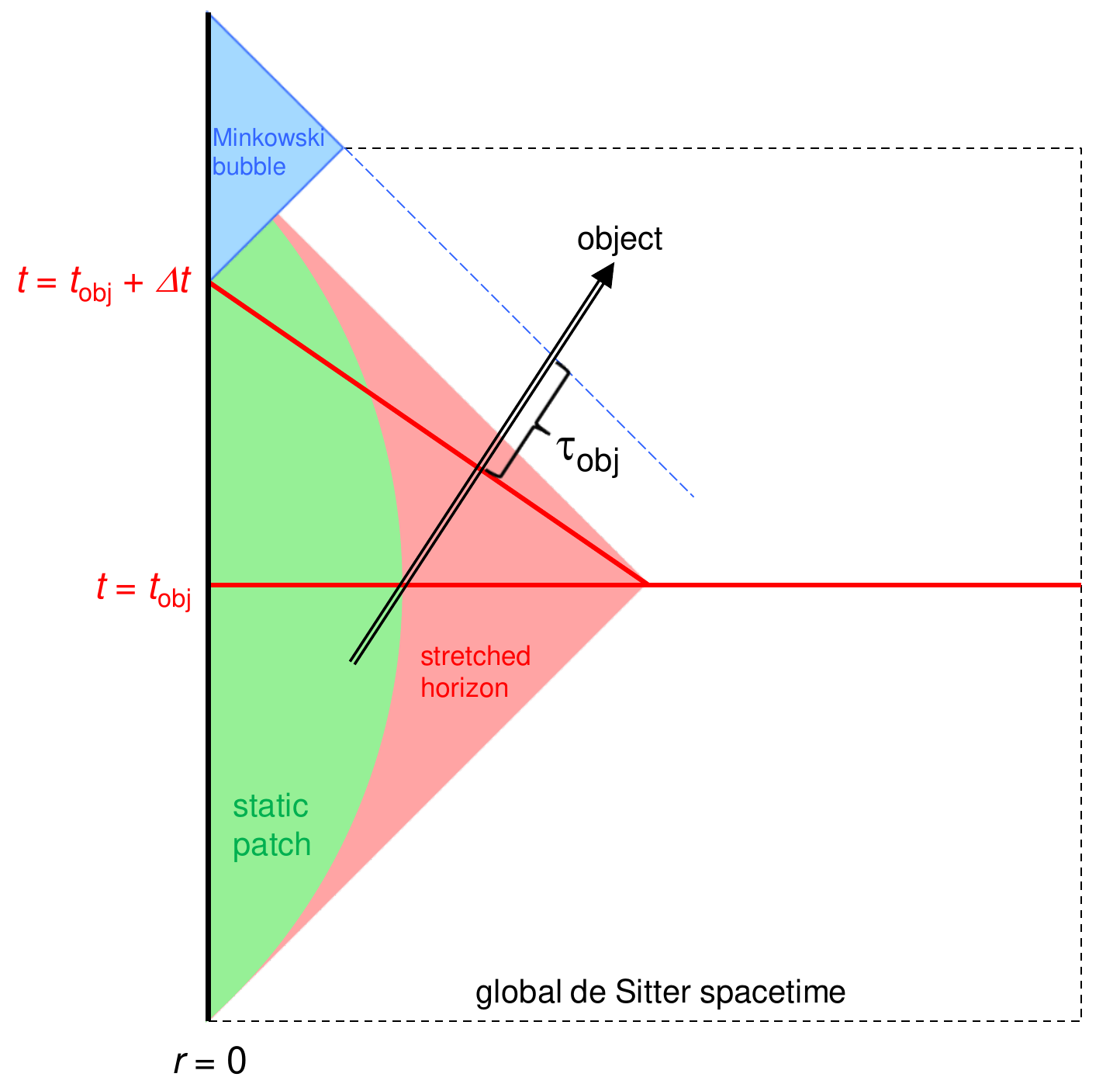}
\vspace{-2mm}
\end{center}
\caption{
 An object falling into the stretched horizon at $t = t_{\rm obj}$ can try to send a physical signal to the observer at $r = 0$ who enters a Minkowski bubble nucleated at $r = 0$ at $t = t_{\rm obj} + \varDelta t$.
 The holographic static-patch description of de~Sitter spacetime suggests that this is not possible if $\varDelta t$ is larger than the scrambling time $t_{\rm scr}$.
 This expectation is consistent with the semiclassical picture, since for $\varDelta t > t_{\rm scr}$ the object would have to send the signal within proper time $\tau_{\rm obj} \lesssim l_{\rm P}$, which is not possible.}
\label{fig:dS-retrieve}
\end{figure}

In Fig.~\ref{fig:dS-retrieve}, we depict the Penrose diagram of the situation in which an object hits the stretched horizon at $t_{\rm obj}$, and a Minkowski bubble is nucleated at the location of the observer, $r = 0$, at $t = t_{\rm obj} + \varDelta t$.
We expect that if $\varDelta t > t_{\rm scr}$, the observer cannot receive any information from the object, since then the effective theory erected at $t = t_{\rm obj} + \varDelta t$ will not have an excitation in the other hemisphere of the effective global de~Sitter spacetime, and so there is nothing that can send a signal to the observer.
From the viewpoint of the object, this implies that it cannot send any signal to the observer after it crosses the $t = t_{\rm obj} + \varDelta t$ hypersurface.

This is indeed consistent with what semiclassical theory implies.
We can show that in order for the signal to reach the Minkowski bubble, and hence the observer, the object must send it within the proper time
\begin{equation}
  \tau_{\rm obj} \lesssim l_{\rm P}
\label{eq:dS-cond}
\end{equation}
after it crosses the $t = t_{\rm obj} + \varDelta t$ hypersurface, but this is not possible because the holographic principle does not allow for the object to send physical information within $l_{\rm P}$.
Incidentally, this conclusion does not change even if the object is accelerated toward $r = 0$ after it crosses the stretched horizon, as long as the invariant acceleration is smaller than $O(1/l_{\rm s})$.%
\footnote{
 Note that here we have considered the case in which the system was in a meta-stable de~Sitter phase before entering a Minkowski phase.
 We expect that the situation is different if the earlier, de~Sitter phase is only approximate in that there is noticeable time dependence of background spacetime, as in the case of slow-roll inflation.
 A different analysis will be needed for such cases.
}

\subsubsection*{Schwarzschild de~Sitter spacetime}

Our analysis applies straightforwardly to a Schwarzschild de~Sitter spacetime which has a black hole of radius $r_{\rm bh}$ (centered at $r=0$) in de~Sitter spacetime of radius $r_{\rm dS}$, as long as $r_{\rm bh}$ is sufficiently smaller than $r_{\rm dS}$.
(The case in which $r_{\rm bh}$ and $r_{\rm dS}$ are comparable is analogous to the case of a near extremal black hole, which we do not consider in this paper.)

Specifically, if $r_{\rm bh} \ll r_{\rm dS}$, the black hole can be viewed as a small excitation for the purpose of erecting the effective theory outside the de~Sitter horizon, so that the construction described so far goes through without significant modifications.
The situation is similar for the effective theory of the black hole interior.
A notable thing is that if the black hole is old enough that it is maximally entangled and that the majority of emitted Hawking radiation has reached the stretched de~Sitter horizon, then operators describing the black hole interior must involve degrees of freedom associated with the de~Sitter horizon.

\section{Intrinsically Extended Spacetimes}
\label{sec:extended}

So far, we have been considering ``realistic,'' single-sided systems and seen how the effective two-sided pictures emerge as collective phenomena associated with the stretched horizon degrees of freedom.
However, there is nothing theoretically wrong in considering ``intrinsically two-sided'' systems which involve two copies of holographic theories discussed so far.
In fact, such systems have been considered in various contexts in black hole~\cite{Maldacena:2001kr,Shenker:2013pqa,Susskind:2014rva,Papadodimas:2015xma,Brown:2015bva,Almheiri:2019yqk,Bouland:2019pvu,Neuenfeld:2021bsb,Verlinde:2021jwu,Leutheusser:2021qhd,Leutheusser:2021frk,Witten:2021unn} and de~Sitter~\cite{Alishahiha:2004md,Alishahiha:2005dj,Susskind:2021dfc,Susskind:2021esx,Shaghoulian:2021cef,Shaghoulian:2022fop,Strominger:2001pn,Aalsma:2020aib,Geng:2020kxh,Geng:2021wcq,Aalsma:2021kle,Kames-King:2021etp} physics.

In this section, we discuss how such intrinsically two-sided systems can be understood in the framework described so far.
In particular, we will see how these and single-sided systems lead to similar semiclassical physics, despite the fact that states in the two cases have rather different structures at the microscopic level.
While we focus here on the two-sided case for a black hole, we expect it is relatively straightforward to extend it to black holes with more than two exterior regions~\cite{Balasubramanian:2014hda}.

\subsection{Two-sided eternal black hole}
\label{subsec:2-sided_BH}

Let us consider a static, two-sided eternal black hole.
Strictly speaking, for a finite black hole mass, this system exists only as a large AdS black hole, but we can imagine that a two-sided black hole in asymptotically flat spacetime also behaves in approximately the same manner if it is sufficiently large compared with the scale that we are interested in.

We specifically consider a thermofield double state which is prepared by the Euclidean path integral over a half of the time compactified on a circle of length $\beta$~\cite{Maldacena:2001kr}.
In the context of the AdS/CFT correspondence, this is a thermally entangled state of two CFTs.
Similar to the case of a single-sided black hole, we divide modes in each side of the black hole into zone and horizon modes at the state preparation time, which we refer to as $t=0$.%
\footnote{
 For a large AdS black hole, we define zone modes to be all the modes other than horizon modes defined as in Section~\ref{subsec:UV}.
}
We then have four classes of modes, i.e.\ zone and horizon modes in both (right and left) sides of the black hole.

Assuming that the two sides of the black hole have the same microscopic structure (which corresponds to the case that the two CFTs are the same), the black hole vacuum state at $t=0$ as viewed from the exterior is given by
\begin{equation}
  \ket{\Psi(M)} = \frac{1}{\sqrt{\sum_{\mu'} e^{-\beta E_{\mu'}}}} \sum_\mu e^{-\frac{\beta}{2}E_\mu} \ketR{\mu} \ketL{\mu},
\label{eq:two-sided-eigen}
\end{equation}
where $\ketR{\mu}$ represent energy eigenstates of the holographic theory describing the right side with energy $E_\mu$ and similarly for $\ketL{\mu}$.
The mass $M$ of the black hole is related to $\beta$ by a smoothness condition for spacetime~\cite{Gibbons:1976ue}:
\begin{equation}
  M = \frac{r_+}{2l_{\rm P}^2}\left( 1 + \frac{r_+^2}{L^2}\right),
\qquad
  \beta = \frac{4\pi r_+ L^2}{3r_+^2 + L^2},
\end{equation}
where $L$ is the AdS length.
In the context of holography, this relation can be viewed as arising from the requirement that the boundary spacetime can be smoothly extended to the bulk spacetime~\cite{Harlow:2020bee}.

We note that the state $\ket{\Psi(M)}$ has exactly zero energy under the two-sided (modular) Hamiltonian $H = H_{\rm R} - H_{\rm L}$, which is possible because the system is static under time evolution generated by $H$.%
\footnote{
 We can extend the state in Eq.~\eqref{eq:two-sided-eigen} to the case in which two CFTs are different.
 In this case, the state must have some, though exponentially small, energy uncertainty, reflecting the fact that the spectra of the two CFTs are not identical.
}
This time evolution corresponds to Schwarzschild time evolution in a single-sided black hole; see Eq.~\eqref{eq:H_HR-HL}.

In general, the energy eigenstates $\ketR{\mu}$ do not agree with the number eigenstates of the zone and horizon modes for the right side (and similarly for the left side).
However, to describe the dynamics of a zone mode that has energy $E \gtrsim 1/\beta = T_{\rm H}$, this issue can be ignored.
Consider, for example, a zone mode of energy $\sim E$ localized at $\bar{r}_*$ in the zone and having a Gaussian tale of the form $\sim e^{-O(1) E^2 (r_* - \bar{r}_*)^2}$.
The correction arising from the existence of the stretched horizon to the quantum theory of matter on classical black hole background is then suppressed by the exponential of $E^2 (\bar{r}_* - r_{*,{\rm s}})^2$.
This implies that the effect of the deviation of zone-mode states from the thermofield double form coming from their interactions with the stretched horizon is negligible compared with the factor of $e^{-\beta E/2}$ if the mode is sufficiently hard ($E \gg 1/\beta$) or away from the stretched horizon ($\bar{r}_* - r_{*,{\rm s}} \gg \beta$).

We now take a subset of the zone modes on the right side as right-side hard modes, whose state we denote by $\ketR{n}$.
Here, $n = \{ n_\alpha \}$ represents the set of occupation numbers of these modes.
The corresponding zone modes on the left side are then left-side hard modes, whose state is $\ketL{n}$.
The rest of the modes on the right and left sides are right- and left-side soft modes, respectively, whose states we denote by $\ketR{\psi_i}$ and $\ketL{\psi_i}$.
Here, the index $i$ runs over all soft-mode states, not just those in a specific energy window, since we no longer impose the energy constraint for modes on one side.

The state in Eq.~\eqref{eq:two-sided-eigen} can then be approximated, for the purpose of describing the dynamics of hard modes, as
\begin{equation}
  \ket{\Psi(M)} \approx \frac{1}{\sqrt{Z}} \sum_n \sum_i e^{-\frac{\beta}{2}(E_n+E_i)} \ketR{n} \ketR{\psi_i} \ketL{n} \ketL{\psi_i},
\label{eq:two-sided}
\end{equation}
where $E_n$ and $E_i$ are the energies carried by $\ketR{n}$ and $\ketR{\psi_i}$ defined with respect to $H$ (or $H_{\rm R}$), and
\begin{equation}
  Z = \sum_n \sum_i e^{-\beta (E_n+E_i)}.
\label{eq:Z}
\end{equation}
We can then build an effective theory of the interior on the vacuum microstate of Eq.~\eqref{eq:two-sided}.%
\footnote{
 In this particular case, we can take the hard modes to be the entire zone modes, though this is not the case in general.}

It is important to realize that the microscopic entanglement structure of $\ket{\Psi(M)}$ in Eq.~\eqref{eq:two-sided} is different from that of a microstate of a single-sided black hole, which takes the form of Eq.~\eqref{eq:sys-state_dS} (ignoring the entanglement with far modes).
In particular, in Eq.~\eqref{eq:two-sided}, the hard modes on the right side, which we identify as ``our'' side, are entangled directly with those on the left side, while in Eq.~\eqref{eq:sys-state_dS} they are entangled with the soft modes on the same side.

The way in which the thermal nature is introduced is also different in two setups.
In Eq.~\eqref{eq:two-sided}, the Boltzmann factors are introduced already at the microscopic level, while in Eq.~\eqref{eq:sys-state_dS} the coefficients take random values, and the Boltzmann factors for the hard modes arise only statistically after tracing out the soft modes; see Eq.~\eqref{eq:rho_vac}.
In the case of the two-sided black hole, tracing out the left side gives
\begin{equation}
  \rho_{\rm R} = \frac{1}{Z} \sum_n \sum_i e^{-\beta(E_n+E_i)} \ketR{n} \ketR{\psi_i}\, \braR{n} \braR{\psi_i}.
\label{eq:rho_R}
\end{equation}
Further tracing out the soft modes lead to
\begin{equation}
  \rho_{\rm R,hard} = \frac{1}{z} \sum_n e^{-\beta E_n} \ketR{n} \braR{n},
\label{eq:rho_Rzone}
\end{equation}
where $z = \sum_n e^{-\beta E_n}$.
At the microscopic level, the two-sided black hole is a model of a single-sided black hole only in the sense that Eq.~\eqref{eq:rho_Rzone} takes the same form as Eq.~\eqref{eq:rho_vac}.

Excited states in which there are objects in the right exterior (our side) of the black hole are obtained by acting with the annihilation/creation operators
\begin{align}
  b_{{\rm R}\gamma} &= \sum_n \sqrt{n_\gamma}\, \ketR{\{ n_\alpha - \delta_{\alpha\gamma} \}} \braR{\{ n_\alpha \}},
\label{eq:ann-R}\\*
  b_{{\rm R}\gamma}^\dagger &= \sum_n \sqrt{n_\gamma + 1}\, \ketR{\{ n_\alpha + \delta_{\alpha\gamma} \}} \braR{\{ n_\alpha \}}
\label{eq:cre-R}
\end{align}
on the vacuum state in Eq.~\eqref{eq:two-sided}.
As in the case of a single-sided black hole, describing the interior requires time evolution operator other than that generated by $H = H_{\rm R} - H_{\rm L}$.
The appropriate generator $\tilde{H}$ can be constructed using annihilation/creation operators acting on left-side hard modes
\begin{align}
  b_{{\rm L}\gamma} &= \sum_n \sqrt{n_\gamma}\, \ketL{\{ n_\alpha - \delta_{\alpha\gamma} \}} \braL{\{ n_\alpha \}},
\label{eq:ann-L}\\*
  b_{{\rm L}\gamma}^\dagger &= \sum_n \sqrt{n_\gamma + 1}\, \ketL{\{ n_\alpha + \delta_{\alpha\gamma} \}} \braL{\{ n_\alpha \}}
\label{eq:cre-L}
\end{align}
as
\begin{equation}
  \tilde{H} = \sum_\xi \Omega_\xi a_\xi^\dagger a_\xi + \tilde{H}_{\rm int}\bigl( \{ a_\xi \}, \{ a_\xi^\dagger \} \bigr),
\label{eq:tilde-H-2sided}
\end{equation}
where
\begin{align}
  a_\xi &= \sum_\gamma \bigl( \alpha_{\xi\gamma} b_{{\rm R}\gamma} + \beta_{\xi\gamma} b_{{\rm R}\gamma}^\dagger + \zeta_{\xi\gamma} b_{{\rm L}\gamma} + \eta_{\xi\gamma} b_{{\rm L}\gamma}^\dagger \bigr),
\label{eq:a_xi-2sided}\\*
  a_\xi^\dagger &= \sum_\gamma \bigl( \beta_{\xi\gamma}^* b_{{\rm R}\gamma} + \alpha_{\xi\gamma}^* b_{{\rm R}\gamma}^\dagger + \eta_{\xi\gamma}^* b_{{\rm L}\gamma} + \zeta_{\xi\gamma}^* b_{{\rm L}\gamma}^\dagger \bigr)
\label{eq:a_xi-dag-2sided}
\end{align}
are infalling mode operators with the coefficients $\alpha_{\xi\gamma}$, $\beta_{\xi\gamma}$, $\zeta_{\xi\gamma}$, and $\eta_{\xi\gamma}$ taking the same values as those in Eqs.~\eqref{eq:a_xi} and \eqref{eq:a_xi-dag} for near horizon modes.
The reason why the construction here need not involve soft modes is that the black hole vacuum state given in Eq.~\eqref{eq:two-sided}, in fact, factors into hard-mode and soft-mode parts:
\begin{equation}
  \ket{\Psi(M)} \approx \biggl( \frac{1}{\sqrt{z}} \sum_n e^{-\frac{\beta}{2} E_n} \ketR{n} \ketL{n} \biggr) \Biggl( \frac{1}{\sqrt{\sum_j e^{-\beta E_j}}} \sum_i e^{-\frac{\beta}{2} E_i} \ketR{\psi_i} \ketL{\psi_i} \Biggr),
\label{eq:two-sided-fac}
\end{equation}
so that the soft-mode piece can simply be ignored.
Note that unlike the construction in Refs.~\cite{Leutheusser:2021qhd,Leutheusser:2021frk}, we do not claim that the generator $\tilde{H}$ of infalling time evolution can be represented purely on the right operators.%
\footnote{
 In the large $N$ limit construction of Refs.~\cite{Leutheusser:2021qhd,Leutheusser:2021frk}, operators $b_{{\rm R}\gamma}$, $b_{{\rm R}\gamma}^\dagger$, $b_{{\rm L}\gamma}$, and $b_{{\rm L}\gamma}^\dagger$ generate a type~III$_1$ von~Neumann algebra (which becomes type~II$_\infty$ at the next order in $1/N$~\cite{Witten:2021unn}), so that Hilbert spaces for $\ketR{n}$, $\ketL{n}$, $\ketR{\psi_i}$, and $\ketL{\psi_i}$ are not defined.
 This is not the case for finite $N$ as envisioned here.
}
The evolution with $\tilde{H}$ in Eq.~\eqref{eq:tilde-H-2sided} allows us to describe the semiclassical physics in the domain of dependence of the union of zones on both sides at $t = 0$.

This erection of the effective theory without involving soft modes is essentially nothing other than the construction of the semiclassical theory from the beginning.
The unknown UV physics, including the effect of ignoring horizon modes, is reflected in the choice of the vacuum in the effective theory.
For example, for a large AdS black hole in thermal equilibrium with the ambient AdS spacetime, the correct choice for typical soft-mode states is the Hartle-Hawking vacuum.
This choice, however, cannot be derived from the low energy consideration alone.

We note that our starting point of Eq.~\eqref{eq:two-sided-eigen} involves a choice of time.
Suppose we evolve the state $\ket{\Psi(M)}$ under the full, microscopic Hamiltonian $H_{\rm R} + H_{\rm L}$:%
\footnote{
 It is important that this evolution is performed using the full, microscopic Hamiltonian (or equivalently performed in the boundary theory), since the analogous operator $H_{\rm R} + H_{\rm L}$ in the bulk field theory is singular at the bifurcation surface.
}
\begin{equation}
  \ket{\Psi(M)} \,\rightarrow\, e^{-i(H_{\rm R} + H_{\rm L}) t} \ket{\Psi(M)} = \frac{1}{\sqrt{\sum_{\mu'} e^{-\beta E_{\mu'}}}} \sum_\mu e^{-\left( \frac{\beta}{2} + 2i t \right)E_\mu} \ketR{\mu} \ketL{\mu}.
\label{eq:two-sided_phases}
\end{equation}
In general, this state is not equivalent to $\ket{\Psi(M)}$; the phases on the right-hand side have a physical meaning.
For example, as $t$ increases from $0$, the maximal interior volume of a spatial surface anchored to the boundaries at time $t$ grows, although for $t \gtrsim e^{S_{\rm BH}}$ the mixing between the hard and soft modes becomes so important that the naive semiclassical picture of interior volume grow ceases to apply~\cite{Iliesiu:2021ari}.

In the two-sided state in Eq.~\eqref{eq:two-sided}, the von~Neumann entropy of the right side is given by
\begin{align}
  S_{\rm R} &= -\Tr\left[\rho_{\rm R} \ln\rho_{\rm R}\right]
\nonumber\\
  &= \sum_n \sum_i \frac{e^{-\beta(E_n+E_i)}}{Z} \left[ \beta(E_n+E_i) + \ln Z \right],
\end{align}
where $\rho_{\rm R}$ is given by Eq.~\eqref{eq:rho_R}.
Approximating that the density of states of the soft modes does not depend on $E_i$ and denoting it by $S_{\rm soft}(M)$, we have up to $O(1)$ coefficients
\begin{equation}
  \sum_i \rightarrow \beta\, \int\! dE_i\, e^{S_{\rm soft}(M)},
\end{equation}
where $\beta$ is used to match the dimension.
This gives
\begin{equation}
  Z \sim \sum_n e^{S_{\rm soft}(M) - \beta E_n} = z\, e^{S_{\rm soft}(M)},
\end{equation}
and
\begin{equation}
  S_{\rm R} = \ln\left[ z\, e^{S_{\rm soft}(M)} \right] + \beta \vev{E_n}_{\rm R} + O(1).
\end{equation}
Here,
\begin{equation}
  \vev{E_n}_{\rm R} = \frac{1}{z} \sum_n E_n e^{-\beta E_n}
\end{equation}
is the thermal energy of the hard modes in the right exterior, as measured by $H$.
Thus, up to the thermal contribution of the hard modes (and an unimportant $O(1)$ term), the entropy $S_{\rm R}$ agrees with $S_{\rm sys}$ given by Eqs.~\eqref{eq:S_sys} and \eqref{eq:def-z}, which is the entropy of a single-sided black hole.
Namely, for a two-sided black hole in the state of Eq.~\eqref{eq:two-sided}, the entropy of the black hole is given at the leading order by the entanglement entropy between the two sides (i.e.\ two CFTs in the AdS/CFT context).

\subsubsection*{Evaporation and the destruction of the wormhole}

Let us now consider coupling the two-sided black hole discussed so far to a reservoir so that the black hole can radiate into it.
We do this on the right side (i.e.\ our side) of the black hole and see its effect on an object falling into the black hole from the same side.

Before coupling the two systems, the state is given by the product of the black hole state in Eq.~\eqref{eq:two-sided} and the ground state of the reservoir system $\ket{\phi_0}$:
\begin{equation}
  \ket{\Psi_0} = \frac{1}{\sqrt{Z}} \sum_m \sum_j e^{-\frac{\beta}{2}(E_m+E_j)} \ketR{m} \ketR{\psi_j} \ketL{m} \ketL{\psi_j} \ket{\phi_0},
\label{eq:2-sided_init}
\end{equation}
where $Z$ is given by Eq.~\eqref{eq:Z}.
After the coupling of the two systems, this state is no longer an energy eigenstate, so it starts evolving.
Specifically, the coupling injects positive energy shock waves into the two systems, after which the Hawking emission process begins.
This process backreacts and produces a superposition of black holes having different masses and momenta (at different locations) in the right-side space.
As discussed in Section~\ref{subsec:vacuum}, we focus on a branch in which the black hole has a well-defined mass and (vanishing) momentum, within the minimum uncertainty required by quantum mechanics.
We can then say that after time $t$ is passed, the state for each $(m,j)$ changes as%
\footnote{
 We regard the left states $\ketL{m} \ketL{\psi_j}$ as (approximate) eigenstates of the left boundary Hamiltonian so that they do not evolve.
}
\begin{equation}
  \ketR{m} \ketR{\psi_j} \ket{\phi_0} \,\,\xrightarrow[\scriptsize{+\mbox{ projection}}]{\scriptsize{\mbox{time evolution}}}\,\, \sum_n \sum_{i_n = 1}^{e^{S_{\rm bh}(M_{\rm R}-E_n)}} \sum_a\, c^{m j}_{n i_n a} \ketR{n} \ketR{\psi^{(n)}_{i_n}} \ket{\phi^{(m,j)}_a},
\label{eq:right-evap}
\end{equation}
where
\begin{equation}
  \sum_n \sum_{i_n = 1}^{e^{S_{\rm bh}(M_{\rm R}-E_n)}} \sum_a\, |c^{m j}_{n i_n a}|^2 = 1,
\end{equation}
$M_{\rm R}$ is the mass of the black hole as viewed from the right side, and $\ket{\phi^{(m,j)}_a}$ represents the state of the reservoir.
Here, $\ketR{n} \ketR{\psi^{(n)}_{i_n}}$ and $\ket{\phi^{(m,j)}_a}$ carry energies $M_{\rm R}$ and $E_m + E_j - M_{\rm R}$, respectively (within the uncertainty), and we have assumed that $t$ is not too large that the structure of the black hole is dramatically altered (for example, that it is fully evaporated).
The index of $\ketR{\psi^{(n)}_{i_n}}$ now carries subindex $n$ because the energy of $\ketR{\psi^{(n)}_{i_n}}$ is correlated with that of $\ketR{n}$.

Substituting Eq.~\eqref{eq:right-evap} into Eq.~\eqref{eq:2-sided_init}, we obtain the state at time $t$
\begin{equation}
  \ket{\Psi_t} = \frac{1}{\sqrt{Z}} \sum_n \sum_m \sum_{i_n = 1}^{e^{S_{\rm bh}(M_{\rm R}-E_n)}} \sum_j \sum_a e^{-\frac{\beta}{2}(E_m+E_j)} c^{m j}_{n i_n a} \ketR{n} \ketR{\psi^{(n)}_{i_n}} \ketL{m} \ketL{\psi_j} \ket{\phi^{(m,j)}_a}.
\label{eq:2-sided_t}
\end{equation}
We assume that for $t > t_{\rm scr}$, the coefficients $c^{m j}_{n i_n a}$ take Gaussian random values across $(n,i_n,a)$ for each $(m,j)$, as in Eq.~\eqref{eq:c-distr-BH}, although this assumption is less justified than that for a single-sided black hole.
Now, consider the (non-normalized) states multiplying $\ketR{n}$ in the sum over $n$ in Eq.~\eqref{eq:2-sided_t}.
\begin{equation}
  \| n \rrangle_{{\rm\scriptscriptstyle R},nn} = \frac{1}{\sqrt{Z}} \sum_m \sum_{i_n = 1}^{e^{S_{\rm bh}(M_{\rm R}-E_n)}} \sum_j \sum_a e^{-\frac{\beta}{2}(E_m+E_j)} c^{m j}_{n i_n a} \ketR{\psi^{(n)}_{i_n}} \ketL{m} \ketL{\psi_j} \ket{\phi^{(m,j)}_a}.
\label{eq:ketcR_nn}
\end{equation}
With the assumption stated above, the norms of these states are given by
\begin{align}
  \,{}_{{\rm\scriptscriptstyle R},nn}\!\llangle n \| n \rrangle_{{\rm\scriptscriptstyle R},nn} &= \frac{1}{Z} \sum_m \sum_{i_n = 1}^{e^{S_{\rm bh}(M_{\rm R}-E_n)}} \sum_j \sum_a e^{-\beta(E_m+E_j)} |c^{m j}_{n i_n a}|^2
\nonumber\\
  &= \frac{1}{(\sum_n e^{-\beta_{\rm R} E_n})}\, e^{-\beta_{\rm R} E_n},
\label{eq:n_nn-norm}
\end{align}
up to corrections exponentially suppressed in $S_{\rm bh}(M_{\rm R})$.
Here, in the last line we have used
\begin{align}
  & \vev{|c^{m j}_{n i_n a}|^2} = \frac{1}{\sum_n \sum_a e^{S_{\rm bh}(M_{\rm R}-E_n)}} \\
  & \qquad\Longrightarrow\quad \sum_{i_n = 1}^{e^{S_{\rm bh}(M_{\rm R}-E_n)}} \sum_a\, |c^{m j}_{n i_n a}|^2 = \frac{e^{S_{\rm bh}(M_{\rm R}-E_n)}}{(\sum_n e^{S_{\rm bh}(M_{\rm R}-E_n)})} = \frac{1}{(\sum_n e^{-\beta_{\rm R} E_n})}\, e^{-\beta_{\rm R} E_n},
\label{eq:2side-stat}
\end{align}
and
\begin{equation}
  \beta_{\rm R} = \left.\frac{\partial S_{\rm bh}(E)}{\partial E}\right|_{E = M_{\rm R}}
\end{equation}
is the temperature of the black hole as viewed from the right side, which is different from $\beta$.
The state $\| n \rrangle_{{\rm\scriptscriptstyle R},nn}$ is thus related to the corresponding normalized state $\ketcR{n}$ by
\begin{equation}
  \| n \rrangle_{{\rm\scriptscriptstyle R},nn} = \frac{1}{\sqrt{\sum_n e^{-\beta_{\rm R} E_n}}}\, e^{-\frac{\beta_{\rm R}}{2} E_n} \ketcR{n}.
\label{eq:ketcR}
\end{equation}

From Eqs~\eqref{eq:ketcR_nn} and \eqref{eq:ketcR}, we find that the state in Eq.~\eqref{eq:2-sided_t} can be written in the thermofield double form
\begin{equation}
  \ket{\Psi_t} = \frac{1}{\sqrt{\sum_n e^{-\beta_{\rm R} E_n}}} \sum_n e^{-\frac{\beta_{\rm R}}{2} E_n} \ketR{n} \ketcR{n}.
\label{eq:2side-eff}
\end{equation}
This allows us to erect the effective theory of the interior, following the construction described in Section~\ref{sec:extension}.
In particular, it implies that an object located in the zone of the right side at time $t$ will smoothly pass through the horizon (of the black hole of mass $M_{\rm R}$) from the right side.%
\footnote{
 We can also erect an effective theory as viewed from the left exterior, i.e.\ the opposite to the side from which the black hole evaporates.
 The construction is analogous to that described here---we identify $\ketcL{m}$ as the state that comes with $\ketL{m}$, leading to the state in the effective theory
 \begin{equation}
   \ket{\Psi'_t} = \frac{1}{\sqrt{\sum_n e^{-\beta E_n}}} \sum_n e^{-\frac{\beta}{2} E_n} \ketL{n} \ketcL{n},
 \end{equation}
 where $\beta = \partial S_{\rm bh}(E)/\partial E|_{E=M}$.
 The analysis presented below also applies to this case with similar conclusions.
}

It is interesting to consider what happens if the initial state has an excitation in the second exterior, i.e.\ the exterior on the left side.
Suppose that the state at the time of the coupling, $t = 0$, has an excitation in the zone, e.g.,
\begin{equation}
  \ket{\Psi_{\rm init}} = \prod_{i=1}^{N} \left( \sum_\gamma f^{(i)}_\gamma b_{{\rm L}\gamma}^\dagger \right) \ket{\Psi_0},
\end{equation}
where $\ket{\Psi_0}$ is given by Eq.~\eqref{eq:2-sided_init}.
In this case, the state after $t > t_{\rm scr}$ is given by Eq.~\eqref{eq:2-sided_t} with the replacement
\begin{equation}
  e^{-\frac{\beta}{2}E_m} c^{m j}_{n i_n a} \ket{\phi^{(m,j)}_a} \,\rightarrow\, \sum_p e^{-\frac{\beta}{2}E_p} U_p^m c^{p j}_{n i_n a} \ket{\phi^{(p,j)}_a},
\label{eq:L-modif}
\end{equation}
where $U_p^m$ is a unitary matrix which has the indices $p = \{ p_\alpha \}$ and $m = \{ m_\alpha \}$ and depends on $f^{(i)}_\gamma$.
Similarly, if there is an excitation on the left side that fell into the stretched horizon at an earlier time, then the state of the left-side soft modes at $t = 0$ deviates from that in Eq.~\eqref{eq:2-sided_init}, causing the change of the state at time $t$
\begin{equation}
  e^{-\frac{\beta}{2}E_j} c^{m j}_{n i_n a} \ket{\phi^{(m,j)}_a} \,\rightarrow\, \sum_k e^{-\frac{\beta}{2}E_k} V_k^j c^{m k}_{n i_n a} \ket{\phi^{(m,k)}_a},
\label{eq:L-modif-2}
\end{equation}
where $V_k^j$ is a unitary matrix acting on the space of the left-side soft states.
With the changes in Eqs.~\eqref{eq:L-modif} and \eqref{eq:L-modif-2}, we can define $\| n \rrangle_{{\rm\scriptscriptstyle R},nn}$ analogously to Eq.~\eqref{eq:ketcR_nn} and calculate its norm
\begin{align}
  \,{}_{{\rm\scriptscriptstyle R},nn}\!\llangle n \| n \rrangle_{{\rm\scriptscriptstyle R},nn} &= \frac{1}{Z} \sum_m \!\sum_{i_n = 1}^{e^{S_{\rm bh}(M_{\rm R}-E_n)}}\! \sum_j \sum_{p,p'} \sum_{k,k'} \sum_{a,a'} e^{-\frac{\beta}{2}(E_p + E_{p'} + E_k + E_{k'})} U_p^{m*} U_{p'}^m V_k^{j*} V_{k'}^j c^{p k *}_{n i_n a} c^{p' k'}_{n i_n a'} \inner{\phi^{(p,k)}_a}{\phi^{(p',k')}_{a'}}
\nonumber\\
  &= \frac{1}{(\sum_n e^{-\beta_{\rm R} E_n})}\, e^{-\beta_{\rm R} E_n},
\end{align}
which we find is the same as Eq.~\eqref{eq:n_nn-norm} up to exponentially suppressed corrections.
Here, we have assumed that there is no intricate (and unexpected) cancellation between $V_k^{j*} V_{k'}^j$ and $c^{p k *}_{n i_n a} c^{p' k'}_{n i_n a'}$ in the sum over $(k, k')$, which would jeopardize the scaling in the last line.%
\footnote{
 A similar assumption is not needed for $U_p^{m*} U_{p'}^m$, since the number of degrees of freedom of semiclassical excitations is too small to jeopardize this scaling anyway.
}

With this assumption, we thus find that the state of the effective theory is still given by Eq.~\eqref{eq:2side-eff}.
This implies that the existence of an excitation on the left exterior at $t = 0$ cannot affect the effective theory erected at $t > t_{\rm scr}$ after evaporation began on the right side.
In other words, the Einstein-Rosen bridge between the two sides is broken by Hawking radiation for $t > t_{\rm scr}$, although an object falling into the black hole from either side sees smooth spacetime when it crosses the horizon.

At the technical level, this occurs because the energy constraint in Eq.~\eqref{eq:right-evap} (i.e.\ the condition that the mass of the black hole as viewed from the right side is $M_{\rm R}$) breaks entanglement between the right and left side modes necessary to have a bridge~\cite{Nomura:2018kia}.
To see this, we can trace out the soft and far modes in the state of Eq.~\eqref{eq:2-sided_t}
\begin{equation}
  \Tr_{\scriptscriptstyle\rm soft+far} \ket{\Psi_t} \bra{\Psi_t} = \frac{1}{\left( \sum_{m'} e^{-\beta_{\rm R} E_{m'}} \right) \left( \sum_{n'} e^{-\beta E_{n'}} \right)} \sum_m \sum_n e^{-\beta_{\rm R} E_m} e^{-\beta E_n} \ketR{m} \ketL{n} \braR{m} \braL{n}
\end{equation}
and find that it takes a different form than
\begin{equation}
  \Tr_{\scriptscriptstyle\rm soft+far} \ket{\Psi_0} \bra{\Psi_0} = \frac{1}{\sum_{n'} e^{-\beta E_{n'}}} \sum_m \sum_n e^{-\frac{\beta}{2}(E_m+E_n)} \ketR{m} \ketL{m} \braR{n} \braL{n},
\end{equation}
obtained from the thermofield double state which has a connected Einstein-Rosen bridge.
In particular, we see that the hard modes on the right side are mostly entangled with the right-side soft modes and far modes in $\ket{\Psi_t}$, and not with modes on the left side as in $\ket{\Psi_0}$.

Physically, this breaking of the Einstein-Rosen bridge occurs due to decoherence caused by Hawking emission from the right side.
After the Hawking emission, black holes of different masses (as viewed from the right side) can be semiclassically discriminated, so that they should be viewed as living in different branches of the wavefunction.
Thus, unless a falling object somehow preserves coherence among these different branches,%
\footnote{
 It is not not clear to what extent this is possible because the falling object necessarily feels gravity generated by the black hole in question.
}
it cannot see any signal sent from the left exterior inside the black hole.
In general, the Einstein-Rosen bridge---or wormhole---prepared by the thermofield double state in Eq.~\eqref{eq:two-sided} is fragile under a realistic physical process occurring to it.%
\footnote{
 If the decoherence is imperfect, the two sides can be ``weakly connected,'' e.g.\ connected in the interior but with a physical domain wall between the two horizons.
}

\subsubsection*{Entangled black holes}

For a reason similar to the case above, entangled black holes of the form
\begin{equation}
  \ket{\Psi(M_1,M_2)} = \sum_{A=1}^{e^{S_{\rm sys}(M_1)}} \sum_{B=1}^{e^{S_{\rm sys}(M_2)}} \eta_{AB} \ket{\Psi_A(M_1)} \ket{\Psi_B(M_2)}
\end{equation}
generically do not have an Einstein-Rosen bridge connecting them; an object falling into one of the black holes sees a smooth horizon but cannot receive any signal sent in from the other black hole~\cite{Nomura:2018kia}.
The same comment applies to more than two entangled black holes.

\subsubsection*{Interior holography?}

\begin{figure}[t]
\begin{center}
  \includegraphics[width=0.5\textwidth]{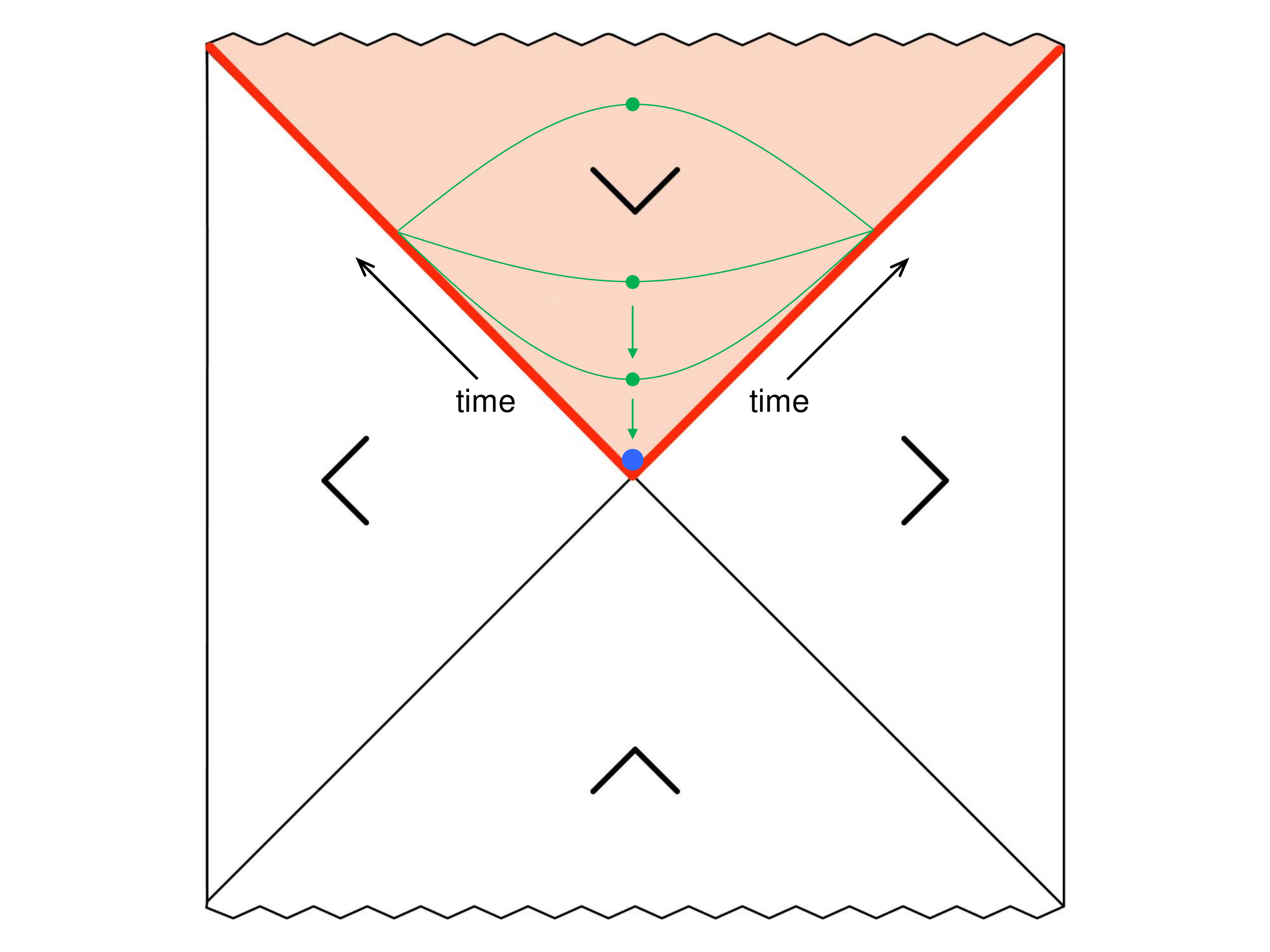}
\vspace{-2mm}
\end{center}
\caption{
 Holographic theory of the interior may be constructed by distributing the interior degrees of freedom over the horizon $H(t)$, which consists of two marginally trapped surfaces as indicated by the wedges.
 Time $t$ evolves toward the future on both components of the horizon.
 Entanglement entropy between the two sides is computed by finding the maximin surface, obtained by minimizing the area of a surface homologous to a component of the horizon on a Cauchy surface anchored to $H(t)$ (green dots) and then maximizing it over possible Cauchy surfaces, which leads to the bifurcation surface (blue dot).
 The entanglement entropy is thus given by the Bekenstein-Hawking entropy of the black hole.}
\label{fig:BH-int-holo}
\end{figure}

We now discuss the possibility of formulating a ``holographic theory of the interior.''
In the discussion so far, the fundamental degrees of freedom are assumed to live on boundaries (e.g.\ holographic screens or AdS boundaries) outside the black hole horizon.
However, discussion in Section~\ref{sec:holo} suggests that a holographic theory may be able to live on an apparent horizon, the surface on which one of the (quantum) expansions vanishes.
We may, therefore, speculate that a holographic theory for the interior can be formulated by picking out the degrees of freedom relevant for describing the interior and distributing them on the horizon; see Fig.~\ref{fig:BH-int-holo}.

As seen in Fig.~\ref{fig:BH-int-holo}, the time evolution of this theory occurs toward the future on both arms of the horizon, so that the holographic space at a given time consists of two disconnected two-spheres.
One can then calculate entanglement entropy between the left and right sides of the black hole by finding a maximin surface~\cite{Wall:2012uf} homologous to one of the two-spheres, which turns out to be the bifurcation surface (see Fig.~\ref{fig:BH-int-holo}).
The entanglement entropy is thus given by
\begin{equation}
  S = \frac{{\cal A}_{\rm bh}}{4G_{\rm N}},
\end{equation}
where ${\cal A}_{\rm bh}$ is the area of the horizon.
The proposed theory, therefore, passes one of the simplest checks for consistency.

It is not clear if the theory described here is fully consistent or useful.
For example, the continuous renormalization procedure in Refs.~\cite{Nomura:2018kji,Murdia:2020iac} cannot be used here to describe the interior, since the procedure requires the bulk to be normal (while it is trapped here).
We have discussed this nonetheless, since it is related to similar proposals in de~Sitter spacetime~\cite{Susskind:2021dfc,Susskind:2021esx,Shaghoulian:2021cef,Shaghoulian:2022fop}, which we will address later.

\subsection{Global de~Sitter spacetime}
\label{subsec:global-dS}

We now discuss a possible description of global de~Sitter spacetime analogous to a static, two-sided black hole.
As in the case of a two-sided black hole in asymptotically flat spacetime, we consider it to be an approximate description of a sufficiently long-lived meta-stable de~Sitter spacetime.

Specifically, we assume that global de~Sitter spacetime at $t=0$ can be described by a state of a thermofield double-like form between two holographic theories each of which describes a static patch.
Following the case of a two-sided black hole, we may divide modes in each theory into ``zone'' and horizon modes.
We then take a subset of the zone modes to be hard modes while leaving all the other modes as soft modes.
In the case of de~Sitter spacetime, however, we expect that gravity is not decoupling in the holographic theories (because the boundary is not in an asymptotic region; see Section~\ref{subsec:holo-dS}), so that the two theories are interacting through it.%
\footnote{
 By gravity we mean the full dynamics associated with gravity at short and long distances.
 Note that in 2+1 dimensions, there is no massless propagating graviton.
}
We thus denote by $\ketN{n}$, $\ketN{\psi_i}$, $\ketS{n}$, and $\ketS{\psi_i}$ the states of the hard and soft modes in the limit that gravity is turned off in these holographic theories.
Here, the subscripts ${\rm N}$ and ${\rm S}$ specify the theory under consideration (referring to the north and south hemispheres, respectively).

\begin{figure}[t]
\centering
  \subfloat[\centering Global de~Sitter spacetime]{{\includegraphics[width=0.35\textwidth]{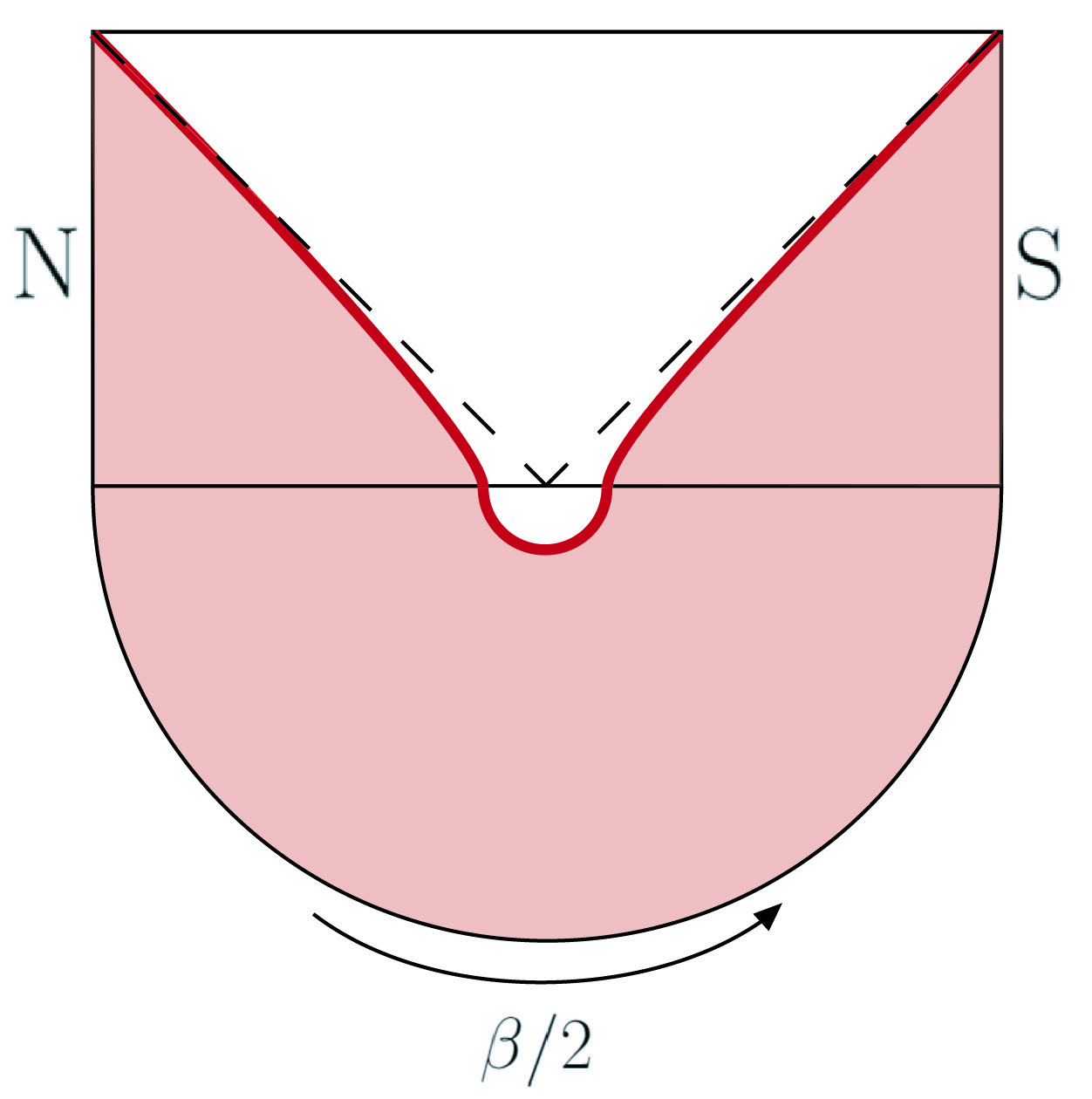} }}
\hspace{1.5cm}
  \subfloat[\centering Two-sided black hole]{{\includegraphics[width=0.35\textwidth]{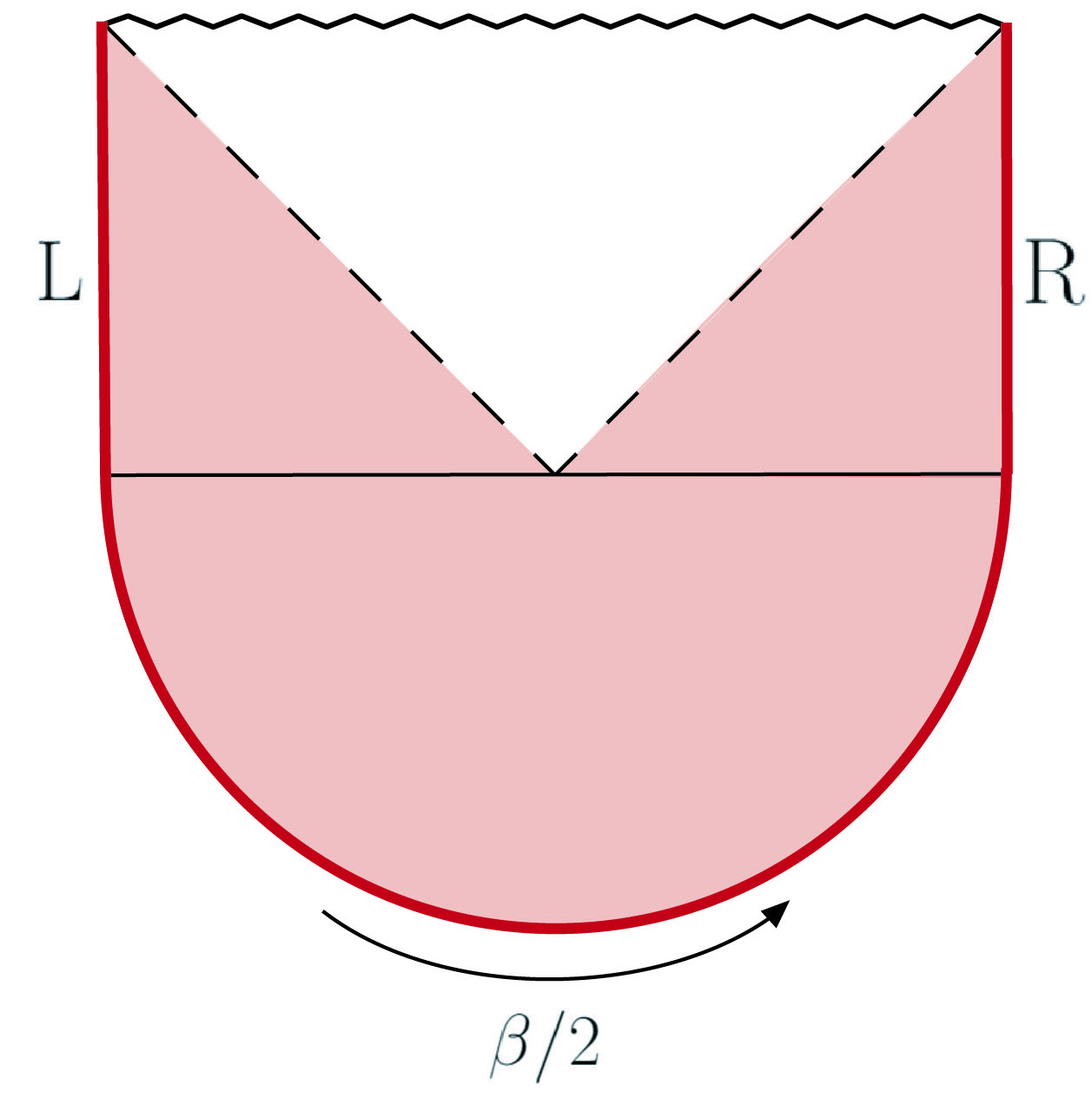} }}
\caption{
 Preparation of a holographic state of the thermofield double form by Euclidean path integral.
 In each panel, the thick red line represents the location at which the holographic theory resides, while the shaded region corresponds to the bulk region that can be simply reconstructed, e.g., by the procedure of continuously pulling in the boundary.}
\label{fig:state-prep}
\end{figure}

Assuming that the two theories have the same microscopic structure, the relevant thermofield double state can be written as
\begin{equation}
  \ket{\Psi'(E)} \approx \frac{1}{\sqrt{Z}} \sum_n \sum_i e^{-\frac{\beta}{2}(E_n+E_i)} \ketN{n} \ketN{\psi_i} \ketS{n} \ketS{\psi_i},
\quad
  Z = \sum_n \sum_i e^{-\beta(E_n+E_i)},
\label{eq:dS-two-sided}
\end{equation}
in the limit that gravity is turned off in holographic theories.
Here, $E = \alpha/l_{\rm P}^2$, and $E_n$ and $E_i$ are the energies carried by the hard and soft modes in the north hemisphere defined with respect to the Hamiltonian $H = H_{\rm N} - H_{\rm S}$.
We expect that the state in Eq.~\eqref{eq:dS-two-sided} can be prepared by the Euclidean path integral over a half of the time compactified on a circle of length $\beta$ (in the limit that gravity is turned off); see Fig.~\ref{fig:state-prep}(a).
While the location of the boundary is now reversed compared with the case of a black hole, depicted in Fig.~\ref{fig:state-prep}(b), we assume that the periodicity $\beta$ of the time direction is still related to the temperature of de~Sitter spacetime as $\beta = 1/T_{\rm H} = 2\pi\alpha$, as in Ref.~\cite{Gibbons:1976ue}.

When gravity is turned on in holographic theories, the Euclidean path integral preparing the state is expected to receive extra contributions suppressed by the gravitational coupling which are not diagonal in the space spanned by $\ket{n} \ket{\psi_i}$.
This gives the correction to the state of the form
\begin{equation}
  \ket{\Psi'(E)} \,\,\longrightarrow\,\, \ket{\Psi(E)} = \ket{\Psi'(E)} + \sum_{m,n} \sum_{j,i} \varepsilon_{m,n;j,i}\, \ketN{m} \ketN{\psi_j} \ketS{n} \ketS{\psi_i},
\label{eq:dS-two-sided-2}
\end{equation}
where $\varepsilon_{m,n;j,i}$ are coefficients of order $l_{\rm P}^2/\alpha^2$.
We suspect that these off-diagonal parts are related to the fact that, unlike the case of a black hole, a positive energy shockwave in de~Sitter spacetime gives a traversable ``wormhole'' between the two hemispheres~\cite{Aalsma:2020aib,Aalsma:2021kle}.

Excited states in which there are objects in the north and south hemispheres of the global de~Sitter spacetime can be obtained by acting annihilation/creation operators, given by Eqs.~\eqref{eq:ann-R}--\eqref{eq:cre-L} with ${\rm R} \rightarrow {\rm N}$ and ${\rm L} \rightarrow {\rm S}$, on the vacuum state in Eq.~\eqref{eq:dS-two-sided-2}.
Describing the region outside the horizons of the two static patches requires time evolution operator other than that generated by $H = H_{\rm N} - H_{\rm S}$.
The appropriate generator $\tilde{H}$ can be constructed as
\begin{equation}
  \tilde{H} = \sum_\xi \Omega_\xi a_\xi^\dagger a_\xi + \tilde{H}_{\rm int}\bigl( \{ a_\xi \}, \{ a_\xi^\dagger \} \bigr),
\label{eq:tilde-H-global-dS}
\end{equation}
where
\begin{align}
  a_\xi &= \sum_\gamma \bigl( \alpha_{\xi\gamma} b_{{\rm R}\gamma} + \beta_{\xi\gamma} b_{{\rm R}\gamma}^\dagger + \zeta_{\xi\gamma} b_{{\rm L}\gamma} + \eta_{\xi\gamma} b_{{\rm L}\gamma}^\dagger \bigr),
\label{eq:a_xi-2sided-dS}\\*
  a_\xi^\dagger &= \sum_\gamma \bigl( \beta_{\xi\gamma}^* b_{{\rm R}\gamma} + \alpha_{\xi\gamma}^* b_{{\rm R}\gamma}^\dagger + \eta_{\xi\gamma}^* b_{{\rm L}\gamma} + \zeta_{\xi\gamma}^* b_{{\rm L}\gamma}^\dagger \bigr)
\label{eq:a_xi-dag-2sided-dS}
\end{align}
are mode operators with the coefficients $\alpha_{\xi\gamma}$, $\beta_{\xi\gamma}$, $\zeta_{\xi\gamma}$, and $\eta_{\xi\gamma}$ determined by semiclassical calculation.

Since the $t=0$ hypersurface is a Cauchy surface, this effective theory describes the entire global de~Sitter spacetime.
The theory, however, is intrinsically semiclassical.
The choice of the vacuum cannot be derived from the first principle, although we expect that the correct choice corresponding to typical soft-mode states is the Bunch–Davies vacuum~\cite{Bunch:1978yq}.
The description of field fluctuations on the de~Sitter background is also only statistical.
To go beyond this, e.g.\ to describe the details of a state depending also on the horizon microstate, we must resort to the microscopic theory.

In the $\ket{\Psi(E)}$ state, the von~Neumann entropy of the north hemisphere is
\begin{equation}
  S_{\rm N} = \sum_n \sum_i \frac{e^{-\beta(E_n+E_i)}}{Z} \left[ \beta(E_n+E_i) + \ln Z \right] + O\biggl(\frac{l_{\rm P}^2}{\alpha^2}\biggr).
\end{equation}
Assuming that the density of states of the soft modes does not depend on $E_i$ and denoting it by $S_{\rm soft}(E)$, we have
\begin{equation}
  S_{\rm N} = \ln\left[ z\, e^{S_{\rm soft}(E)} \right] + \beta \vev{E_n}_{\rm N} + O(1),
\end{equation}
where $z = \sum_n e^{-\beta E_n}$, and $\vev{E_n}_{\rm N} = \sum_n E_n e^{-\beta E_n} / z$ is the thermal energy of the hard modes in the north hemisphere, as measured by $H$.
Thus, the entropy $S_{\rm N}$ agrees with $S_{\rm sys}$ given by Eqs.~\eqref{eq:S_sys} and \eqref{eq:def-z} at the leading order.
Namely, in the global state of Eq.~\eqref{eq:dS-two-sided-2}, the entropy of de~Sitter spacetime can be interpreted as the entanglement entropy between the two hemispheres, at the leading order.

\subsubsection*{Relation to the DS/dS correspondence}

The theory we are considering consists of two holographic systems located on the stretched horizons of two static patches, each of which covers one half of the spatial section of global de~Sitter spacetime at $t = 0$, the global time at which the spatial volume becomes minimal.
These two systems, each of which is expected to be strongly coupled, are weakly interacting through gravity.
This structure is reminiscent of that in the DS/dS correspondence~\cite{Alishahiha:2004md,Alishahiha:2005dj}, a proposed holographic description of de~Sitter spacetime.

\begin{figure}[t]
\begin{center}
  \includegraphics[width=0.37\textwidth]{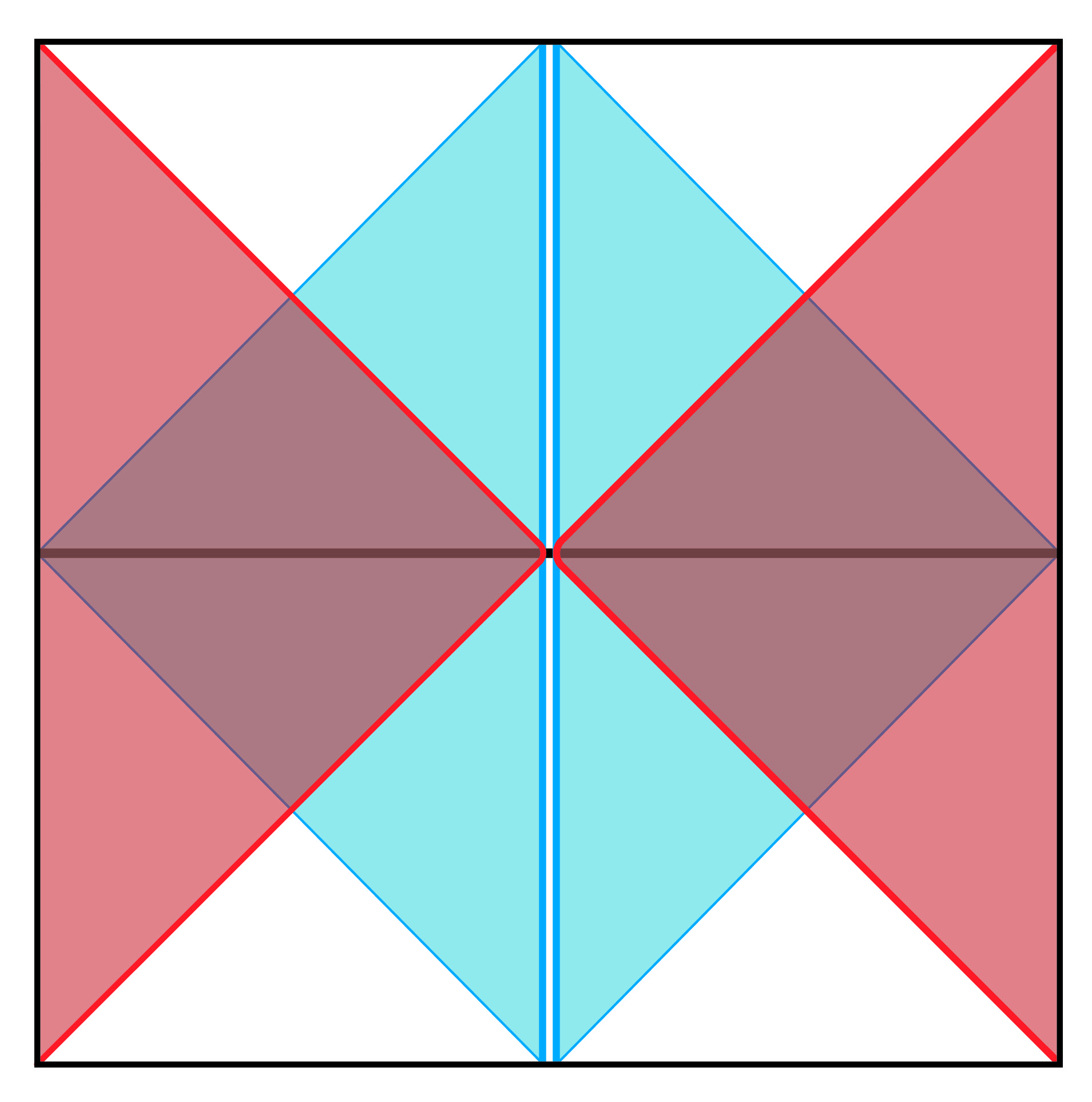}
\vspace{-2mm}
\end{center}
\caption{
The holographic theory based on static patches consists of two weakly interacting systems located on the stretched horizons of two static patches, with the simply reconstructed region staying within the two patches (red).
In the DS/dS correspondence, each of the holographic systems is located by itself on de~Sitter spacetime with one lower dimensions, leading to the simply reconstructed region called the DS/dS patch (blue).
The structures of the two theories are similar at $t = 0$.}
\label{fig:dS-dS}
\end{figure}

While the two theories have similar structures at $t = 0$, they can be different at $t \neq 0$.
In particular, if we evolve our theory with $H_{\rm N} + H_{\rm S}$, the places where the holographic systems are located move toward the future along the stretched horizons of the two static patches, as indicated by the red drawings in Fig.~\ref{fig:dS-dS}.
This makes simply reconstructed regions stay within the static patches.
On the other hand, in the DS/dS correspondence, each of the holographic systems is completed itself into de~Sitter spacetime of one lower dimensions, leading to the simply reconstructed region called a DS/dS patch, which is depicted by the blue drawings in the figure.

One might speculate that the two theories represent the same system evolved differently in holographic space.
It will be interesting to study this possible relation, but we leave it for the future.

\subsubsection*{Relation to the Shaghoulian-Susskind proposal}

\begin{figure}[t]
\begin{center}
  \includegraphics[width=0.5\textwidth]{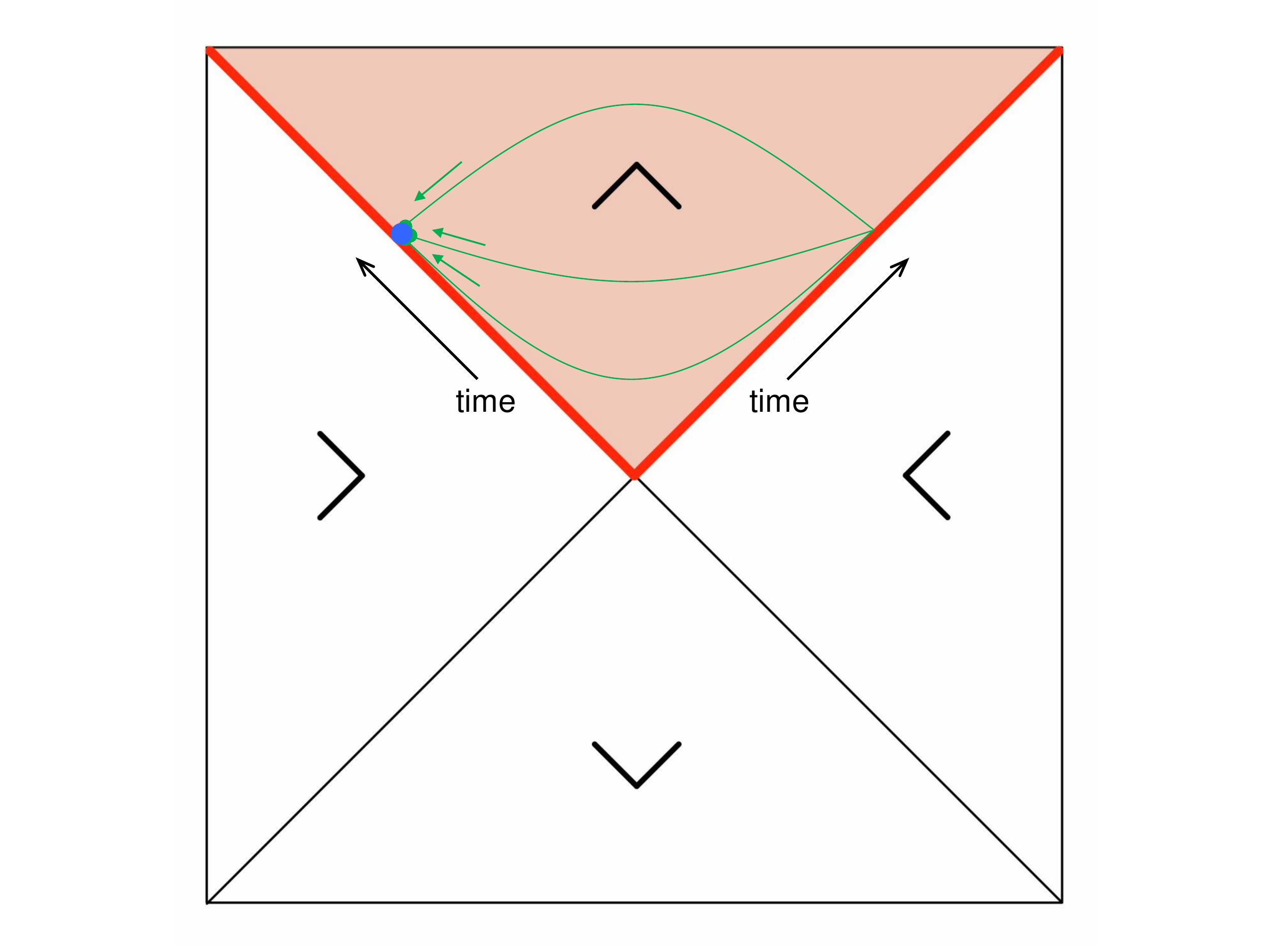}
\vspace{-2mm}
\end{center}
\caption{
 Holographic theory placed on the de~Sitter horizons of two static patches can describe the region outside the horizons (as viewed by polar observers).
 Entanglement entropy between the two sides is computed by finding the maximin surface, which is nothing other than one of the horizons (blue dot).
 The entanglement entropy, therefore, is given by the Gibbons-Hawking entropy.}
\label{fig:dS-int-holo}
\end{figure}

At the end of Section~\ref{subsec:2-sided_BH}, we discussed the possibility of having a holographic theory of the black hole interior by placing the relevant degrees of freedom on the apparent horizon.
We can consider an analogous situation for de~Sitter spacetime, in which case the degrees of freedom relevant for describing the region outside the horizons of the two static patches are placed on the horizon; see Fig.~\ref{fig:dS-int-holo}.

As in the case of the black hole, we can test the consistency of this picture by computing the entanglement entropy between the two patches, which can be done using the maximin procedure.
By minimizing a surface homologous to one of the horizons on a spacelike surface bounded by the two horizons at a fixed time, we find the resulting surface is nothing other than one of the horizons.
The subsequent maximization over spacelike surfaces thus gives us the same surface, so the de~Sitter horizon is the maximin surface.
This gives the entanglement entropy
\begin{equation}
  S = \frac{{\cal A}_{\rm dS}}{4G_{\rm N}},
\end{equation}
where ${\cal A}_{\rm dS}$ is the area of the de~Sitter horizon, which is consistent with the picture that the evolution of the system is unitary.
This seems to be what is proposed in Refs.~\cite{Susskind:2021dfc,Susskind:2021esx} as a holographic theory of de~Sitter spacetime (referred to as the monolayer proposal in Ref.~\cite{Shaghoulian:2022fop}), although our interpretation here says that the theory describes only the shaded region in Fig.~\ref{fig:dS-int-holo}; in particular, it does not describe the regions inside the two static patches.

In our picture, the degrees of freedom used for the theory described above are a subset of the degrees of freedom of holographic theories describing the static patches.
Therefore, the former are not really independent of the latter at the microscopic level.
However, we may treat them independent for the purpose of describing semiclassical physics in the bulk.
If this is the case, we can regard the holographic theory as consisting of two layers of degrees of freedom on the horizons, one describing the outside and the other describing the interiors of the horizons.
This seems to be the proposal of Ref.~\cite{Shaghoulian:2021cef}, called the bilayer proposal in Ref.~\cite{Shaghoulian:2022fop}.

\section{Gravitational Path Integral}
\label{sec:path-int}

The approach we have described so far is based on the canonical formalism of quantum mechanics.
In particular, we have assumed the existence of a Hilbert space factor representing states of the horizon degrees of freedom, with which we could construct an effective theory describing a spacetime region behind the horizon---the interior in the case of a black hole---using only the low energy input that the dynamics of these degrees of freedom is maximally chaotic (and fast scrambling).

Quantum mechanics, however, can also be formulated using path integrals, and we expect that the same physical conclusions would be obtained from this formalism.
In this section, we discuss what the picture looks like in this case.
We will see that the corresponding picture is that of Refs.~\cite{Penington:2019kki,Almheiri:2019qdq}, in which aspects of unitary evolution can be reproduced by gravitational path integral that fixes the ``boundary condition'' based on the quantity of interest and then integrates over all possible semiclassical geometries consistent with it, including those with nontrivial topologies (in particular, replica wormholes).
This allows us to relate the framework described so far to the treatment based on the quantum extremal surface prescription~\cite{Penington:2019npb,Almheiri:2019psf,Almheiri:2019hni}.
While we focus on the case of a black hole here, we suspect that a similar story can be developed for de~Sitter spacetime as well.

The picture presented here was outlined in Refs.~\cite{Langhoff:2020jqa,Nomura:2020ewg}.
The ensemble nature of the semiclassical description arising from an ensemble of microstates was also discussed in Refs.~\cite{Pollack:2020gfa,Belin:2020hea,Liu:2020jsv,Freivogel:2021ivu,Cotler:2022rud}.
The understanding of the Page curve presented here is based on the picture developed in Refs.~\cite{Bousso:2020kmy,Engelhardt:2020qpv,Renner:2021qbe,Qi:2021oni,Almheiri:2021jwq,Blommaert:2021fob,Chandra:2022fwi}.

\subsection{Ensemble from coarse graining}

The starting point for the path integral formalism is very different from that of the canonical formalism.
Specifically, in the present context it should start from a collection of classical field configurations on classical geometries, which will then be integrated over.
We are interested in a ``low energy'' framework in which the detailed microscopic knowledge is not necessary to understand the physics.
This implies that a black hole must be treated as a (semi)classical object in which the detailed microscopic structure cannot be discriminated.

This treatment, in fact, is required by quantum mechanics.
As we have argued in Sections~\ref{sec:QFT}~and~\ref{sec:stretched}, microstates of a black hole can be regarded as independent quantum states arising from superposing energy eigenstates in a small energy window, e.g.\ of order $T_{\rm H}$, around $M$.
Thus, to discriminate these microstates, one would need an exponentially long measurement time $\sim e^{S_{\rm bh}}/T_{\rm H}$.
The black hole, however, would already have evaporated by the time such a measurement would be completed (or the state of the black hole will be altered significantly for a large AdS black hole thermalized with the environment).
One therefore cannot operationally discriminate different microstates by performing a measurement on the black hole.%
\footnote{
 One can, of course, {\it infer} the microstate of the black hole if we prepare a specific state of initial collapsing matter and then use the microscopic theory to simulate its time evolution.
}

\begin{figure}[t]
\begin{center}
  \includegraphics[width=0.5\textwidth]{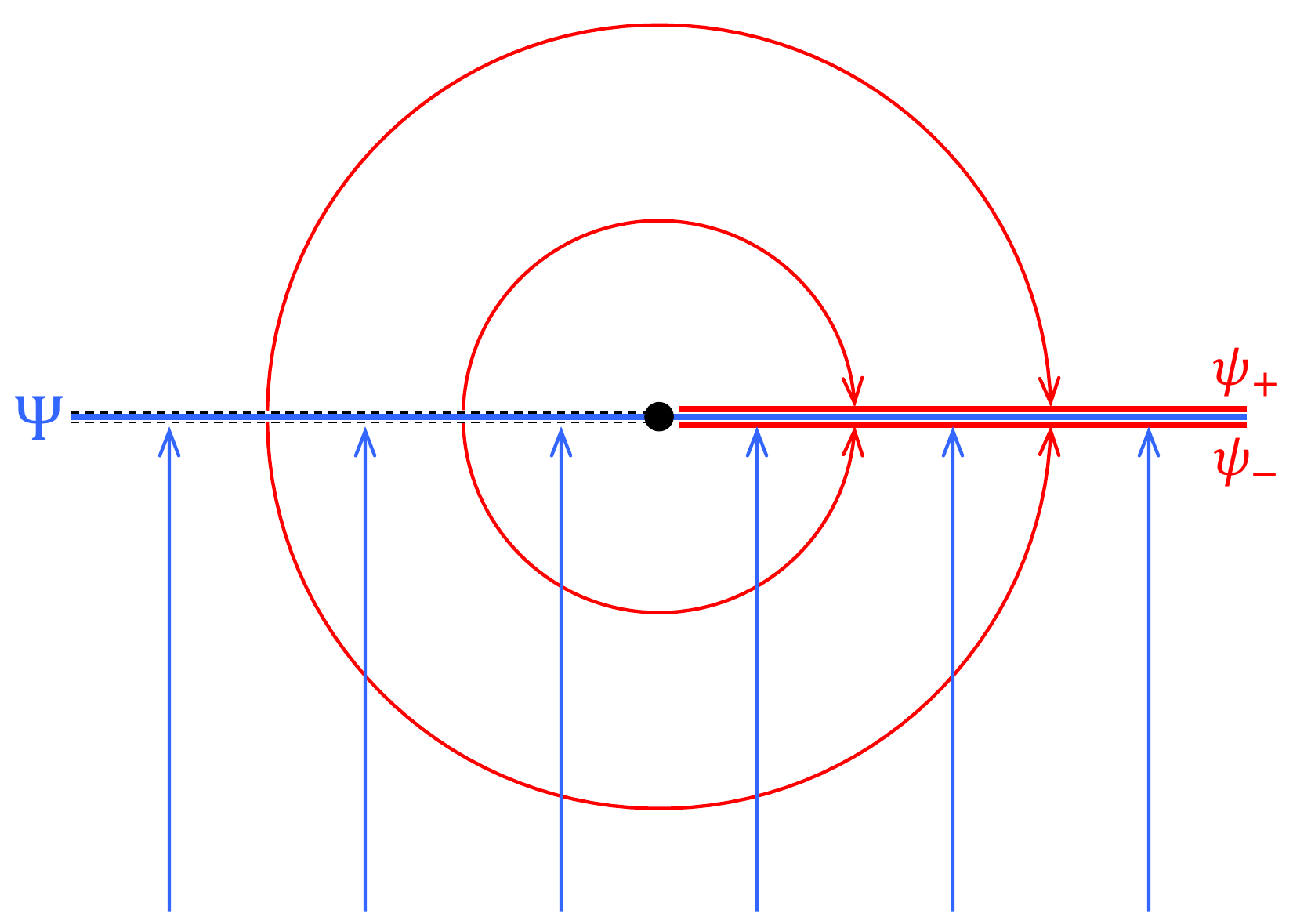}
\vspace{-3mm}
\end{center}
\caption{
 The density matrix of the exterior region, whose element is specified by the field configurations $\psi_+$ and $\psi_-$, can be computed by Euclidean path integral, which can be interpreted to receive equal contributions from all the black hole microstates.
 Euclidean path integral can also be used to prepare a two-sided state $\Psi$, which is pure and unique.}
\label{fig:Euc-PI}
\end{figure}

We thus have to regard a black hole spacetime appearing in gravitational path integral as representing a maximally mixed ensemble of the microstates consistent with the classical specification of the black hole~\cite{Langhoff:2020jqa,Nomura:2020ewg}.
Note that this picture is associated with a single-sided (or a single-sided description) of a black hole.
To see this, one can imagine preparing a black hole state using Euclidean path integral in the approximation that the black hole is static.
As represented in Fig.~\ref{fig:Euc-PI}, the ``$\psi_+$-$\psi_-$ component'' of the density matrix of the exterior region can then be obtained by the path integral in which the spatial field configurations are fixed to be $\psi_+$ and $\psi_-$ above and below the cut corresponding to the exterior region of the black hole.

At the semiclassical level, this simply gives the thermal density matrix which we regard as the semiclassical black hole vacuum state.
We can, however, imagine following the same procedure at the microscopic level to get the corresponding density matrix
\begin{equation}
  \rho_{\rm micro} \propto \sum_n e^{-\frac{\beta}{2} \varDelta H} \ket{n} \bra{n} e^{-\frac{\beta}{2} \varDelta H},
\label{eq:Eucl-PI}
\end{equation}
where $\{ \ket{n} \}$ is a complete set of black hole vacuum microstates consistent with the background used in the semiclassical path integral, $\beta = 1/T_{\rm H}$ is the periodicity of Euclidean time in the angular direction, and $\varDelta H$ is the microscopic exterior (boost) Hamiltonian.
Here, the zero of $\varDelta H$ is chosen to be a typical energy associated with the space ${\cal H}_{\rm vac}$ spanned by $\ket{n}$'s.%
\footnote{
 The index $n$ here corresponds to the index $A$ in the notation in Sections~\ref{sec:stretched}~and~\ref{sec:extension}.
}

Since the matrix elements of the $\beta \varDelta H/2$ operator in ${\cal H}_{\rm vac}$ are of order or smaller than $1$, we find that many microstates contribute to $\rho_{\rm micro}$.
In particular, there are exponentially many microstates that contribute dominantly and almost equally; these are the microstates among which the matrix elements $\beta \varDelta H/2$ is much smaller than $1$.
The fact that semiclassical theory cannot resolve the detailed structure of microstates implies that it perceives $\rho_{\rm micro}$ to be the maximally mixed state%
\footnote{
 The contributions to $\rho_{\rm micro}$ from states with $|\beta \varDelta H| \gg 1$ are negligible either because of the Boltzmann suppression (for $\varDelta H > 0$) or because the number of states is too small (for $\varDelta H < 0$).
 The chaotic nature of the black hole dynamics then implies that the remaining contribution can be well approximated to come from the maximally mixed ensemble of microstates with $|\beta \varDelta H| \lesssim 1$.
}
\begin{equation}
  \rho_{\rm sc} \propto \sum_{n=1}^{e^{S_{\rm bh}}} \ket{n} \bra{n},
\label{eq:max-mixed}
\end{equation}
where the sum runs over a sufficiently large set of vacuum microstates $\ket{n}$, whose precise specification is not important as discussed in Section~\ref{sec:stretched} (or as in standard statistical mechanics).
The number of elements of this set is of order $e^{S_{\rm bh}}$, which we have already indicated in Eq.~\eqref{eq:max-mixed}.

This interpretation indeed reproduces many features which are attributed to the ensemble nature of holographic {\it theories} in lower dimensional quantum gravity~\cite{Penington:2019kki} in terms of an ensemble of microscopic {\it states}~\cite{Langhoff:2020jqa} (see also Refs.~\cite{Pollack:2020gfa,Belin:2020hea,Liu:2020jsv,Freivogel:2021ivu,Cotler:2022rud}).
Incidentally, this should be contrasted with the situation in which a global (two-sided) state is prepared by Euclidean path integral.
In this case, the generated state is a pure state
\begin{equation}
  \ket{\Psi_{\rm micro}} \propto \lim_{\tau \rightarrow \infty} e^{-\varDelta\tilde{H} \tau} \ket{i}
\end{equation}
obtained by evolving some initial state $\ket{i}$ at $\tau = -\infty$ to $\tau = 0$.
Here, $\ket{i}$ is chosen in the space of black hole (not necessarily vacuum) microstates ${\cal H}_{\rm bh}$, and $\varDelta \tilde{H}$ is the microscopic ``infalling'' (inertial) Hamiltonian with its zero chosen to be a typical energy of vacuum microstates.
This leads to the unique, lowest energy eigenstate, determined by the choice of ${\cal H}_{\rm bh}$; see also discussion around Eq.~\eqref{eq:two-sided-eigen}.

\subsection{Replica method, entanglement island, and the Page curve}

With the understanding that the black hole spacetime appearing in gravitational path integral represents the (maximally mixed) ensemble of black hole microstates, the results of Refs.~\cite{Penington:2019kki,Almheiri:2019qdq} can be understood in a simple manner.

\subsubsection*{Euclidean gravitational path integral}

Let us first consider a setup similar to Ref.~\cite{Almheiri:2019qdq} in which a black hole living in a gravitational region is coupled to a nongravitational region.%
\footnote{
 The nongravitational region is a proxy for a weakly gravitational region.
 The distinction between gravitational and nongravitational regions is necessary to correctly identify semiclassical degrees of freedom in the latter region, in particular Hawking radiation quanta, of which the von~Neumann entropy is calculated.
 Without this separation, the Page curve of our interest cannot be computed using the method adopted here~\cite{Geng:2020fxl}, although we expect that there is a way to reproduce the same result without involving an artificial separation of regions, at least in principle.
}
The black hole then radiates into the nongravitational region.
Our interest is to compute the von~Neumann entropy of the emitted radiation.

\begin{figure}[t]
\begin{center}
  \includegraphics[width=0.75\textwidth]{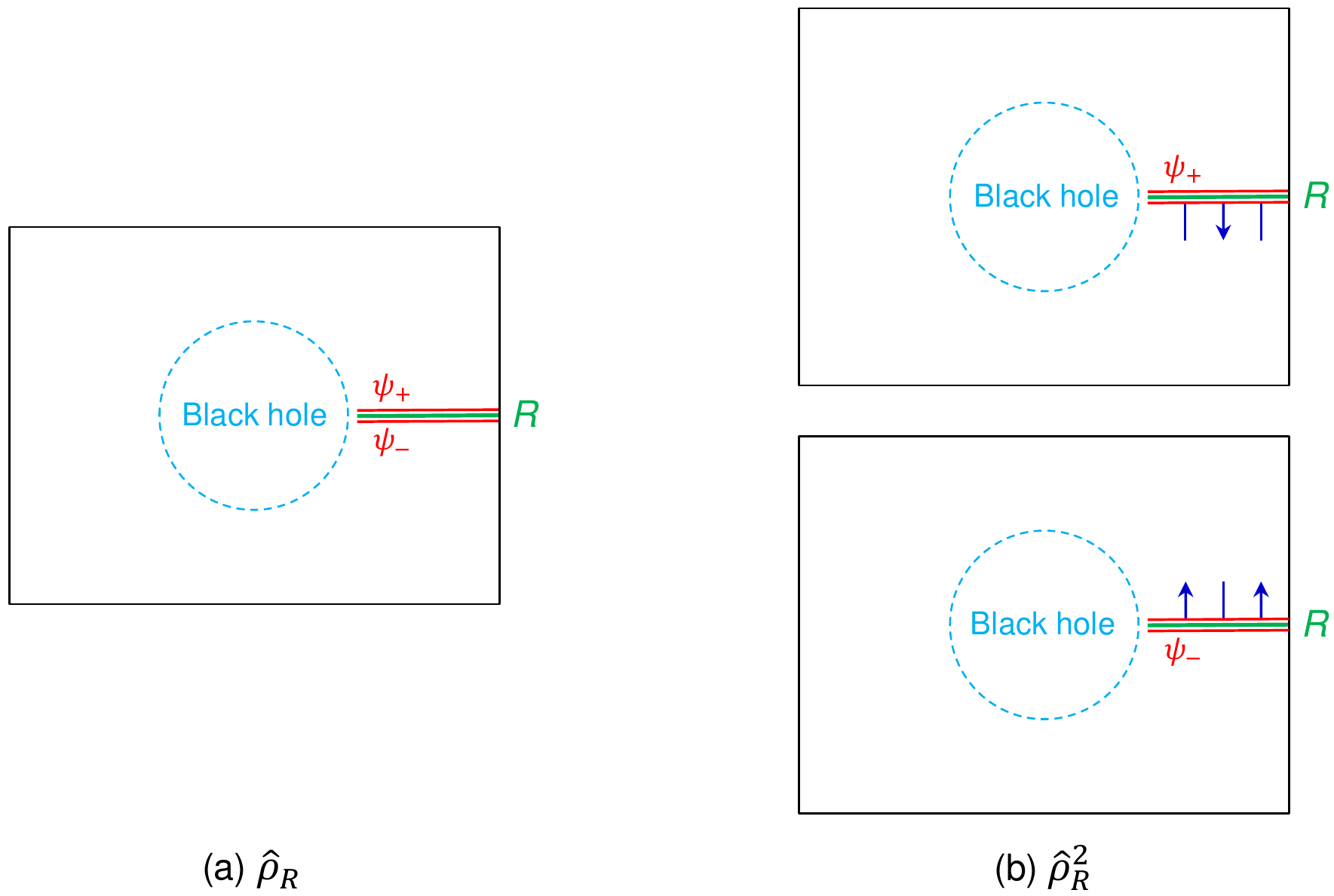}
\vspace{-3mm}
\end{center}
\caption{
 (a) The (nonnormalized) density matrix $\hat{\rho}_R$ of radiation in nongravitational region $R$ can be calculated by performing path integral with the boundary condition fixing the spatial field configurations above and below the cut along $R$.
 (b) The square of the density matrix, $\hat{\rho}_R^2$ is calculated using the replica method, which replicates the spacetime into two and imposes the boundary condition such that one side of the cuts along $R$ on the two sheets are appropriately sewn together (as represented by blue arrows).}
\label{fig:density-mat}
\end{figure}

Since the radiation lives in a nongravitational region, we must be able to describe its state semiclassically.
In particular, assuming that the system can be viewed as quasi-static at each instance of time, we expect that the density matrix of the radiation in a region $R$ can be calculated using a Euclidean path integral by specifying its element by spatial configurations of the radiation field above and below the cut on $R$; see Fig.~\ref{fig:density-mat}(a).
The obtained density matrix is not normalized, which we denote by $\hat{\rho}_R$ with the hat over $\rho_R$ indicating that the density matrix is not normalized.%
\footnote{
 The density matrix $\rho_R$ here refers to the fine-grained (reduced) density matrix of fields in $R$.
 In the language of Ref.~\cite{Almheiri:2019yqk}, the radiation here is the ``radiation in boldface.''
}

The path integral performed, however, includes the gravitational region in which a black hole resides.
Thus, the obtained density matrix, in fact, involves an ensemble average over the black hole microstates in the sense of Eq.~\eqref{eq:max-mixed}
\begin{equation}
  \overline{\hat{\rho_R}} = \frac{1}{e^{S_{\rm bh}}} \sum_{n=1}^{e^{S_{\rm bh}}} \hat{\rho}_{R,n},
\label{eq:rho_PI}
\end{equation}
where $\hat{\rho}_{R,n} = \mathrm{Tr}_{R^c} \ket{n} \bra{n}$ represents the (nonnormalized) density matrix of $R$ for the microstate $\ket{n}$ with $R^c$ being the complement of $R$.
Defining the normalized density matrix by
\begin{equation}
  \overline{\rho_R} = \frac{\overline{\hat{\rho_R}}}{\Tr\overline{\hat{\rho_R}}},
\label{eq:rho_PI-norm}
\end{equation}
we can compute the von~Neumann entropy of $R$ as
\begin{equation}
  S_R^{({\rm sc})} = -\Tr\left[ \overline{\rho_R} \ln \overline{\rho_R} \right].
\label{eq:S_R-Hawking}
\end{equation}
Since this involves coarse-graining, i.e.\ the ensemble average over microstates, the quantity $S_R^{({\rm sc})}$ is well approximated by the thermal entropy of the radiation, which increases monotonically in time (until the black hole is fully evaporated, if it is not eternal).
This is Hawking's result~\cite{Hawking:1976ra} showing the apparent violation of unitarity in the black hole evaporation process.

However, what we really want to understand is the behavior of
\begin{equation}
  S_R^{({\rm micro})} = -\Tr\left[ \rho_{R,n} \ln \rho_{R,n} \right]
\end{equation}
(for each microstate $\ket{n}$), which must show the Page behavior~\cite{Page:1993wv} if the evolution of the system is unitary.
In particular, $S_R^{({\rm micro})}$ must go down to zero when the black hole is fully evaporated and all the emitted radiation is included in $R$.
A question is how (or if) we can see this behavior in a semiclassical analysis.

The idea is that while the semiclassical description necessarily involves the ensemble average over black hole microstates, it still allows for calculating the ensemble average of many different quantities.
For example, we can calculate the ensemble average of the square of the density matrix for the radiation using the replica method~\cite{Lewkowycz:2013nqa,Callan:1994py,Calabrese:2004eu}, i.e.\ by replicating the spacetime into two copies and imposing the boundary condition for the path integral such that (one side of) the cuts along $R$ on two sheets are appropriately sewn together; see Fig.~\ref{fig:density-mat}(b).
Performing path integral with this boundary condition gives
\begin{equation}
  \overline{\hat{\rho}_R^2} = \frac{1}{e^{S_{\rm bh}}} \sum_{n=1}^{e^{S_{\rm bh}}} \hat{\rho}_{R,n}^2.
\label{eq:rhosq_PI}
\end{equation}
An important point is that the performed path integral must involve all possible geometries in the gravitational region, including those having a topology that geometrically connects the two sheets (the replica wormhole)~\cite{Penington:2019kki,Almheiri:2019qdq}.
This is because the gravitational path integral should not predetermine the geometry on which quantum fields are integrated over.%
\footnote{
 As elucidated, e.g., in Ref.~\cite{Blommaert:2021fob}, this prescription calculates $\overline{\hat{\rho}_R^2}$ with exponential accuracy, taking into account the detailed microscopic structure of radiation (radiation in boldface in the language of Ref.~\cite{Almheiri:2019yqk}).
 If we instead calculate $\overline{\hat{\rho}_R^2}$ including only trivial topology, i.e.\ the disconnected contribution, then we would fail to capture the effect from the exponentially complex microscopic structure, yielding $\overline{\hat{\rho}_R^2}$ as computed in the semiclassical theory (i.e.\ the squared density matrix of non-boldface radiation).
 This latter prescription would lead to Hawking's result~\cite{Hawking:1976ra} when we compute the von~Neumann entropy using $S_R = - \lim_{n \rightarrow 1} \partial (\Tr \overline{\rho_R^n})/\partial n$, where $\overline{\rho_R^n} = \overline{\hat{\rho}_R^n}/(\Tr \overline{\hat{\rho}_R})^n$.
}
In general, the inclusion of geometries with nontrivial topologies makes Eq.~\eqref{eq:rhosq_PI} different from the square of Eq.~\eqref{eq:rho_PI}:\ $\overline{\hat{\rho}_R^2} \neq \bigl( \overline{\hat{\rho}_R} \bigr)^2$.

\begin{figure}[t]
\begin{center}
  \includegraphics[width=0.9\textwidth]{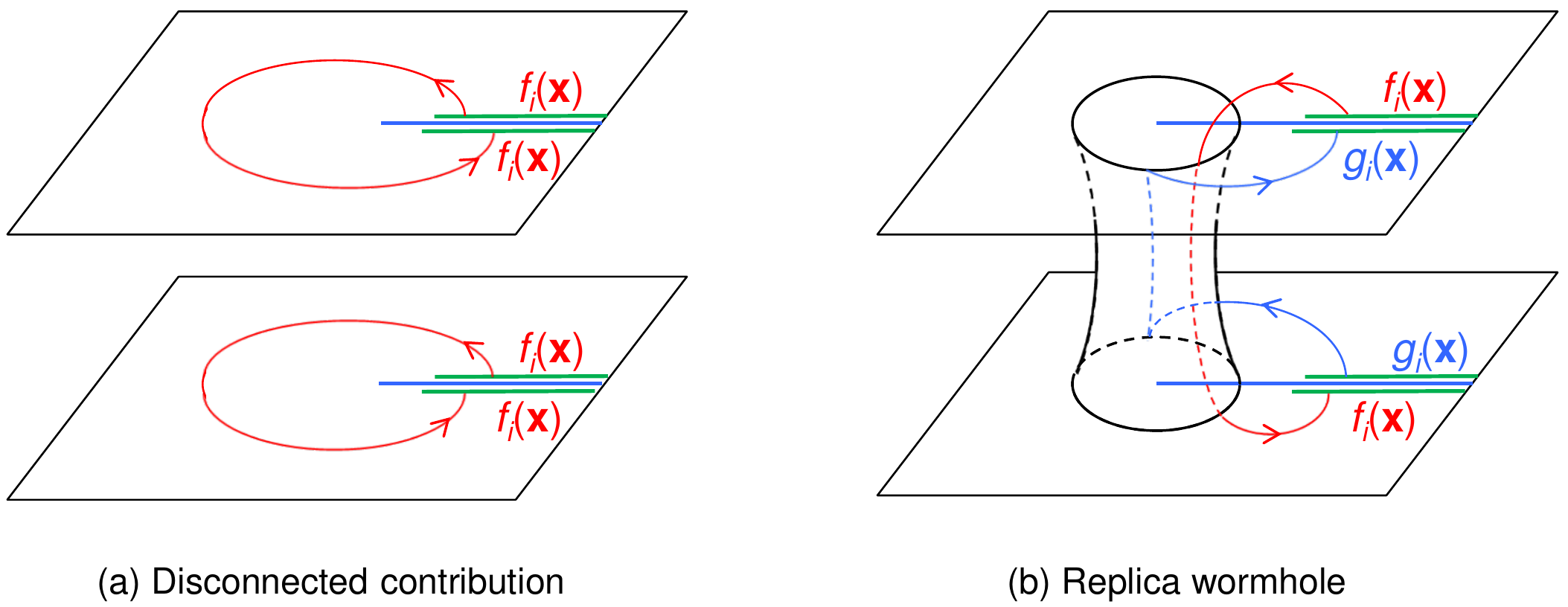}
\vspace{-2mm}
\end{center}
\caption{
 Two contributions to replica method calculation of $\overline{\Tr \hat{\rho}_R^2}$.
 (a) The normal contribution in which the gravitational region is filled separately on each sheet.
 Integration of semiclassical matter gives $e^{S_{\rm rad}}$ coming from Euclidean evolution relating all 4~field configurations above and below the cuts on two sheets (red arrows), while gravitational path integral gives the $e^{S_{\rm bh}}$ factor for each sheet.
 (b) For the replica wormhole configuration, the gravitational regions on two sheets are geometrically connected.
 The contribution from gravitation path integral is thus only $e^{S_{\rm bh}}$, while the contribution from matter integration gives $e^{2S_{\rm rad}}$ since there are two independent cycles for evolution (red and blue arrows).}
\label{fig:replica-2}
\end{figure}

In Fig.~\ref{fig:replica-2}, we illustrate the calculation of the trace of the square of the nonnormalized density matrix $\hat{\rho}_R$, which we denote by $Z_2 = \overline{\Tr \hat{\rho}_R^2}$.
As found in Refs.~\cite{Penington:2019kki,Almheiri:2019qdq}, there are two contributions to this.
The first is the one in which the gravitational region is filled separately for two sheets; Fig.~\ref{fig:replica-2}(a).
Restricting the radiation configuration to those resembling the emitted Hawking radiation at a coarse-grained level, the integral of semiclassical fields provides a factor of $e^{S_{\rm rad}}$, since the relevant four spatial configurations of the fields (both sides of the two cuts) are all related by Euclidean evolution as indicated by the red arrows in the figure.
On the other hand, the gravitational path integral gives $e^{S_{\rm bh}}$ for each sheet, so that this contribution is given by
\begin{equation}
  Z_2^{({\rm disconnected})} \sim e^{2 S_{\rm bh} + S_{\rm rad}}.
\end{equation}
The other contribution is the one coming from a replica wormhole; Fig.~\ref{fig:replica-2}(b).
In this case, the radiation contribution is $e^{2 S_{\rm rad}}$, since we now have two independent cycles for the evolution, so that the four relevant configurations are related only pairwise (the red and blue arrows).
On the other hand, the gravitational contribution is now $e^{S_{\rm bh}}$ because it comes from a single, connected component.
This therefore gives
\begin{equation}
  Z_2^{({\rm wormhole})} \sim e^{S_{\rm bh} + 2 S_{\rm rad}}.
\end{equation}
Adding the two contributions together, we find that $Z_2 \sim {\rm max}\{ e^{2 S_{\rm bh} + S_{\rm rad}}, e^{S_{\rm bh} + 2 S_{\rm rad}} \}$.

\begin{figure}[t]
\centering
  \subfloat[\centering Disconnected contribution]{{\includegraphics[width=0.4\textwidth]{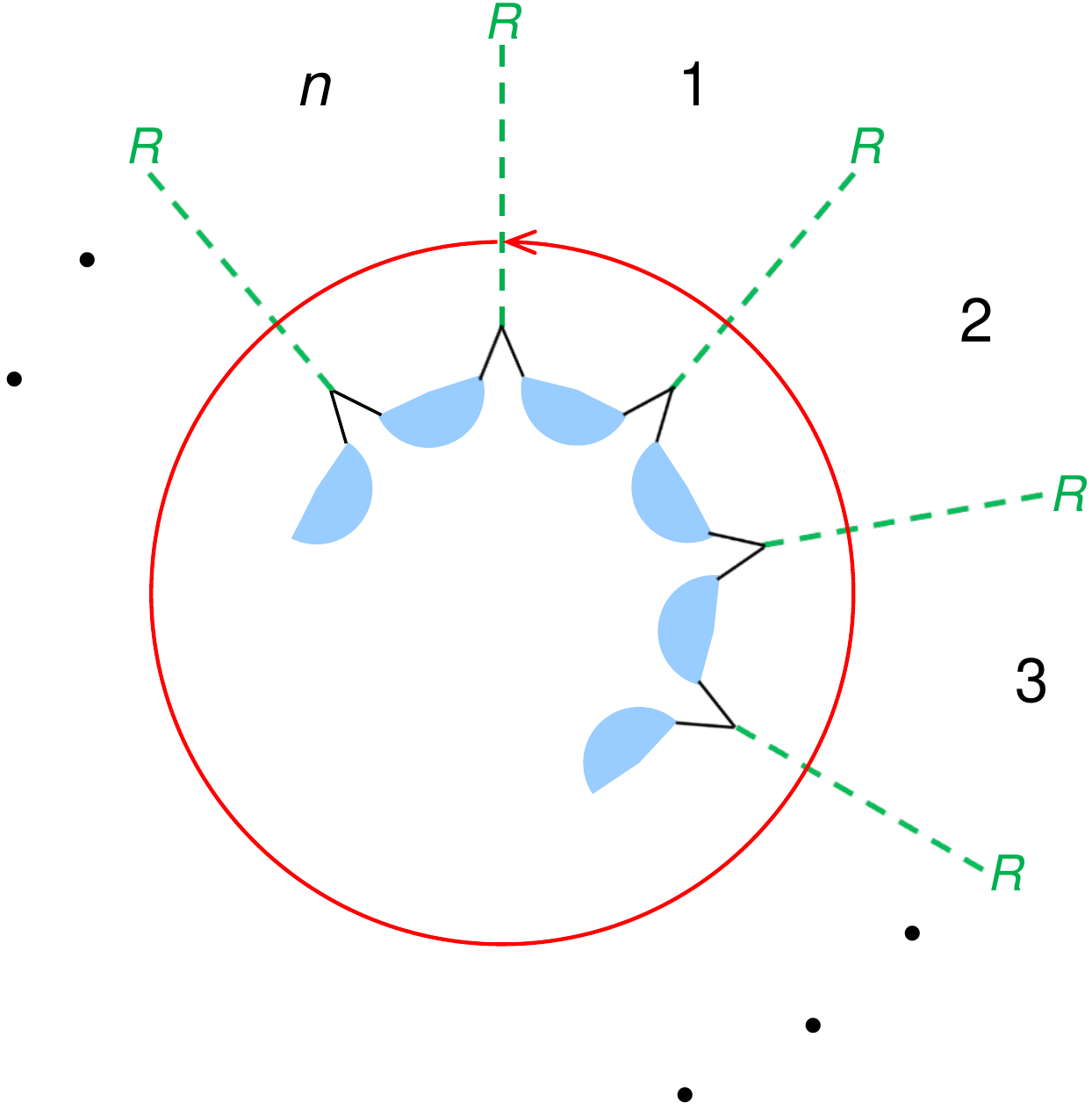} }}
\hspace{1.5cm}
  \subfloat[\centering Replica wormhole]{{\includegraphics[width=0.4\textwidth]{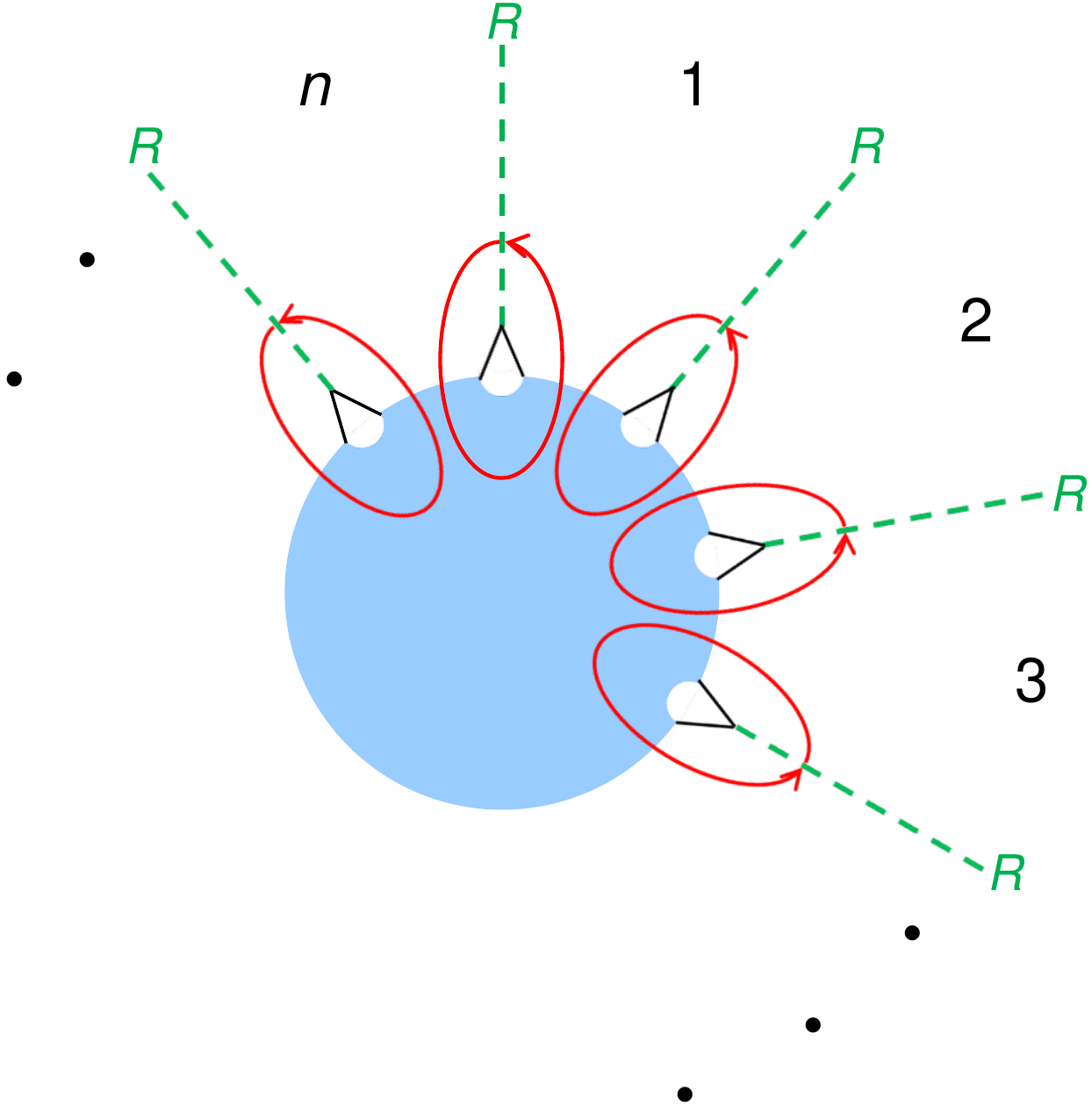} }}
\caption{
 (a) The contribution to $Z_n$ with the gravitational region filled separately (light blue) on each sheet.
 All semiclassical matter configurations above and below the cuts $R$ (green dashed) are related by Euclidean evolution (red arrow), giving the factor of $e^{S_{\rm rad}}$.
 (b) For the replica symmetric wormhole configuration, the gravitational regions on $n$ sheets are geometrically connected.
 This allows for $n$ independent cycles for evolution (red arrows), giving the contribution from matter integration of order $e^{nS_{\rm rad}}$.}
\label{fig:replica-n}
\end{figure}

A similar analysis can be performed for $Z_n$ for $n \in \mathbb{N}$.
Assuming that the replica symmetric wormhole connecting all $n$ sheets dominates for $S_{\rm rad} > S_{\rm bh}$, we find
\begin{equation}
  Z_n \sim \begin{cases} 
    e^{n S_{\rm bh} + S_{\rm rad}} & \mbox{for } S_{\rm bh} > S_{\rm rad}\\
    e^{S_{\rm bh} + n S_{\rm rad}} & \mbox{for } S_{\rm rad} > S_{\rm bh}.
  \end{cases}
\end{equation}
This is illustrated in Fig.~\ref{fig:replica-n}.
Since the density matrix $\rho_R$ is given by $\hat{\rho}_R/\Tr \hat{\rho}_R$, we thus find
\begin{equation}
  \overline{\Tr \rho_R^n} = \overline{\left( \frac{\Tr \hat{\rho}_R^n}{(\Tr \hat{\rho}_R)^n} \right)} \approx \frac{\overline{\Tr \hat{\rho}_R^n}}{\bigl( \overline{\Tr \hat{\rho}_R} \bigr)^n} = \frac{Z_n}{Z_1^n} \sim \begin{cases} 
    e^{(1-n) S_{\rm rad}} & \mbox{for } S_{\rm bh} > S_{\rm rad}\\
    e^{(1-n) S_{\rm bh}}  & \mbox{for } S_{\rm rad} > S_{\rm bh},
  \end{cases}
\end{equation}
where in the second equation we have used the fact that the standard deviation of $\Tr \hat{\rho}_R^n$ is smaller than its typical size in the ensemble, which we can easily be convinced.
By analytically continuing this result in $n$, we can now obtain the ensemble average of the von~Neumann entropy of the radiation
\begin{equation}
  \overline{S_R} = - \lim_{n \rightarrow 1} \frac{\partial}{\partial n} \overline{\Tr \rho_R^n} \sim 
  \begin{cases} 
    S_{\rm rad} & \mbox{for } S_{\rm bh} > S_{\rm rad}\\
    S_{\rm bh}  & \mbox{for } S_{\rm rad} > S_{\rm bh},
  \end{cases}
\end{equation}
which reproduces the Page curve.
This is because we have calculated the ensemble average of the von~Neumann entropy (which obeys the Page curve for all members of the ensemble), and not the von~Neumann entropy of the averaged density matrix as in Eq.~\eqref{eq:S_R-Hawking}.

Note that this observation is similar to that in Ref.~\cite{Bousso:2020kmy}, but here we do not consider any ensemble of holographic theories.
Instead, the relevant ensemble arises from the coarse graining of black hole microstates, which is {\it forced} on us if we adopt any formalism involving the semiclassical picture of the black hole, such as gravitational path integral.
See also Refs.~\cite{Blommaert:2021fob,Chandra:2022fwi} for related discussion.

\subsubsection*{Entanglement island and the Lorentzian picture}

It was shown in Ref.~\cite{Almheiri:2019qdq}, building on the technique developed in Refs.~\cite{Lewkowycz:2013nqa,Dong:2016hjy,Dong:2017xht}, that the prescription using the replica method in gravitational path integral is equivalent to the quantum extremal surface prescription~\cite{Engelhardt:2014gca} in the original unreplicated spacetime.%
\footnote{
 Strictly speaking, this was shown only in the limit that the contribution from the graviton to the semiclassical field integration is negligible compared with those from other quantum fields.
 Below we assume that the equivalence of the two prescriptions persists beyond this limit.
}
Going to the Lorentzian signature, this therefore leads to the following picture~\cite{Penington:2019npb,Almheiri:2019psf,Almheiri:2019hni}.

In order to calculate the von~Neumann entropy of radiation $S_{\rm rad}$ in some region $R$, we need to find an entanglement island $I$ which extremizes the following quantity%
\footnote{
 The regions $R$ and $I$ can be viewed either as spatial regions or causal regions associated with them.
}
\begin{equation}
  S_{\rm gen}(R,I) = \frac{{\cal A}(\partial I)}{4 G_{\rm N}} - \Tr\left[\rho_{\rm sc}(R \cup I) \ln\rho_{\rm sc}(R \cup I)\right],
\end{equation}
where ${\cal A}(\partial I)$ is the area of the boundary of $I$, and $\rho_{\rm sc}(X)$ is the reduced density matrix of the region $X$ in the semiclassical theory.
Note that $I$ can be a null region.
In general, there can be multiple such $I$'s, and the entropy $S_{\rm rad}$ is given by the minimum of $S_{\rm gen}(R,I)$'s associated with all such $I$'s:
\begin{equation}
  S_{\rm rad} = {\rm min} \stackbin[I]{}{\rm ext} S_{\rm gen}(R,I).
\end{equation}
This entropy is the same as that calculated by the replica method in path integral
\begin{equation}
  S_{\rm rad} = \overline{S_R},
\end{equation}
so that it obeys the Page curve.

It is important that this extremization procedure is performed on a global spacetime of general relativity.
In particular, for a black hole spacetime, it must be performed on the whole spacetime including the interior of the black hole.
This, therefore, gives a complementary picture to that described in Sections~\ref{sec:QFT}~--~\ref{sec:extended}.
In the picture described here, the existence of the black hole interior is obvious---in fact, the framework assumes it---while to understand the unitary nature of black hole evolution, one needs to resort to a method that appropriately incorporates nonperturbative effects of quantum gravity, such as replica wormholes.
On the other hand, in the framework of Sections~\ref{sec:QFT}~--~\ref{sec:extended}, the unitarity of the evolution is an assumption and, as a consequence, the existence of the interior is not manifest---the interior emerges only effectively as a collective phenomenon involving horizon degrees of freedom, which are subject to universally chaotic and fast-scrambling dynamics.

\subsubsection*{Manifestly unitary vs global spacetime descriptions}

We can, in fact, understand the existence of the two frameworks discussed above---based on the manifestly unitary and global spacetime pictures, respectively---from the viewpoint of gauge symmetries of the underlying theory~\cite{Langhoff:2020jqa,Nomura:2020ewg}.
As emphasized in Ref.~\cite{Marolf:2020xie}, a theory of quantum gravity has nonperturbative gauge redundancies much larger than the standard diffeomorphism.
A particular manifestation of this is the apparent violation of the Bekenstein-Hawking entropy bound in the semiclassical description of a black hole~\cite{Langhoff:2020jqa,Nomura:2020ewg,Chakravarty:2020wdm}.
This violation arises from huge spatial volume inside the black hole~\cite{Christodoulou:2014yia,Christodoulou:2016tuu} (including the so-called bags-of-gold configurations), which leads to the number of independent quantum states exceeding the Bekenstein-Hawking bound.
Many of these semiclassically independent states, however, are equivalent under the nonperturbative gauge symmetries, making the number of truly independent states satisfy the Bekenstein-Hawking bound.

The two frameworks describe this phenomenon in very different, though equivalent, ways.
In the framework discussed in Sections~\ref{sec:QFT}~--~\ref{sec:extended}, the nonperturbative gauge redundancies (as well as a part of the standard diffeomorphism) are explicitly fixed by employing a Schwarzschild-like time foliation, which is motivated by holography.
In this framework, making unitarity manifest, the number of independent black hole states is simply $e^{S_{\rm BH}}$, and the apparently much larger number of independent interior states in semiclassical theory comes from the fact that the Hilbert space spanned by the $e^{S_{\rm BH}}$ states admit $e^{e^{S_{\rm BH}}}$ approximately orthogonal states, between which inner products are of order $e^{-S_{\rm BH}/2}$ or smaller.
Since the semiclassical theory cannot detect such small inner products, it appears to accommodate independent quantum states larger than the Bekenstein-Hawking bound in the interior of a black hole.

On the other hand, in the framework making the interior manifest, the argument runs in the other direction.
In this case, the starting point is the global spacetime picture, which is highly redundant.
In the canonical formulation of quantum mechanics, this implies the existence of over-entropic semiclassical interior states $\ket{\psi_i}$.
After including nonperturbative effects of gravity, however, these semiclassically orthogonal states develop small overlaps, $\inner{\psi_i}{\psi_j} \sim e^{-S_{\rm BH}/2}$, in a such way that the rank of the matrix $M_{ij} \equiv \inner{\psi_i}{\psi_j}$ is reduced drastically to $e^{S_{\rm BH}}$.
Such a large number of null states are a manifestation of the large nonperturbative gauge redundancies~\cite{Marolf:2020xie,McNamara:2020uza}, which relate even spaces with different topologies~\cite{Marolf:2020xie,Jafferis:2017tiu}.
In the path integral formulation, this reduction of the semiclassical Hilbert space to the physical one is achieved by including additional contributions to the path integral (e.g.\ replica wormholes) which projects states onto those invariant under the relevant gauge symmetries as we have seen in this section.

\section{Conclusion}
\label{sec:concl}

In quantum gravity, there has been a difficulty in reconciling fundamental principles of physics in the presence of a black hole (or other) horizon, particularly the unitarity of quantum mechanics and the equivalence principle of general relativity~\cite{Hawking:1976ra}.
This difficulty is, in fact, an avatar of the conventional difficulty of describing UV physics at the Planck scale, albeit in a redshifted form.
Because of a huge gravitational redshift between the horizon and asymptotic regions, the degrees of freedom represented by the Bekenstein-Hawking entropy~\cite{Bekenstein:1973ur,Hawking:1974sw}, which obey an intrinsically ``stringy'' dynamics, appear to have exponentially degenerate states that cannot be discriminated operationally by performing a measurement on the black hole.
A low energy description of quantum gravity treats these states in a thermal way, so that unitarity appears to be lost.

While a semiclassical theory, as a low energy description of quantum gravity, cannot describe all the microscopic dynamics of the fundamental theory, it can still be used to obtain a coarse-grained understanding of how unitarity and the equivalence principle can coexist in a black hole system, with only a few inputs from the UV theory.
This is what we have explored in this paper.

One way of doing this is simply to postulate that the evolution of a black hole is unitary when viewed from the exterior~\cite{tHooft:1984kcu,tHooft:1990fkf,Susskind:1993if,Page:1993wv}.
This is a view motivated by holography~\cite{Maldacena:1997re}.
In this case, the existence of the black hole interior appears to be in jeopardy~\cite{Almheiri:2012rt}.
We have seen, however, that with the assumption that the dynamics of horizon degrees of freedom is chaotic and fast scrambling across all low energy species, the interior emerges at a semiclassical level as a collective phenomenon involving the horizon, and possibly other, degrees of freedom~\cite{Nomura:2018kia,Nomura:2019qps,Nomura:2019dlz,Langhoff:2020jqa,Nomura:2020ska}.
An important point is that for a black hole with minimal uncertainties (i.e.\ a specific black hole at the semiclassical level), the assumption of Gaussian randomness of the coefficients for microstates allows us to learn many features associated with the construction of the effective theory of the interior.
We do not need any further details about the microscopic theory to figure these things out.
A similar construction applies to de~Sitter spacetime, which we have also elaborated in this paper.

The compatibility between unitarity and the equivalence principle can alternatively be seen by starting from the global spacetime of general relativity.
In this case, the existence of the interior is manifest, but at the apparent expense of unitarity as Hawking's original calculation indicates~\cite{Hawking:1976ra}.
However, while a semiclassical picture necessarily involves an average over microstates, we can directly calculate the average of the quantity we are interested in, e.g.\ the von~Neumann entropy of emitted Hawking radiation, rather than the quantity in the averaged state.
This is what is done in Refs.~\cite{Penington:2019npb,Almheiri:2019psf,Almheiri:2019hni,Penington:2019kki,Almheiri:2019qdq} by employing the replica method in gravitational path integral, or equivalently the quantum extremal surface prescription.
In this way, one can see the unitarity of the underlying theory without knowing its detailed dynamics.

Despite the fact that the two frameworks described above appear very different, they give the same physical conclusions.
In particular, a black hole evolves unitarily and has a smooth horizon.
Viewed in this way, one can conclude that historical confusions about black hole physics come from the fact that only one of these features can be made manifest in a given low energy description; the other one appears in a highly nontrivial manner.
We have seen that the description making unitarity (quantum mechanics) manifest comes more naturally with the canonical/Hamiltonian formulation of quantum mechanics, while the one making the interior (general relativity) manifest is associated with the path integral/Lagrangian formulation.
It remains to be seen if there is a microscopic formulation of quantum gravity in which both these features are manifest at the same time.

\section*{Acknowledgments}

We would like to thank Chris Akers, Sinya Aoki, Vijay Balasubramanian, Arjun Kar, Lampros Lamprou, Adam Levine, Juan Maldacena, Masamichi Miyaji, Dominik Neuenfeld, Edgar Shaghoulian, Tomonori Ugajin, Shreya Vardhan, Zixia Wei, and Zhenbin Yang for useful discussions on this and related subjects.
This work was supported in part by the Department of Energy, Office of Science, Office of High Energy Physics under QuantISED Award DE-SC0019380 and contract DE-AC02-05CH11231 and in part by MEXT KAKENHI grant number JP20H05850, JP20H05860.

\end{document}